\documentclass[12pt]{article}
\usepackage{eqsection,subeqnarray,indent,amsfonts,amssymb,epsf}
\usepackage{epic,eepic}

 \def\subdominantcolor{pink}         

\begin{document}

\bibliographystyle{plain}

\date{April 26, 2002 \\[1mm] revised March 21, 2003}

\title{\vspace*{-2cm}
       Transfer Matrices and Partition-Function Zeros \\
       for Antiferromagnetic Potts Models \\[5mm]
       \large\bf III.~Triangular-lattice chromatic polynomial}

\author{
  \\
  {\small Jesper Lykke Jacobsen}                            \\[-0.2cm]
  {\small\it Laboratoire de Physique Th\'eorique et Mod\`eles Statistiques}
                                                            \\[-0.2cm]
  {\small\it Universit\'e Paris-Sud}                        \\[-0.2cm]
  {\small\it B\^atiment 100}                                \\[-0.2cm]
  {\small\it F-91405 Orsay, FRANCE }        \\[-0.2cm]
  {\small\tt JACOBSEN@IPNO.IN2P3.FR}                        \\[5mm]
  {\small Jes\'us Salas}                                    \\[-0.2cm]
  {\small\it Departamento de F\'{\i}sica Te\'orica}         \\[-0.2cm]
  {\small\it Facultad de Ciencias, Universidad de Zaragoza} \\[-0.2cm]
  {\small\it Zaragoza 50009, SPAIN}                         \\[-0.2cm]
  {\small\tt JESUS@MELKWEG.UNIZAR.ES}     \\[5mm]
  {\small Alan D.~Sokal}                  \\[-0.2cm]
  {\small\it Department of Physics}       \\[-0.2cm]
  {\small\it New York University}         \\[-0.2cm]
  {\small\it 4 Washington Place}          \\[-0.2cm]
  {\small\it New York, NY 10003 USA}      \\[-0.2cm]
  {\small\tt SOKAL@NYU.EDU}               \\[-0.2cm]
  {\protect\makebox[5in]{\quad}}  
  \\
}

\maketitle
\thispagestyle{empty}   

\begin{abstract}
We study the chromatic polynomial $P_G(q)$ for $m\times n$
triangular-lattice strips of widths $m \le 12_{\rm P}, 9_{\rm F}$
(with periodic or free transverse boundary conditions, respectively)
and arbitrary lengths $n$ (with free longitudinal boundary conditions).
The chromatic polynomial gives the zero-temperature limit of
the partition function for the $q$-state Potts antiferromagnet.
We compute the transfer matrix for such strips in the 
Fortuin--Kasteleyn representation and obtain the 
corresponding accumulation sets of chromatic zeros
in the complex $q$-plane in the limit $n\to\infty$.
We recompute the limiting curve obtained by Baxter
in the thermodynamic limit $m,n\to\infty$
and find new interesting features with possible physical consequences.
Finally, we analyze the isolated limiting points
and their relation with the Beraha numbers. 
\end{abstract}


\clearpage

\newcommand{\be}{\begin{equation}}
\newcommand{\ee}{\end{equation}}
\newcommand{\<}{\langle}
\renewcommand{\>}{\rangle}
\newcommand{\widebar}{\overline}
\def\reff#1{(\protect\ref{#1})}
\def\spose#1{\hbox to 0pt{#1\hss}}
\def\ltapprox{\mathrel{\spose{\lower 3pt\hbox{$\mathchar"218$}}
 \raise 2.0pt\hbox{$\mathchar"13C$}}}
\def\gtapprox{\mathrel{\spose{\lower 3pt\hbox{$\mathchar"218$}}
 \raise 2.0pt\hbox{$\mathchar"13E$}}}
\def\textprime{${}^\prime$}
\def\proof{\par\medskip\noindent{\sc Proof.\ }}
\def\qed{\hbox{\hskip 6pt\vrule width6pt height7pt depth1pt \hskip1pt}\bigskip}
\def\proofof#1{\bigskip\noindent{\sc Proof of #1.\ }}
\def\half{ {1 \over 2} }
\def\third{ {1 \over 3} }
\def\twothird{ {2 \over 3} }
\def\smfrac#1#2{\textstyle{#1\over #2}}
\def\smhalf{ \smfrac{1}{2} }
\newcommand{\real}{\mathop{\rm Re}\nolimits}
\renewcommand{\Re}{\mathop{\rm Re}\nolimits}
\newcommand{\imag}{\mathop{\rm Im}\nolimits}
\renewcommand{\Im}{\mathop{\rm Im}\nolimits}
\newcommand{\sgn}{\mathop{\rm sgn}\nolimits}
\newcommand{\tr}{\mathop{\rm tr}\nolimits}
\newcommand{\diag}{\mathop{\rm diag}\nolimits}
\newcommand{\Gal}{\mathop{\rm Gal}\nolimits}
\newcommand{\mycup}{\mathop{\cup}}
\newcommand{\Arg}{\mathop{\rm Arg}\nolimits}
\def\hboxscript#1{ {\hbox{\scriptsize\em #1}} }
\def\zhat{ {\widehat{Z}} }
\def\phat{ {\widehat{P}} }
\def\qtilde{ {\widetilde{q}} }
\newcommand{\mod}{\mathop{\rm mod}\nolimits}
\renewcommand{\emptyset}{\varnothing}

\def\scra{\mathcal{A}}
\def\scrb{\mathcal{B}}
\def\scrc{\mathcal{C}}
\def\scrd{\mathcal{D}}
\def\scrf{\mathcal{F}}
\def\scrg{\mathcal{G}}
\def\scrl{\mathcal{L}}
\def\scro{\mathcal{O}}
\def\scrp{\mathcal{P}}
\def\scrq{\mathcal{Q}}
\def\scrr{\mathcal{R}}
\def\scrs{\mathcal{S}}
\def\scrt{\mathcal{T}}
\def\scrv{\mathcal{V}}
\def\scrz{\mathcal{Z}}

\def\q{{\sf q}}

\def\Z{{\mathbb Z}}
\def\R{{\mathbb R}}
\def\C{{\mathbb C}}
\def\Q{{\mathbb Q}}

\def\T{{\mathsf T}}
\def\H{{\mathsf H}}
\def\V{{\mathsf V}}
\def\D{{\mathsf D}}
\def\J{{\mathsf J}}
\def\P{{\mathsf P}}
\def\QQ{{\mathsf Q}}
\def\RR{{\mathsf R}}

\def\bsigma{\mbox{\protect\boldmath $\sigma$}}
\def\bone{{\mathbf 1}}
\def\vv{{\bf v}}
\def\uu{{\bf u}}
\def\w{{\bf w}}

\newtheorem{theorem}{Theorem}[section]
\newtheorem{proposition}[theorem]{Proposition}
\newtheorem{lemma}[theorem]{Lemma}
\newtheorem{corollary}[theorem]{Corollary}
\newtheorem{conjecture}[theorem]{Conjecture}


\newenvironment{sarray}{
          \textfont0=\scriptfont0
          \scriptfont0=\scriptscriptfont0
          \textfont1=\scriptfont1
          \scriptfont1=\scriptscriptfont1
          \textfont2=\scriptfont2
          \scriptfont2=\scriptscriptfont2
          \textfont3=\scriptfont3
          \scriptfont3=\scriptscriptfont3
        \renewcommand{\arraystretch}{0.7}
        \begin{array}{l}}{\end{array}}

\newenvironment{scarray}{
          \textfont0=\scriptfont0
          \scriptfont0=\scriptscriptfont0
          \textfont1=\scriptfont1
          \scriptfont1=\scriptscriptfont1
          \textfont2=\scriptfont2
          \scriptfont2=\scriptscriptfont2
          \textfont3=\scriptfont3
          \scriptfont3=\scriptscriptfont3
        \renewcommand{\arraystretch}{0.7}
        \begin{array}{c}}{\end{array}}

%
%
\section{Introduction}   \label{sec1}

The antiferromagnetic $q$-state Potts model
\cite{Wu_82,Wu_84,Baxter_82,Martin_91,%
Baxter_70_TRI,Baxter_82b,WSK_90,Saleur_90,Saleur_91,Adler_95,%
Salas_97,Salas_98,Ferreira_99,Cardy_01}
exhibits unusual behavior not found in ferromagnets.
Indeed, its phase diagram and critical behavior
depend crucially on the lattice structure,
unlike the ferromagnetic case where the concept of universality applies.
Specifically, for each lattice ${\cal L}$
there exists a number $q_c({\cal L})$ such
that for all $q>q_c$ the model is disordered at any temperature,
including zero temperature \cite{Salas_97}.
Exactly at $q=q_c({\cal L})$ the system is disordered at
all positive temperatures and has a zero-temperature critical point.
The zero-temperature limit of the antiferromagnetic Potts model
is particularly interesting because its partition function
on a finite graph $G$ coincides with the chromatic polynomial $P_G(q)$,
which counts the number of ways of coloring the vertices of $G$
using $q$ colors subject to the constraint that adjacent vertices
always receive different colors \cite{Read_88}.

In Refs.~\cite{transfer1,transfer2} we undertook a study of the
zeros of the chromatic polynomial when the parameter $q$ is allowed to
take complex values (see \cite{transfer1} for detailed references
to the previous literature).
In those papers we studied strips of the square lattice;
here we extend that work to the triangular lattice.
The triangular-lattice case is of particular interest
because the path-breaking work of Baxter \cite{Baxter_86,Baxter_87}
provides a conjectured exact solution in the thermodynamic limit.

The study of the complex zeros of the chromatic polynomial is
inspired by the Yang--Lee picture of phase transitions
\cite{Yang-Lee_52}.  We study families of graphs $G_n$ for which the
chromatic polynomial can be expressed via a transfer matrix of fixed
size $M\times M$:
\begin{subeqnarray}
   P_{G_n}(q)  & = &   \tr[ A(q) \, T(q)^n ]  \\[2mm]
               & = &   \sum\limits_{k=1}^{M}
                           \alpha_k(q) \, \lambda_k(q)^n  \;,
\label{general_form_P}
\end{subeqnarray}
where the transfer matrix $T(q)$ and the boundary-condition matrix
$A(q)$ are polynomials in $q$, so that the eigenvalues $\{ \lambda_k \}$
of $T$ and the amplitudes $\{ \alpha_k \}$ are algebraic functions of $q$.
Rather than using $T(q)$ to compute the zeros of the
chromatic polynomial for a {\em finite}\/ strip $m\times n$,
we have focussed on the direct calculation of
their accumulation points in the limit $n\rightarrow\infty$,
i.e.~for the case of an {\em semi-infinite}\/ strip
\cite{Beraha_79,Beraha_80,Shrock_97a,Shrock_98a,transfer1,transfer2}.
According to the Beraha--Kahane--Weiss theorem
\cite{BKW_75,BKW_78,Sokal_chromatic_roots},
the accumulation points of zeros when $n\rightarrow\infty$ can either
be isolated limiting points (when the amplitude associated to the
dominant eigenvalue vanishes, or when all eigenvalues vanish simultaneously)
or belong to a limiting curve $\scrb$ (when
two dominant eigenvalues cross in modulus). By studying the limiting
curves for different values of the strip width $m$, we hope to learn new
features of the thermodynamic limit $m\rightarrow\infty$.

To determine the isolated limiting points, we shall take advantage of the
following simple result \cite{Beraha_80}:
\be
 \det D = \prod_{k=1}^M \alpha_k \prod_{1 \le i \le j \le M}
 (\lambda_j - \lambda_i)^2  \;,
 \label{def_Det_D}
\ee
where $D$ is the $M\times M$ matrix with entries
$D_{ij} = \sum_{k=1}^M \alpha_k (\lambda_k)^{i+j-2} = P_{G_{i+j-2}}$.

An important feature of the limiting curve ${\cal B}$ is that it
typically crosses the positive real axis at a point $q_0(m)$.\footnote{
   If there is more than one such crossing,
   we define $q_0(m)$ to be the {\em smallest}\/ such crossing.
   When no such crossing occurs, the limiting curve often
   includes a pair of complex-conjugate endpoints rather close to the
   positive real $q$-axis. In these cases, we define $q_0(m)$ to be the
   point closest to that axis with positive imaginary part.
}
Physically,
$q_0(m)$ corresponds to a point of non-analyticity of the ground-state
degeneracy per site \cite{Shrock_97a,Shrock_97b}.
As the strip width $m$ grows, this crossing point $q_0(m)$ increases
and presumably tends to a limiting value $q_0(\infty)$.
On the other hand, as $m\to\infty$ the curve $\scrb = \scrb_m$
presumably tends to a thermodynamic-limit curve $\scrb_\infty$.
We define $q_c$ to be the largest value where ${\cal B}_\infty$ crosses  
the real $q$-axis:
please note that $q_c$ may or may not correspond to crossings of the
real axis for any finite $m$.
In general the value $q_0(\infty)$ is smaller than $q_{\rm c}$, 
although for some lattices they may coincide
(this depends on the shape of the curves ${\cal B}_m$ and ${\cal B}_\infty$).
Indeed, in Ref.~\cite{transfer2} evidence was presented suggesting that
for the square lattice $q_0(\infty) = q_{\rm c} = 3$.

The crucial role of $q_0$ is further emphasized by studying the relation
between chromatic polynomials and the so-called {\em Beraha numbers}
\be
  B_n \,=\, 4 \cos^2 {\pi \over n} \,=\, 2 + 2 \cos {2\pi \over n}
  \qquad \mbox{for }n=2,3,\ldots \;.
 \label{def_Bn}
\ee
It has been found in a number of cases \cite{Beraha_unpub,Beraha_79,Beraha_80}
that chromatic roots tend to accumulate at some of the Beraha numbers.
In the case of the {\em square lattice}\/, we have found empirically
\cite{transfer1,transfer2} that on a strip of width $m$ with either free or
periodic transverse boundary conditions\footnote{
   Let $m$ (resp.~$n$) denote the number of sites in the
   transverse (resp.~longitudinal) direction of the strip,
   and let F (resp.~P) denote free (resp.~periodic) boundary conditions
   in a given direction. Then we use the terminology:
   free ($m_{\rm F} \times n_{\rm F}$),
   cylindrical ($m_{\rm P} \times n_{\rm F}$),
   cyclic ($m_{\rm F} \times n_{\rm P}$),
   and toroidal ($m_{\rm P} \times n_{\rm P}$).
   In this paper we consider free and cylindrical boundary conditions,
   as well as a new type of boundary condition that we shall call
   ``zig-zag'' (Section~\ref{sec_Z}).
},
there is at least one vanishing amplitude $\alpha_i(q)$ at each of the
first $m$ Beraha numbers $B_2,\ldots,B_{m+1}$ (but not higher ones).
It thus appears that in the limit $m \to\infty$ all the Beraha numbers
will be zeros of some amplitude.\footnote{
   As we shall see, a similar (but not identical)
   statement appears to hold true also for the triangular lattice:
   see Section~\ref{sec8.2}.
}
Moreover, we found that the
vanishing amplitude corresponds to the eigenvalue obtained by analytic
continuation in $q$ from the one that is dominant at small real $q$,
in agreement with a conjecture of Baxter \cite[p.~5255]{Baxter_87}.
Thus, the first few Beraha numbers --- namely, those (up to at most
$B_{m+1}$) that lie below the point $q_0(m)$ 
--- correspond to the vanishing of a dominant amplitude and hence,
by the Beraha--Kahane--Weiss theorem, to a limit point of chromatic
roots, while the remaining Beraha numbers do not.  As the strip width
$m$ grows, $q_0(m)$ tends to $q_0(\infty)$, and the limiting points of
chromatic roots are thus constrained to be the points
$B_2,B_3,\ldots,B_{p} < q_0(\infty)$.  This scenario for the
accumulation of chromatic roots at {\em some}\/ of the Beraha numbers
was set forth by Baxter \cite{Baxter_87} and elaborated by Saleur
\cite{Saleur_90}; further references can be found in \cite{transfer1}.

In the present publication we shall be concerned with the
antiferromagnetic Potts model on the {\em triangular lattice}\/. For
this case, Baxter and collaborators
\cite{Baxter_78,Baxter_86,Baxter_87} have determined the exact free
energy (among other quantities) on two special curves in the
$(q,v)$-plane:
\begin{eqnarray}
   v^3 + 3v^2 - q   & = &  0             \label{eqDOB1.3}  \\[1mm]
   v                & = &  -1            \label{eqDOB1.4}
\end{eqnarray}
The uppermost branch ($v \ge 0$) of curve \reff{eqDOB1.3} is known to
correspond to the ferromagnetic critical point
\cite{Baxter_78,Baxter_82}, and Baxter \cite{Baxter_86} initially
conjectured (following a hint of Nienhuis \cite{Nienhuis_82}) that
the zero-temperature antiferromagnetic model \reff{eqDOB1.4} is critical
in the interval $0 \le q \le 4$. This prediction is known to be correct
for $q=2$ \cite{Stephenson_64,Blote_82b,Nienhuis_84b} and is believed
to be correct also for $q=4$
\cite{Baxter_70_TRI,Henley_unpublished,Salas_TRI4state}. On the other
hand, for $q=3$ the conjecture contradicts the rigorous result
\cite{vEFS_unpub}, based on Pirogov-Sinai theory, that there is a
low-temperature phase with long-range order and small correlation
length.\footnote{ A Monte Carlo study of the $q=3$ model found strong
evidence for a first-order transition to an ordered phase at $\beta J
\approx -1.594$ \cite{Adler_95}.}
In any case, for $q>4$ we expect that the triangular-lattice Potts model
is noncritical even at zero temperature;
this has recently been confirmed by Monte Carlo simulation
of the models with $q=5,6$ \cite{Salas-Sokal_TRI56}.
We therefore expect that for the triangular lattice $q_c = 4$.
%
%
%

For the model \reff{eqDOB1.4}, Baxter \cite{Baxter_86} used a Bethe Ansatz
to compute three
different expressions $g_i(q)$ [$i=1,2,3$] that he argued
correspond to the dominant eigenvalues of the transfer matrix in
different regions $\scrd_i$ of the complex $q$-plane; in a second
paper \cite{Baxter_87} he provided corrected estimates for the precise
locations of $\scrd_1,\scrd_2,\scrd_3$. 
Using these formulae, he determined the value of $q_0(\infty)$ as
\be
 q_0(\infty,{\rm tri}) \approx 3.81967 \qquad\hbox{(Baxter)}
 \label{eq.q0.Baxter}
\ee
One important outcome of the present paper (see Section~\ref{sec6.2})
is that the ``phase diagram'' predicted by Baxter's formulae
is actually more complicated than what Baxter found.
In particular, it now appears that
the correct value of $q_0(\infty,{\rm tri})$
coming from Baxter's eigenvalues $g_i$ is
slightly smaller than Baxter's value \reff{eq.q0.Baxter},
the corrected value being
\be
 q_0(\infty,{\rm tri}) = B_{12} = 2 + \sqrt{3} \approx 
         3.73205 \qquad\hbox{(this paper)}
 \label{eq.q0.thispaper}
\ee
We therefore conjecture that the isolated limiting chromatic roots
of the infinite-size triangular lattice are
$B_2,\ldots,B_{11}$ and possibly $B_{12}$,
rather than $B_2,\ldots,B_{14}$ as conjectured by Baxter.

To study the approach to the thermodynamic limit,
we have computed the transfer matrix for triangular-lattice strips
of widths $2 \le m \le 9$ with free boundary conditions
and $2 \le m \le 12$ with cylindrical boundary conditions,
and we have determined the corresponding limiting curves $\scrb$.
We have also undertaken a detailed comparison of our finite-lattice
eigenvalues with Baxter's eigenvalues $g_i$,
and have found a surprising fact (Section~\ref{sec6.3}):
in the region of the $q$-plane corresponding to Baxter's $\scrd_3$,
we find no evidence in our finite-lattice data of any eigenvalue
corresponding to $g_1$.
As a consequence, the ``extra'' branches of the phase diagram
found in Section~\ref{sec6.2} appear to be absent after all!
There seem to be two possibilities:
\begin{itemize}
   \item[(a)]  The eigenvalue $g_1$ really is present in the region $\scrd_3$,
       but only for strip widths much larger than those we have studied.
       In this case, the limiting curve $\scrb_\infty$ really would
       exhibit all the complexities found in Section~\ref{sec6.2},
       and the correct value of $q_0$ would be given by
       \reff{eq.q0.thispaper} rather than \reff{eq.q0.Baxter}.
   \item[(b)]  For some reason, the eigenvalue $g_1$ is not present
       in this region (though it is clearly present elsewhere).
       In this case, the limiting curve $\scrb_\infty$ would be
       similar to that depicted by Baxter,
       and the correct value of $q_0$ would be
       given by \reff{eq.q0.Baxter} after all.
\end{itemize}
We discuss these issues further in Section~\ref{sec6.4}.

Our finite-lattice data also serve as a testing ground for the general
conjectures on Beraha numbers as stated above (see Ref.~\cite{transfer1}
for further details).
We discuss this further in Section~\ref{sec8.2}.

Previous studies using a similar transfer-matrix approach have been made
notably by Shrock and collaborators.
In particular, they have considered triangular-lattice strips
of width $m \le 5$ for free and cylindrical boundary conditions 
\cite{Shrock_98a,Shrock_98c,Shrock_98f,Shrock_99g,Shrock_00c}.\footnote{
  The case $m=4$ with cylindrical boundary conditions was first done by 
  Beraha and Kahane \protect\cite{Beraha_79}. 
  The case $m=5$ with cylindrical boundary conditions was first done by 
  Beraha, Kahane and Weiss \protect\cite{Beraha_80}. 
}
They have also considered other boundary conditions for the same lattice  
\cite{Shrock_00c,Shrock_00a,Shrock_01a,Shrock_01b}.\footnote{
 The case $2_{\rm F} \times n_{\rm P}$ (i.e., cyclic boundary conditions) 
 was first done by Beraha, Kahane and Weiss \protect\cite{Beraha_80}.
} 
Generalizations to nonzero temperature
for several boundary conditions have been 
carried out in Refs.~\cite{Shrock_00e,Shrock_inprep}.
Finally, Refs.~\cite{Shrock_01b,Shrock_02a} discuss
some general structural properties 
of the Potts-model partition function and chromatic polynomial
on square-lattice and triangular-lattice strips.

This paper is laid out as follows:
In Section~\ref{sec_prelim} we discuss some brief preliminaries.
In Section~\ref{sec_F} we give our numerical results for free transverse
boundary conditions, and in Section~\ref{sec_P} for
periodic transverse (cylindrical) boundary conditions.
We have also examined a third type of boundary conditions,
called ``zig-zag'', which we introduce and motivate in Section~\ref{sec_Z}.
In Section~\ref{secBaxter} we analyze Baxter's \cite{Baxter_86,Baxter_87}
exact solution for the thermodynamic limit
and carefully recompute his phase diagram,
finding interesting new features with possible physical consequences;
we also compare his predictions for the dominant eigenvalues
with our finite-lattice data 
and comment on the agreements and discrepancies.
Finally, in Section~\ref{sec8} we present our conclusions.

%
%
\section{Preliminaries}  \label{sec_prelim}

The general theory of the transfer-matrix method used here has been explained
in Ref.~\cite{transfer1}, and the implementational details of an improved
algorithm have been given in Ref.~\cite{transfer2}.  Suffice it here to 
mention that we have used the Fortuin--Kasteleyn representation
\cite{Kasteleyn_69,Fortuin_72} of the Potts model
in the computation of the transfer matrix;
therefore, all quantities are expressed as polynomials in $q$.

To compute the limiting curves $\scrb$ we have used two different techniques:
the resultant method and a direct-search method.
These techniques have been described in \cite[Section 4.1]{transfer1},
and we use here the same conventions and notation.

Let us briefly mention a few improvements/additions to our methodology:

\medskip

{\em Computation of T points.}\/  We have adopted an improved method
for locating T points, based on applying a numerical minimization
algorithm (e.g.\ {\sc Mathematica}'s {\tt FindMinimum})
to the function
\be
   F(q)  \;=\;
   (|\lambda_1(q)| - |\lambda_2(q)|)^2  \,+\,
   (|\lambda_1(q)| - |\lambda_3(q)|)^2  \,+\,
   (|\lambda_2(q)| - |\lambda_3(q)|)^2
   \;,
\ee
where $\lambda_1,\lambda_2,\lambda_3$ are the three eigenvalues
of the transfer matrix of largest modulus.  At any given $q$
these eigenvalues can easily be computed numerically
by finding the roots of the characteristic polynomial of the transfer matrix.
Using this method, we are able to locate T points far more precisely
than in our previous work.

\medskip

{\em Fixed zeros.}\/
When $q$ is an integer and the graph $G$ is not $q$-colorable,
we have $P_G(q) = 0$.  For this reason, certain small integers $q$
can be ``fixed'' zeros of the zero-temperature partition function,
independent of the strip length $n$.
In particular, $q=0,1$ are roots for all widths $m \ge 2$
and lengths $n \ge 1$.
Furthermore, $q=2$ is a root for all triangular-lattice strips
of widths $m \ge 2$ and lengths $n \ge 2$,
because the triangular lattice is not bipartite.
Finally, $q=3$ is a root for all {\em cylindrical}\/ triangular-lattice strips
of widths that are not multiples of 3 (with lengths $n \ge 2$),
because these graphs are not 3-colorable.

It is interesting to see how these behaviors come about
from the point of view of the transfer-matrix formalism.
The partition function on a lattice of length $n$ has the form
\be
   Z_n   \;=\;
   \sum\limits_{k=1}^{M} \alpha_k(q) \, \lambda_k(q)^{n-1}
   \;,
\ee
where the $\{ \lambda_k \}$ are the eigenvalues of the transfer matrix
and the $\{ \alpha_k \}$ are the corresponding amplitudes.
A particular value $q$ can then be a ``fixed'' zero of $Z$
for any of three reasons:
\begin{itemize}
   \item[1)]  All the amplitudes $\alpha_k$ vanish at $q$.
      Then $Z_n(q) = 0$ for all $n \ge 1$.
   \item[2)]  All the eigenvalues $\lambda_k$ vanish at $q$.
      Then $Z_n(q) = 0$ for all $n \ge 2$.
   \item[3)]  ``Mixed case'':  Neither all the amplitudes nor
      all the eigenvalues vanish at $q$,
      but for each $k$ either $\alpha_k$ or $\lambda_k$ vanishes at $q$
      (or both).
      Then $Z_n(q) = 0$ for all $n \ge 2$.
\end{itemize}
As we shall see, the points $q=0$ and $q=1$ will be fixed roots
belonging to Case 1:
all the amplitudes vanish due to an overall prefactor $q(q-1)$.
The point $q=2$ will be a fixed root
belonging sometimes to Case 1 and sometimes to Case 3
(and to Case 2 when $m=2$ for all boundary conditions).
For cylindrical strips where the width is not a multiple of 3,
the point $q=3$ will be a fixed root
belonging to Case 2 for $m=4_{\rm P}$ and to Case 3 for $m\ge 5_{\rm P}$.
We shall endeavor to explain in each case
the mechanism underlying the fixed zeros;
these results will be summarized in Section~\ref{sec8.3}.

\medskip

{\em Computation of isolated limiting points.}\/
To find the isolated limiting points, we first compute symbolically
the determinant $\det D(q)$ defined by \reff{def_Det_D};
this determinant is a polynomial in $q$ with integer coefficients,
typically of very high degree.
We then compute numerically the zeros of this polynomial,
using the MPSolve 2.1.1 package \cite{Bini_package,Bini-Fiorentino};
these zeros correspond to points $q$ where at least one amplitude
$\alpha_k(q)$ vanishes.
Finally, we test numerically each of these zeros to see whether the amplitude
corresponding to the {\em dominant}\/ eigenvalue is vanishing;
if it is, then the point in question is an isolated limiting point.
This method is guaranteed to discover {\em all}\/ of the isolated
limiting points.
We shall not bother to report here all the zeros of $\det D(q)$,
but only (a) the isolated limiting points
and (b) the Beraha numbers $B_n$ that are zeros
of some subdominant amplitude.
Please note that whenever $B_n$ is a zero of $\det D(q)$,
so are all the primitive generalized Beraha numbers
\be
  B_n^{(k)} \;=\; 4 \cos^2 {k\pi \over n} \,=\, 2 + 2 \cos {2\pi k \over n}
 \label{def_Bnk}
\ee
where $k$ is relatively prime to $n$, since they have the same
minimal polynomial $p_n(q)$ \cite[Section~2.3]{transfer1}.

Unfortunately, in some cases the matrix $D(q)$ is so large that
we have been unable to compute symbolically its determinant.
In these cases, it is more convenient to compute numerically the eigenvalues 
$\{\lambda_j\}$ and their corresponding amplitudes $\{\alpha_j\}$ and check 
(a) whether any of the amplitudes vanish and (b) whether the 
amplitude $\alpha^\star$ associated to the dominant eigenvalue vanishes. 
We have restricted our search to certain ``candidate''
values of $q$ (or neighborhoods in the complex $q$-plane),
namely (a) the Beraha numbers $B_n$ for $n \le 50$,
and (b) any real or complex values of $q \notin {\cal B}$
where zeros of $Z_n$ seem to be accumulating as $n$ gets large.
When there is a exact candidate (such as the Beraha numbers $B_n$), we have  
computed the amplitudes with high-precision arithmetic (200 digits of precision
at least). We considered that an amplitude is zero when its absolute value
is less than (for instance) $10^{-190}$. 
When we do not have an exact candidate, we tried to minimize the 
dominant amplitude $|\alpha^\star|$ around a region where zeros of $Z_n$
tend to accumulate as $n$ grows. This situation occurred only for the strips of 
widths $8_{\rm Z}$ and $10_{\rm Z}$ (``zig-zag''
boundary conditions, see Section~\ref{sec_Z}). 
The condition $|\alpha^\star|\ltapprox 10^{-52}$ holds for all
the cases reported here. 

In all the cases where we are unable to compute $\det D(q)$ symbolically,
we are able to assert that certain points are indeed
isolated limiting points, but we cannot claim with confidence that
we have found {\em all}\/ of the isolated limiting points.

\medskip

%
%
\section{Numerical Results for the Triangular-Lattice
   \hfill\break Chromatic Polynomial:
   \hfill\break Free Boundary Conditions}
   \label{sec_F}

We have computed the transfer matrix $\T(m_{\rm F})$ and the limiting 
curves $\scrb$ for triangular-lattice strips of widths $2 \leq m \leq 9$
with free boundary conditions in both directions.
We also write $L_x$ as a synonym for the strip width $m$.

As explained in Ref.~\cite{transfer1}, the chromatic polynomial for this
family of strip lattices can be written as
\be
Z(m_{\rm F} \times n_{\rm F}) \;=\; \uu^{\rm T} \H \T(m_{\rm F})^{n-1} 
                                \vv_{\rm id}
\label{chromatic_free}
\ee
where $\uu$ and $\vv_{\rm id}$ are certain vectors,
and $\T(m_{\rm F}) = \V \H$ is the transfer matrix.
Here $\H$ (resp.\ $\V$) corresponds to adding one new layer of
horizontal (resp.\ vertical and diagonal) bonds:
see Figure~\ref{Figure_transfer}(a).
The matrices $\H$, $\V$ and $\T(m_{\rm F})$
act on the space of connectivities of sites in the top layer,
whose basis elements $\vv_{\scrp}$ are indexed by partitions $\scrp$ of 
the single-layer vertex set $\{1,2,\ldots,m\}$.
In particular, $\vv_{\rm id} = \vv_{ \{\{1\},\{2\},\ldots,\{m\}\} }$.
Since the strip lattices we are dealing with are planar,
only {\em non-crossing}\/ partitions $\scrp$ can occur.

In the particular case of the chromatic polynomial
(i.e.\ the {\em zero-temperature}\/ antiferromagnet),
the horizontal operator $\H$ is a projection ($\H^2 = \H$),
and we can work in its image subspace by using the 
modified transfer matrix $\T'(m_{\rm F}) = \H \V \H$ in place of 
$\T(m_{\rm F}) = \V \H$, and using the basis vectors 
\be
 \w_{\scrp}   \;=\; \H \vv_{\scrp}
 \label{def_wP}
\ee
in place of $\vv_{\scrp}$.
Then we can rewrite \reff{chromatic_free} as
\be
Z(m_{\rm F} \times n_{\rm F}) \;=\; \uu^{\rm T} \T'(m_{\rm F})^{n-1}
                                \w_{\rm id}
\label{chromatic_free_final}
\ee
where $\w_{\rm id} = \H \vv_{\rm id}$.
Please note that $\w_{\scrp} = \H \vv_{\scrp} = 0$ for any 
partition $\scrp$ that includes nearest-neighbor sites in the same block,
so we can ignore all such partitions. 
The dimension of the transfer matrix $\T'(m_{\rm F})$
is therefore equal to the number of
non-crossing non-nearest-neighbor partitions of the set $\{1,2,\ldots,m\}$, 
which is given by the Motzkin number $M_{m-1}$ \cite{transfer1}. 
To simplify the notation, we will drop the prime in $\T'(m_{\rm F})$
and denote the basis elements $\w_{\scrp}$ by a shorthand
using Kronecker delta functions:
for instance, $\w_{ \{\{1,3\},\{2\},\{4,6\},\{5\}\} }$
will be written $\delta_{13} \delta_{46}$.
We denote the set of basis elements for a given strip as 
${\bf P} = \{ \w_{\scrp} \}$.
For instance, the basis for  $m=3$ is ${\bf P} = \{ 1, \delta_{13} \}$.

We have checked the self-consistency of our finite-lattice results
using the trivial identity
\be
  Z(m_{\rm F} \times n_{\rm F})  \;=\;   Z(n_{\rm F} \times m_{\rm F})
\ee
for all pairs $2 \le m,n \le 9$.

%
%
\subsection{$L_x = 2_{\rm F}$} \label{sec2F}

This case is trivial, as the transfer matrix is one-dimensional:
\be
Z(2_{\rm F} \times n_{\rm F}) = q(q-1) (q-2)^{2(n-1)} 
\ee
Since there is only one eigenvalue, there is obviously no crossing, hence
${\cal B} = \emptyset$. However, there are zeros for all $n$ at 
$q=0,1$ and for $n\geq 2$ at $q=2$.
The fixed zeros at $q=0,1$ arise from a vanishing amplitude,
and the fixed zero at $q=2$ arises from a vanishing eigenvalue.

%
%
\subsection{$L_x = 3_{\rm F}$} \label{sec3F}

The transfer matrix is two-dimensional. In the basis  
${\bf P} = \{ 1, \delta_{13} \}$ it can be written as
\be
T(3_{\rm F}) \;=\; \left( \begin{array}{cc} 
              q^3 - 7q^2 + 17q - 14  & q^2 - 6q + 9  \\
                           - q +  2  &        q - 3 
              \end{array} \right)
\ee
and the partition function is equal to
\be
Z( 3_{\rm F} \times n_{\rm F} ) \;=\; 
      q(q-1) \left( \begin{array}{c} 
                        q-1 \\
                          1 
                    \end{array}\right)^{\!\rm T} \cdot 
                  T(3_{\rm F})^{n-1} \cdot  
             \left( \begin{array}{c}  1 \\ 0 \end{array} \right)  
 \label{eq.Z.3F}
\ee

The limiting curve ${\cal B}$ (see Figure~\ref{Figure_tri_3FxInftyF}) contains 
three disconnected pieces and it crosses the real axis at 
$q_0 \approx 2.5698402910$. There are six endpoints:
\begin{subeqnarray}
q &\approx& 1.2047381150 \pm 1.1596169599\,i \\
q &\approx& 2.3930361082 \pm 0.2538745688\,i \\
q &\approx& 3.4022257768 \pm 0.5865084714\,i
\end{subeqnarray}
These results were previously obtained by Ro\v{c}ek {\em et al}.\ 
\cite{Shrock_98a}. 

The determinant $\det D(q)$ has the form
\be
\det D(q) = q^2 (q-1)^2 (q-2)^3 
\ee
Thus, it vanishes at the first three Beraha numbers $q=0,1,2$.
At those points
the dominant amplitude vanishes, hence they are isolated limiting points. 

In fact, the partition function vanishes at $q=0,1$ for all $n$,
and at $q=2$ for all $n \ge 2$.
Obviously, at $q=0,1$ both amplitudes vanish,
due to the prefactor $q(q-1)$ in \reff{eq.Z.3F}.
[This happens for {\em all}\/ strips of width $m \ge 2$;
 we will henceforth call these zeros ``trivial''.]
At the fixed zero $q=2$,
there is one nonzero eigenvalue ($\lambda^\star = -1$)
with a vanishing amplitude
and one zero eigenvalue with a nonvanishing amplitude;
we are therefore in Case 3 described in Section~\ref{sec_prelim}.
The fourth real zero (see Table~\ref{table_zeros_free}) converges at an 
approximate $1/n$ rate to the value $q_0 \approx 2.5698402910$. 

%
%
\subsection{$L_x = 4_{\rm F}$} \label{sec4F}

The transfer matrix is four-dimensional. In the basis 
${\bf P} = \{1, \delta_{13},\delta_{24},\delta_{14} \}$, it takes 
the form 
\be
T(4_{\rm F}) \;=\; \left( \begin{array}{cccc}
  T_{11}         & T_{12}        & T_{13}        & T_{14}        \\
  - q^2 + 5q - 6 & q^2 - 5q + 6  & -q + 3        &     -2(q - 3) \\ 
  - q^2 + 5q - 6 & q^2 - 5q + 6  & q^2 - 6q + 9  & q^2 - 8q + 15 \\
           q - 2 & q^2 - 5q + 6  &       -q + 3  & q^2 - 7q + 13
              \end{array} \right)
\ee
where 
\begin{subeqnarray}
T_{11} &=& q^4 - 10q^3 + 39q^2 - 70q + 48 \\ 
T_{12} &=&         q^3 -  9q^2 + 26q - 24 \\ 
T_{13} &=&         q^3 -  9q^2 + 28q - 30 \\ 
T_{14} &=&         q^3 - 10q^2 + 36q - 45  
\end{subeqnarray}
The partition function is equal to
\be
Z( 4_{\rm F} \times n_{\rm F} ) \;=\;
      q(q-1) \left( \begin{array}{c}
                   (q-1)^2 \\ 
                    q-1    \\ 
                    q-1    \\ 
                    q-2 
                    \end{array}\right)^{\!\rm T} \cdot
                  T(4_{\rm F})^{n-1} \cdot
             \left( \begin{array}{c}  1 \\ 0 \\ 0 \\ 0 \end{array} \right)
\ee

The limiting curve ${\cal B}$ (see Figure~\ref{Figure_tri_4FxInftyF})
contains two complex-conjugate disconnected pieces
that do not cross the real axis.
The closest points to the real axis are 
$q_0 \approx 2.7592502040 \pm 0.1544431251\,i$. 
There are six endpoints:
\begin{subeqnarray}
q &\approx& 0.8164709452 \pm 1.2804094073\,i \\
q &\approx& 2.7592502040 \pm 0.1544431251\,i \\
q &\approx& 3.6398304896 \pm 0.5986827987\,i
\end{subeqnarray}
There are T points at $q \approx 3.3341785562 \pm 0.8829730283\,i$.
These results were previously obtained by Ro\v{c}ek {\em et al}.\ 
\cite{Shrock_98a,Shrock_98c}.

The determinant $\det D(q)$ has the form
\be
\det D(q) = -q^4 (q-1)^4 (q-2)^{13} (q^2-3q+1) (q-3)^6 
            (q^4 - 11q^3 + 46q^2 -86q+61)^2 
 \label{eq.4F.detD}
\ee
We recognize, as factors in \reff{eq.4F.detD},
the first five minimal polynomials $p_k(q)$
for the Beraha numbers $B_k$ \cite[Table 1]{transfer1};
therefore $\det D(q)$ vanishes at the first five Beraha numbers $q=0,1,2,B_5,3$.
The dominant amplitude vanishes at $q=0,1,2,B_5$,
so these are isolated limiting points.
At $q=3$ only two subdominant amplitudes vanish,
so this is not an isolated limiting point.
Similarly, all of the zeros of the last factor in \reff{eq.4F.detD}
correspond to the vanishing of subdominant amplitudes only,
so none of them is an isolated limiting point.  

In fact, the partition function vanishes at $q=0,1$ for all $n$,
and at $q=2$ for all $n \ge 2$.
The fixed zeros at $q=0,1$ are trivial.
At $q=2$, there are three zero eigenvalues and a unique leading eigenvalue 
$\lambda^\star=4$ with zero amplitude.
Notice that the transfer matrix is not diagonal for $q=2$: there is a 
$2\times2$ nontrivial Jordan block corresponding to $\lambda=0$ and 
whose contribution to the partition 
function is zero for all $n$. The amplitude corresponding to the 
other $\lambda=0$ eigenvalue is 2; we are therefore in Case 3 described 
in Section~\ref{sec_prelim}.
Finally, the fourth real zero converges exponentially fast to $B_5$
(see Table~\ref{table_zeros_free}).

Please note that for this strip there is a vanishing subdominant
amplitude at $q=B_k$ for a Beraha number $k$ {\em greater than}\/ $m+1$
(namely, $B_6 = 3$).
As we shall see, this occurs frequently for the triangular lattice,
and contrasts with the behavior observed for the square lattice
\cite{transfer1,transfer2}.

%
%
\subsection{$L_x = 5_{\rm F}$} \label{sec5F}

The transfer matrix is nine-dimensional; it can be found in the 
{\sc Mathematica} file {\tt transfer3.m} available as part of
the electronic version of this paper in the {\tt cond-mat} archive.
This strip has been previously studied by
Chang and Shrock \cite{Shrock_00c};
but they did not compute the limiting curve. 

The limiting curve ${\cal B}$ is connected 
(see Figure~\ref{Figure_tri_5FxInftyF}). It crosses the real
axis at $q_0=3$. There are six endpoints:
\begin{subeqnarray}
q &\approx& 0.5586170364 \pm 1.2816149610\,i \\
q &\approx& 3.0474871745 \pm 0.8171660680\,i \\
q &\approx& 3.7782975917 \pm 0.5699779858\,i
\end{subeqnarray}

The topology of the limiting curve is rather involved. It has 12 T points: 
$q \approx 3.1572589261 \pm 0.7951215102\,i$,
$q \approx 3.1251751109 \pm 0.8152460413\,i$,
$q \approx 3.2093444343 \pm 0.9296294663\,i$,
$q \approx 3.3452062643 \pm 0.9758086833\,i$,
$q \approx 3.3248793469 \pm 0.9987588766\,i$, and
$q \approx 3.2492362818 \pm 1.1185073809\,i$.
These points define four closed regions. The first five T points are the
vertices of two complex-conjugate regions which look approximately like
rectangular bands. The third, fifth and sixth T points above define two 
complex-conjugate triangular-like regions. 

The determinant $\det D(q)$ is given by 
\be
\det D(q) \;=\; q^9 (q-1)^9 (q-2)^{66} (q^2 -3q +1)^4 (q-3)^{61} P(q)^2 
\ee
where the polynomial $P(q)$ can be found in the {\sc Mathematica} file
{\tt transfer3.m}. The factors appearing in $\det D(q)$ are the first 
five polynomials $p_k(q)$ given in \cite[Table 1]{transfer1};
therefore $\det D(q)$ vanishes at $q=0,1,2,B_5,3$.
The dominant amplitude vanishes at the first four of them (but not at $q=3$),
so that $q=0,1,2,B_5$ are isolated limiting points.  
All of the zeros of $P(q)$ correspond to the vanishing of
subdominant amplitudes only, so none of them is an isolated limiting point.  

The first two real zeros $q=0,1$ are trivial ones. At $q=2$ 
there are two nonzero eigenvalues with zero amplitudes 
and seven zero eigenvalues. The transfer matrix is not diagonalizable:
there are two Jordan blocks of dimension 3 and 2, respectively, associated to 
the eigenvalue $\lambda=0$. The contribution of these blocks to the 
partition function vanishes for all $n$. In addition, the amplitude 
corresponding to the other two zero eigenvalues are 0 and 2; we are 
therefore in Case 3 described in Section~\ref{sec_prelim}.  
The fourth real zero converges exponentially fast to the value $q=B_5$
(see Table~\ref{table_zeros_free});
and the fifth real zero converges at an approximate $1/n$ rate to the value 
$q_0 =3$. This agrees with the fact that $q=3$ is a regular limiting point
and not an isolated limiting point. 

%
%
\subsection{$L_x = 6_{\rm F}$} \label{sec6F}

The transfer matrix is 21-dimensional; it can be found in the
{\sc Mathematica} file {\tt transfer3.m}. 

The limiting curve ${\cal B}$ is connected
(see Figure~\ref{Figure_tri_6FxInftyF}). It crosses the real
axis at $q_0\approx 3.1609256737$. There are six endpoints:
\begin{subeqnarray}
q &\approx& 0.3796307748 \pm 1.2450702104\,i \\
q &\approx& 2.9641235697 \pm 1.1179839989\,i \\
q &\approx& 3.8664092416 \pm 0.5329463088\,i
\end{subeqnarray}
There are four T points at 
$q \approx 3.3081144403 \pm 1.2171494282\,i$, and  
$q \approx 3.5005856709 \pm 0.9442298756\,i$.

The determinant $\det D(q)$ is given by
\be
\det D(q) \;=\; q^{21} (q-1)^{21} (q-2)^{360} (q^2 - 3q + 1)^{13} 
                (q-3)^{503} (q^3 - 5q^2 + 6q - 1) P(q)^2
\ee
where the polynomial $P(q)$ can be found in the {\sc Mathematica} file
{\tt transfer3.m}. The factors appearing in $\det D(q)$ are the first
six polynomials $p_k(q)$ given in \cite[Table 1]{transfer1}.
Thus, $\det D(q)$ vanishes at $q=0,1,2,B_5,3,B_7$.
The dominant amplitude vanishes at the first five of them (but not at $B_7$),
so that $q=0,1,2,B_5,3$ are isolated limiting points.
All of the zeros of $P(q)$ correspond to the vanishing of
subdominant amplitudes only, so none of them is an isolated limiting point.  

The first two real zeros $q=0,1$ are trivial ones. At $q=2$ we get 
six nonzero eigenvalues with zero amplitudes and 15 zero eigenvalues. 
We find again that the transfer matrix is not diagonalizable for $q=2$: there 
are five nontrivial Jordan blocks (one of dimension 3 and four of dimension 2) 
corresponding 
to the eigenvalue $\lambda=0$, and whose contribution to the partition 
function always vanishes. The amplitudes of the other four zero eigenvalues are
(0,0,0,2). 
This combination seems to be the generic case for free boundary conditions:
all amplitudes vanish except one corresponding to a zero eigenvalue. 
When the transfer matrix is not diagonalizable, then the nontrivial Jordan 
blocks correspond to $\lambda=0$ and their contribution is always zero.
The fourth and fifth 
real zeros converge exponentially fast to the values $q=B_5$ and $q=3$,
respectively (see Table~\ref{table_zeros_free});
the sixth real zero converges at an approximate $1/n$ rate to the value 
$q_0 \approx 3.1609256737$.

%
%
\subsection{$L_x = 7_{\rm F}$} \label{sec7F}

The transfer matrix is 51-dimensional; it can be found in the
{\sc Mathematica} file {\tt transfer3.m}. In this case we have been unable
to compute symbolically the resultant, hence the computation of the 
limiting curve has been performed using the direct-search method. 

The limiting curve ${\cal B}$ is connected
(see Figure~\ref{Figure_tri_7FxInftyF}). It crosses the real
axis at $q_0\approx 3.2764013231$. There are six endpoints:
\begin{subeqnarray}
q &\approx& 0.250538 \pm 1.196864\,i  \\
q &\approx& 3.925804 \pm 0.496672\,i \\
q &\approx& 2.878928 \pm 1.343851\,i
\end{subeqnarray}
There are four T points at 
$q \approx 3.6146786603 \pm 0.9081562491\,i$ and
$q \approx 3.2704141478 \pm 1.5194310419\,i$.

We have been unable to compute the determinant $\det D(q)$. However, we
computed the amplitudes numerically at each of the Beraha numbers $B_n$
up to $B_{50}$ and determined in particular whether it is an isolated
limiting point or not.  As always, $q=0,1$ are trivial
isolated limiting points where all the amplitudes vanish. The dominant 
amplitude also vanishes at $q=2,B_5,3,B_7$, so they are isolated
limiting points too. Finally, a subdominant amplitude vanishes at $q=B_8$.

The first two real zeros $q=0,1$ are trivial ones. 
At $q=2$, all amplitudes vanish except one corresponding to a zero
eigenvalue. In particular, there are 12 nonzero eigenvalues with zero 
amplitudes, 8 zero eigenvalues with zero amplitudes, and one zero eigenvalue 
with nonzero amplitude. We also find 12 nontrivial Jordan blocks corresponding 
to $\lambda=0$, whose contribution to the partition function is always zero.
The fourth, fifth
and sixth real zeros converge exponentially fast to the values
$B_5$, 3 and $B_7$, respectively (see Table~\ref{table_zeros_free_bis});
however, the convergence to $B_7 \approx 3.246979603717$ is slowed
by its nearness to the regular limiting point $q_0\approx 3.2764013231$.
For lengths $n\gtapprox 77$, a seventh real zero appears:
it converges (at an approximate $1/n$ rate) to $q_0$.

%
%
\subsection{$L_x = 8_{\rm F}$} \label{sec8F}

The transfer matrix is 127-dimensional; it can be found in the
{\sc Mathematica} file {\tt transfer3.m}. Again we have used the direct-search
method to locate the points of the limiting curve ${\cal B}$.  

The limiting curve ${\cal B}$ is connected
(see Figure~\ref{Figure_tri_8FxInftyF}). It crosses the real
axis at $q_0\approx 3.3610599515$. There are six endpoints:
\begin{subeqnarray}
q &\approx& 0.154432 \pm 1.146669\,i \\
q &\approx& 2.793496 \pm 1.521468\,i \\
q &\approx& 3.967566 \pm 0.463648\,i 
\end{subeqnarray}
There are four T points at
$q\approx 3.2555859898 \pm 1.7000353877\,i$ 
and $q\approx 3.6703287722 \pm 0.8845072864\,i$.

We were unable to compute the determinant $\det D(q)$. However, we
computed the amplitudes numerically at each of the Beraha numbers $B_n$
up to $B_{50}$ and determined in particular whether it is an isolated
limiting point or not.
As always, $q=0,1$ are trivial
isolated limiting points where all the amplitudes vanish. The dominant
amplitude vanishes also at $q=2,B_5,3,B_7$, so they are isolated
limiting points too. Finally, some subdominant amplitudes vanish at 
$q=B_8,B_9$; they are not isolated limiting points. 

The first two real zeros $q=0,1$ are trivial ones. 
At the third real zero $q=2$, all amplitudes vanish except one 
corresponding to a zero eigenvalue.
At $q=2$ the transfer matrix is not diagonalizable: we find 30 nontrivial 
Jordan blocks corresponding to $\lambda=0$, but their contribution to the
partition function vanishes for all $n\geq 1$. We also get 33 nonzero 
eigenvalues with zero amplitudes, 20 zero eigenvalues with zero amplitudes,
and one zero eigenvalue with nonzero amplitude. The fourth, fifth
and sixth real zeros converge exponentially fast to the values $B_5$, $3$
and $B_7$, respectively (see Table~\ref{table_zeros_free_bis}).
We also expect a seventh real zero converging (at an approximate $1/n$ rate)
to $q_0\approx 3.3610599515$. Such zero does not appear up to lengths
$n=96$ (see Table~\ref{table_zeros_free_bis}). We would need to go to 
very large $n$ to observe this additional zero.
 
%
%
\subsection{$L_x = 9_{\rm F}$} \label{sec9F}

The transfer matrix is 323-dimensional; it can be found in the
{\sc Mathematica} file {\tt transfer3.m}. The size of this transfer matrix 
prevented us from computing the limiting curve ${\cal B}$.  
However, we were able to compute the point where the limiting curve crosses the 
real axis: it is $q_0\approx 3.4251304673$. 
In Figure~\ref{Figure_tri_9FxInftyF} we show the zeros of $Z$
for the finite lattices
$9_{\rm F}\times 45_{\rm F}$ and $9_{\rm F}\times 90_{\rm F}$.

The first two real zeros $q=0,1$ are trivial ones. We have checked numerically
that the dominant amplitude vanishes at $q=2,B_5,3,B_7,B_8$
(and at no other Beraha numbers up to $B_{50}$);
these are therefore isolated limiting points.
By inspection of Figure~\ref{Figure_tri_9FxInftyF} we do not 
find any candidate for a non-Beraha real isolated limiting point
or for a complex isolated limiting point. 
A subdominant amplitude vanishes for $q=B_9$ and $B_{10}$.

At $q=2$ the transfer matrix is not diagonalizable: we find 76 nontrivial 
Jordan blocks (up to dimension $5\times 5$) corresponding to $\lambda=0$; 
none of these blocks contribute to
the partition function for any $n\ge1$. We also get 74 nonzero eigenvalues 
with zero amplitudes, 50 zero eigenvalues with zero amplitudes,
and one zero eigenvalue with nonzero amplitude.
The fourth, fifth, and sixth real zeros converge exponentially fast to the 
values $B_5$, $3$ and $B_7$, respectively 
(see Table~\ref{table_zeros_free_bis}). 
We expect that the seventh real zero will converge exponentially fast to
$B_8 = 2 + \sqrt{2}$ and that
there will be (for large enough length $n$) an additional real zero
that will eventually converge to the value 
$q \approx 3.4251304673 > B_8$ at an approximate $1/n$ rate.
However, we would probably need to go
to very large lengths $n$ in order to observe this behavior.

%
%
\section{Numerical Results for the Triangular-Lattice
   \hfill\break Chromatic Polynomial:
   \hfill\break Cylindrical Boundary Conditions}
   \label{sec_P}

We have also computed the transfer matrix $\T(m_{\rm P})$
and the limiting curves $\scrb$
for triangular-lattice strips of widths $2 \leq m \equiv L_x \leq 12$
with cylindrical boundary conditions,
i.e.\ periodic b.c.\  in the transverse direction
and free b.c.\  in the longitudinal direction.

The partition function can be written analogously to
\reff{chromatic_free_final} as
\be
Z(m_{\rm P} \times n_{\rm F}) = \uu^{\rm T} \T(m_{\rm P})^{n-1}
                                \w_{\rm id}
\label{chromatic_cyl_final}
\ee
where $\T(m_{\rm P}) = \H \V \H$.
(See \cite[Section 3.1]{transfer1} for some 
remarks about the actual computation of this transfer matrix.)
Since $\T(m_{\rm P})$ commutes with translations
(due to the periodic b.c.)\ and the vectors $\uu$ and $\w_{\rm id}$
are translation-invariant,
we can restrict attention to the translation-invariant subspace.
The dimension of $\T(m_{\rm P})$ is therefore given
by the number of equivalence classes modulo translation
of non-crossing non-nearest-neighbor partitions.
This number is denoted by TriCyl($m$) in \cite[Table 2]{transfer1}. A general
analytic expression for TriCyl($m$) is not known, although such a formula
has been obtained for prime values of $m$ \cite[Theorem 3]{Tutte_sq}. 

We have checked our results for widths $2 \le m_{\rm P} \le 12$
and lengths $n_{\rm F} = 2,3,4$ by comparing to the results of
Beraha--Kahane--Weiss \cite{Beraha_80}
(resp.\ Chang--Shrock \cite{Shrock_00c})
for {\em width}\/ $n_{\rm F} = 2$ (resp.\ $n_{\rm F} = 3,4$)
and {\em length}\/ $m_{\rm P}$
(i.e.\ cyclic boundary conditions), using the trivial identity
\be
   Z(m_{\rm P} \times n_{\rm F})  \;=\;  Z(n_{\rm F} \times m_{\rm P})
   \;.
\ee
This is a highly non-trivial check of the correctness of our results.

The triangular lattice with cylindrical boundary conditions possesses
a curious reflection symmetry that we shall now explain.
Note first that the triangular lattice is {\em not}\/ invariant
under reflection in the transverse direction\footnote{
   More precisely, if the strip width is even, one can choose
   to reflect either through a pair of lattice sites
   or through a pair of bisectors (sites with half-integer coordinates);
   the two choices differ by a subsequent translation.
   If the strip width is odd, then every reflection axis
   passes through one lattice site and one bisector.
},
since reflection changes the direction of the diagonal bonds.
Nevertheless, {\em in the translation-invariant subspace}\/
the transfer matrix does commute with reflection,
because by translating the upper row (of a pair of rows) by one unit,
one can change$\!$
\setlength{\unitlength}{3mm}
\begin{picture}(1.3,1)(-0.2,0)
   \put(0,1){\line(0,-1){1}}
   \put(0,1){\line(1,-1){1}}
   \put(0,0){\circle*{0.2}}
   \put(0,1){\circle*{0.2}}
   \put(1,0){\circle*{0.2}}
\end{picture}
into
\begin{picture}(1.3,1)(-0.2,0)
   \put(0,0){\line(1,1){1}}
   \put(1,0){\line(0,1){1}}
   \put(0,0){\circle*{0.2}}
   \put(1,0){\circle*{0.2}}
   \put(1,1){\circle*{0.2}}
\end{picture}
, thereby converting the triangular lattice into a reflected
triangular lattice!
Because the transfer matrix (in the translation-invariant subspace)
commutes with reflection,
we can pass to a new basis consisting of connectivities
that are either even or odd under reflection.\footnote{
   Let us call $\{v_j\}_{j=1}^M$ the connectivity basis in
   the translation-invariant subspace.
   Some of these basis elements are invariant under reflection;
   the rest can be grouped into pairs $(v_\alpha,v_\beta)$
   that map into each other under reflection.
   A basis for the reflection-even (i.e.\ reflection-invariant) subspace
   is then given by the basis elements in the first set together with the
   combinations $v_\alpha+v_\beta$ from the second set.
   A basis for the reflection-odd subspace is given by the
   combinations $v_\alpha-v_\beta$ from the second set.
}
In this new basis, the transfer matrix $\T(m_{\rm P})$
is block-diagonal:
\be
\T(m_{\rm P}) \;=\; \left( \begin{array}{cc}
                               \T_+(m_{\rm P})  &  0  \\
                               0                &  \T_-(m_{\rm P})
                            \end{array} \right)
   \;\,,
\label{transfer_general_diag}
\ee
where $\T_+$ (resp.\ $\T_-$) corresponds to the
reflection-even (resp.\ reflection-odd) subspace.
Furthermore, the reflection-odd components of the start vector
$\w_{\rm id} = \H \vv_{\rm id}$ are identically zero,
since both $\vv_{\rm id}$ and $\H$ are manifestly reflection-invariant.
Likewise, the reflection-odd components of the final vector $\uu$
are identically zero, since the definition of $\uu$
involves only $\H$ and not $\V$.
Therefore (for either of these two reasons),
the amplitudes $\alpha_k(q)$ corresponding to the reflection-odd subspace
are all identically vanishing;
these eigenvalues make no contribution whatsoever to the partition function.
It follows that we can work entirely within the
reflection-invariant subspace,
which has dimension $\hbox{SqCyl}(m)$ \cite[Table 2]{transfer1}.
For strip widths $m \ge 8$, $\hbox{SqCyl}(m)$
is strictly smaller than $\hbox{TriCyl}(m)$,
so that the matrix $\T_-$ is nontrivial.

%
%
\subsection{$L_x = 2_{\rm P}$} \label{sec2P}

This case is trivial, as the transfer matrix is one-dimensional:
\be
Z(2_{\rm P} \times n_{\rm F}) \;=\; q(q-1) [(q-2)(q-3)]^{n-1}  \;.
\ee
Please note that the triangular-lattice strip $2_{\rm P} \times n_{\rm F}$ is 
{\em not}\/ equivalent to the strip $2_{\rm F} \times n_{\rm F}$
for any length $n \ge 2$.

Since there is only one eigenvalue, there is obviously no crossing, hence
${\cal B} = \emptyset$. However, there are zeros for all $n$ at
$q=0,1$ and for $n\geq 2$ at $q=2,3$.

%
%
\subsection{$L_x = 3_{\rm P}$} \label{sec3P}

This case is also trivial, as the transfer matrix is again one-dimensional:
\be
Z(3_{\rm P} \times n_{\rm F}) \;=\;
 q(q-1)(q-2) (q^3 - 9q^2 + 29q -32)^{n-1}  \;.
\ee
Again ${\cal B} = \emptyset$, and the amplitude
vanishes at $q=0,1,2$ (which are the first three Beraha numbers). For 
$n\geq 2$ there are additional fixed zeros at $q\approx 2.5466023485$
and $q\approx 3.2266988258 \pm 1.4677115087\,i$,
where the eigenvalue vanishes. 
This strip was studied by Ro\v{c}ek {\em et al}.\ \cite{Shrock_98c}.

%
%
\subsection{$L_x = 4_{\rm P}$} \label{sec4P}

The transfer matrix is two-dimensional. In the basis
${\bf P} = \{ 1, \delta_{13}+\delta_{24} \}$ it can be written as
\be
T(4_{\rm P}) \;=\; \left( \begin{array}{cc}
    q^4 - 12q^3 + 58q^2 - 135q + 126 & 2(q^3 - 10q^2 + 33q - 36) \\
                 -2(q^2 -   7q + 12) & 2(        q^2 -  6q +  9)  
              \end{array} \right)
\ee
and the partition function is equal to
\be
Z( 4_{\rm P} \times n_{\rm F} ) \;=\;
      q(q-1) \left( \begin{array}{c}
                        q^2 -3q + 3 \\
                        2(q-1)
                    \end{array}\right)^{\!\rm T} \cdot
                  T(4_{\rm P})^{n-1} \cdot
             \left( \begin{array}{c}  1 \\ 0 \end{array} \right)
\ee

The limiting curve ${\cal B}$ (see Figure~\ref{Figure_tri_4PxInftyF}) contains
three pieces: two complex-conjugate arcs and a self-conjugate loop.
In addition, at the point $q=3$ both eigenvalues vanish simultaneously
(i.e.\ the transfer matrix $T(4_{\rm P})$ itself vanishes);
this is a special (degenerate) species of isolated limiting point
\cite{Sokal_chromatic_roots}.
(One of the amplitudes does not vanish at $q=3$, but that is irrelevant.)
The self-conjugate loop-like component 
crosses the real axis at $q_0 \approx 3.4814056002$ and $q = 4$. 
There are four endpoints:
\begin{subeqnarray}
q &\approx& 1.3705340683 \pm 2.7508526144\,i\\
q &\approx& 3.6294659317 \pm 0.6958422780\,i 
\end{subeqnarray}
This limiting curve was first obtained in the pioneering paper of
Beraha and Kahane \cite{Beraha_79} (see also \cite{Shrock_98c}).
They drew the important conclusion that $q=4$ is a limiting point of
(complex) chromatic roots for the sequence $4_{\rm P} \times n_{\rm F}$
of {\em planar}\/ graphs --- hence the wonderful title of their paper,
``Is the Four-Color Conjecture Almost False?''\footnote{
   Ironically, by the time that this article was published,
   the Four-Color Conjecture had become the Four-Color Theorem.
   The Beraha--Kahane article was submitted in 1976
   but not published until 1979.
}

The determinant $\det D(q)$ has the form
\be
\det D(q) = 8 q^2 (q-1)^2 (q-2) (q^2 - 3 q + 1) (q-3)^2 (q-4)^2 
\ee
We recognize the first five minimal polynomials $p_k(q)$ given in
\cite[Table 1]{transfer1}. Hence, the determinant vanishes at the
first five Beraha numbers $q=0,1,2,B_5,3$. It also vanishes at $q=4$, which
corresponds to $B_\infty$. The dominant amplitude vanishes only at 
$q=0,1,2,B_5$; these values of $q$ correspond to isolated limiting points.

The fixed zeros at $q=0,1$ are trivial ones. 
At the fixed zero $q=2$,
there is one nonzero eigenvalue ($\lambda^\star = 10$)
with a vanishing amplitude
and one zero eigenvalue with a nonvanishing amplitude ($\alpha=2$);
we are therefore in Case 3 described in Section~\ref{sec_prelim}.
The fourth real zero converges exponentially fast to $B_5$
(see Table~\ref{table_zeros_cyl}),
in agreement with the fact that this is an isolated limiting point.
The fifth real zero, at $q=3$, is a fixed zero where both eigenvalues vanish
(one of the amplitudes is 0 and the other is 18);
we are therefore on Case 2 described in Section~\ref{sec_prelim}. 
The fact that $q=3$ is a fixed zero is due to the width not being a
multiple of 3. 
Finally, $q=4$ is a crossing point where both eigenvalues take the value 
$\lambda=2$ and one of the amplitudes vanishes. The sixth real zero in 
Table~\ref{table_zeros_cyl} converges at an approximate $1/n$ rate to the 
value $q_0 \approx 3.4814056002$.  

For this strip there is a vanishing subdominant
amplitude at $q=B_6=3$, which is {\em greater than}\/ $B_{m+1} = B_5$,
in contrast 
with the behavior observed for the square lattice \cite{transfer1,transfer2}.

%
%
\subsection{$L_x = 5_{\rm P}$} \label{sec5P}

The transfer matrix is two-dimensional. In the basis
${\bf P} = \{ 1, \delta_{13}+\hbox{\rm perm.} \}$ it can be written as
\be
T(5_{\rm P}) \;=\; \left( \begin{array}{cc}
  T_{11} &  
   5(q^4 - 14q^3 + 76q^2 - 187q + 174)           \\ 
         -   q^3 + 11q^2 -  43q + 58             &
            3q^3 - 35q^2 + 132q  -162 
              \end{array} \right)
\ee
where 
\be
T_{11} \;=\; q^5 - 15q^4 + 95 q^3 - 320 q^2 + 579 q -452  \;, 
\ee
and the partition function is equal to
\be
Z( 5_{\rm P} \times n_{\rm F} ) \;=\;
      q(q-1)(q-2) \left( \begin{array}{c}
                        q^2 - 2q + 2\\ 
                        5(q-1) 
                    \end{array}\right)^{\!\rm T} \cdot
                  T(5_{\rm P})^{n-1} \cdot
             \left( \begin{array}{c}  1 \\ 0 \end{array} \right)
 \label{eq.Z.5P}
\ee

The limiting curve ${\cal B}$ (see Figure~\ref{Figure_tri_5PxInftyF}) contains
three disconnected pieces: two complex-conjugate arcs and a self-conjugate 
loop-like arc. This latter piece crosses the real axis at 
$q_0 \approx 3.2072219810$ and at $q \approx 3.2847747616$.
There are four endpoints:
\begin{subeqnarray}
q &\approx& 0.4772525688 \pm 2.5694937945\,i\\
q &\approx& 3.5227474312 \pm 0.1876729035\,i
\end{subeqnarray}
The limiting curve was first obtained in \cite{Shrock_00c}. \footnote{
   Two decades earlier, Beraha, Kahane and Weiss \cite{Beraha_80}
   computed the transfer matrix and reported the crossings of the
   limiting curve ${\cal B}$ with the real axis.
   But they did not show a plot of the limiting curve.
}

The determinant $\det D(q)$ has the form
\be
\det D(q) = 5 q^2 (q-1)^2 (q-2)^2 (q^2 - 3q + 1) (q-3) 
            (q^3 - 11q^2 + 43q -58)^2 
\ee
We recognize the first five minimal polynomials $p_k(q)$ given in
\cite[Table 1]{transfer1}. Hence, the determinant vanishes at the
first five Beraha numbers $q=0,1,2,B_5,3$. Furthermore, the dominant amplitude
vanishes at all these points; hence they are isolated limiting points. 
The last factor of the determinant does not provide additional isolated
limiting points.

The fixed zeros at $q=0,1,2$ are trivial ones where all amplitudes
vanish due to the prefactor $q(q-1)(q-2)$ in \reff{eq.Z.5P}.
The fourth real zero converges exponentially fast to $B_5$
(see Table~\ref{table_zeros_cyl}).
The fifth real zero, at $q=3$, is a fixed zero where
there is one nonzero eigenvalue ($\lambda^\star = -2$)
with a vanishing amplitude
and one zero eigenvalue with a nonvanishing amplitude ($\alpha=30$);
we are therefore in Case 3 described in Section~\ref{sec_prelim}.
The fact that $q=3$ is a fixed zero is due to the width not being a
multiple of 3. 
Finally, the sixth real zero
in Table~\ref{table_zeros_cyl} converges at an approximate $1/n$ rate to the 
value $q_0 \approx 3.2072219810$.  

%
%
\subsection{$L_x = 6_{\rm P}$} \label{sec6P}

The transfer matrix is five-dimensional; it can be found in the 
{\sc Mathematica} file {\tt transfer3.m}. 
This strip has been previously studied by
Chang and Shrock \cite{Shrock_00c};
but they did not compute the limiting curve. 

The limiting curve ${\cal B}$ is connected
(see Figure~\ref{Figure_tri_6PxInftyF}). It crosses the real
axis at $q_0\approx 3.2524186216$. There are four endpoints:
\begin{subeqnarray}
q \approx 0.0207708231 \pm 2.2756742729\,i\\
q \approx 4.2838551928 \pm 0.6111544521\,i 
\end{subeqnarray}
There are T points at 
$q \approx  3.9766954928 \pm 0.9167681670\,i$.

The determinant $\det D(q)$ has the form
\begin{eqnarray}
\det D(q) &=& 1769472 q^5 (q-1)^5 (q-2)^6 (q^2 - 3 q + 1)^3  
            (q-3)^{15} (q^3 - 5q^2 + 6q -1) \nonumber \\ 
          & & \qquad \times (q^2 - 5q + 5)^2 (q-4)^8 P(q)^2 
 \label{eq.detD.6P}
\end{eqnarray} 
where the polynomial $P(q)$ can be found in the file {\tt transfer3.m}. 
The first six factors in \reff{eq.detD.6P}
are the first six minimal polynomials given in \cite[Table 1]{transfer1};
hence the determinant vanishes at the first six Beraha numbers
$q=0,1,2,B_5,3,B_7$.
It also vanishes at the Beraha number $B_{10}$,
whose minimal polynomial is $q^2 -5q + 5$, and at $q=4$.
The dominant amplitude vanishes only at the first six Beraha numbers,
so these values are the only isolated limiting points.  
The polynomial $P(q)$ does not provide additional isolated
limiting points. 

The first two zeros $q=0,1$ are trivial ones. 
At $q=2$ we have three nonzero eigenvalues with vanishing amplitudes, 
one zero eigenvalue with zero amplitude, and one zero eigenvalue with
nonzero amplitude;  we therefore fall in Case 3 of Section~\ref{sec_prelim}.
The fourth and fifth real zeros converge exponentially fast to
$q=B_5,3$ (see Table~\ref{table_zeros_cyl}).
The next Beraha number $B_7\approx 3.2469796037$ is very close to the value
$q_0 \approx 3.2524186216$ where the limiting curve ${\cal B}$
crosses the real axis.
This explains why the convergence rate to the sixth real zero in 
Table~\ref{table_zeros_cyl} is not as fast as expected
(empirically the convergence is roughly $\sim n^{-1.9}$);
but we expect that it will be ultimately exponential (for very large $n$).
We also expect a seventh real zero for large enough $n$; this zero is 
expected to converge (at an approximate $1/n$ rate) to the value 
$q_0 \approx 3.2524186216$. We would need to go to very large $n$ to 
observe this additional zero.  

Finally, for this strip there is a vanishing 
subdominant amplitude at $q=B_{10}$, which is {\em greater than}\/
$B_{m+1} = B_7$, in contrast with the behavior observed for the square lattice 
\cite{transfer1,transfer2}.

%
%
\subsection{$L_x = 7_{\rm P}$} \label{sec7P}

The transfer matrix is six-dimensional; it can be found in the
{\sc Mathematica} file {\tt transfer3.m}.

The limiting curve ${\cal B}$ is connected
(see Figure~\ref{Figure_tri_7PxInftyF}). It crosses the real
axis at $q_0\approx 3.4790022937$ and $q\approx 3.6798199576$.
It enters the half-plane $\real(q) < 0$, and there are four endpoints 
\begin{subeqnarray}
q \approx           -0.2279183274 \pm  2.0134503491\,i\\
q \approx \phantom{-}3.9930118897 \pm  0.6273386181\,i
\end{subeqnarray}
There are four T points at 
$q \approx 3.6222949363 \pm 0.1398555812\,i$ and 
$q \approx 3.9816630253 \pm 0.8993269516\,i$. 
Finally, it is worth noticing that the 
limiting curve encloses a small region around  
$3.479 \ltapprox \real(q) \ltapprox 3.680$ and
$|\imag(q) | \ltapprox 0.14$. 

The determinant $\det D(q)$ has the form
\begin{eqnarray}
\det D(q) &=& 68841472 q^6 (q-1)^6 (q-2)^6 (q^2 - 3q + 1)^4 (q-3)^{21} 
              (q^3 - 5q^2 + 6q -1) \nonumber  \\
          & & \qquad \times (q^2 - 4q + 2)(q^2 - 5q + 5)^2 P(q)^2 
 \label{eq.detD.7P}
\end{eqnarray}
where the polynomial $P(q)$ can be found in the file {\tt transfer3.m}.
The first seven factors in \reff{eq.detD.7P}
are precisely the first minimal polynomials given
in \cite[Table 1]{transfer1}. The next one ($q^2 -5q + 5$) is the tenth 
minimal polynomial in \cite[Table 1]{transfer1}. Hence, the determinant 
$\det D(q)$ vanishes at the Beraha numbers $q=0,1,2,B_5,3,B_7,B_8,B_{10}$. 
However, the dominant amplitude vanishes only at the first seven Beraha
numbers $q=0,1,2,B_5,3,B_7,B_8$; these values are the only isolated 
limiting points for this strip. The polynomial $P(q)$ does not provide 
additional isolated limiting points.

The first three zeros $q=0,1,2$ are trivial ones.
The fifth real zero, at $q=3$, is a fixed zero where there are
2 nonzero eigenvalues with zero amplitudes,
one zero eigenvalue with zero amplitude,
and 3 zero eigenvalues with nonzero amplitudes.
The fourth, sixth and seventh real zeros converge exponentially fast to
$q=B_5,B_7,B_8$ (see Table~\ref{table_zeros_cyl}).
The eighth real zero seems to converge at an approximate $1/n$ rate to the 
value $q_0 \approx 3.4790022937$. 

In contrast with the behavior observed for the square lattice,  
we find a vanishing subdominant amplitude at $q=B_{10}$, which is greater 
than $B_{m+1} = B_8$. 

%
%
\subsection{$L_x = 8_{\rm P}$} \label{sec8P}

The transfer matrix $T(8_{\rm P})$ is 15-dimensional.
As discussed at the beginning of this section,
the transfer matrix can be brought into block-diagonal form
\be
       T(8_{\rm P}) \;=\; \left( \begin{array}{cc}
                                    T_+(8_{\rm P}) & 0 \\
                                    0              & T_-(8_{\rm P})
                                 \end{array} \right)
\label{transfer_8P_diag}
\ee
where the block $T_+$ (resp.\ $T_-$)
is 14-dimensional (resp.\ 1-dimensional)
and corresponds to the subspace of reflection-invariant
(resp.\ reflection-odd) connectivities.
Moreover, the amplitude corresponding to the reflection-odd eigenvalue
is identically vanishing;
this eigenvalue therefore makes no contribution to the partition function.
The reduced transfer matrix $T_+(8_{\rm P})$
can be found in the {\sc Mathematica} file {\tt transfer3.m}.
The one-dimensional block is $T_-(8_{\rm P}) = -q^3 + 6q^2 - 8q - 3$.

There are, however, two further curious features
for which we have, as yet, no explanation.
First of all, we find {\em another}\/ eigenvalue
$\lambda = -q^3 + 6q^2 - 8q - 3$,
this time inside the reflection-invariant subspace.
Secondly (and even more mysteriously),
this eigenvalue too has an identically vanishing amplitude.
(We have checked this fact numerically.)
The pair of eigenvalues $\lambda = -q^3 + 6q^2 - 8q - 3$
can be observed explicitly by forming
the characteristic polynomial of the transfer matrix,
which can be factored as\footnote{
  In most of the previously studied cases with
  cylindrical boundary conditions
  (namely, triangular-lattice strips of widths
   $4_{\rm P} \le m \le 7_{\rm P}$
   and square-lattice strips of widths $4_{\rm P} \le m \le 9_{\rm P}$),
  the characteristic polynomial associated to the transfer matrix
  cannot be factored as in \protect\reff{char_poly_8P}.
  In other words, none of the eigenvalues $\lambda$ is a polynomial in $q$.
  The cases with $m\leq3_{\rm P}$ are trivial as the transfer matrix
  is one-dimensional: there is a single eigenvalue,
  which is indeed a polynomial in $q$.
}
\be
\det[ T(8_{\rm P}) - \lambda {\bf 1} ]  \;=\;
  (\lambda + q^3 - 6q^2 + 8q + 3)^2   \, Q_2(q,\lambda)
\label{char_poly_8P}
\ee
where $Q_2(q,\lambda)$ is a polynomial in $q$ and $\lambda$
(it is obviously of degree 13 in $\lambda$).
Unfortunately, we have been unable to find a further change of basis
to make $T_+(8_{\rm P})$ block-diagonal and thereby bring out explicitly
the eigenvalue $\lambda = -q^3 + 6q^2 - 8q - 3$ lying inside that subspace.

In order to compute the limiting curve ${\cal B}$ we have mainly used the 
resultant method. The existence of a double eigenvalue
$\lambda = -q^3 + 6q^2 - 8q - 3$ in the full transfer 
matrix $T(8_{\rm P})$ makes the resultant identically zero
for $\theta=0$.
However, this problem does not arise if we consider the reduced
matrix $T_+(8_{\rm P})$.
Nevertheless, the existence of an identically zero amplitude
within the reduced subspace
makes the computation of the limiting curve ${\cal B}$
not completely straightforward,
as only those eigenvalues with non-identically-vanishing amplitudes
should be taken into account in computing ${\cal B}$.
A simple solution is devised by noting that the resultant method uses only
the characteristic polynomial of the transfer matrix.
Therefore, we can drop the factor $(\lambda + q^3 - 6q^2 + 8q + 3)^2$
in \reff{char_poly_8P} and compute the resultant using the polynomial
$Q_2(q,\lambda)$.
In this way, we obtain a nonzero resultant,
into which the zero-amplitude eigenvalues do not enter.\footnote{
   We also checked --- though this is not relevant to computing ${\cal B}$
   for the boundary conditions being considered here --- that the
   zero-amplitude eigenvalue $\lambda= - q^3 + 6q^2 - 8q - 3$ is not dominant 
   at any of the zeros of our resultant.
   If the zero-amplitude eigenvalue $\lambda= - q^3 + 6q^2 - 8q - 3$ were
   in fact dominant somewhere in the complex $q$-plane, then by modifying the 
   top and bottom endgraphs (as shown in \protect\cite{Shrock_98c})
   it might be possible to make that eigenvalue
   (in either the reflection-even or reflection-odd sector or both)
   contribute to the chromatic polynomial 
   and thereby obtain a {\em different}\/ limiting curve ${\cal B}$
   for the different choice of endgraphs.
}  

The limiting curve ${\cal B}$ is connected
(see Figure~\ref{Figure_tri_8PxInftyF}). It crosses the real
axis at $q_0\approx 3.5147694243$.
It enters the half-plane $\real(q) < 0$, and there are four endpoints
\begin{subeqnarray}
q &\approx&           -0.3713655472 \pm 1.7983425919\,i\\ 
q &\approx& \phantom{-}4.0496984440 \pm 0.7359317819\,i 
\end{subeqnarray}
There are cusp-like structures around $q \approx 4.04 \pm 0.74\,i$.
A closer look shows that these structures are in fact T points located at 
$q \approx 4.0428606488 \pm 0.7417105390\,i$,
from which there emerge very short branches terminating
at the endpoints $q \approx 4.0496984440 \pm 0.7359317819\,i$.

We form a matrix $D(q)$ of dimension 13 rather than 15;
in this way we can avoid the two identically vanishing amplitudes.\footnote{
  If we were to form a 15-dimensional or 14-dimensional matrix $D(q)$,
  its determinant would be identically zero.
  This tells us (in case we did not know it already)
  that two of the amplitudes are identically zero.
}
Its determinant has the form   
\begin{eqnarray}
\det D(q) &=& {\rm const.} \times 
               q^{13} (q-1)^{13} (q-2)^{24} (q^2 - 3q + 1)^{10} 
              (q-3)^{191}  \nonumber \\  
          & & \quad \times (q^3 - 5q^2 + 6q -1)^4 (q^2 - 4q + 2) 
              (q^3 - 6 q^2 + 9q -1) \nonumber \\ 
          & & \quad \times (q^2 - 5q + 5)^{14}  (q^3 - 7q^2 + 14q -7)^2 
              (q-4)^{54}  \nonumber \\
          & & \quad \times 
  (12q^6 - 196q^5 +1355q^4 - 5126q^3 + 11337q^2 - 14086q +7755) \nonumber \\
          & & \quad \times P_1(q)^2
 \label{eq.detD.8P}
\end{eqnarray}
where the polynomial 
$P_1(q)$ can be found in the file {\tt transfer3.m}.
The first nine factors in \reff{eq.detD.8P}
are precisely the first minimal polynomials given
in \cite[Table 1]{transfer1};  therefore, $\det D(q)$ vanishes at the Beraha 
numbers $q=B_2,\ldots,B_{10}$. The next factor is the square of the polynomial 
$q^3 - 7q^2 + 14q -7$, which is $p_{14}(q)$ \cite[Table 1]{transfer1}, 
so that $\det D(q)$  also vanishes at $q=B_{14}$. Finally, the determinant
also vanishes at $q=4=B_\infty$.
Unlike what we have seen for cylindrical
strips of smaller width, in this case the remaining part of 
$\det D(q)$ is {\em not}\/ the square of a polynomial
with integer coefficients;  rather, there is the additional degree-6 factor
preceding $P_1(q)^2$.
The dominant amplitude vanishes only at the first seven Beraha
numbers $q=0,1,2,B_5,3,B_7,B_8$, so these values are the only isolated
limiting points for this strip.
The degree-6 factor and the polynomial $P_1(q)$ do not provide
additional isolated limiting points.

The first two zeros $q=0,1$ are trivial ones. At $q=2,3$, all amplitudes
vanish except for a few corresponding to 
zero eigenvalues;  we are thus in Case 3 of Section~\ref{sec_prelim}.
At $q=2$, there are 10 nonzero eigenvalues with zero amplitudes,
3 zero eigenvalues with zero amplitudes, and one zero eigenvalue with 
nonzero amplitude. At $q=3$ the transfer matrix is not diagonalizable:
there are 2 nonzero eigenvalues with zero amplitudes, one $2\times2$ 
nontrivial Jordan block corresponding to $\lambda=0$ with no contribution 
at all to the partition function for all $n$, and 10 zero eigenvalues 
with nonzero amplitudes.
The fourth, sixth and seventh real zeros converge exponentially fast to
$q=B_5,B_7,B_8$ (see Table~\ref{table_zeros_cyl}).
The eighth real zero seems to converge at an approximate $1/n$ rate to the
value $q_0 \approx 3.5147694243$.

Finally, we find vanishing subdominant amplitudes at $q=B_{10}, B_{14}$,
which are greater than $B_{m+1} = B_9$.

%
%
\subsection{$L_x = 9_{\rm P}$} \label{sec9P}

The transfer matrix $T(9_{\rm P})$ is 28-dimensional.
It can be brought into block-diagonal form
\be
       T(9_{\rm P}) \;=\; \left( \begin{array}{cc}
                                    T_+(9_{\rm P}) & 0 \\
                                    0              & T_-(9_{\rm P})
                                 \end{array} \right)
  \;\,,
\label{transfer_9P_diag}
\ee
where the block $T_+$ (resp.\ $T_-$)
is 22-dimensional (resp.\ 6-dimensional)
and corresponds to the subspace of reflection-invariant
(resp.\ reflection-odd) connectivities.
Moreover, all the amplitudes corresponding to the reflection-odd subspace
are identically vanishing;
this subspace therefore makes no contribution to the partition function.
The reduced transfer matrix $T_+(9_{\rm P})$
can be found in the {\sc Mathematica} file {\tt transfer3.m}.

Mysteriously, all of the eigenvalues in the reflection-odd subspace
have ``copies'' in the reflection-even subspace.
This can be seen by computing the characteristic polynomial
associated to the transfer matrix $T(9_{\rm P})$,
which can be factored as follows:
\be
\det[ T(9_{\rm P}) - \lambda {\bf 1} ] \;=\;
 Q_1(q,\lambda)^2 \, Q_2(q,\lambda)  
\label{char_tri_9P}
\ee
where $Q_1(q,\lambda)$ and $Q_2(q,\lambda)$ are polynomials
in $q$ and $\lambda$;
here $Q_1$ (resp.\ $Q_2$) is of degree 6 (resp. 16) in $\lambda$.
More specifically, $Q_1$ (resp.\ $Q_1 Q_2$)
is the characteristic polynomial of $T_-$ (resp.\ $T_+$);
the fact that $Q_1$ appears as a factor in the characteristic polynomial
of $T_+$ is direct proof of the just-mentioned ``copying'' of eigenvalues.
Even more mysteriously, our numerical checks suggest that
all the eigenvalues coming from $Q_1$ have
identically zero amplitudes --- not only in the reflection-odd subspace
(where this is well understood) but also in the reflection-invariant
subspace.
We thus find 6 pairs of equal eigenvalues
with identically vanishing amplitudes.

In order to be able to use the resultant method in this case, we proceed as
in the previous subsection: we drop the polynomial $Q_1(q,\lambda)^2$ (which 
contains the zero-amplitude eigenvalues) from the 
characteristic polynomial associated to $T(9_{\rm P})$ and compute 
the resultant with the polynomial $Q_2(q,\lambda)$ (which contains the 
eigenvalues with nonzero amplitudes). We have computed the points with 
$\theta=0$ with the resultant method; the points with other values of $\theta$ 
have been computed using the direct-search method.

The limiting curve ${\cal B}$ is connected
(see Figure~\ref{Figure_tri_9PxInftyF}). It crosses the real
axis at $q_0\approx 3.5270636990$.
It enters the half-plane $\real(q) < 0$, and there are four endpoints
\begin{subeqnarray}
q &\approx&           -0.4576020413 \pm 1.6238415411\,i\\ 
q &\approx& \phantom{-}4.2828643197 \pm 0.3823491910\,i
\end{subeqnarray}
There are T points located at 
$q \approx 4.0160853030 \pm  0.7870153859\,i$. 

We form a matrix $D(q)$ of dimension 16 rather than 28,
in order to avoid the 12 identically zero amplitudes.
Its determinant has the form
\begin{eqnarray}
\det D(q) &=& {\rm const} \times  
              q^{16} (q-1)^{16} (q-2)^{16} (q^2 - 3q + 1)^{13}
              (q-3)^{173} \nonumber \\
          & & \qquad \times (q^3 - 5q^2 + 6q -1)^5 (q^2 - 4q + 2)^4 
              (q^3 - 6q^2 + 9q -1)        \nonumber \\
          & & \qquad \times (q^2 - 5q + 5)^7 (q^3 - 7q^2 + 14q -7)^2 P(q)   \;,
\end{eqnarray}
where the polynomial
$P(q)$ can be found in the file {\tt transfer3.m}. We find the same
``Beraha factors'' as in the triangular-lattice strip of width $8_{\rm P}$ 
(Section~\ref{sec8P}). Thus, $\det D(q)$ vanishes at the Beraha
numbers $q=B_2,\ldots,B_{10},B_{14}$. The polynomial $P(q)$ is not the 
square of any polynomial with integer coefficients; rather it can be written
as $P_1(q) P_2(q)^2$ where $P_1$ and $P_2$ are integer-coefficient polynomials.
The dominant amplitude vanishes only at the first seven Beraha
numbers $q=0,1,2,B_5,3,B_7,B_8$;  thus, these values are the only isolated
limiting points for this strip. The polynomial $P(q)$ does not provide
additional isolated limiting points.

The first three zeros $q=0,1,2$ are trivial ones (i.e., all amplitudes
vanish identically).
The fourth, fifth, sixth and seventh real zeros converge exponentially fast to
$q=B_5,3,B_7,B_8$ (see Table~\ref{table_zeros_cyl_bis}).
The eighth real zero seems to converge at an approximate $1/n$ rate to the
value $q_0 \approx 3.5270636990$.

We again find a vanishing subdominant amplitude 
at $q=B_{14}$, which is greater than $B_{m+1} = B_{10}$.

%
%
\subsection{$L_x = 10_{\rm P}$} \label{sec10P}

The transfer matrix $T(10_{\rm P})$ is 67-dimensional.
It can be brought into block-diagonal form
\be
       T(10_{\rm P}) \;=\; \left( \begin{array}{cc}
                                    T_+(10_{\rm P}) & 0 \\
                                    0              & T_-(10_{\rm P})
                                 \end{array} \right)
  \;\,,
\label{transfer_10P_diag}
\ee
where the block $T_+$ (resp.\ $T_-$)
is 51-dimensional (resp.\ 16-dimensional)
and corresponds to the subspace of reflection-invariant
(resp.\ reflection-odd) connectivities.
All the amplitudes corresponding to the reflection-odd subspace
are identically vanishing;
this subspace therefore makes no contribution to the partition function.
The reduced transfer matrix $T_+(10_{\rm P})$
can be found in the {\sc Mathematica} file {\tt transfer3.m}.

The characteristic polynomial of the transfer matrix $T(10_{\rm P})$
obviously factors as
$\det[ T(10_{\rm P}) - \lambda {\bf 1} ] = Q_1(q,\lambda)\, Q(q,\lambda)$,
where $Q_1$ (resp.\ $Q$)
is the characteristic polynomial of $T_-$ (resp.\ $T_+$).
Numerically we have found, once again,
that all the eigenvalues in the reflection-odd subspace
have ``copies'' in the reflection-even subspace.
Therefore, the polynomial $Q(q,\lambda)$ should have $Q_1(q,\lambda)$
as a factor, so that
\be
\det[ T(10_{\rm P}) - \lambda {\bf 1} ] \;=\;
   Q_1(q,\lambda)^2 \, Q_2(q,\lambda)
\label{char_tri_10P}
\ee
where $Q_2(q,\lambda)$ is a polynomial of degree 35 in $\lambda$.
Unfortunately, we have been unable to compute the characteristic polynomial 
$Q(q,\lambda)$ of the reduced transfer matrix $T_+(10_{\rm P})$
and verify the conjectured factorization \reff{char_tri_10P}.

Once again, we have found numerically that the ``copied'' eigenvalues
have identically vanishing amplitudes.
We thus find 16 pairs of equal eigenvalues
with identically vanishing amplitudes.

We have used the direct-search method in the computation of ${\cal B}$.
The limiting curve ${\cal B}$ is connected
(see Figure~\ref{Figure_tri_10PxInftyF}). It crosses the real
axis at $q_0\approx 3.6348299654$.
It enters the half-plane $\real(q) < 0$, and there are four endpoints
\begin{subeqnarray}
q &\approx&           -0.510807 \pm 1.481233 \,i\\
q &\approx& \phantom{-}4.113231 \pm 0.492835 \,i
\end{subeqnarray}
There are T points located at $q \approx 4.0632619066 \pm 0.8803786140\,i$.

We have numerically checked that the dominant amplitude vanishes at the first 
nine Beraha numbers $q=0,1,2,B_5,3,B_7,B_8,B_9, B_{10}$
(and at no others);
therefore, these values are the only isolated limiting points for this strip.
We have found no evidence of complex isolated limiting points from the
zeros of the finite-length strips.  
 
The first two zeros $q=0,1$ are trivial ones. 
At $q=2,3$ we are in Case 3 described in Section~\ref{sec_prelim}.
At $q=2$, there are 38
nonzero eigenvalues with zero amplitudes, 12 zero eigenvalues with 
zero amplitudes and one zero eigenvalue with a nonzero amplitude. 
At $q=3$ the transfer matrix is not diagonalizable: we find 4 nonzero
eigenvalues with zero amplitudes, one $2\times 2$ nontrivial Jordan block 
corresponding to 
$\lambda=-3$ which does not contribute to the partition function for any $n$,
40 zero eigenvalues with nonzero amplitudes, and two nontrivial Jordan blocks
(of dimensions 3 and 2 respectively) corresponding to $\lambda =0$. 
The contribution of these later blocks is zero except for $n=1$.
The fourth, sixth, seventh, eight and ninth real zeros
converge exponentially fast to
$q=B_5, B_7, B_8, B_9, B_{10}$
(see Table~\ref{table_zeros_cyl_bis}). 
The tenth real zero seems to converge at an approximate $1/n$ rate to the
value $q_0 \approx 3.6348299654$.

Finally, we have also checked that there are vanishing amplitudes (in addition 
to the trivial 32 identically zero amplitudes) for $q=B_{11}$, $B_{14}$, and
$B_{18}$. In all these cases, the vanishing amplitude corresponds to 
a subdominant eigenvalue; thus, none of these points is an isolated limiting 
point. Please note that the last two values (namely, $B_{14}$ and 
$B_{18}$) are greater than $B_{m+1}=B_{11}$.

%
%
\subsection{$L_x = 11_{\rm P}$} \label{sec11P}

The transfer matrix $T(11_{\rm P})$ is 145-dimensional.
It can be brought into block-diagonal form
\be
       T(11_{\rm P}) \;=\; \left( \begin{array}{cc}
                                    T_+(11_{\rm P}) & 0 \\
                                    0              & T_-(11_{\rm P})
                                 \end{array} \right)
  \;\,,
\label{transfer_11P_diag}
\ee
where the block $T_+$ (resp.\ $T_-$)
is 95-dimensional (resp.\ 50-dimensional)
and corresponds to the subspace of reflection-invariant
(resp.\ reflection-odd) connectivities.
All the amplitudes corresponding to the reflection-odd subspace
are identically vanishing;
this subspace therefore makes no contribution to the partition function.
The reduced transfer matrix $T_+(11_{\rm P})$
can be found in the {\sc Mathematica} file {\tt transfer3.m}.

The characteristic polynomial of the transfer matrix $T(11_{\rm P})$
obviously factors as
$\det[ T(11_{\rm P}) - \lambda {\bf 1} ] = Q_1(q,\lambda)\, Q(q,\lambda)$,
where $Q_1$ (resp.\ $Q$)
is the characteristic polynomial of $T_-$ (resp.\ $T_+$).
Numerically we have found, once again,
that all the eigenvalues in the reflection-odd subspace
have ``copies'' in the reflection-even subspace.
Therefore, the polynomial $Q(q,\lambda)$ should have $Q_1(q,\lambda)$
as a factor, so that
\be
\det[ T(11_{\rm P}) - \lambda {\bf 1} ] \;=\;
   Q_1(q,\lambda)^2 \, Q_2(q,\lambda)
\label{char_tri_11P}
\ee
where $Q_2(q,\lambda)$ is a polynomial of degree 45 in $\lambda$.
Unfortunately, we have been unable to compute the characteristic polynomial
$Q(q,\lambda)$ of the reduced transfer matrix $T_+(11_{\rm P})$
and verify the conjectured factorization \reff{char_tri_11P}.

Once again, we have found numerically that the ``copied'' eigenvalues
have identically vanishing amplitudes.
We thus find 50 pairs of equal eigenvalues
with identically vanishing amplitudes.

As in the preceding subsection, we used the direct-search method in 
the computation of ${\cal B}$. This curve crosses the real $q$-axis at
$q_0\approx 3.6441399017$ (see Figure~\ref{Figure_tri_11PxInftyF}).
It enters the half-plane $\real(q) < 0$, and there are four endpoints
\begin{subeqnarray}
q &\approx&           -0.543988 \pm 1.363241 \,i\\
q &\approx& \phantom{-}4.156093 \pm 0.529420 \,i
\end{subeqnarray}
There are T points located at
$q \approx 4.0425923021 \pm 0.6927608569\,i$.

We have numerically checked that the dominant amplitude vanishes at the first
nine Beraha numbers $q=0,1,2,B_5,3,B_7,B_8,B_9,B_{10}$ (and at no others),
so that these values are the only real isolated limiting points for this strip.
We inspected the zeros of the finite-length strips and
found no evidence of complex isolated limiting points.

The first three zeros $q=0,1,2$ are trivial ones, as all amplitude vanish. 
At $q=3$ the transfer matrix is not diagonalizable. There are four nontrivial
Jordan blocks corresponding to four nonzero eigenvalues. Furthermore, the 
contribution of these Jordan blocks to the partition function vanishes for
all $n\geq 1$. We also find 8 nonzero eigenvalues with zero amplitudes,  
30 zero eigenvalues with nonzero amplitudes, and 30 
zero eigenvalues with zero amplitudes.
The fourth, sixth, seventh, eighth and ninth real zeros
converge exponentially fast to $q=B_5,B_7, B_8, B_9, B_{10}$
(see Table~\ref{table_zeros_cyl_bis}).
The tenth real zero seems to converge at an approximate $1/n$ rate to the
value $q_0 \approx 3.6441399017$.

Finally, we have also checked that there are vanishing amplitudes (in addition
to the trivial 100 identically zero amplitudes) for
$q=B_{11}, B_{12}, B_{14}, B_{18}$.
In all these cases, the vanishing amplitude is subdominant;
thus, none of these points is an isolated limiting point.
Again, the values $B_{14}$ and $B_{18}$ 
are greater than $B_{m+1} = B_{12}$.

%
%
\subsection{$L_x = 12_{\rm P}$} \label{sec12P}

The transfer matrix $T(12_{\rm P})$ is 368-dimensional.
It can be brought into block-diagonal form
\be
       T(12_{\rm P}) \;=\; \left( \begin{array}{cc}
                                    T_+(12_{\rm P}) & 0 \\
                                    0              & T_-(12_{\rm P})
                                 \end{array} \right)
  \;\,,
\label{transfer_12P_diag}
\ee
where the block $T_+$ (resp.\ $T_-$)
is 232-dimensional (resp.\ 136-dimensional)
and corresponds to the subspace of reflection-invariant
(resp.\ reflection-odd) connectivities.
All the amplitudes corresponding to the reflection-odd subspace
are identically vanishing;
this subspace therefore makes no contribution to the partition function.
The reduced transfer matrix $T_+(12_{\rm P})$
can be found in the {\sc Mathematica} file {\tt transfer3.m}.

The characteristic polynomial of the transfer matrix $T(12_{\rm P})$
obviously factors as
$\det[ T(12_{\rm P}) - \lambda {\bf 1} ] = Q_1(q,\lambda)\, Q(q,\lambda)$,
where $Q_1$ (resp.\ $Q$)
is the characteristic polynomial of $T_-$ (resp.\ $T_+$).
Numerically we have found, once again,
that all the eigenvalues in the reflection-odd subspace
have ``copies'' in the reflection-even subspace.
Therefore, the polynomial $Q(q,\lambda)$ should have $Q_1(q,\lambda)$
as a factor, so that
\be
\det[ T(12_{\rm P}) - \lambda {\bf 1} ] \;=\;
   Q_1(q,\lambda)^2 \, Q_2(q,\lambda)
\label{char_tri_12P}
\ee
where $Q_2(q,\lambda)$ is a polynomial of degree 96 in $\lambda$.
Unfortunately, we have been unable to compute the characteristic polynomial
$Q(q,\lambda)$ of the reduced transfer matrix $T_+(12_{\rm P})$
and verify the conjectured factorization \reff{char_tri_12P}.

Once again, we have found numerically that the ``copied'' eigenvalues
have identically vanishing amplitudes.
We thus find 136 pairs of equal eigenvalues
with identically vanishing amplitudes.

Due to the large dimension of the transfer matrix, we have been unable to
compute the limiting curve. However, we have managed using the direct-search
method to compute the point where ${\cal B}$ crosses the real $q$-axis:
$q_0\approx 3.6431658979$. 
We have also computed the position of the pair of complex-conjugate T points 
that are obvious in Figure~\ref{Figure_tri_11PxInftyF}; the result is 
$q \approx 4.05713658 \pm 0.73432479\,i$. 

We have numerically checked that the dominant amplitude vanishes at the first
nine Beraha numbers $q=0,1,2,B_5,3,B_7,B_8,B_9$, and $B_{10}$.
These values are the only isolated limiting points for this strip. We
have found no evidence of complex isolated limiting points by inspecting the
zeros of the finite-length strips. 

The first two zeros $q=0,1$ are trivial ones. At $q=2$ there are 164 
nonzero eigenvalues with zero amplitudes, 2 zero eigenvalues with 
nonzero amplitudes, and 66 zero eigenvalues with zero amplitudes. 
The convergence to $q=B_5,3,B_7$, $B_8$, $B_9$, $B_{10}$ is exponentially fast
(see Table~\ref{table_zeros_cyl_bis}).
The tenth real zero seems to converge at an approximate $1/n$ rate to the
value $q_0 \approx 3.6431658979$. 

Finally, we have also checked that there are vanishing amplitudes (in addition
to the trivial 272 identically zero amplitudes) for
$q=B_{11}, B_{12}, B_{13}, B_{14}, B_{18}, B_{22}$.
In all these cases, the vanishing amplitude is subdominant;
thus, none of these points is an isolated limiting point.

In this case we find three vanishing subdominant amplitudes 
($B_{14}$, $B_{18}$, and $B_{22}$) that are greater than the value 
$B_{m+1} = B_{13}$.

%
%
\section{Numerical Results for the Triangular-Lattice
   \hfill\break Chromatic Polynomial:
   \hfill\break Zig--Zag Boundary Conditions}
   \label{sec_Z}

Until now, we have built up the triangular lattice by transferring
along a direction that is perpendicular to one of the principal
directions of the lattice. An alternative choice, of course, would be
to transfer along a direction that is {\em parallel}\/ to a principal
direction. When periodic boundary conditions are imposed
across the strip, these two constructions are inequivalent:
they yield different finite graphs. We shall
refer to this alternative construction, with periodic boundary
conditions in the transversal direction and free boundary conditions
in the longitudinal direction, as ``zig--zag'' boundary conditions,
and it will be denoted by the subscript Z.
Note that the lattice width $m$ must be {\em even}\/.

For zig--zag boundary conditions,
the transfer matrix is not given by the formulae
of Ref.~\cite{transfer1}.
Rather, as is evident from Figure~\ref{Figure_transfer}(b),
the transfer matrix now takes the following form
\be
\T(m_{\rm Z}) = \H \V_{\rm even} \H \V_{\rm odd}   \;,
\label{def_tm_zigzag}
\ee
where $\V_{\rm even}$ (resp.~$\V_{\rm odd}$) is the product of the 
matrices associated to the vertical bonds located at even (resp.~odd) sites
of the lattice. With this definition, all the formulae applied in the 
previous sections hold.  

Our original motivation for introducing this new construction
was the following:
It is clear from Sections~\ref{sec_F} and \ref{sec_P} that the limiting
curves for the strips with free and cylindrical boundary conditions
differ qualitatively by the existence of a small additional inward-pointing
branch for the case of free b.c., which is absent for cylindrical b.c.
In the limit of infinite width, one might wonder whether this branch
extends to $q=2$, as the
triangular-lattice Ising model is known to have a zero-temperature
critical point \cite{Stephenson_64}.
We found it interesting to examine whether we can
recover such a branch by imposing zig--zag boundary conditions.
The answer turns out to be negative;
but it seems to us that zig--zag b.c.\ are interesting in their own right,
irrespective of this initial motivation.

We have computed the transfer matrix $T(m_{\rm Z})$
and the limiting curves $\scrb$
for triangular-lattice strips of even widths $2 \leq m \equiv L_x \leq 10$. 
It is interesting to note that the trick discussed in
\cite[Sections 3.1 and 3.3]{transfer1} 
for the standard construction of a cylindrical triangular-lattice strip
is not necessary here. 
On the other hand, the dimension of the transfer matrix $T(m_{\rm Z})$
is in general different from TriCyl($m$),
because the invariances are different:
for zig-zag boundary conditions,
the system is invariant under translations of {\em even}\/ (but not odd)
length and under reflections.  

%
%
\subsection{$L_x = 2_{\rm Z}$} \label{sec2Z}

This case is trivial, as the transfer matrix is one-dimensional:
\be
Z(2_{\rm Z} \times n_{\rm F}) = q(q-1) (q-2)^{2(n-1)}
\ee
Please note that the strip $2_{\rm Z} \times n_{\rm F}$ is
equivalent to $2_{\rm F} \times n_{\rm F}$.
Since there is only one eigenvalue, there is obviously no crossing, hence
${\cal B} = \emptyset$. However, there are zeros for all $n$ at
$q=0,1$ and for $n\geq 2$ at $q=2$.

%
%
\subsection{$L_x = 4_{\rm Z}$} \label{sec4Z}

The transfer matrix is three-dimensional. In the basis
${\bf P} = \{ 1, \delta_{13},\delta_{24} \}$ it can be written as
\be
T(4_{\rm Z}) \;=\; \left( \begin{array}{ccc}
   T_{11}     &  T_{12}                & T_{13} \\ 
   2q -5      & q^2 - 4q + 4           & 1 \\
   T_{31}     & q^2 - 4q + 4           & 3q^2 -17q + 25 
              \end{array} \right)
\ee
where 
\begin{subeqnarray}
T_{11}  &=& q^4 - 12q^3 + 56q^2 - 121q + 101 \\
T_{12}  &=&         q^3 -  8q^2 +  20q -  16\\
T_{13}  &=&         q^3 - 10q^2 +  34q -  40\\
T_{31}  &=&        2q^3 - 17q^2 +  50q -  50
\end{subeqnarray}
and the partition function is equal to
\be
Z( 4_{\rm P} \times n_{\rm Z} ) \;=\;
      q(q-1) \left( \begin{array}{c}
                        q^2 - 3q + 3 \\
                        q-1 \\
                        q-1
                    \end{array}\right)^{\!\rm T} \cdot
                  T(4_{\rm Z})^{n-1} \cdot
             \left( \begin{array}{c}  1 \\ 0 \\ 0 \end{array} \right)
\ee

The limiting curve ${\cal B}$ (see Figure~\ref{Figure_tri_4PxInftyZ}) contains
two complex-conjugate arcs, which do not cross the real $q$-axis.
There are four endpoints:
\begin{subeqnarray}
q &\approx& 2.0991442518 \pm 2.5589234827\,i\\ 
q &\approx& 2.7371672817 \pm 0.1723332852\,i 
\end{subeqnarray}
We have found a complex-conjugate pair of double zeros of the resultant 
for $\theta=0$ (see \cite[Section~4.1.1]{transfer1}) at 
$q \approx 3.7718445063 \pm 1.1151425080\,i$. At these values the 
limiting curve appears at first glance to be singular
(see Figure~\ref{Figure_tri_4PxInftyZ}).
However, a closer look reveals that this is {\em not}\/ the case;
in fact, the limiting curve is perfectly analytic
around these two points.\footnote{
  In the notation of \protect\cite[Section 4.2]{transfer1}, the 
  characteristic polynomial of $T(4_{\rm Z})$ can be expanded as 
$$
P(\lambda,q) \;=\; c(\lambda-\lambda_0)^2 + d (q-q_0)^2 + 
               e (q-q_0)(\lambda-\lambda_0) + \ldots 
$$
  around the points $q_0 \approx 3.7718445063 \pm 1.1151425080\,i$ and 
  the dominant (double) eigenvalue 
  $\lambda_0 = -0.5237532362 + 4.5580089825\,i$
  (the linear terms in $\lambda-\lambda_0$ and $q-q_0$ vanish).
  This expansion leads to
  analytic eigenvalues $\lambda_\pm(q)$ around $q=q_0$ and 
  to an analytic equimodular locus around $q=q_0$.
} 

The determinant $\det D(q)$ has the form
\begin{eqnarray}
\det D(q) &=& - q^3 (q-1)^3 (q-2)^2 (q^2 - 3q + 1) (q-3)^2 (2q -5)^4
          \nonumber \\
          & & \qquad \times (q^3 -10q^2 + 34q - 38)^2 
\end{eqnarray}
We recognize the first five minimal polynomials $p_k(q)$ given in
\cite[Table 1]{transfer1}. Hence, the determinant vanishes at the
first five Beraha numbers $q=0,1,2,B_5,3$. The dominant amplitude vanishes 
at all these points except at $q=3$; therefore, $q=0,1,2,B_5$ are 
isolated limiting points.
It is interesting to note that the dominant amplitude vanishes also at 
$q=5/2$, so that this too is an isolated limiting point. This is the first 
time we have found a real isolated limiting point that is not a Beraha number.
(For the square lattice, we did not find any non-Beraha real 
isolated limiting point \cite{transfer1,transfer2}.)
There are no additional isolated limiting points coming from the last
factor $q^3 -10q^2 + 34q - 38$,
as the vanishing amplitudes in question all correspond to
subdominant eigenvalues.

The first two real zeros $q=0,1$ are trivial as all the amplitudes vanish. 
The third real zero $q=2$ belongs to Case 3 of Section~\ref{sec_prelim}: 
the two nonzero eigenvalues have zero amplitude, and there is an 
additional zero eigenvalue with a nonzero amplitude. Finally, the 
fourth and fifth real zeros converge exponentially fast to the 
values $5/2$ and $B_5$, respectively.  
In summary, we find five isolated limiting points $q=0,1,2,5/2$ and $B_5$. 
This is in agreement with Table~\ref{table_zeros_zig}.

Please note that there is a vanishing subdominant amplitude at
$q=B_{6}=3$. This value is greater than $B_{m+1}=B_5$, in contrast with the 
observed behavior for the square lattice \cite{transfer1,transfer2}. 

%
%
\subsection{$L_x = 6_{\rm Z}$} \label{sec6Z}

The transfer matrix is seven-dimensional; it can be found in the
{\sc Mathematica} file {\tt transfer3.m}. 

The limiting curve ${\cal B}$ is connected
(see Figure~\ref{Figure_tri_6PxInftyZ}). It crosses the real
axis at $q\approx 3.1752579126$. There are four endpoints:
\begin{subeqnarray}
q &\approx& 0.3618461880 \pm 2.5093731708\,i\\
q &\approx& 4.2589504182 \pm 0.7015734543\,i
\end{subeqnarray}
There are T points at $q \approx 3.8395346820 \pm 1.1149959335\,i$.

The determinant $\det D(q)$ has the form
\be
\det D(q) = 81 q^7 (q-1)^7 (q-2)^{18} (q^2 - 3 q + 1)^4
            (q-3)^{44} (q^3 -5q^2 + 6q -1) P(q)^2
\ee 
where the polynomial $P(q)$ can be found in the file {\tt transfer3.m}.
The first six polynomial are the first six minimal polynomials given
in \cite[Table 1]{transfer1};  hence the determinant vanishes at the
first six Beraha numbers $q=0,1,2,B_5,3,B_7$. 
The dominant amplitude vanishes at the first five Beraha
numbers $q=0,1,2,B_5,3$ as well as at three of the zeros of $P(q)$, namely
$q\approx 2.7226328355$ and $q\approx 3.6696077451 \pm 0.9506864736\,i$. 
This is the first triangular-lattice strip where we find complex 
isolated limiting points. In the square-lattice case, complex isolated
limiting points were quite common: we found such limiting points for 
$L \geq 6$ with both free and cylindrical boundary conditions 
\cite{transfer1,transfer2}. 

The first two real zeros $q=0,1$ are trivial ones. The third real zero
$q=2$ falls in Case 3 of Section~\ref{sec_prelim}: there are four nonzero 
eigenvalues with zero amplitudes, two zero eigenvalues with nonzero
amplitudes, and one zero eigenvalue with zero amplitude. 
The convergence of the fourth, fifth and sixth real zeros to their 
corresponding limiting values (namely, $B_5,2.7226328355$, and $3$) 
is exponentially fast, as shown in Table~\ref{table_zeros_zig}. 
Finally, the seventh real zero converges at an approximate $1/n$ rate 
to the value $q\approx 3.1752579126$. 

In summary, we find six real isolated limiting points at
$q=0,1,2,B_5,2.7226328355,3$ and $B_7$ (see Table~\ref{table_zeros_zig})
and two complex isolated limiting points at 
$q\approx 3.6696077451 \pm 0.9506864736\,i$. 
However, since the complex isolated limiting points are very near ${\cal B}$,
it is very difficult to observe the convergence to them
as distinct from the convergence to ${\cal B}$
(see Figure~\ref{Figure_tri_6PxInftyZ}).

%
%
\subsection{$L_x = 8_{\rm Z}$} \label{sec8Z}

The transfer matrix is 24-dimensional; it can be found in the
{\sc Mathematica} file {\tt transfer3.m}.

The limiting curve ${\cal B}$ is connected
(see Figure~\ref{Figure_tri_8PxInftyZ}). It crosses the real
axis at $q\approx 3.3941047539$. There are four endpoints:
\begin{subeqnarray}
q &\approx&           -0.2143469947 \pm 2.0301412598\,i \\ 
q &\approx& \phantom{-}4.2899063418 \pm 0.5046183096\,i
\end{subeqnarray}
There are T points at $q \approx 4.0055796610 \pm 0.8830638824\,i$.

In this case we were unable to compute the determinant $\det D(q)$. However,
we checked numerically whether any of the amplitudes vanishes at the Beraha 
numbers $B_n$, and if this occurs, whether the vanishing amplitude is the 
leading one or not. We have made this check up to $n=50$.  
We have found that at least one amplitude vanishes at
the Beraha numbers $q=0,1,2,B_5,3,B_7,B_8,B_9$. The dominant amplitude
vanishes only at the first six (namely, $q=0,1,2,B_5,3,B_7$),
so that these latter numbers are isolated limiting points 
(see Table~\ref{table_zeros_zig}). In Table~\ref{table_zeros_zig} we also
notice an additional isolated zero at $q\approx 2.8214204955$. 
We have numerically confirmed that this point is indeed an isolated 
limiting point by minimizing the absolute value of the dominant amplitude
$\alpha^\star$ in a neighborhood of that point.

The first two real zeros $q=0,1$ are trivial ones. At $q=2$, there are 13
nonzero eigenvalues with zero amplitudes, 2 zero eigenvalues with nonzero
amplitudes, and 9 zero eigenvalues with zero amplitudes.
Finally, the other real zeros
in Table~\ref{table_zeros_zig} converge exponentially fast to their
corresponding limiting values (namely, $B_5,2.8214204955,3$, and $B_7$).  
We expect, for sufficiently larger lengths $n$,
an additional real zero larger than $B_7$ and converging to 
$q \approx 3.3941047539$;  but we apparently need to go beyond $n=80$
to see it.
In summary, there are seven real isolated limiting points at
$q=0$, $1$, $2$, $B_5$, $2.8214204955$, $3$, and $B_7$. 

By minimizing the absolute value of the dominant amplitude $\alpha^\star$, 
we have found a pair of complex-conjugate isolated limiting points 
at $q\approx 3.8327415674 \pm 0.73050211595\,i$ 
(See Figure~\ref{Figure_tri_8PxInftyZ}). Again, we are not sure that we have
found all complex isolated limiting points for this strip. 

%
%
\subsection{$L_x = 10_{\rm Z}$} \label{sec10Z}

The transfer matrix is 87-dimensional; it can be found in the
{\sc Mathematica} file {\tt transfer3.m}.

In this case we were unable to compute the limiting curve.
However, we were able 
to compute the value of $q$ where that curve crosses the real axis:
$q \approx 3.5204366907$. 

The matrix $D(q)$ is too large for us to compute its determinant.  
Instead, we have checked numerically
the eigenvalues and amplitudes of the transfer
matrix at the Beraha numbers $B_n$ with $2\leq n \leq 50$.  
We have found that at least one amplitude vanishes at
the Beraha numbers $q=0,1,2,B_5,3,B_7,B_8,B_9,B_{10},B_{11}$. 
The dominant amplitude vanishes only at the first seven 
(namely, $q=0,1,2,B_5,3,B_7,B_8$), so that these latter numbers are 
isolated limiting points (see Table~\ref{table_zeros_zig}). 
In Table~\ref{table_zeros_zig} we also
notice two additional isolated zeros at $q\approx 2.8737312493$ and 
$q\approx 3.3831285312$. We have numerically checked that in both cases
the leading amplitudes vanish, so they too are isolated limiting points. 

The first two real zeros $q=0,1$ are trivial. At $q=2$, there are 35 nonzero
eigenvalues with zero amplitudes, 31 zero eigenvalues with zero amplitudes,
and 21 zero eigenvalues with nonzero amplitudes. 
The convergence of the fourth through ninth real zeros
to their corresponding limiting values 
(namely, $B_5,2.8737312493,3,B_7,3.3831285312,B_8$) is exponentially
fast as shown in Table~\ref{table_zeros_zig}. Finally, the tenth real zero
converges at an approximate $1/n$ rate to the value
$q\approx 3.5204366907$.

In summary, there are nine real isolated limiting points at
$q=0$, $1$, $2$, $B_5$, $2.8737312493$, $3$, $B_7$, $3.3831285312$, and $B_8$. 
We have been unable to say whether or not there are any
complex isolated limiting points;  but we do not see 
any candidate in Figure~\ref{Figure_tri_10PxInftyZ}.  

%
%
\section{Thermodynamic Limit} \label{secBaxter}

In this section we will review the Bethe-Ansatz solution found by Baxter 
\cite{Baxter_86,Baxter_87} for the thermodynamic limit of the zero-temperature 
triangular-lattice Potts antiferromagnet (Section~\ref{sec6.1}),
and carefully recalculate Baxter's predictions
for the limiting curve $\scrb_\infty$
where the chromatic roots are expected to accumulate (Section~\ref{sec6.2}).
The resulting picture will be substantially similar to that
set forth by Baxter \cite{Baxter_87},
but with a few important qualitative differences.
Next we will compare Baxter's predictions for the dominant eigenvalues
with our finite-lattice data (Section~\ref{sec6.3}),
and comment on the agreements and discrepancies (Section~\ref{sec6.4}).

\subsection{Baxter's solution} \label{sec6.1}

In terms of the variables $x$ and $\theta$ defined by
\be
q   \;=\;  2 - x - x^{-1}  \;=\; 2 + 2 \cos \theta
  \label{def_theta}
\ee
with $|x| < 1$ and $0 < \real \theta < \pi$,
Baxter defined three functions (eigenvalues) $g_i(q)$ as follows:
\begin{subeqnarray}
g_1(q) &=& -{1\over x}\prod\limits_{j=1}^\infty 
            {(1-x^{6j-3})(1-x^{6j-2})^2(1-x^{6j-1}) \over 
             (1-x^{6j-5})(1-x^{6j-4})(1-x^{6j})(1-x^{6j+1}) }
\slabel{def_g1} \\[4mm]
\log g_2(q) &=& \int_{-\infty}^\infty {\rm d}k \, {\sinh k\theta \over 2k} 
           \left( {\sinh[ k(\pi-2\theta)/2] \over 
                  (2\cosh k\theta -1)\sinh(\pi k/2) } \right. \nonumber \\
\slabel{def_g2}
            & & \qquad\qquad\qquad\qquad\qquad \left. - 
                  {\cosh[ k(\pi-2\theta)/2] \over
                  (2\cosh k\theta +1)\cosh(\pi k/2) } \right) \\[4mm]
\log g_3(q) &=& \int_{-\infty}^\infty {\rm d}k \, 
                {\sinh k\theta \, [\sinh k(\pi-\theta)] \over 
                 k \sinh \pi k \, [2\cosh k(\pi-\theta)-1] }
\slabel{def_g3}
\label{def_gi}
\end{subeqnarray}
These formulae were obtained in \cite{Baxter_86}, but the corresponding 
ranges of validity were established only in \cite{Baxter_87}. 
In particular, Baxter found that the complex $q$-plane can be divided
into three domains ${\cal D}_i$ [$i=1,2,3$], in each of which the dominant
eigenvalue is $g_i$.
According to Baxter \cite{Baxter_87},
the intersections of these regions with the real axis are as follows:
\begin{subeqnarray}
{\cal D}_1 \cap \R & = & \{q> 4\} \cup \{q<0\}  \\  
{\cal D}_2 \cap \R & = & \{q_0<q<4\}   \\  
{\cal D}_3 \cap \R & = & \{0<q<q_0\}
  \label{baxter_D123}
\end{subeqnarray}
Baxter therefore determined the parameter $q_0$ by solving the equation   
\be 
g_2(q_0) = g_3(q_0) 
\label{def_q0_integrals}
\ee
via Newton's method and numerical integration of \reff{def_g2}/\reff{def_g3},
and found $q_0 \approx 3.81967$. 
We have refined this computation using the same method, and find
\be
q_0(\hbox{Baxter})  \,\approx\, 3.819671731239719 \;.
\label{def_q0}
\ee 
This point is labelled F in
Figures~\ref{Figure_Baxter_t} and \ref{Figure_Baxter_q_zoom}
($\theta_{\rm F} \approx 0.427907971348122$).
We have also plotted the eigenvalues $g_2$ and $g_3$
over the entire range $0 < \theta < \pi$
and verified that there is only one crossing point, namely \reff{def_q0}.
However, as we will argue later,
the intersections ${\cal D}_i \cap \R$ are in fact more complicated
than what is claimed in \reff{baxter_D123},
so that \reff{def_q0} is {\em not}\/ in fact the correct value of $q_0$
{\em if Baxter's three eigenvalues are in fact the dominant ones}\/.
(See Sections~\ref{sec6.3} and \ref{sec6.4} for further discussion.)

In order to obtain the limiting curve $\scrb_\infty$ in the complex $q$-plane,
Baxter \cite{Baxter_87} took advantage of the following simpler expressions
for the ratios of eigenvalues:
\begin{subeqnarray}
{g_2(q) \over g_1(q) } &=& \prod\limits_{j=1}^\infty \left( 
              {1-\omega p^{2j-1} \over 1-\omega p^{2j} } \right)^3 
              {1-p^{6j} \over 1 - p^{6j-3} } 
\slabel{def_g2_over_g1} \\[2mm]
{g_3(q) \over g_1(q) } &=& \prod\limits_{j=1}^\infty \left( 
              {1+\omega^2 y^{j} \over 1-\omega^2 y^{j} } \right)^3 
              {1-y^{3j} \over 1 + y^{3j} } 
\slabel{def_g3_over_g1}
\label{def_ratios_g}
\end{subeqnarray}  
where 
\begin{subeqnarray}
 p &=& - e^{i \pi^2 /3\theta } \slabel{def_p} \\ 
 y &=&   e^{-2 i \pi^2 /3(\pi-\theta) } \slabel{def_y} \\ 
 \omega &=&   e^{2\pi i/3} \slabel{def_omega} 
\end{subeqnarray}
In these equations we require that $|p|<1$ and $|y|<1$ so that the products 
converge;  this corresponds to $\imag \theta < 0$.

As Baxter \cite{Baxter_87} noted, the parameter $\theta$ enters
into the products (\ref{def_ratios_g}a,b) only via $p$ or $y$, respectively,
and these two variables are invariant under the transformations
\begin{subeqnarray}
{\pi \over \theta} &\rightarrow& {\pi \over \theta} + 6 k \qquad
 \quad\;\,\Longrightarrow\qquad
 p \rightarrow p
\slabel{def_symmetries_ratio21} \\[4mm]
 {\pi \over \pi-\theta} &\rightarrow& {\pi \over \pi-\theta} + 3 k' \;\,
 \quad\Longrightarrow\qquad
 y \rightarrow y
\slabel{def_symmetries_ratio31}
\label{def_symmetries_ratios}
\end{subeqnarray}
for any integers $k,k'$.
Thus, each solution of $|g_2/g_1| = 1$ in the complex $p$-plane
corresponds to an infinite family of solutions in the complex $\theta$-plane:
these can be thought of as a ``primary'' solution
(namely, the one with largest $\real\theta$ contained in the
 physical region $0 < \real\theta < \pi$)
and its ``images'' under the transformation \reff{def_symmetries_ratio21}
with $k \ge 1$.
As $k \to \infty$ these ``image'' curves converge to $\theta = 0$ ($q=4$).
Likewise, each solution of $|g_3/g_1| = 1$ in the complex $y$-plane
corresponds to an infinite family of solutions in the complex $\theta$-plane:
a ``primary'' solution
(the one with smallest $\real\theta$ contained in the physical region)
and its ``images'' under the transformation \reff{def_symmetries_ratio31}
with $k' \ge 1$.
As $k' \to \infty$ these ``image'' curves converge to $\theta = \pi$ ($q=0$).
It is important to note that the transformations \reff{def_symmetries_ratios}
do {\em not} tell anything about the dominant or subdominant character of 
the equimodular curve at the transformed value of $\theta$;
this property has to be checked by other means.

Let us emphasize that the symmetries \reff{def_symmetries_ratios}
play no essential logical role in our analysis:
one can, in principle, discover all the equimodular curves by direct search,
without any reference to these symmetries.
But it is {\em useful}\/ to know, once one has discovered one equimodular
curve, that in certain circumstances there must exist others as well
(and to know their exact location).

{\bf Important Remark.}
Neither the eigenvalues (\ref{def_gi}b,c)
nor the eigenvalue ratios (\ref{def_ratios_g}a,b)
or $g_3/g_1 =$ (\ref{def_ratios_g}a)/(\ref{def_ratios_g}b)
--- nor even their absolute values ---
are invariant under the transformation $\theta \to \theta + 2\pi k''$
that leaves $q$ invariant.
Therefore, we {\em must}\/ require $0 < \real\theta < \pi$
when using these formulae.

\bigskip 

\noindent
{\bf Remarks.}
1.  The formula $q = 2 - x - x^{-1}$ maps the disc $|x| < 1$
one-to-one onto the $q$-plane cut along the interval $[0,4]$.
Therefore (as Baxter observed \cite{Baxter_87}),
when $q$ is real and $q>4$ or $q<0$, the parameter $x$ is also real,
as is the eigenvalue $g_1$ defined in \reff{def_g1}.

2.  The formula $q = 2 + 2\cos\theta$ maps the strip $0 < \real \theta < \pi$
one-to-one onto the $q$-plane cut along the intervals
$(-\infty,0]$ and $[4,\infty)$.
In particular, when $q$ is real and $0 < q < 4$, the parameter $\theta$
is real ($0 < \theta < \pi$), as are the eigenvalues 
$g_2$ and $g_3$ defined in \reff{def_g2}/\reff{def_g3}. 

3.  The definitions
\reff{def_g1}, \reff{def_g2_over_g1} and \reff{def_g3_over_g1}
cannot be applied directly {\em on}\/ the real $q$-axis for $0 < q < 4$;
rather, one must consider a limit in which $\imag\theta \uparrow 0$
and hence $|x|, |p|, |y| \uparrow 1$.

\subsection{Recomputation of Baxter's phase diagram} \label{sec6.2}

In this subsection we will carefully recompute
the limiting curve ${\cal B}_\infty$
{\em under the tentative hypothesis that Baxter's three eigenvalues
  are in fact the dominant ones}\/.
(In Section~\ref{sec6.3} we will test this hypothesis
 against our finite-lattice data.)

\subsubsection{Computation of limiting curves}

Our goal is to compute the locus of points where
two or more of the eigenvalues $g_i$ are equimodular,
and to determine at each such point whether these 
equimodular eigenvalues are dominant or subdominant.
We carry out this procedure as follows:
\begin{itemize}
   \item[1)]  We use \reff{def_g2_over_g1} to compute the locus $|g_2/g_1| = 1$
in the complex $p$-plane (Figure~\ref{Figure_Baxter_p});
we then transform the resulting plot to the complex $\theta$-plane
using \reff{def_p}.
As noted above, each curve in the $p$-plane corresponds to
an infinite family of curves in the $\theta$-plane.
Along each of these latter curves, we compute $|g_3/g_1|$
and classify the curve (or portions of it) as dominant or subdominant.
   \item[2)]  We use \reff{def_g3_over_g1} to compute the locus $|g_3/g_1| = 1$
in the complex $y$-plane (Figure~\ref{Figure_Baxter_y});
we then transform the resulting plot to the complex $\theta$-plane
using \reff{def_y}.
Each curve in the $y$-plane corresponds to
an infinite family of curves in the $\theta$-plane.
Along each of these latter curves, we compute $|g_2/g_1|$
and classify the curve (or portions of it) as dominant or subdominant.
   \item[3)]  We use (\ref{def_ratios_g}a,b) to compute the locus
$|g_3/g_2| = 1$ directly in the complex $\theta$-plane and
to determine dominance or subdominance.
   \item[4)]  We combine the three families of equimodular curves
into a single $\theta$-plane plot (Figure~\ref{Figure_Baxter_t}).
   \item[5)]  Finally, we transform the resulting curves to
the complex $q$-plane using \reff{def_theta}.
The resulting ``phase diagram'' is shown in Figure~\ref{Figure_Baxter_q};
a detailed view near the point $q=4$ is shown in
Figure~\ref{Figure_Baxter_q_zoom}.
\end{itemize}

Despite the explicit formulae \reff{def_ratios_g},
these computations are far from straightforward,
due to the slow convergence of the products when $|p|$ or $|y|$ is near 1
(i.e.\ when $q$ is near the interval $0 < q < 4$ of the real axis)
and to the need for very high numerical precision in intermediate
stages of the computation.
We discuss these technical points in the Appendix.

The equimodular curves $|g_2/g_1| = 1$ in the complex $p$-plane
are shown in Figure~\ref{Figure_Baxter_p}.
Each equimodular curve $C_n$ has exactly two endpoints. As we approach the 
circle $|p| = 1$, more smaller equimodular curves appear. In order 
to disentangle the larger curves from these new smaller curves, we have  
followed each equimodular curve carefully as it approaches the 
$|p| = 1$ limit. In Figure~\ref{Figure_Baxter_p} we have shown all 
equimodular curves that intersect the circle $|p| = 0.99$. 

The principal feature is a curve $C_1$ running from
point G ($p=-1$, $\theta=\pi/6$ and images)
via the origin ($p=0$, $\theta = -i0$)
to point H ($p=-i$, $\theta = 2\pi/3$ and images).
The next-longest curve ($C_2$) runs from
point I ($p=i$, $\theta=2\pi/9$ and images)
to point J ($p=e^{3\pi i/7}$, $\theta=7\pi/30$ and images).
The third-longest curve ($C_3$) runs from
point K ($p=e^{-\pi i/5}$, $\theta=5\pi/12$ and images)
to point L ($p=e^{-\pi i/4}$, $\theta=4\pi/9$ and images).
Some further equimodular curves and their endpoints are shown on
Figure~\ref{Figure_Baxter_p}
and enumerated in Table~\ref{table_fits_p}. 
In this table we have first shown the curves $C_n$ for which both endpoints
are reasonably well determined (see below). Then we have listed some other 
well-determined endpoints whose counterparts could not be estimated with
sufficient accuracy; these points are grouped into the category ``Others''. 

It is curious that all these endpoints appear to lie at $p = e^{i\varphi}$
where $\varphi$ is a rational multiple of $\pi$ (with a small denominator).
In order to test this conjecture,
we have performed detailed fits as follows:
For each endpoint we first obtained ten nearby equimodular points 
$p$ with $|p| = 0.990,0.991,\ldots,0.999$.
Then we performed a least-squares fit of the data\footnote{
   We have made the computations with data truncated to eight decimal digits,
   hence the error of the input data is $10^{-8}$. 
   Note, however, that the deviations from \reff{Ansatz_fits_p}
   are {\em not}\/ statistical fluctuations;  rather, they are
   ``corrections to scaling'', i.e.\ due to neglected higher-order terms
   in \reff{Ansatz_fits_p}.
   A large value of the $\chi^2$ is thus a signal that we need to include 
   higher-order terms in our Ansatz.
}
using the polynomial Ansatz
\be
 {\Arg p \over \pi} \;\,=\;\, {\Arg p_{\rm Fit} \over \pi}  \,+\, 
                       \sum\limits_{k=1}^8 a_k (1 - |p|)^k 
 \label{Ansatz_fits_p}
\ee
in order to estimate the phase $\varphi = \Arg p_{\rm Fit}$ at the endpoint.
If the equimodular curve is smooth close to $|p|=1$, this Ansatz is expected
to work well. We have chosen an eighth-order polynomial in order
to take into account as many data points as possible
while allowing a little freedom (we have one degree 
of freedom in the fits). As a check, we have repeated this computation with 
lower-degree polynomial Ans\"atze and dropping the data with the smallest
values of $|p|$ (in order to have at least one degree of freedom in the fit).
We have used the stability of the estimates for $\Arg p_{\rm Fit}$ as 
a guideline to decide whether a fit is good or not (see below).

We next asked whether the estimated value of $\Arg p_{\rm Fit}/\pi$
is or is not close to a rational number with a small denominator.
We have used the following basic criterion:
the real number $x$ is ``close'' to the rational number $m/n$ whenever the
\be
  \hbox{``discrepancy''}  \;\equiv\; |n x - m| 
 \label{def_discrepancy}
\ee
is sufficiently small.
In order to make precise what we mean by ``sufficiently small'',
let us first define
\be
   d_n(x)  \;=\;  \min_{m \in \Z} |nx - m|   \;.
\ee
Clearly, for {\em every}\/ real number $x$ and {\em every}\/ integer $n$,
we have $d_n(x) \le 1/2$;
so, if we are to have good evidence that $x=m/n$,
we will need to insist, at the very least,
that the ``discrepancy'' be $\ll 1/2$.
But this requirement is, in fact, nowhere near stringent enough:
the trouble is that, in our procedure,
the denominator $n$ is not fixed in advance;
rather, we start from the number $x$ and we ask whether there {\em exists}\/
a (small) integer $n$ such that $d_n(x) \ll 1$.
And the occurrence of {\em some}\/ integers $n$ such that $d_n(x) \ll 1$
is by no means a rare event.
Indeed, a theorem of Hurwitz (1891) asserts that
for {\em every}\/ real number $x$, {\em there exist}\/ infinitely many
integers $n$ such that $d_n(x) < 1/(\sqrt{5} \, n)$.\footnote{
   The constant $\sqrt{5}$ is best possible, as is shown by
   $x = (1+\sqrt{5})/2$.
   For a proof of Hurwitz's theorem, see 
   \cite[Chapter 1]{Niven_63}, \cite[Chapters 5 and 6]{Rademacher_77}
   or \cite[Chapter I]{Cassels_57}.
}
So we will need to demand that $d_n(x) \ll 1/n$.
This suggests using the criterion $d_n(x) \le \delta/n^s$
for some $\delta > 0$ and some $s > 1$.

Suppose, in fact, that $x$ is a {\em randomly chosen}\/ real number
(from a uniform distribution).
Then it is not hard to show that
\begin{subeqnarray}
   {\rm Prob}(\hbox{there exists an integer $n$ with } d_n(x) \le \delta/n^s)
   & \le &  \sum\limits_{n=1}^\infty {2\delta \phi(n) \over n^{s+1}}  \qquad
       \slabel{eq.number.1a} \\[2mm]
   & = &   {2\delta \zeta(s) \over \zeta(s+1)}  \;,
       \slabel{eq.number.1b}
       \label{eq.number.1}
\end{subeqnarray}
where $\phi(n)$ is the Euler totient function
(i.e.\ the number of integers $1 \le k \le n$ that are
 relatively prime to $n$)
and $\zeta(s)$ is the Riemann zeta function.\footnote{
   {\sc Proof of \reff{eq.number.1}:}
   Let $x$ be a uniform random number in $[0,1]$;
   we do all our arithmetic in the real numbers modulo 1.
   For any fixed $m$ and $n$, we clearly have
   ${\rm Prob}(|nx-m| \le \delta/n^s) = \min[2\delta/n^{s+1},\, 1]$.
   Now we need to sum over pairs $(m,n)$ with $1 \le m \le n$;
   in doing this, it suffices to consider only pairs $(m,n)$ that are
   {\em relatively prime}\/, since if $m/n = m'/n'$ with $n' < n$,
   then the interval $|nx-m| \le \delta/n^s$ is entirely contained in
   the larger interval $|n'x-m'| \le \delta/{n'}^s$.
   This proves \reff{eq.number.1a}.
   The equality with \reff{eq.number.1b} is a standard identity
   in analytic number theory: see e.g.\ \cite[p.~371]{Graham_94}.

   Of course, \reff{eq.number.1} is still an overestimate,
   because it neglects the overlaps between intervals
   whose denominators $n$ and $n'$ do not divide each other.
   It would be interesting to know what the true behavior
   of this probability is.
}
Therefore, if we insist that
\be
  \hbox{``discrepancy''}  \;\le\;
  {\epsilon \zeta(s+1) \over 2\zeta(s) n^s}  \;,
 \label{eq.number.2}
\ee
we are following a procedure whose probability of yielding
a ``false positive signal'' is less than $\epsilon$
(if $x$ is chosen uniformly at random).
Otherwise put, we can define a ``significance level''
\be
   \epsilon  \;=\;
   {2\zeta(s) n^s \over \zeta(s+1)} \,\times\, \hbox{``discrepancy''}
   \;.
   \label{eq.number.3}
\ee
For concreteness, we have chosen $s = 3/2$.
Roughly speaking, we consider the equality $x = m/n$ to be
very strongly supported if $\epsilon \ltapprox 0.001$,
strongly supported if $\epsilon \ltapprox 0.01$,
and a plausible guess if $\epsilon \ltapprox 0.05$.

We have been able to fit the data corresponding to 42 endpoints.
(We have numerically located additional endpoints,
 but we have not included here those endpoints that correspond
 to very small equimodular curves close to the circle $|p|=1$.) 
Of these 42 endpoints, we have obtained a reasonably good fit for the 
28 points displayed in Table~\ref{table_fits_p}:
here a ``reasonably good fit'' is defined as one for which
$\epsilon \leq 0.05$ and for which the values of $\Arg p_{\rm Fit}$
and $\epsilon$ are stable under changes in the
degree of the polynomial Ansatz \reff{Ansatz_fits_p}.
We find that 18 points have $\epsilon \le 10^{-3}$
(eight of them even have $\epsilon \le 10^{-4}$),
while 10 points have $10^{-3} < \epsilon \le 0.05$.

We have found several smaller equimodular curves that are hard to see  
on Figure~\ref{Figure_Baxter_p}, and we suspect that they too have rational 
endpoints.
Indeed, we would not be surprised to learn that such rational endpoints
are {\em dense}\/ in the unit circle
(though we have no idea how to prove this conjecture).

The equimodular curves $|g_3/g_1| = 1$ in the complex $y$-plane
are shown in Figure~\ref{Figure_Baxter_y}.
Again, each equimodular curve $D_n$ has exactly two endpoints, and 
smaller equimodular curves appear as we approach the $|y| = 1$ limit.
We have shown in Figure~\ref{Figure_Baxter_p} all equimodular curves 
that intersect the circle $|y| = 0.99$. 

The principal feature is again a curve $D_1$ running from
point G ($y=e^{-4\pi i/5}$, $\theta=\pi/6$ and images)
via the origin ($p=0$, $\theta = \pi-i0$)
to point H ($y=1$, $\theta = 2\pi/3$ and images).
The next-longest curve ($D_2$) runs from
point K ($y=e^{6\pi i/7}$, $\theta=5\pi/12$ and images)
to point L ($y=e^{4\pi i/5}$, $\theta=4\pi/9$ and images).
Finally, a small curve ($D_4$) runs from
point G${}'$ ($y=e^{-8\pi i/11}$, $\theta=\pi/12$ and images)
to point H${}'$ ($y=e^{-10\pi i/13}$, $\theta=2\pi/15$ and images);
we call these points G${}'$ and H${}'$ because they correspond
to $\theta$ values that are images of the points G and H
under the $p$-plane transformation \reff{def_symmetries_ratio21}
[though they are not images in the $y$-plane].

Once again, all the endpoints appear to lie at $y = e^{i\psi}$
where $\psi$ is a rational multiple of $\pi$ with small denominator.
Even more curiously, many of these endpoints correspond to $\theta$ values 
that are also observed as endpoints in the $|g_2/g_1| = 1$ plot (e.g., all 
points with a label in Tables~\ref{table_fits_p} and \ref{table_fits_y}).
We performed fits to those endpoints in the same manner
as just explained for the $p$-plane.
We found 44 endpoints, and of these we obtained reasonably good fits for the 
23 endpoints displayed in Table~\ref{table_fits_y}.
All these points
satisfy the criterion $\epsilon \leq 0.05$, and both 
$\Arg p_{\rm Fit}$ and $\epsilon$ are rather stable as we vary
the degree of the polynomial Ansatz \reff{Ansatz_fits_p}.
Of these 23 endpoints, we find 12 satisfying $\epsilon \leq 10^{-3}$
(six of them satisfy the stronger condition $\epsilon \leq 10^{-4}$),
while the other 11 endpoints satisfy $10^{-3} < \epsilon \le 0.05$.

We have also found more smaller equimodular curves that are hard to
see on Figure~\ref{Figure_Baxter_y}; we conjecture that they have 
rational endpoints
and we wonder whether these rational endpoints are dense in the unit circle.

\bigskip 

\noindent
{\bf Remark}. It is important to note that the two main points in 
Figures~\ref{Figure_Baxter_t}--\ref{Figure_Baxter_q} (namely, G and H) are  
extremely well determined in both the $p$- and $y$-planes. In particular, their 
$\epsilon$ values range from $4 \times 10^{-5}$ down to $6\times 10^{-8}$. 
Thus, we can trust that the values of $\theta$ for these two points are given 
{\em exactly}\/ by $\theta_G = \pi/6$ and $\theta_H = 2\pi/3$ (see 
Tables~\ref{table_fits_p}--\ref{table_fits_y}).

\subsubsection{Summary of $\theta$-plane phase diagram}

Let us now describe the resulting zero-temperature ``phase diagram'' in the 
complex $\theta$-plane (Figure~\ref{Figure_Baxter_t}) and discuss the 
agreements and discrepancies with respect to Baxter \cite{Baxter_87}.  
For simplicity we have labelled the points by the same letters as in
Figure~5 of \cite{Baxter_87}.
The portions of curves where the equimodular eigenvalues are dominant
(resp.\ subdominant) are depicted in black (resp.\ \subdominantcolor).

The curve A--B (resp.\ B--C) corresponds to the dominant equimodularity 
of $g_3$ (resp.\ $g_2$) and $g_1$. These two curves together with the 
real $\theta$ axis enclose the rest of the dominant curves. The position of
these points is
\begin{subeqnarray}
\theta_{\rm A} &=& \pi                      \label{def_t_A} \\
\theta_{\rm B} &\approx& 0.508588719845180 - 0.625516375803391\,i
    \label{def_t_B}  \\
\theta_{\rm C} &=& 0                        \label{def_t_C}
\end{subeqnarray} 

The point B is triply equimodular (i.e.\ a T point),
so that three equimodular curves cross there:
\begin{itemize}
   \item[1)]  C--B--H, which corresponds to $|g_1| = |g_2|$
     (it is dominant along C--B and subdominant along B--H);  and
   \item[2)]  A--B--Q--G, which corresponds to $|g_1| = |g_3|$
     (it is dominant along A--B, subdominant along B--Q,
      and dominant again along Q--G);
   \item[3)]  R--B--Q--F, which corresponds to $|g_2| = |g_3|$
     (it is subdominant along R--B, dominant along B--Q,
      and subdominant again along Q--F). Point R corresponds to 
      $\theta=-i\infty$.
\end{itemize}
This last result contradicts \cite{Baxter_87},
where the entire curve B--Q--F is claimed to be dominant. The 
position of these points is\footnote{
   Despite appearances, the point F in Figure~\ref{Figure_Baxter_t}
   is {\em not}\/ triply equimodular:
   the dominant equimodular curve $|g_1| = |g_2|$ meets the real axis
   at H${}'$ ($\theta = 2\pi/15 \approx 0.41888$),
   i.e.\ slightly {\em below}\/ point F;
   this is discussed below at \reff{def_theta_H_k_12} ff.
   This splitting is somewhat more visible
   on Figure~\ref{Figure_Baxter_q_zoom}.
}
\begin{subeqnarray}
\theta_{\rm Q} &\approx& 0.440568708859061 - 0.235993788540783 \,i
   \slabel{def_t_Q} \\
\theta_{\rm F} &\approx& 0.427907971348122  \slabel{def_t_F}
\end{subeqnarray}

The point Q is also  triply equimodular,
so that three equimodular curves cross there.
Two of them have just been discussed:
A--B--Q--G and R--B--Q--F.
The third is C--Q--G, which corresponds to $|g_1| = |g_2|$;
it is dominant along C--Q and subdominant along Q--G.
Please note that the subdominant curve Q--G lies always to the
right of (but very close to) the dominant curve Q--G.

There are infinitely many equimodular curves $|g_1| = |g_2|$
terminating at point C ($q=4$) and converging to it:
they are images under \reff{def_symmetries_ratio21}
of the two curves C--B--H and C--Q--G.
The dominant eigenvalue alternates between $g_1$ and $g_2$
as these curves are crossed.
For simplicity, we have shown in Figure~\ref{Figure_Baxter_t}
only the first few of these image curves.
The endpoints of these curves
can be obtained easily using the values of $\theta_G = \pi/6$ and
$\theta_H=2\pi/3$ (see Table~\ref{table_fits_p}) and
transformation \reff{def_symmetries_ratio21}. The result is
\begin{subeqnarray}
\slabel{def_theta_G_k_12}
\theta_{G,k} & = & { \pi \over 6(1+k)} \\
\theta_{H,k} & = & {2\pi \over 3(1+ 4 k)} 
\slabel{def_theta_H_k_12}
\end{subeqnarray}
According to \reff{def_Bn}/\reff{def_theta}, they correspond to Beraha
numbers: $\theta_{G,k}$ is $B_{12+12k}$, and $\theta_{H,k}$ is 
$B_{3+12k}$. 

Likewise, there are infinitely many equimodular curves $|g_1| = |g_3|$
terminating at point A ($q=0$) and converging to it:
they are images under \reff{def_symmetries_ratio31}
of the two curves A--B--Q--G and A--H.
The dominant eigenvalue alternates between $g_1$ and $g_3$
as these curves are crossed.
Once again, we have shown in Figure~\ref{Figure_Baxter_t}
only the first few of these image curves. The endpoints of these curves
can be obtained easily using the values of $\theta_G = \pi/6$ and 
$\theta_H=2\pi/3$ (see Table~\ref{table_fits_y}) and 
transformation \reff{def_symmetries_ratio31}. The result is
\begin{subeqnarray}
\slabel{def_theta_G_k_13}
\theta_{G,k'} & = & {1 + 15k' \over 6 + 15 k' } \pi \\
\theta_{H,k'} & = & {2 + 3 k' \over 3 + 3 k'}   \pi
\slabel{def_theta_H_k_13}
\end{subeqnarray}
In this case, none of them corresponds to a Beraha number.

There are also many small equimodular curves lying near the real $\theta$-axis,
some of which are dominant;
they arise from the curves I--J, K--L, etc.\ in
Figures~\ref{Figure_Baxter_p} and \ref{Figure_Baxter_y}.
The number of these curves grows rapidly as $|\imag\theta| \to 0$,
so we cannot possibly compute all of them;
moreover, the computation becomes increasingly difficult
as $|\imag\theta| \to 0$, since $|p|$ and $|y|$ are tending to 1.
We have, in any case, shown in Figure~\ref{Figure_Baxter_t}
all those curves in the range $0.2 \leq \real \theta \leq 2.3$ that
intersect the half-plane $\imag \theta \leq -0.01$. 
Again, one can obtain the endpoints of these curves by applying 
transformations \reff{def_symmetries_ratios} to the corresponding 
$\theta$ values in Tables~\ref{table_fits_p} and \ref{table_fits_y}. 
These endpoints are not in general Beraha numbers. Only point I 
($\theta_I = 2\pi/9$) and its transformed values under 
\reff{def_symmetries_ratio21},
\be
\theta_{I,k} = {2\pi \over 9+12 k}   \;,
\label{def_theta_I_k_12}
\ee
are Beraha numbers (namely, $B_{9+12k}$).

It follows from Figure~\ref{Figure_Baxter_t}
that the point $q_0$ should not be identified with F
(as Baxter \cite{Baxter_87} did), but rather with the point G at position
\be
\theta_{\rm G} = {\pi\over 6} \approx 0.523598775598299 \label{def_t_G} 
\ee
(see Tables~\ref{table_fits_p} and \ref{table_fits_y}).

\subsubsection{Summary of $q$-plane phase diagram}

In Figure~\ref{Figure_Baxter_q} we show the above ``phase diagram'' in 
the $q$-plane (for clarity, only the dominant equimodular curves
have been depicted).
This is quite similar to Figure~5 of \cite{Baxter_87},  
except for four issues:

1)
The ``phase diagram'' around point C ($q=4$) is richer than the one found 
by Baxter
(see Figure~\ref{Figure_Baxter_q_zoom} for a detailed plot of this region). 
The largest components of the region ${\cal D}_2$ (where $g_2$ is dominant)
are bounded by C--B--Q--C and its complex-conjugate counterpart 
C--E--Q${}'$--C.
The points B and Q take the values
\begin{subeqnarray}
q_{\rm B} &\approx&  4.099903170634857 + 0.649694690705481\,i
   \label{def_q_B} \\
q_{\rm Q} &\approx&  3.859627688708099 + 0.203154495450945\,i
   \label{def_q_Q}
\end{subeqnarray}
However, there are additional components of ${\cal D}_1$ and ${\cal D}_2$
near point C;  indeed, as we approach point C the dominant 
eigenvalue alternates between $g_1$ and $g_2$.
Thus, $g_1$ is dominant on the region lying between C--Q--G--Q${}'$--C
and C--$B_{15}$--C;
$g_2$ is dominant on the region between the curve C--$B_{15}$--C and the next 
curve we find towards C, etc. Baxter found [via the transformation 
\reff{def_symmetries_ratio21}] only half of the curves around 
point C, namely, those curves that are images of C--B--H and past through
the Beraha numbers $B_{3+12k}$.
(In Figure~\ref{Figure_Baxter_q_zoom} we have 
shown two of these curves, corresponding to $B_{15}$ and $B_{27}$.) 

2) If we define $q_0$ as the point on the real $q$-axis where the region 
${\cal D}_3$ ends, then the above discussion implies that 
$q_0$ is not given by $q_{\rm F} \approx 3.819671731239719$
as Baxter believed [cf.\ \reff{def_q0}], but rather by
\be
q_0 = q_{\rm G} = B_{12} = 2 + \sqrt{3} \approx 3.732050807568877
   \label{def_q_G}
\ee
[cf.\ \reff{def_t_G}].

3)
The ``phase diagram'' around point A ($q=0$) is also richer than the one found 
by Baxter. The analytic structure is similar to the one already discussed 
for the point C, except for the fact that now the dominant eigenvalue 
alternates between $g_1$ and $g_3$. Again, Baxter found 
[via \reff{def_symmetries_ratio31}]
only some of the equimodular curves in this region. The physical meaning 
of these new curves is not clear to us.  

4)
We also find many new dominant equimodular curves lying very close
to the real $q$-axis between points A and C.
Some of them (lying between H and G)
have been depicted in Figure~\ref{Figure_Baxter_q}.
These curves were missed by Baxter and their physical meaning is unclear.  

Let us stress that Baxter was aware of some (though not all)
of the ``extra'' equimodular curves discussed here.
He said \cite[p.~5252]{Baxter_87}:
\begin{quote}
\small
   We have ignored those alternative curves as the finite lattice
   calculations give no evidence of there being zeros on them,
   and the full set of complex zeros just fits onto BAE, BCE, BFE.
\end{quote}
It is an open question whether ignoring these curves is indeed
the correct thing to do;
we shall address this question more fully in Sections~\ref{sec6.3}
and \ref{sec6.4}.

On Figure~\ref{Figure_Baxter_q} we have also superposed the limiting curve 
for $L=11_{\rm P}$ (see Figure~\ref{Figure_tri_11PxInftyF}).\footnote{
   $L=12_{\rm P}$ would be substantially similar, had we been able
   to compute the limiting curve for it
   (compare Figures~\ref{Figure_tri_11PxInftyF} and
    \ref{Figure_tri_12PxInftyF}).
}
This curve lies quite close to the main parts of the
infinite-strip-width limiting curve obtained in this section. 
The point $q_0$ for the $L=11_{\rm P}$ strip
(as well as for the rest of the strips considered in this paper:
 see Table~\ref{table_summary})
lies below both Baxter's prediction \reff{def_q0} and our somewhat lower
prediction \reff{def_q_G}.
So our results in Sections~\ref{sec_F}--\ref{sec_Z}
are compatible with both predictions;
unfortunately our strips are not yet wide enough to distinguish between them.
On the other hand, it is precisely around 
points A and C that the finite-strip limiting curve is not defined, so 
our transfer-matrix results do not give any clue as to whether the additional
dominant equimodular curves we have found there can be neglected
(as Baxter did) or not.  

Another way of discovering whether the true value of $q_0$ is given by 
point F \reff{def_q0} or by point G \reff{def_q_G} is to consider the 
isolated limiting points of wide triangular-lattice strips. We expect that
all {\em real} isolated limiting points are smaller than $q_0$. Thus, if we
find any isolated limiting point larger than \reff{def_q_G} [and smaller 
than \reff{def_q0}], then we should conclude that our prediction is false. 
In Tables~\ref{table_zeros_free}, \ref{table_zeros_cyl}, 
\ref{table_zeros_cyl_bis} and \ref{table_zeros_zig} we do not find any such 
zero. The largest real zero we have found is 
$\approx 3.6345747 < q_{\rm G} = B_{12} \approx 3.7320508\ldots$.
On the other hand, for free and cylindrical boundary conditions, all 
real isolated limiting points are expected to be Beraha numbers.
Thus, if $q_0$ were given by F, as Baxter predicts,
then the largest real isolated limiting point would be
$B_{14} \approx 3.801938$. On the other hand, if $q_0$ is given by G, 
as we predict, then the largest real isolated limiting point would be
$B_{11} \approx 3.682507$
(or perhaps $B_{12} \approx 3.732051$).
In Figure~\ref{Figure_Baxter_q_zoom} we have 
also depicted the position of the Beraha numbers $B_{11},\ldots,B_{16}$ to
make easier the comparison with the two alternative values of $q_0$.

\subsection{Comparison of Baxter's eigenvalues with finite-lattice data}
   \label{sec6.3}

Let us now compare Baxter's predictions for the leading eigenvalues
$g_i(q)$ [$i=1,2,3$] with our finite-lattice data
on strips up to width $L=12$ (with cylindrical boundary conditions).
We shall consider slices of fixed $\imag q$,
and plot $\log |g_i(q)|$ as a function of $\real q$.
Simultaneously, we shall also plot $f_k(q,L) = L^{-1} \log |\lambda_k(q)|$
for the five largest (in modulus) eigenvalues of the transfer matrix.
Our goal is to see whether analogues of Baxter's eigenvalues
$g_1,g_2,g_3$ can be observed in our finite-lattice data.

Let us begin at large $\imag q$, where the behavior is simplest.
A plot for $\imag q = 2$ is shown in Figure~\ref{fig_Imq=2}.
Baxter's eigenvalues $g_1,g_2,g_3$ are shown as a solid blue line,
a dashed black line, and a dashed-dotted red line, respectively.
The eigenvalue $g_1$ is dominant for $\real q \ltapprox -0.52$,
the eigenvalue $g_3$ is dominant for $-0.52 \ltapprox \real q \ltapprox 3.82$,
and the eigenvalue $g_1$ is once again dominant for $\real q \gtapprox 3.82$.
The eigenvalue $g_2$ is everywhere strongly subdominant.
Comparing these curves with our finite-lattice eigenvalues,
we see that:
\begin{itemize}
   \item In the region $\real q \ltapprox -0.52$,
      the leading eigenvalue agrees almost perfectly with $g_1$,
      and the second eigenvalue agrees very closely with $g_3$.
   \item In the region $-0.52 \ltapprox \real q \ltapprox 3.82$,
      the leading eigenvalue agrees quite well (though not perfectly)
      with $g_3$.
      Moreover, in a small part of this region near its two extremes,
      we can identify a subleading eigenvalue that appears to correspond
      to $g_1$.  However, in most of this region we are unable to
      identify an eigenvalue corresponding to $g_1$ --- conceivably it exists
      but is hidden under too many other subleading eigenvalues.
   \item In the region $\real q \gtapprox 3.82$,
      the leading eigenvalue almost perfectly with $g_1$.
      However, we are unable to identify any eigenvalue corresponding
      to $g_3$, except perhaps for $\real q$ just above the
      crossover point $\approx 3.82$.
\end{itemize}

We have also extrapolated to the infinite-volume limit $L\to\infty$
the values of the free energy $f_1(L)$ corresponding to
the leading eigenvalue.
At values of $q$ where the theory is critical,
conformal field theory predicts that the finite-size corrections
to the bulk free energy should be of the form
\be
f_1(L) \; = \; f_1(\infty) \,+\, {\pi G \over 6} \, {c \over L^2}  \,+\, \cdots
\label{Ansatz_fk}
\ee
where $c$ is the system's conformal charge,
$G$ is a geometrical factor that equals $\sqrt{3}/2$ on the triangular
lattice, and the dots stand for higher-order corrections.
At values of $q$ where the theory is noncritical,
the finite-volume effects should decay exponentially as $L\to\infty$,
so one may expect that the Ansatz \reff{Ansatz_fk} will again work well
(and simply yield a very small number for $c$).
We therefore estimated the free energy $f_1(\infty)$
by inserting the data points $L=11,12$ into the Ansatz \reff{Ansatz_fk}.
The results are shown as solid green dots on Figure~\ref{fig_Imq=2}.

For $\real q \ltapprox -0.52$ and $\real q \gtapprox 3.82$,
the extrapolated free energy $f_1(\infty)$ coincides almost exactly
with the value for $L=12$ (as well as with Baxter's eigenvalue $g_1$),
in accordance with our expectation that the region outside the curve ABCEA
is noncritical and hence has exponentially rapid convergence to the
infinite-volume limit.
In the intermediate regime $-0.52 \ltapprox \real q \ltapprox 3.82$,
the extrapolated free energy coincides almost exactly
with Baxter's eigenvalue $g_3$
(except near the boundaries of this regime, where the extrapolation
 behaves less well);
indeed, it coincides with $g_3$ substantially better than $L=12$ points did,
suggesting that the theory in this region is indeed critical
and satisfies \reff{Ansatz_fk} with $c \neq 0$.

The behavior is similar at $\imag q = 1.5$ (Figure~\ref{fig_Imq=1.5})
and $\imag q = 1$ (Figure~\ref{fig_Imq=1});
the main difference is that the subleading eigenvalue in the region
$\real q \ltapprox -0.5$ is more distant from $g_3$ than it was
for $\imag q = 2$.
In particular, for $\imag q = 1$ the two leading finite-volume eigenvalues
do not even cross;  this reflects the fact that $\imag q = 1$ lies below
the leftmost endpoint of the equimodular curve
(see Figure~\ref{Figure_tri_12PxInftyF}).
However, this is almost certainly an effect that will go away
for larger strip widths $L$, as the equimodular curve closes up towards $q=0$
(compare Figures~\ref{Figure_tri_allP} and \ref{Figure_Baxter_q}).
The leading finite-volume eigenvalue agrees almost perfectly with $g_1$
for $\real q \ltapprox -0.5$
and agrees quite well with $g_3$ for $-0.5 \ltapprox \real q \ltapprox 4$,
despite the lack of crossing;
moreover, the agreement with $g_3$ in the latter interval is almost
perfect (except near the endpoints) after extrapolation to $L=\infty$.
All this is exactly the behavior one would expect when a curve such as
Figure~\ref{Figure_tri_12PxInftyF} is trying to approximate that of
Figure~\ref{Figure_Baxter_q}.
The agreement in the region $\real q \ltapprox -0.5$
of the two leading eigenvalues with $g_1$ and $g_3$
must therefore be reckoned as excellent.
Please note that all these slices lie well above the point B and the curve AH.

Consider now the slice $\imag q = 0.5$,
which crosses the curves BQ and BC and twice crosses the curve AH;
the corresponding eigenvalues are plotted in Figure~\ref{fig_Imq=0.5}.
Here the eigenvalue $g_1$ is dominant for
$\real q \ltapprox -0.56$;
the eigenvalue $g_3$ becomes ever-so-slightly dominant
for $-0.56 \ltapprox \real q \ltapprox -0.09$;
the eigenvalue $g_1$ again becomes dominant in the interval
$-0.09 \ltapprox \real q \ltapprox 0.95$
(between the two crossings of the curve AH),
where it has a sharp peak;
the eigenvalue $g_3$ is dominant for
$0.95 \ltapprox \real q \ltapprox 4.01$;
the eigenvalue $g_2$ is briefly dominant in the interval
$4.01 \ltapprox \real q \ltapprox 4.17$;
and finally, the eigenvalue $g_1$ is dominant for $\real q \gtapprox 4.17$.
But very little of this complexity is seen in our finite-lattice eigenvalues:
\begin{itemize}
   \item In the region $\real q \ltapprox -0.56$,
      the leading eigenvalue agrees almost perfectly with $g_1$.
   \item In the entire region $-0.56 \ltapprox \real q \ltapprox 4.01$,
      we see no sign whatsoever of $g_1$;
      in particular, {\em we see no sign of the sharp peak in the interval}\/
      $-0.09 \ltapprox \real q \ltapprox 0.95$.
      Rather, the leading eigenvalue everywhere agrees closely
      (though not perfectly) with $g_3$.
   \item The region $\real q \approx 4$ is very hard to disentangle.
   \item For $\real q \gtapprox 4.17$, the leading eigenvalue
      agrees almost perfectly with $g_1$.
      There may conceivably be one or more eigenvalues corresponding to $g_2$;
      this is uncertain.
\end{itemize}

When we extrapolate the leading eigenvalue to $L\to\infty$,
there is essentially no change in the regions $\real q \ltapprox -0.56$
and $\real q \gtapprox 4.17$ (where the agreement with $g_1$ was
essentially perfect anyway);
on the other hand, we obtain an even better agreement with $g_3$
in the intermediate region $-0.56 \ltapprox \real q \ltapprox 4.01$,
except near the endpoint $\real q \approx 4$
where the extrapolation goes haywire.

The behavior is similar at $\imag q = 0.35$ (Figure~\ref{fig_Imq=0.35}).
The peak in $g_1$ is even sharper, and once again no sign of it
is seen in the finite-lattice data.
Please note that the slice at $\imag q = 0.35$
intersects only the outermost curve AH, but it is very close
to the next curve AH${}'$
(the maximum value of  $\imag q$ along this second curve is $\approx 0.3435$).

Likewise at $\imag q = 0.2$ (Figure~\ref{fig_Imq=0.2}):
this slice intersects {\em two}\/ of the curves (AH and AH${}'$),
corresponding to a sharp dip in $g_1$ near $\real q \approx 0.36$,
followed by an even sharper rise near $\real q \approx 0.89$
and then another precipitous dip.
Once again, we see no sign of this behavior in the finite-lattice eigenvalues
or in the extrapolated leading eigenvalue.
The same situation holds for $\imag q = 0.1$ (Figure~\ref{fig_Imq=0.1}):
there are more and sharper dips and rises in $g_1$,
including some near $\real q \approx 4$; but there is no sign of 
this abrupt behavior in the finite-size data or the extrapolated data.

Finally, let us consider $\imag q = 0$ (Figure~\ref{fig_Imq=0}).
Here $g_1$ is well-defined only for $q < 0$ and $q > 4$ [cf.\ \reff{def_g1}],
while $g_2$ and $g_3$ are well-defined only for $0 < q < 4$
[cf.\ (\ref{def_gi}b,c)].
The eigenvalues $g_2$ and $g_3$ cross at the point
$q_{\rm F} \approx 3.819671731239719$ found by Baxter [cf.\ \reff{def_q0}].
The leading eigenvalue $f_1$ for $L=12_{\rm P}$ reproduces well Baxter's 
prediction for the leading eigenvalue nearly everywhere
(namely, $g_1$ for $q \leq 0$ and $q \ge 4$,
$g_3$ for $0\leq q \leq q_{\rm F}$,
and  $g_2$ for $q_{\rm F}\leq q \leq 4$).
The main discrepancy occurs around $q = q_{\rm F}$:
we find that our finite-size data agree with $g_2$
in the interval $q_0(L=12_{\rm P}) \approx 3.64317 < q < 4$,
which is somewhat larger than the interval 
$q_{\rm F} \approx 3.81967 < q < 4$ predicted by Baxter.
But this is perfectly understandable as a finite-size effect
due to the fact that $q_0(L=12_{\rm P})$ is smaller than $q_{\rm F}$.
If we look at the extrapolation to infinite volume of the 
leading eigenvalue, we observe an even better agreement with Baxter's
prediction in the interval $0\ltapprox \real q \ltapprox 3$;
but in a wide interval around $q=q_{\rm F}$ the extrapolation looks horrible.
This is probably due to the large finite-size effects
that characterize that region where three eigenvalues
are closely competing for dominance and there is a complicated
pattern of level crossings.

In summary, our finite-lattice data are in excellent agreement
with Baxter's predictions, {\em except on one major point}\/:
we find no evidence of an eigenvalue corresponding to $g_1$
in the region of the $q$-plane enclosed by the curve ABCEA.
As a consequence, we find no evidence of the equimodular curves
in this region that should correspond to the crossings of $g_1$ and $g_3$
(e.g.\ AHA and its images, LK, etc.)\ 
or of $g_1$ and $g_2$ (e.g.\ CQGQ${}'$C and its images, etc.).
There seem to be two possibilities:
\begin{itemize}
   \item[(a)]  The eigenvalue $g_1$ really is present in this region,
       but only for strip widths $L$ larger than those we have studied
       (possibly {\em much}\/ larger).
       In this case, the limiting curve $\scrb_\infty$ really would
       exhibit all the complexities shown in
       Figures~\ref{Figure_Baxter_q} and \ref{Figure_Baxter_q_zoom},
       and the correct value of $q_0$ would be given by
       $q_{\rm G} = 2 + \sqrt{3} \approx 3.73205$ rather than by
       $q_{\rm F} \approx 3.81967$ as predicted by Baxter.
   \item[(b)]  For some reason, the eigenvalue $g_1$ is not present
       in this region (though it is clearly present elsewhere).
       In this case, the limiting curve $\scrb_\infty$ would be given
       by ABCEA and CBQFQ${}'$EC only,
       and the correct value of $q_0$ would be $q_{\rm F} \approx 3.81967$
       after all.
\end{itemize} 
At present our numerical data are insufficient to distinguish definitively
between these two scenarios.  But some relevant theoretical considerations
will be presented in the next subsection.

\subsection{Critical discussion}  \label{sec6.4}

Baxter \cite[p.~5242]{Baxter_87} stresses that
\begin{quote}
\small
we expect the results to be exact:
the only way they could be wrong would be if the domain structure
were more complicated.  For instance, we cannot rule out the
existence of a fourth domain within which $W(q)$
[the ``partition function per site'']
has some yet different form.  All we can say is that we have seen
no sign of it.
\end{quote}
But the trouble seems to be not that a fourth dominant eigenvalue appears
in some part of the complex plane --- we see no sign of that either ---
but rather that in part of the complex plane
one of Baxter's three eigenvalues (namely, $g_1$) does \emph{not} appear!

One possible explanation may have been hinted at by Baxter in his first paper,
where he points out \cite[p.~2823]{Baxter_86} that the equivalence
between Nienhuis' loop model and the triangular-lattice chromatic polynomial
``should be treated with caution'';  indeed he observes that
it gives manifestly {\em wrong}\/ answers on a finite lattice
when $q=1$ or 3.
He then gives the following explanation for this discrepancy,
which we now find strikingly prescient:
\begin{quotation}
\small
   We can resolve this apparent contradiction by repeating Nienhuis'
   argument for a finite lattice.  We start by writing the triangular
   Potts model as a Kagom\'e lattice six-vertex model, as has been done
   by Baxter {\em et al}\/ (1976).\footnote{
      The reference being cited here is
      Baxter, Kelland and Wu \cite{Baxter-Kelland-Wu}.
}
   We interpret this as an SOS model
   by placing spins on the faces of the Kagom\'e lattice, adjacent spins
   differing by ${1 \over 2}$, the greater being to the left of the
   intervening arrow.  Choosing the hexagonal (triangular) faces to have
   integer (half-integer) spins and taking $J = -\infty$,
   we can sum over the half-integer spins to obtain Nienhuis' SOS form
   of the loop model, {\em but with special boundary weights}\/ 
   [emphasis added].
   For $0 < q < 4$ these boundary weights are complex, so it is possible
   for them to modify the bulk behaviour.
   Obviously this is happening in the $q=1$ and 3 cases mentioned above.

   In terms of the loop model transfer matrices, these boundary conditions
   may mean that not all eigenvalues contribute to the partition function.
   (If an eigenvector is `orthogonal to the boundary vector', then it
   never enters the calculation of $Z_{\rm Loop}$.)
   This must be happening in the $q=1$ and 3 cases,
   so is presumably a general phenomenon.
   (Similar problems with using the six-vertex form of the Potts model
   have previous[ly] been observed (see Baxter 1982b).\footnote{
      The reference being cited here is \cite{Baxter_JSP82}.
})

   Thus not only is care necessary to select the largest eigenvalue of
   the loop model transfer matrix but one should also verify that it is
   a contributing eigenvalue.
\end{quotation}

Indeed, it is quite possible that Baxter's solution of the
hexagonal-lattice loop model corresponds to some unusual boundary condition
for the triangular-lattice chromatic polynomial ---
different from the free and cylindrical boundary conditions studied here ---
in which one or more eigenvalues present in the latter are
entirely absent in the former (or vice versa).
The presence or absence of such eigenvalues can, of course,
radically change the limiting curve $\scrb$
in case the ``missing'' eigenvalue becomes dominant
somewhere in the complex $q$-plane.\footnote{
   Here is an example of such a situation:
   In computing the triangular-lattice chromatic polynomial
   with cylindrical boundary conditions,
   one has the choice of working entirely
   within the translation-invariant subspace (as we have done)
   or else working in a larger space that includes non-translation-invariant
   connectivities (as Ro\v{c}ek {\em et al.}\/ \cite{Shrock_98c} have done).
   If one takes the latter approach, one will find that
   one of the eigenvalues has an identically vanishing amplitude
   (provided that translation-invariant endgraphs are used);
   when this non-contributing eigenvalue is ignored (as it should be),
   both approaches will give the same limiting curve $\scrb$.
   On the other hand, the latter approach allows one to consider
   also the use of {\em non-}\/translation-invariant endgraphs;
   and it can sometimes be arranged for the ``extra'' eigenvalue
   to contribute and indeed be dominant in some region of the
   complex $q$-plane, with the result that the limiting curve $\scrb$
   is {\em different}\/ from what it was without that eigenvalue.
   Compare, for example, \cite[Figures 4(a) and (b)]{Shrock_98c},
   which show the limiting curves $\scrb$ for the triangular lattice
   of width $L=4_{\rm P}$ with different endgraphs.
}

But this explanation has a severe defect:
if true, it would suggest that the ``missing'' eigenvalue
should be absent {\em throughout}\/ the complex $q$-plane
(corresponding to an {\em identically}\/ vanishing amplitude).
In our case, by contrast, the eigenvalue $g_1$ is unambiguously observed
outside the curve ABCEA;  it is only inside this curve that the
eigenvalue $g_1$ ``disappears''.
It is hard to see how such a $q$-dependent effect could be
caused by boundary conditions.

There remains, of course, the other possibility noted earlier:
namely, that the eigenvalue $g_1$ really is present inside the curve ABCEA
as well, but only for strip widths $L$ larger than those we have studied.
Indeed, it is not so surprising that on lattices $L \le 12$
we have failed to see the gyrations of $g_1$ that give rise to
the equimodular curves AH, AH${}'$, \ldots:
after all, in this same neighborhood ($q \approx 0$)
we also fail to see the crossing of $g_1$ and $g_3$
that corresponds to the completion of the equimodular curve BA at $q=0$,
and there is little doubt that {\em this}\/ feature will be restored
in the limit $L \to\infty$.

Light could perhaps be shed on this issue by a more detailed
examination of Baxter's Bethe Ansatz for finite-width strips \cite{Baxter_86}
and a comparison of his finite-width results with our own.
It could also be useful to study boundary conditions that are periodic
in the longitudinal direction (e.g.\ cyclic or toroidal),
as these seem to lead to curves ${\cal B}$ that close at $q=0$
already for finite $L$ \cite{Shrock_00a,Shrock_01a,Shrock_01b}.
The trouble is that these studies are restricted, at present,
to very small $L$ (namely, $L \le 4$).

\section{Summary and Outlook}   \label{sec8}

\subsection{Behavior of dominant-eigenvalue-crossing curves $\scrb$}
  \label{sec8.1}

In this paper we have computed the transfer matrix for triangular-lattice 
strips of width $3 \leq L_x \leq 9$ for free boundary conditions,  
$4\leq L_x \leq 12$ for cylindrical boundary conditions, and $L_x=4,6,8,10$
for ``zig--zag'' boundary conditions. The transfer matrix
allows the computation of the chromatic zeros for strips of arbitrary length
$L_y$. As the length $L_y$ tends to infinity (for fixed width $L_x$),
the chromatic zeros accumulate along
certain curves (limiting curves ${\cal B}$)
and around certain points (isolated limiting points)
according to the Beraha--Kahane--Weiss theorem
\cite{BKW_75,BKW_78,Sokal_chromatic_roots}. 
For all the above strips except $L_x=10_{\rm F}, 12_{\rm P}, 10_{\rm Z}$,
we have been able to compute the limiting curves ${\cal B}$.
The exact computation of all the isolated limiting points has been 
carried out for $L_x\leq 6_{\rm F}$, $L_x\leq 9_{\rm P}$
and $L_x\leq 6_{\rm Z}$;
for the larger strips we were able to check that certain values 
of $q$ are isolated limiting points,
but we cannot be certain that we have found all of them.
By studying the behavior of the limiting curves
and isolated limiting points
as a function of the strip width $L_x$ (and boundary conditions), we hope
to shed light on the thermodynamic limit $L_x,L_y\to\infty$. 

The basic properties of both limiting curves and isolated limiting points 
are summarized in Table~\ref{table_summary}. In all cases the identity 
\begin{eqnarray}
\hbox{\rm endpoints} &=& 
 \hbox{\rm (2 $\times$ components) + (2 $\times$ double points) + (T points)}  
 \nonumber \\
 & & \qquad -\, \hbox{\rm (2 $\times$ enclosed regions)} 
\end{eqnarray}
holds. This identity can be derived by simple topological/graph-theoretic 
arguments.  

By inspection of Table~\ref{table_summary}, we observe some regularities
when $L_x$ becomes large. 
For all three boundary conditions,
the curve ${\cal B}$ appears to become connected (\#C = 1)
when $L_x$ is large enough.
For all $L_x$, the number of endpoints (\#E)
is 6 for free boundary conditions
and 4 for the other two boundary conditions.
(Note, however, that for $L_x \geq 7_{\rm F}$, $L_x \geq 10_{\rm P}$
 and $L_x = 8_{\rm Z}$ our counts on the number of endpoints
 are only lower bounds:  we may have missed some.)
When $L_x$ is large enough,
the number of T points (\#T) is 4 for free boundary conditions
and 2 for the other boundary conditions.
We have found no evidence of double points for any of the strips considered.
Finally, the number of enclosed regions is zero except for
$L_x = 5_{\rm F}$, $L_x\leq 5_{\rm P}$ and  $L_x=7_{\rm P}$.
These regularities are in sharp contrast with the 
square-lattice case \cite{transfer1,transfer2},
where the number of connected components and endpoints
seems to grow with the strip width.
In particular, we have not found in the triangular-lattice limiting curves
any trace of the small gaps and bulb-like regions
that are so common in the square-lattice case.
It therefore seems that the thermodynamic limit may be achieved in a 
smoother way for the triangular lattice than for the square lattice.
Finally, it is worth mentioning 
that in all cases except $L_x = 4_{\rm F}$ and $4_{\rm Z}$,
the limiting curve ${\cal B}$ crosses the real $q$-axis, thus defining $q_0$.
By contrast, for the square lattice,
$q_0$ is well-defined only for odd widths;
for even widths with both free and cylindrical boundary conditions,
we found either that ${\cal B}$ fails to intersect the real axis
or that it contains a segment of the real axis passing through a
double point \cite{transfer1,transfer2}.

Let us also note that,
as in the square-lattice case \cite{transfer1,transfer2},
we find chromatic zeros with $\real q < 0$.
Indeed, for $L_x \geq 7_{\rm P}$ and $L_x \geq 8_{\rm Z}$
we find that the limiting curve ${\cal B}$ intersects 
the half-plane $\real q < 0$.
For free boundary conditions,
none of our limiting curves ($L_x \le 8_{\rm F}$)
reach this half-plane;
but from Table~\ref{table_summary} we can see that $\min \real q$
is decreasing and we expect that it will be $< 0$ for $L_x \ge 10_{\rm F}$
(and possibly already for $9_{\rm F}$).

The regularities exhibited by the limiting curves become clearer when we 
superpose them all (with fixed boundary conditions).
This is done in  Figures~\ref{Figure_tri_allF}, \ref{Figure_tri_allP}
and \ref{Figure_tri_allZ}
for free, cylindrical and zig-zag b.c., respectively.
We find an overall behavior similar to that found 
for the square lattice \cite{transfer1,transfer2}.
For free boundary conditions (Figure~\ref{Figure_tri_allF}),
we find a monotonic behavior with the width $L_x$:
both the leftmost arcs and the rightmost arcs move outwards as $L_x$
is increased
(see also the columns labelled 
$\min \real q$ and $\max \real q$ in Table~\ref{table_summary}).
The value of $q_0$ (or $\real q_0$ for $L_x = 4_{\rm F}$)
is also monotonically increasing in $L_x$.
The overall shape of the limiting curves is similar to the expected limiting
curve in the thermodynamic limit (Figure~\ref{Figure_Baxter_q}).  
We expect that as $L_x$ grows, the leftmost endpoints will 
tend towards $q=0$, while the rightmost endpoints will go to 
$q = q_c({\rm tri})=4$. The crossing point $q_0$ will eventually go
to either point F [cf.\ \reff{def_q0}] or point G [cf.\ \reff{def_q_G}] in 
Figure~\ref{Figure_Baxter_q}. Unfortunately, our numerical data are not good
enough to tell unambiguously the true limit.   
There is additional one feature of the limiting curves with 
free boundary conditions that does {\em not}\/ correspond
to any feature of the predicted thermodynamic-limit curve:
namely, a pair of small complex-conjugate branches emerging from T points
and pointing inwards. From Figure~\ref{Figure_tri_allF}, it seems that
the size of these branches does not go to zero as $L_x$ is increased
(at least up to 8 or 9);  rather their size stays more or less constant.
We are unable to say whether these branches will get shorter
for larger $L_x$ and ultimately disappear in the limit $L_x\to\infty$.  

In Figure~\ref{Figure_tri_allP} we superpose all the limiting curves
with cylindrical boundary conditions.
As in the square-lattice case, the behavior of
the leftmost part of these curves seems to be monotonic:
the arcs move outwards as $L_x$ is increased.
In particular, $\min \real q$ decreases monotonically
with the strip width (see Table~\ref{table_summary}).
However, the behavior on the right side of the plot is clearly not monotonic:
there are differences depending on the quantity $L_x$ mod 3.
This is to be expected, since with periodic boundary conditions in
the transversal direction, strip widths that are not multiples of 3 are 
somewhat {\em unnatural} as they introduce frustration in the 3-state Potts 
antiferromagnet. Thus, the dependence on $L_x$ in the interval 
$3\ltapprox \real q \ltapprox 4.5$ is not a surprise (the same feature is 
present in the square-lattice case, where we find an even-odd dependence 
on the limiting curves \cite{transfer1,transfer2}).
For fixed values of $L_x$ mod 3,
we find that $q_0$ is monotonic in $L_x$:
for $L_x = 1$ mod 3 it decreases,
while for $L_x = 0$ or 2 mod 3 it increases.
The shape of the limiting curves for $L_x = 0$ mod 3
is very similar to the expected thermodynamic limit 
(Figure~\ref{Figure_Baxter_q}), while in the other cases we find 
stronger finite-size effects that we expect to disappear
in the limit $L_x \to\infty$.
Finally, it is worth mentioning the absence of the extra 
branches that appear for free boundary conditions.

In Figure~\ref{Figure_tri_allZ} we superpose the limiting curves
for ``zig-zag'' boundary conditions.
In this case we only have three curves, so we are unable to
extract any definitive conclusion.
We can only confirm the monotonic behavior
on the leftmost side of the curves and the absence of any extra branches. 
Again, the overall shape is similar to the expected thermodynamic limit
depicted in Figure~\ref{Figure_Baxter_q}.

If we compare the limiting curves for different boundary conditions, we
see that the thermodynamic limit is achieved faster for  
cylindrical and ``zig-zag'' boundary conditions than for free 
boundary conditions (due to existence of ``surface'' effects in the later). 
This is similar to the behavior observed in the square-lattice case.

The main unsolved problem is, of course, whether the extra equimodular
curves predicted by Baxter's formulae (Figure~\ref{Figure_Baxter_q})
are really there for large enough $L$.
Unfortunately, the evidence from strip widths $L \le 12$
is inconclusive (Sections~\ref{sec6.3} and \ref{sec6.4}).

\bigskip

\noindent
{\bf Remark}. In the computation of the limiting curves ${\cal B}$
for triangular-lattice strips with {\em cylindrical}\/ boundary conditions,
we have found a curious behavior involving identically vanishing amplitudes.
(This is important, because the eigenvalues corresponding to the
 identically vanishing amplitudes must be {\em excluded}\/
 from the computation of the equimodular curves.
 For square-lattice strips, by contrast,
 we have not observed any identically vanishing amplitudes
 \cite{transfer1,transfer2}.)
As explained in the introduction to Section~\ref{sec_P},
the transfer matrix can be written (after a change of basis)
in the block-diagonal form
\be
\T(m_{\rm P}) \;=\; \left( \begin{array}{cc}
                           \T_+(m_{\rm P}) & 0 \\
                             0                      & \T_-(m_{\rm P})
                             \end{array} \right)
   \;,
\label{transfer_mP_diag}
\ee
where the matrix $\T_+(m_{\rm P})$ lives on the subspace of
reflection-invariant connectivities and has dimension SqCyl($m$)
[i.e., the dimension of the transfer matrix for a square-lattice
strip of width $m$ with cylindrical boundary conditions],
while the matrix $\T_-(m_{\rm P})$ lives on the subspace of reflection-odd
connectivities and has dimension ${\rm TriCyl}(m) - {\rm SqCyl}(m)$.
For $m \geq 8_{\rm P}$ we have ${\rm TriCyl}(m) > {\rm SqCyl}(m)$
and this decomposition becomes nontrivial.
Now, simple symmetry arguments (see Section~\ref{sec_P})
explain why all the eigenvalues in the reflection-odd subspace
should have identically vanishing amplitudes.
But what is curious and mysterious is that each of these eigenvalues
has an identical ``partner'' in the reflection-even subspace,
also with identically vanishing amplitude.
This means that the characteristic polynomial
associated to the transfer matrix $\T(m_{\rm P})$ can be factored as
\be
\det[ \T(m_{\rm P}) - \lambda {\bf 1} ] \; = \;
                     Q_1(q,\lambda)^2 \, Q_2(q,\lambda)  \;,
\ee
where the zero-amplitude eigenvalues are those coming from the factor
$Q_1(q,\lambda)^2$.
In particular, the number of eigenvalues with zero amplitude (\# VA)
is always even, and it equals twice the dimension of the
reflection-odd subspace:
\be
    {\rm \# VA}(m) \;=\;
    2 \times \left[ {\rm TriCyl}(m) - {\rm SqCyl}(m) \right]
\label{number_zero_amplitudes}
\ee
(see Table~\ref{table_dimensions}).
This is, at any rate, what we have found for $8_{\rm P} \le m \le 12_{\rm P}$
(see Sections~\ref{sec8P}--\ref{sec12P});
we conjecture that it holds for larger widths as well.
It follows that
the effective dimension of the transfer matrix TriCyl${}'$($m$) is given by
\be
     {\rm TriCyl}'(m)  \;=\;  2 {\rm SqCyl}(m) - {\rm TriCyl}(m)
     \;.
\label{effective_number_eigenvalues}
\ee
Numerical values for all these quantities are displayed in
Table~\ref{table_dimensions};
of course, the values of \# VA and TriCyl${}'$($m$)
for $m=13,14$ are {\em conjectures}\/.

Numerical values of TriCyl($m$) and SqCyl($m$) were first reported in
\cite[Table 2]{transfer1}. An analytic formula of TriCyl($m$) for prime
values of $m$ has been obtained in \cite[Theorem 3]{Tutte_sq}.
This paper also contains a conjecture for SqCyl($m$) with prime $m\geq 3$
\cite[Conjecture 2]{Tutte_sq}.
Finally, an analytic formula for TriCyl${}'$($m$)
has been conjectured for arbitrary values of $m$
\cite[Conjecture 1]{Tutte_sq}.

\subsection{Behavior of amplitudes and the Beraha conjecture}  \label{sec8.2} 

Let us now discuss the isolated limiting points and the role of the Beraha 
numbers in the triangular-lattice strips.  Our results show that
the number of isolated limiting 
points is a non-decreasing function of the strip width $L_x$ (for each  
boundary condition), at least up to the maximum $L_x$ we have been
able to investigate.
For free and cylindrical boundary conditions we did not find any
complex isolated limiting points (see Table~\ref{table_summary}).
For ``zig-zag'' boundary conditions we find a pair
of complex-conjugate isolated limiting points for $L_x=6_{\rm Z}$, 
and we have evidence of the existence of another pair of
complex-conjugate isolated limiting points for $L_x=8_{\rm Z}$.

Concerning the real isolated limiting points, most of them are Beraha numbers  
\reff{def_Bn}.   It is only for ``zig-zag'' boundary conditions
that we find real isolated limiting points that are {\em not}\/
Beraha numbers: for $L_x=4_{\rm Z}$ we
find $q=5/2$; for $L_x=6_{\rm Z}$, $q\approx 2.722633$;  
for $L_x=8_{\rm Z}$, $q\approx 2.821420$;
and for $L_x=10_{\rm Z}$, we find two such points,
$q\approx 2.873731$ and $q\approx 3.383129$.
It is not clear to us how these non-Beraha real isolated limiting points
behave as $L_x\to\infty$, e.g., whether their number is bounded or unbounded.

For all the lattices we have studied, we observed empirically that there
is at least one vanishing amplitude $\alpha_i(q)$ at each of the Beraha 
numbers up to $B_{L+1}$ (see Table~\ref{table_zeros_D}).  
It is reasonable to conjecture that this holds for all $L$
(in agreement with a similar conjecture for the square lattice 
 \cite[Conjecture~7.1]{transfer1}):

\begin{conjecture} \label{conjecture1}
For a triangular-lattice strip of width $L$ with free, cylindrical or 
``zig-zag'' boundary conditions, at each Beraha number $q=B_2,\ldots,B_{L+1}$
there is at least one vanishing amplitude $\alpha_i(q)$. That is, 
$\det D(q)=0$ for $q=B_2,\ldots,B_{L+1}$.
\end{conjecture} 

In contrast with the square-lattice case
\cite[Conjectures~7.2 and 7.3]{transfer1},
however, we find that there is a vanishing amplitude [hence $\det D(q) = 0$]
also at some Beraha numbers {\em larger}\/ than $B_{L+1}$.
Indeed, we find examples for each boundary condition
(see Table~\ref{table_zeros_D}): 

\begin{itemize}

\item
For free boundary conditions,
$q=B_6$ is a zero of $\det D(q)$ for $L=4_{\rm F}$.

\item
For cylindrical boundary conditions,
$q=B_6$ is a zero of $\det D(q)$ for $L=4_{\rm P}$;
$q=B_{10}$ is a zero for $L=6_{\rm P},7_{\rm P},8_{\rm P}$;
$q=B_{14}$ is a zero for
   $L=8_{\rm P},9_{\rm P},10_{\rm P},11_{\rm P},12_{\rm P}$;
$q=B_{18}$ is a zero for $L=10_{\rm P},11_{\rm P},12_{\rm P}$;
and finally, $q=B_{22}$ is a zero for $L=12_{\rm P}$.  

\item 
For ``zig-zag'' boundary conditions, $q=B_6$ is 
a zero of $\det D(q)$ for $L=4_{\rm Z}$. 

\end{itemize}

\noindent
We have systematically checked all Beraha numbers up to $B_{50}$ to make
this list.   

Please note that for free and ``zig-zag'' boundary conditions,
we have found only {\em one}\/ case each in which a Beraha number
beyond $B_{L+1}$ is a zero of $\det D(q)$,
namely the relatively small value of $L=4$.
It is conceivable that for all larger $L$
there are no such Beraha zeros, as is conjectured for all $L$
for the square lattice \cite[Conjecture~7.2]{transfer1}.
However, this conjecture clearly cannot be true for
cylindrical boundary conditions. 
There is presumably some pattern that tells us
{\em which}\/ Beraha numbers $q=B_k$ with $k>L+1$
can be zeros of $\det D(q)$.
Thus far only a few Beraha numbers 
($B_6, B_{10}, B_{14}, B_{18}, B_{22}$) have appeared on that list. 
Indeed, we conjecture that the pattern is the following:

\begin{conjecture} \label{conjecture2}
For a triangular-lattice strip of width $L$ with cylindrical 
boundary conditions, the Beraha numbers where $\det D(q)$ vanishes is 
given by the union of the sets $\{ B_2,B_3,...,B_{L+1} \}$ and  
$\{ B_{4k-2} \; | \; k=1,2,...,\lfloor L/2 \rfloor \}$, 
the upper limit on $k$ being the integer part of $L/2$. 
\end{conjecture}

\subsection{Nature of the fixed zeros}  \label{sec8.3}

In Section~\ref{sec_prelim}
we discussed the ``fixed'' zeros that occur
at small integers $q$ (here $q=0,1,2,3$)
when the graph fails to be $q$-colorable.
{}From the point of view of the transfer-matrix formalism,
these fixed zeros can arise in either of three ways:
\begin{itemize}
   \item[1)]  All the amplitudes $\alpha_k$ vanish at $q$.
      Then $Z_n(q) = 0$ for all lengths $n \ge 1$.
   \item[2)]  All the eigenvalues $\lambda_k$ vanish at $q$.
      Then $Z_n(q) = 0$ for all $n \ge 2$.
   \item[3)]  ``Mixed case'':  Neither all the amplitudes nor
      all the eigenvalues vanish at $q$,
      but for each $k$ either $\alpha_k$ or $\lambda_k$ vanishes at $q$
      (or both).
      Then $Z_n(q) = 0$ for all $n \ge 2$.
\end{itemize}
Let us now summarize what we have found concerning
the nature of these fixed zeros for triangular-lattice strips:

\bigskip

{\boldmath$q=0,1.$}
At $q=0,1$ all the amplitudes vanish,
due to the prefactor $q(q-1)$ in the left vector ${\bf u}$.
These points therefore belong to Case 1.

\bigskip
 
{\boldmath$q=2.$}
At $q=2$ the behavior depends on the boundary conditions
and on the strip width $L_x$:
\begin{itemize}
   \item Free boundary conditions:
      \begin{itemize}
         \item $L_x = 2_{\rm F}$:  The one eigenvalue vanishes at $q=2$
            (Case 2).
         \item $L_x = 3_{\rm F}$:  There is one nonzero eigenvalue 
            with a zero amplitude,
            and one zero eigenvalue with a nonzero amplitude (Case 3).
         \item $L_x = 4_{\rm F}$: There is at least one nonzero eigenvalue
             with a zero amplitude and exactly one zero eigenvalue with a 
             nonzero amplitude (Case 3). The transfer matrix at $q=2$
             is not diagonalizable: it has a nontrivial Jordan block 
             corresponding to $\lambda=0$.  
         \item $L_x \ge 5_{\rm F}$:  There is at least one nonzero eigenvalue
            with a zero amplitude,
            at least one zero eigenvalue with a zero amplitude, 
            and exactly one zero eigenvalue with a nonzero amplitude (Case 3).
            We also find that for all $L_x \ge 4_{\rm F}$ the transfer matrix
            at $q=2$ is not diagonalizable, i.e.\ it has nontrivial
            Jordan blocks (all corresponding to eigenvalue $\lambda=0$).
       \end{itemize}
   \item Cylindrical boundary conditions:
      \begin{itemize}
         \item $L_x$ odd:  All the amplitudes vanish,
             due to the prefactor $q(q-1)(q-2)$ in the left vector ${\bf u}$
             (Case 1).
         \item $L_x = 2_{\rm P}$:  The one eigenvalue vanishes at $q=2$
            (Case 2).
         \item $L_x = 4_{\rm P}$:  There is one nonzero eigenvalue
            with a zero amplitude,
            and one zero eigenvalue with a nonzero amplitude (Case 3).
         \item $L_x$ even $\ge 6_{\rm P}$:  
            There is at least one nonzero eigenvalue with a zero amplitude,
            at least one zero eigenvalue with a zero amplitude,
            and at least one zero eigenvalue with a nonzero amplitude (Case 3).
       \end{itemize}
   \item Zig-zag boundary conditions:
      \begin{itemize} 
         \item $L_x = 2_{\rm Z}$:  This is identical to $L_x = 2_{\rm F}$ 
               (Case 2).
         \item $L_x = 4_{\rm Z}$: There are two nonzero eigenvalues
            with zero amplitudes, and one zero eigenvalue with a nonzero 
            amplitude (Case 3).
         \item $L_x$ even $\ge 6_{\rm Z}$: There is at least one nonzero 
            eigenvalue with a zero amplitude,
            at least one zero eigenvalue with a zero amplitude,
            and at least one zero eigenvalue with a nonzero amplitude (Case 3). 
      \end{itemize}
\end{itemize} 

\bigskip
 
{\boldmath$q=3.$}
The point $q=3$ is a fixed zero only for cylindrical boundary conditions
with strip widths $L_x$ that are not a multiple of 3.
There are two distinct situations:
\begin{itemize}
   \item[{}]
\begin{itemize}
   \item $L_x = 4_{\rm P}$:  Both eigenvalues vanish, so that the
      whole transfer matrix vanishes (Case 2).
   \item $L_x = 5_{\rm P}$: There is one nonzero eigenvalue with a zero
      amplitude, and one zero eigenvalue with a nonzero amplitude (Case 3). 
   \item $L_x = 7_{\rm P}, 8_{\rm P}, 10_{\rm P}, 11_{\rm P}$:
      There is at least one nonzero eigenvalue with zero amplitude, 
      at least one zero eigenvalue with a nonzero amplitude, and at least 
      one zero eigenvalue with a zero amplitude [or nontrivial Jordan block  
      corresponding to $\lambda=0$ with no contribution to the partition 
      function for any $n\geq 1$] (Case 3).
      We also find that for $L_x \ge 8_{\rm P}$ the transfer matrix
      at $q=3$ is not diagonalizable, i.e.\ it has nontrivial
      Jordan blocks.
\end{itemize}
\end{itemize}

\section*{Acknowledgments}

We wish to thank Dario Bini for supplying us the MPSolve 2.1.1 package
\cite{Bini_package,Bini-Fiorentino} and for many discussions about its use;
George Andrews and Mireille Bousquet-M\'elou
for useful suggestions concerning the numerical computation
of the products \reff{def_ratios_g};
Hubert Saleur for emphasizing the importance of the Beraha numbers;
Norman Weiss for suggesting that we study the resultant;
and Robert Shrock for many helpful conversations
throughout the course of this work.

We would also like to express our gratitude to an anonymous referee,
whose critical comments on the first version of this paper
led us to make significant improvements in Section~\ref{secBaxter}
and in particular to radically rethink our interpretations.

The authors' research was supported in part
by U.S.\ National Science Foundation grants PHY--9900769 (J.S.\ and A.D.S.),
PHY--0099393 (A.D.S.)\ and PHY--0116590 (A.D.S.)\ 
and by CICyT (Spain) grant FPA2000-1252 (J.S.).

\appendix

\section{Numerical Computation of
         \boldmath$\prod\limits_{n=1}^\infty (1 - tx^n)$}

In this appendix we discuss briefly some of the technical issues
involved in the numerical computation of Baxter's products \reff{def_ratios_g}.
Everything can be expressed in terms of the function
\be
   R(t,x) \;=\; \prod\limits_{n=1}^\infty (1 - tx^n) \;,
 \label{def_R}
\ee
which is defined for complex $t$ and $x$ satisfying $|x| < 1$
and was first studied by Euler \cite{Euler}.
Here we need the cases $t= \pm 1$
and $t = \pm$ a cube root of unity.
A more detailed discussion, including proofs,
can be found in \cite{Sokal_infprod}.

The numerical computation of $R(t,x)$ clearly becomes delicate
when $|x| \uparrow 1$.
In particular, direct use of the product \reff{def_R} gives an algorithm
that is only ``linearly convergent'',
i.e.\ the number of significant digits in the answer
grows linearly with the number of terms taken.
Moreover, the constant of proportionality in this relation
is proportional to $1 - |x|$, and thus deteriorates linearly as
$|x| \uparrow 1$.
Finally, there is severe loss of numerical precision when multiplying
numbers that are very near 1.
An alternative approach can be based on the representation
\be
   \log R(t,x)  \;=\;
   - \sum_{k=1}^\infty {t^k \over k} \, {x^k \over 1-x^k}
   \;,
 \label{eq_logR}
\ee
which is valid whenever $|x| < 1$ and $|tx| < 1$.
This sum is again only linearly convergent,
but the problem of loss of numerical precision
is alleviated by use of the logarithm.

A much more efficient algorithm can be based on the identity
\be
   R(t,x) \;=\;  \sum_{n=0}^\infty
            {(-t)^n x^{n(n+1)/2}  \over (1-x)(1-x^2) \cdots (1-x^n)}
 \label{eq_qseries}
\ee
due to Euler.\footnote{
   For a proof of \reff{eq_qseries},
   see e.g.\ \cite[p.~19, Corollary 2.2]{Andrews_98},
   \cite[p.~34, Lemma 4(a)]{Knopp_70}
   or \cite[pp.~22--23]{Remmert_98}.
}
Because of the $x^{n(n+1)/2}$ factor in the numerator,
this algorithm is ``quadratically convergent'':

\begin{proposition}[\protect\cite{Sokal_infprod}]
   \label{propA.1}
Define
\be
   a_n  \;=\;  {(-t)^n x^{n(n+1)/2}  \over (1-x)(1-x^2) \cdots (1-x^n)}
   \;.
\ee
Then, for $|t| \le 1$ and $|x| \le e^{-\gamma}$ ($\gamma > 0$), we have
\begin{itemize}
   \item[(a)]  $\displaystyle \Delta_N  \,\equiv\,
        \left| \sum\limits_{n=N}^{\infty}  a_n \right|
        \,\le\,
        {e^{\pi^2/6\gamma - N(N+1)\gamma/2}  \over
         1 - e^{-(N+1)\gamma}
        }$
   \item[(b)]  $\displaystyle \delta_N  \,\equiv\,
        {\left| \sum\limits_{n=N}^{\infty}  a_n \right|  \over |R(t,x)|}
        \,\le\,
        {e^{\pi^2/3\gamma - N(N+1)\gamma/2}  \over
         1 - e^{-(N+1)\gamma}
        }$
\end{itemize}
\end{proposition}

\begin{corollary}[\protect\cite{Sokal_infprod}]
   \label{corA.2}
Let $K \ge 0$,
and suppose that $|t| \le 1$ and $|x| \le e^{-\gamma}$ ($\gamma > 0$).
\begin{itemize}
   \item[(a)]  If ${\displaystyle N \,\ge\,
                   \sqrt{ {\pi^2 \over 3\gamma^2} \,+\, {2K \over \gamma} } }$,
        then $\Delta_N \le e^{-K}$.
   \item[(b)]  If ${\displaystyle N \,\ge\,
                   \sqrt{ {2\pi^2 \over 3\gamma^2} \,+\, {2K \over \gamma} } }$,        then $\delta_N \le e^{-K}$.
\end{itemize}
\end{corollary} 

It turns out \cite{Sokal_infprod}
that the {\em a priori}\/ bound of Proposition~\ref{propA.1}(b)
is asymptotically within 9.1\% of being sharp when
$t=1$, $0 < x=e^{-\gamma} < 1$ and $N \gg 1/\gamma$
(moreover, in this case it is asymptotically sharp as $\gamma \downarrow 0$).
But since this bound is overly pessimistic in other cases,
it is of some value to provide an {\em a posteriori}\/ bound
on the truncation error that is more realistic, when $x \notin (0,1)$,
than the {\em a priori}\/ bound.
Here is such a bound, which
can be used a stopping criterion in the numerical algorithm:

\begin{proposition}[\protect\cite{Sokal_infprod}]
   \label{propA.3}
Let $|t| \le 1$, $|x| \le e^{-\gamma}$ ($\gamma > 0$)
and $N > (\log 2)/\gamma$.  Then:
\begin{itemize}
   \item[(a)]  $\displaystyle \Delta_N  \,\equiv\,
        \left| \sum\limits_{n=N}^{\infty}  a_n \right|
        \,\le\,
        |a_{N-1}| \, {e^{-N\gamma} \over 1 - 2e^{-N\gamma}}$
   \item[(b)]  $\displaystyle \delta'_N  \,\equiv\,
        {\left| \sum\limits_{n=N}^{\infty}  a_n \right|  \over |S_N|}
        \,\le\,
        {|a_{N-1}| \over |S_N|} \, {e^{-N\gamma} \over 1 - 2e^{-N\gamma}}$
        where $S_N  \,\equiv\,  \sum\limits_{n=0}^{N-1}  a_n$
        \hfill\break

        [Note also that $\delta_N \le \delta'_N/(1-\delta'_N)$.]
\end{itemize}
In particular, if $N \ge (\log 3)/\gamma$, we have
$\Delta_N \le |a_{N-1}|$ and $\delta'_N \le |a_{N-1}|/|S_N|$.
\end{proposition}

Let us conclude by making some brief remarks about the
numerical precision that is required in intermediate stages of
the calculation based on \reff{eq_qseries}.
It turns out \cite{Sokal_infprod} that the largest term $\max\limits_n |a_n|$
can be as large in magnitude as $e^{\pi^2/12\gamma}$
(and is indeed of this order when $0 < x < 1$),
while the answer $R(t,x)$ can be as small in magnitude as $e^{-\pi^2/6\gamma}$
(and is indeed of this order when $t=1$ and $0 < x < 1$).
It is therefore necessary to maintain,
in intermediate stages of the calculation, approximately
$(\pi^2/4\gamma)/\log 10 \approx 1.07/\gamma$ digits of
working precision beyond the number of significant digits desired
in the final answer.

We used all three algorithms ---
the product \reff{def_R}, the logarithmic sum \reff{eq_logR}
and the quadratically convergent sum \reff{eq_qseries} ---
and carefully cross-checked the value of $R(t,x)$;
we also verified numerically the error bounds of Proposition~\ref{propA.1},
Corollary~\ref{corA.2} and Proposition~\ref{propA.3}.
In order to guarantee that the roundoff error is under control,
we performed all computations using {\sc Mathematica} with 
a working precision of at least 100 digits and often much more
(increasing the working precision until the answer is independent
of the precision used).

%
%
\clearpage
%
%
\begin{table}[t]
\vspace*{-1cm}
\centering
\begin{tabular}{|l|l|l|l|}
\hline\hline
 Lattice & 4th Zero   & 5th Zero   & 6th Zero   \\
\hline\hline
$ 3_{\rm F}\times 3_{\rm F}$ &   &   &   \\
$ 3_{\rm F}\times 6_{\rm F}$ &  2.552816126636  &   &   \\
$ 3_{\rm F}\times 9_{\rm F}$ &   &   &   \\
$ 3_{\rm F}\times 12_{\rm F}$ &  2.562658027317  &   &   \\
$ 3_{\rm F}\times 15_{\rm F}$ &   &   &   \\
$ 3_{\rm F}\times 18_{\rm F}$ &  2.565287184975  &   &   \\
$ 3_{\rm F}\times 21_{\rm F}$ &   &   &   \\
$ 3_{\rm F}\times 24_{\rm F}$ &  2.566507072062  &   &   \\
$ 3_{\rm F}\times 27_{\rm F}$ &   &   &   \\
$ 3_{\rm F}\times 30_{\rm F}$ &  2.567211365497  &   &   \\
\hline
$ 4_{\rm F}\times 4_{\rm F}$ &  2.604661945742  &   &   \\
$ 4_{\rm F}\times 8_{\rm F}$ &  2.618028652707  &   &   \\
$ 4_{\rm F}\times 12_{\rm F}$ &  2.618033986251  &   &   \\
$ 4_{\rm F}\times 16_{\rm F}$ &  2.618033988749  &   &   \\
$ 4_{\rm F}\times 20_{\rm F}$ &  2.618033988750  &   &   \\
$ 4_{\rm F}\times 24_{\rm F}$ &  2.618033988750  &   &   \\
$ 4_{\rm F}\times 28_{\rm F}$ &  2.618033988750  &   &   \\
$ 4_{\rm F}\times 32_{\rm F}$ &  2.618033988750  &   &   \\
$ 4_{\rm F}\times 36_{\rm F}$ &  2.618033988750  &   &   \\
$ 4_{\rm F}\times 40_{\rm F}$ &  2.618033988750  &   &   \\
\hline
$ 5_{\rm F}\times 5_{\rm F}$ &  2.618161303055  &  2.795370504128  &   \\
$ 5_{\rm F}\times 10_{\rm F}$ &  2.618033988749  &   &   \\
$ 5_{\rm F}\times 15_{\rm F}$ &  2.618033988750  &  2.947523648832  &   \\
$ 5_{\rm F}\times 20_{\rm F}$ &  2.618033988750  &   &   \\
$ 5_{\rm F}\times 25_{\rm F}$ &  2.618033988750  &  2.968180058756  &   \\
$ 5_{\rm F}\times 30_{\rm F}$ &  2.618033988750  &   &   \\
$ 5_{\rm F}\times 35_{\rm F}$ &  2.618033988750  &  2.976760450197  &   \\
$ 5_{\rm F}\times 40_{\rm F}$ &  2.618033988750  &   &   \\
$ 5_{\rm F}\times 45_{\rm F}$ &  2.618033988750  &  2.981534673779  &   \\
$ 5_{\rm F}\times 50_{\rm F}$ &  2.618033988750  &   &   \\
\hline
$ 6_{\rm F}\times 6_{\rm F}$ &  2.618033979731  &   &   \\
$ 6_{\rm F}\times 12_{\rm F}$ &  2.618033988750  &  3.001429148693  &  3.054848659601  \\
$ 6_{\rm F}\times 18_{\rm F}$ &  2.618033988750  &  3.000001523178  &  3.100527321592  \\
$ 6_{\rm F}\times 24_{\rm F}$ &  2.618033988750  &  3.000000001785  &  3.118151997375  \\
$ 6_{\rm F}\times 30_{\rm F}$ &  2.618033988750  &  3.000000000002  &  3.127749140385  \\
$ 6_{\rm F}\times 36_{\rm F}$ &  2.618033988750  &  3.000000000000  &  3.133811079422  \\
$ 6_{\rm F}\times 42_{\rm F}$ &  2.618033988750  &  3.000000000000  &  3.137993327670  \\
$ 6_{\rm F}\times 48_{\rm F}$ &  2.618033988750  &  3.000000000000  &  3.141054810628  \\
$ 6_{\rm F}\times 54_{\rm F}$ &  2.618033988750  &  3.000000000000  &  3.143393623378  \\
$ 6_{\rm F}\times 60_{\rm F}$ &  2.618033988750  &  3.000000000000  &  3.145239011028  \\
\hline
\hline
 Beraha &2.618033988750  &   3 & 3.246979603717  \\
\hline
\end{tabular}
\caption{
   Real zeros of the chromatic polynomials of finite triangular-lattice strips
   with free boundary conditions in both directions, to 12 decimal places.
   A blank means that the zero in question is absent.
   The first three real zeros $q=0,1,2$ are exact on all lattices.
   ``Beraha'' indicates the Beraha numbers $B_5 = (3+\sqrt{5})/2$, $B_6=3$,
   and $B_7$.  
}
\protect\label{table_zeros_free}
\end{table}

\clearpage
%
%
\begin{table}[t]
\centering
\begin{tabular}{|l|l|l|l|l|}
\hline\hline
 Lattice & 4th Zero   & 5th Zero   & 6th Zero   & 7th Zero   \\
\hline\hline
$ 7_{\rm F}\times 7_{\rm F}$ &  2.618033988750  &  2.978584823651  &   &   \\
$ 7_{\rm F}\times 14_{\rm F}$ &  2.618033988750  &  3.000000029690  &  3.160410975706  &   \\
$ 7_{\rm F}\times 21_{\rm F}$ &  2.618033988750  &  3.000000000000  &   &   \\
$ 7_{\rm F}\times 28_{\rm F}$ &  2.618033988750  &  3.000000000000  &  3.218685236695  &   \\
$ 7_{\rm F}\times 35_{\rm F}$ &  2.618033988750  &  3.000000000000  &   &   \\
$ 7_{\rm F}\times 42_{\rm F}$ &  2.618033988750  &  3.000000000000  &  3.236121891966  &   \\
$ 7_{\rm F}\times 49_{\rm F}$ &  2.618033988750  &  3.000000000000  &   &   \\
$ 7_{\rm F}\times 56_{\rm F}$ &  2.618033988750  &  3.000000000000  &  3.243833695579  &   \\
$ 7_{\rm F}\times 63_{\rm F}$ &  2.618033988750  &  3.000000000000  &   &   \\
$ 7_{\rm F}\times 70_{\rm F}$ &  2.618033988750  &  3.000000000000  &  3.246633282347  &   \\
$ 7_{\rm F}\times 77_{\rm F}$ &  2.618033988750  &  3.000000000000  &  3.247059872523  &  3.254369173708  \\
$ 7_{\rm F}\times 84_{\rm F}$ &  2.618033988750  &  3.000000000000  &  3.246965843358  &   \\
$ 7_{\rm F}\times 91_{\rm F}$ &  2.618033988750  &  3.000000000000  &  3.246982133140  &  3.258435734303  \\
\hline
$ 8_{\rm F}\times 8_{\rm F}$ &  2.618033988750  &  3.000359693703  &  3.095706393163  &   \\
$ 8_{\rm F}\times 16_{\rm F}$ &  2.618033988750  &  3.000000000000  &  3.229632685380  &   \\
$ 8_{\rm F}\times 24_{\rm F}$ &  2.618033988750  &  3.000000000000  &  3.246928323759  &   \\
$ 8_{\rm F}\times 32_{\rm F}$ &  2.618033988750  &  3.000000000000  &  3.246979586275  &   \\
$ 8_{\rm F}\times 40_{\rm F}$ &  2.618033988750  &  3.000000000000  &  3.246979603712  &   \\
$ 8_{\rm F}\times 48_{\rm F}$ &  2.618033988750  &  3.000000000000  &  3.246979603717  &   \\
$ 8_{\rm F}\times 56_{\rm F}$ &  2.618033988750  &  3.000000000000  &  3.246979603717  &   \\
$ 8_{\rm F}\times 64_{\rm F}$ &  2.618033988750  &  3.000000000000  &  3.246979603717  &   \\
$ 8_{\rm F}\times 72_{\rm F}$ &  2.618033988750  &  3.000000000000  &  3.246979603717  &   \\
$ 8_{\rm F}\times 80_{\rm F}$ &  2.618033988750  &  3.000000000000  &  3.246979603717  &   \\
$ 8_{\rm F}\times 88_{\rm F}$ &  2.618033988750  &  3.000000000000  &  3.246979603717  &   \\
$ 8_{\rm F}\times 96_{\rm F}$ &  2.618033988750  &  3.000000000000  &  3.246979603717  &   \\
\hline
$ 9_{\rm F}\times 9_{\rm F}$ &  2.618033988750  &  2.999999518372  &   &   \\
$ 9_{\rm F}\times 18_{\rm F}$ &  2.618033988750  &  3.000000000000  &  3.246969773686  &   \\
$ 9_{\rm F}\times 27_{\rm F}$ &  2.618033988750  &  3.000000000000  &  3.246979603720  &  3.342943823308  \\
$ 9_{\rm F}\times 36_{\rm F}$ &  2.618033988750  &  3.000000000000  &  3.246979603717  &   \\
$ 9_{\rm F}\times 45_{\rm F}$ &  2.618033988750  &  3.000000000000  &  3.246979603717  &  3.374646284957  \\
$ 9_{\rm F}\times 54_{\rm F}$ &  2.618033988750  &  3.000000000000  &  3.246979603717  &   \\
$ 9_{\rm F}\times 63_{\rm F}$ &  2.618033988750  &  3.000000000000  &  3.246979603717  &  3.387946181123  \\
$ 9_{\rm F}\times 72_{\rm F}$ &  2.618033988750  &  3.000000000000  &  3.246979603717  &   \\
$ 9_{\rm F}\times 81_{\rm F}$ &  2.618033988750  &  3.000000000000  &  3.246979603717  &  3.395349738491  \\
$ 9_{\rm F}\times 90_{\rm F}$ &  2.618033988750  &  3.000000000000  &  3.246979603717  &   \\
\hline
\hline
 Beraha &2.618033988750  &   3 & 3.246979603717  & 3.414213562373  \\
\hline
\end{tabular}
\caption{
   Real zeros of the chromatic polynomials of finite triangular-lattice strips
   with free boundary conditions in both directions, to 12 decimal places.
   We use the same notation as in Table~\protect\ref{table_zeros_free}.
}
\protect\label{table_zeros_free_bis}
\end{table}

\clearpage
%
%
\begin{table}[t]
\vspace*{-1cm}
\centering
\scriptsize
\begin{tabular}{|l|l|l|l|l|l|}
\hline\hline
 Lattice & 4th Zero   & 5th Zero   & 6th Zero   & 7th Zero   & 8th Zero   \\
\hline\hline
$ 4_{\rm P}\times 4_{\rm F}$ &  2.617986010522  &  3   &  3.465246100723  &   &   \\
$ 4_{\rm P}\times 8_{\rm F}$ &  2.618033988740  &  3   &  3.475055224065  &   &   \\
$ 4_{\rm P}\times 12_{\rm F}$ &  2.618033988750  &  3   &  3.477452996799  &   &   \\
$ 4_{\rm P}\times 16_{\rm F}$ &  2.618033988750  &  3   &  3.478536268722  &   &   \\
$ 4_{\rm P}\times 20_{\rm F}$ &  2.618033988750  &  3   &  3.479153472532  &   &   \\
$ 4_{\rm P}\times 24_{\rm F}$ &  2.618033988750  &  3   &  3.479552148708  &   &   \\
$ 4_{\rm P}\times 28_{\rm F}$ &  2.618033988750  &  3   &  3.479830901859  &   &   \\
$ 4_{\rm P}\times 32_{\rm F}$ &  2.618033988750  &  3   &  3.480036768366  &   &   \\
$ 4_{\rm P}\times 36_{\rm F}$ &  2.618033988750  &  3   &  3.480195030232  &   &   \\
$ 4_{\rm P}\times 40_{\rm F}$ &  2.618033988750  &  3   &  3.480320488501  &   &   \\
\hline
$ 5_{\rm P}\times 5_{\rm F}$ &  2.618033990394  &  3   &   &   &   \\
$ 5_{\rm P}\times 10_{\rm F}$ &  2.618033988750  &  3   &  3.196843987850  &   &   \\
$ 5_{\rm P}\times 15_{\rm F}$ &  2.618033988750  &  3   &   &   &   \\
$ 5_{\rm P}\times 20_{\rm F}$ &  2.618033988750  &  3   &  3.202699178454  &   &   \\
$ 5_{\rm P}\times 25_{\rm F}$ &  2.618033988750  &  3   &   &   &   \\
$ 5_{\rm P}\times 30_{\rm F}$ &  2.618033988750  &  3   &  3.204333275156  &   &   \\
$ 5_{\rm P}\times 35_{\rm F}$ &  2.618033988750  &  3   &   &   &   \\
$ 5_{\rm P}\times 40_{\rm F}$ &  2.618033988750  &  3   &  3.205100311429  &   &   \\
$ 5_{\rm P}\times 45_{\rm F}$ &  2.618033988750  &  3   &   &   &   \\
$ 5_{\rm P}\times 50_{\rm F}$ &  2.618033988750  &  3   &  3.205545558020  &   &   \\
\hline
$ 6_{\rm P}\times 6_{\rm F}$ &  2.618033988750  &  3.001033705947  &  3.125892136302  &   &   \\
$ 6_{\rm P}\times 12_{\rm F}$ &  2.618033988750  &  3.000000003803  &  3.198900652620  &   &   \\
$ 6_{\rm P}\times 18_{\rm F}$ &  2.618033988750  &  3.000000000000  &  3.217111179820  &   &   \\
$ 6_{\rm P}\times 24_{\rm F}$ &  2.618033988750  &  3.000000000000  &  3.225649637432  &   &   \\
$ 6_{\rm P}\times 30_{\rm F}$ &  2.618033988750  &  3.000000000000  &  3.230657835149  &   &   \\
$ 6_{\rm P}\times 36_{\rm F}$ &  2.618033988750  &  3.000000000000  &  3.233968503481  &   &   \\
$ 6_{\rm P}\times 42_{\rm F}$ &  2.618033988750  &  3.000000000000  &  3.236327213212  &   &   \\
$ 6_{\rm P}\times 48_{\rm F}$ &  2.618033988750  &  3.000000000000  &  3.238096251767  &   &   \\
$ 6_{\rm P}\times 54_{\rm F}$ &  2.618033988750  &  3.000000000000  &  3.239473538415  &   &   \\
$ 6_{\rm P}\times 60_{\rm F}$ &  2.618033988750  &  3.000000000000  &  3.240576619481  &   &   \\
$ 6_{\rm P}\times 66_{\rm F}$ &  2.618033988750  &  3.000000000000  &  3.241479828709  &   &   \\
$ 6_{\rm P}\times 72_{\rm F}$ &  2.618033988750  &  3.000000000000  &  3.242232528364  &   &   \\
$ 6_{\rm P}\times 78_{\rm F}$ &  2.618033988750  &  3.000000000000  &  3.242868805497  &   &   \\
$ 6_{\rm P}\times 84_{\rm F}$ &  2.618033988750  &  3.000000000000  &  3.243412961909  &   &   \\
$ 6_{\rm P}\times 90_{\rm F}$ &  2.618033988750  &  3.000000000000  &  3.243882786313  &   &   \\
\hline
$ 7_{\rm P}\times 7_{\rm F}$ &  2.618033988750  &  3   &  3.247001348628  &  3.404690481534  &   \\
$ 7_{\rm P}\times 14_{\rm F}$ &  2.618033988750  &  3   &  3.246979603718  &  3.414217072295  &  3.458917430738  \\
$ 7_{\rm P}\times 21_{\rm F}$ &  2.618033988750  &  3   &  3.246979603717  &  3.414213561735  &   \\
$ 7_{\rm P}\times 28_{\rm F}$ &  2.618033988750  &  3   &  3.246979603717  &  3.414213562373  &  3.470544903913  \\
$ 7_{\rm P}\times 35_{\rm F}$ &  2.618033988750  &  3   &  3.246979603717  &  3.414213562373  &   \\
$ 7_{\rm P}\times 42_{\rm F}$ &  2.618033988750  &  3   &  3.246979603717  &  3.414213562373  &  3.473634831556  \\
$ 7_{\rm P}\times 49_{\rm F}$ &  2.618033988750  &  3   &  3.246979603717  &  3.414213562373  &   \\
$ 7_{\rm P}\times 56_{\rm F}$ &  2.618033988750  &  3   &  3.246979603717  &  3.414213562373  &  3.475070205361  \\
$ 7_{\rm P}\times 63_{\rm F}$ &  2.618033988750  &  3   &  3.246979603717  &  3.414213562373  &   \\
$ 7_{\rm P}\times 70_{\rm F}$ &  2.618033988750  &  3   &  3.246979603717  &  3.414213562373  &  3.475899672990  \\
\hline
$ 8_{\rm P}\times 8_{\rm F}$ &  2.618033988750  &  3   &  3.246979601854  &   &   \\
$ 8_{\rm P}\times 16_{\rm F}$ &  2.618033988750  &  3   &  3.246979603717  &  3.414214415195  &  3.472683999084  \\
$ 8_{\rm P}\times 24_{\rm F}$ &  2.618033988750  &  3   &  3.246979603717  &  3.414213562387  &  3.488644630018  \\
$ 8_{\rm P}\times 32_{\rm F}$ &  2.618033988750  &  3   &  3.246979603717  &  3.414213562373  &  3.495735217349  \\
$ 8_{\rm P}\times 40_{\rm F}$ &  2.618033988750  &  3   &  3.246979603717  &  3.414213562373  &  3.499773262291  \\
$ 8_{\rm P}\times 48_{\rm F}$ &  2.618033988750  &  3   &  3.246979603717  &  3.414213562373  &  3.502387969424  \\
$ 8_{\rm P}\times 56_{\rm F}$ &  2.618033988750  &  3   &  3.246979603717  &  3.414213562373  &  3.504221641913  \\
$ 8_{\rm P}\times 64_{\rm F}$ &  2.618033988750  &  3   &  3.246979603717  &  3.414213562373  &  3.505579831565  \\
$ 8_{\rm P}\times 72_{\rm F}$ &  2.618033988750  &  3   &  3.246979603717  &  3.414213562373  &  3.506626776159  \\
$ 8_{\rm P}\times 80_{\rm F}$ &  2.618033988750  &  3   &  3.246979603717  &  3.414213562373  &  3.507458740757  \\
\hline
\hline
 Beraha &2.618033988750  &   3 & 3.246979603717  & 3.414213562373  & 3.532088886238  \\
\hline
\end{tabular}
\caption{
   Real zeros of the chromatic polynomials of finite triangular-lattice strips
   with periodic boundary conditions in the transverse direction
   and free boundary conditions in the longitudinal direction,
   to 12 decimal places. 
   A blank means that the zero in question is absent.
   The first three real zeros $q=0,1,2$ are exact on all lattices.
   ``Beraha'' indicates the Beraha numbers $B_5 = (3+\sqrt{5})/2$, $B_6=3$,
   $B_7$, $B_8=2+\sqrt{2}$, and $B_9$.
}
\protect\label{table_zeros_cyl}
\end{table}

\clearpage
%
%
\begin{table}[t]
\hspace*{-2cm}
\scriptsize
\begin{tabular}{|l|l|l|l|l|l|l|l|}
\hline\hline
 Lattice & 4th Zero   & 5th Zero   & 6th Zero   & 7th Zero   & 8th Zero   & 9th Zero   & 10th Zero   \\
\hline\hline
$ 9_{\rm P}\times 9_{\rm F}$ &  2.618033988750  &  3.000000000000  &  3.246980644227  &  3.382733076359  &   &   &   \\
$ 9_{\rm P}\times 18_{\rm F}$ &  2.618033988750  &  3.000000000000  &  3.246979603717  &  3.414215827400  &  3.467483864312  &   &   \\
$ 9_{\rm P}\times 27_{\rm F}$ &  2.618033988750  &  3.000000000000  &  3.246979603717  &  3.414213562359  &   &   &   \\
$ 9_{\rm P}\times 36_{\rm F}$ &  2.618033988750  &  3.000000000000  &  3.246979603717  &  3.414213562373  &  3.499429426359  &   &   \\
$ 9_{\rm P}\times 45_{\rm F}$ &  2.618033988750  &  3.000000000000  &  3.246979603717  &  3.414213562373  &   &   &   \\
$ 9_{\rm P}\times 54_{\rm F}$ &  2.618033988750  &  3.000000000000  &  3.246979603717  &  3.414213562373  &  3.508825024982  &   &   \\
$ 9_{\rm P}\times 63_{\rm F}$ &  2.618033988750  &  3.000000000000  &  3.246979603717  &  3.414213562373  &   &   &   \\
$ 9_{\rm P}\times 72_{\rm F}$ &  2.618033988750  &  3.000000000000  &  3.246979603717  &  3.414213562373  &  3.513393802382  &   &   \\
$ 9_{\rm P}\times 81_{\rm F}$ &  2.618033988750  &  3.000000000000  &  3.246979603717  &  3.414213562373  &   &   &   \\
$ 9_{\rm P}\times 90_{\rm F}$ &  2.618033988750  &  3.000000000000  &  3.246979603717  &  3.414213562373  &  3.516109505154  &   &   \\
\hline
$ 10_{\rm P}\times 10_{\rm F}$ &  2.618033988750  &  3   &  3.246979603717  &  3.414213601215  &  3.522072913706  &   &   \\
$ 10_{\rm P}\times 20_{\rm F}$ &  2.618033988750  &  3   &  3.246979603717  &  3.414213562373  &  3.532088885496  &   &   \\
$ 10_{\rm P}\times 30_{\rm F}$ &  2.618033988750  &  3   &  3.246979603717  &  3.414213562373  &  3.532088886238  &   &   \\
$ 10_{\rm P}\times 40_{\rm F}$ &  2.618033988750  &  3   &  3.246979603717  &  3.414213562373  &  3.532088886238  &   &   \\
$ 10_{\rm P}\times 50_{\rm F}$ &  2.618033988750  &  3   &  3.246979603717  &  3.414213562373  &  3.532088886238  &   &   \\
$ 10_{\rm P}\times 60_{\rm F}$ &  2.618033988750  &  3   &  3.246979603717  &  3.414213562373  &  3.532088886238  &  3.618274945403  &  3.620352727045  \\
$ 10_{\rm P}\times 70_{\rm F}$ &  2.618033988750  &  3   &  3.246979603717  &  3.414213562373  &  3.532088886238  &  3.618041117772  &  3.623157797032  \\
$ 10_{\rm P}\times 80_{\rm F}$ &  2.618033988750  &  3   &  3.246979603717  &  3.414213562373  &  3.532088886238  &  3.618034257877  &  3.624885448640  \\
$ 10_{\rm P}\times 90_{\rm F}$ &  2.618033988750  &  3   &  3.246979603717  &  3.414213562373  &  3.532088886238  &  3.618033998995  &  3.626138707957  \\
$ 10_{\rm P}\times 100_{\rm F}$ &  2.618033988750  &  3   &  3.246979603717  &  3.414213562373  &  3.532088886238  &  3.618033989140  &  3.627101535574  \\
\hline
$ 11_{\rm P}\times 11_{\rm F}$ &  2.618033988750  &  3   &  3.246979603717  &  3.414213539527  &   &   &   \\
$ 11_{\rm P}\times 22_{\rm F}$ &  2.618033988750  &  3   &  3.246979603717  &  3.414213562373  &  3.532088885575  &   &   \\
$ 11_{\rm P}\times 33_{\rm F}$ &  2.618033988750  &  3   &  3.246979603717  &  3.414213562373  &  3.532088886238  &  3.608601511861  &   \\
$ 11_{\rm P}\times 44_{\rm F}$ &  2.618033988750  &  3   &  3.246979603717  &  3.414213562373  &  3.532088886238  &   &   \\
$ 11_{\rm P}\times 55_{\rm F}$ &  2.618033988750  &  3   &  3.246979603717  &  3.414213562373  &  3.532088886238  &  3.617975980728  &   \\
$ 11_{\rm P}\times 66_{\rm F}$ &  2.618033988750  &  3   &  3.246979603717  &  3.414213562373  &  3.532088886238  &  3.618034451624  &  3.627344614702  \\
$ 11_{\rm P}\times 77_{\rm F}$ &  2.618033988750  &  3   &  3.246979603717  &  3.414213562373  &  3.532088886238  &  3.618033985241  &   \\
$ 11_{\rm P}\times 88_{\rm F}$ &  2.618033988750  &  3   &  3.246979603717  &  3.414213562373  &  3.532088886238  &  3.618033988777  &  3.631979435582  \\
$ 11_{\rm P}\times 99_{\rm F}$ &  2.618033988750  &  3   &  3.246979603717  &  3.414213562373  &  3.532088886238  &  3.618033988750  &   \\
$ 11_{\rm P}\times 110_{\rm F}$ &  2.618033988750  &  3   &  3.246979603717  &  3.414213562373  &  3.532088886238  &  3.618033988750  &  3.634574709990  \\
\hline
$ 12_{\rm P}\times 12_{\rm F}$ &  2.618033988750  &  3.000000000000  &  3.246979603717  &  3.414213593041  &  3.511032635472  &   &   \\
$ 12_{\rm P}\times 24_{\rm F}$ &  2.618033988750  &  3.000000000000  &  3.246979603717  &  3.414213562373  &  3.532088885001  &   &   \\
$ 12_{\rm P}\times 36_{\rm F}$ &  2.618033988750  &  3.000000000000  &  3.246979603717  &  3.414213562373  &  3.532088886238  &   &   \\
$ 12_{\rm P}\times 48_{\rm F}$ &  2.618033988750  &  3.000000000000  &  3.246979603717  &  3.414213562373  &  3.532088886238  &   &   \\
$ 12_{\rm P}\times 60_{\rm F}$ &  2.618033988750  &  3.000000000000  &  3.246979603717  &  3.414213562373  &  3.532088886238  &   &   \\
$ 12_{\rm P}\times 72_{\rm F}$ &  2.618033988750  &  3.000000000000  &  3.246979603717  &  3.414213562373  &  3.532088886238  &  3.618040035384  &  3.624320958404  \\
$ 12_{\rm P}\times 84_{\rm F}$ &  2.618033988750  &  3.000000000000  &  3.246979603717  &  3.414213562373  &  3.532088886238  &  3.618034035926  &  3.627419917635  \\
$ 12_{\rm P}\times 96_{\rm F}$ &  2.618033988750  &  3.000000000000  &  3.246979603717  &  3.414213562373  &  3.532088886238  &  3.618033989120  &  3.629588218978  \\
$ 12_{\rm P}\times 108_{\rm F}$ &  2.618033988750  &  3.000000000000  &  3.246979603717  &  3.414213562373  &  3.532088886238  &  3.618033988753  &  3.631215401061  \\
$ 12_{\rm P}\times 120_{\rm F}$ &  2.618033988750  &  3.000000000000  &  3.246979603717  &  3.414213562373  &  3.532088886238  &  3.618033988750  &  3.632487726562  \\
\hline
\hline
 Beraha &2.618033988750  &   3 & 3.246979603717  & 3.414213562373  & 3.532088886238  & 3.618033988750  & 3.682507065662  \\
\hline
\end{tabular}
\caption{
   Real zeros of the chromatic polynomials of finite triangular-lattice strips
   with periodic boundary conditions in the transverse direction
   and free boundary conditions in the longitudinal direction,
   to 12 decimal places. We use the same notation as in 
   Table~\protect\ref{table_zeros_cyl}. 
}
\protect\label{table_zeros_cyl_bis}
\end{table}

%
%
\begin{table}[t]
\hspace*{-2cm}
\scriptsize
\begin{tabular}{|l|l|l|l|l|l|l|l|}
\hline\hline
 Lattice & 4th Zero   & 5th Zero   & 6th Zero   & 7th Zero   & 8th Zero   & 9th Zero   & 10th Zero   \\
\hline\hline
$ 4_{\rm Z}\times 4_{\rm F}$ &  2.485072022789  &  2.527537649962  &  2.596617094656  &   &   &   &   \\
$ 4_{\rm Z}\times 8_{\rm F}$ &  2.499965989337  &  2.500034085574  &  2.618031965217  &   &   &   &   \\
$ 4_{\rm Z}\times 12_{\rm F}$ &  2.499999937358  &  2.500000062643  &  2.618033988527  &   &   &   &   \\
$ 4_{\rm Z}\times 16_{\rm F}$ &  2.499999999885  &  2.500000000115  &  2.618033988750  &   &   &   &   \\
$ 4_{\rm Z}\times 20_{\rm F}$ &  2.500000000000  &  2.500000000000  &  2.618033988750  &   &   &   &   \\
$ 4_{\rm Z}\times 24_{\rm F}$ &  2.500000000000  &  2.500000000000  &  2.618033988750  &   &   &   &   \\
$ 4_{\rm Z}\times 28_{\rm F}$ &  2.500000000000  &  2.618033988750  &   &   &   &   &   \\
$ 4_{\rm Z}\times 32_{\rm F}$ &  2.500000000000  &  2.618033988750  &   &   &   &   &   \\
$ 4_{\rm Z}\times 36_{\rm F}$ &  2.500000000000  &  2.618033988750  &   &   &   &   &   \\
$ 4_{\rm Z}\times 40_{\rm F}$ &  2.500000000000  &  2.618033988750  &   &   &   &   &   \\
\hline
$ 6_{\rm Z}\times 6_{\rm F}$ &  2.618033988528  &   &   &   &   &   &   \\
$ 6_{\rm Z}\times 12_{\rm F}$ &  2.618033988750  &  3.000017186720  &  3.117917986708  &   &   &   &   \\
$ 6_{\rm Z}\times 18_{\rm F}$ &  2.618033988750  &  3.000000004191  &  3.141107899326  &   &   &   &   \\
$ 6_{\rm Z}\times 24_{\rm F}$ &  2.618033988750  &  3.000000000001  &  3.150834657646  &   &   &   &   \\
$ 6_{\rm Z}\times 30_{\rm F}$ &  2.618033988750  &  3.000000000000  &  3.156227017803  &   &   &   &   \\
$ 6_{\rm Z}\times 36_{\rm F}$ &  2.618033988750  &  2.722632835458  &  3.000000000000  &  3.159661924115  &   &   &   \\
$ 6_{\rm Z}\times 42_{\rm F}$ &  2.618033988750  &  2.722632835458  &  3.000000000000  &  3.162043675850  &   &   &   \\
$ 6_{\rm Z}\times 48_{\rm F}$ &  2.618033988750  &  2.722632835458  &  3.000000000000  &  3.163793040154  &   &   &   \\
$ 6_{\rm Z}\times 54_{\rm F}$ &  2.618033988750  &  2.722632835458  &  3.000000000000  &  3.165132700997  &   &   &   \\
$ 6_{\rm Z}\times 60_{\rm F}$ &  2.618033988750  &  2.722632835458  &  3.000000000000  &  3.166191662980  &   &   &   \\
\hline
$ 8_{\rm Z}\times 8_{\rm F}$ &  2.618033988750  &  3.000000844168  &  3.203925019292  &   &   &   &   \\
$ 8_{\rm Z}\times 16_{\rm F}$ &  2.618033988750  &  3.000000000000  &  3.246976356780  &   &   &   &   \\
$ 8_{\rm Z}\times 24_{\rm F}$ &  2.618033988750  &  3.000000000000  &  3.246979603696  &   &   &   &   \\
$ 8_{\rm Z}\times 32_{\rm F}$ &  2.618033988750  &  2.821420495535  &  3.000000000000  &  3.246979603717  &   &   &   \\
$ 8_{\rm Z}\times 40_{\rm F}$ &  2.618033988750  &  2.821420495535  &  3.000000000000  &  3.246979603717  &   &   &   \\
$ 8_{\rm Z}\times 48_{\rm F}$ &  2.618033988750  &  2.821420495535  &  3.000000000000  &  3.246979603717  &   &   &   \\
$ 8_{\rm Z}\times 56_{\rm F}$ &  2.618033988750  &  2.821420495535  &  3.000000000000  &  3.246979603717  &   &   &   \\
$ 8_{\rm Z}\times 64_{\rm F}$ &  2.618033988750  &  2.821420495535  &  3.000000000000  &  3.246979603717  &   &   &   \\
$ 8_{\rm Z}\times 72_{\rm F}$ &  2.618033988750  &  2.821420495535  &  3.000000000000  &  3.246979603717  &   &   &   \\
$ 8_{\rm Z}\times 80_{\rm F}$ &  2.618033988750  &  2.821420495535  &  3.000000000000  &  3.246979603717  &   &   &   \\
\hline
$ 10_{\rm Z}\times 10_{\rm F}$ &  2.618033988750  &  3.000000000000  &  3.246953122227  &   &   &   &   \\
$ 10_{\rm Z}\times 20_{\rm F}$ &  2.618033988750  &  3.000000000000  &  3.246979603717  &  3.416320582746  &  3.428202969384  &   &   \\
$ 10_{\rm Z}\times 30_{\rm F}$ &  2.618033988750  &  3.000000000000  &  3.246979603717  &  3.414213564771  &  3.470075808656  &   &   \\
$ 10_{\rm Z}\times 40_{\rm F}$ &  2.618033988750  &  2.873731249334  &  3.000000000000  &  3.246979603717  &  3.414213562373  &  3.484585415185  &   \\
$ 10_{\rm Z}\times 50_{\rm F}$ &  2.618033988750  &  2.873731249334  &  3.000000000000  &  3.246979603717  &  3.414213562373  &  3.492475877808  &   \\
$ 10_{\rm Z}\times 60_{\rm F}$ &  2.618033988750  &  2.873731249334  &  3.000000000000  &  3.246979603717  &  3.414213562373  &  3.497477600415  &   \\
$ 10_{\rm Z}\times 70_{\rm F}$ &  2.618033988750  &  2.873731249334  &  3.000000000000  &  3.246979603717  &  3.383128531235  &  3.414213562373  &  3.500942791087  \\
$ 10_{\rm Z}\times 80_{\rm F}$ &  2.618033988750  &  2.873731249334  &  3.000000000000  &  3.246979603717  &  3.383128531235  &  3.414213562373  &  3.503489509874  \\
$ 10_{\rm Z}\times 90_{\rm F}$ &  2.618033988750  &  2.873731249334  &  3.000000000000  &  3.246979603717  &  3.383128531235  &  3.414213562373  &  3.505442176204  \\
$ 10_{\rm Z}\times 100_{\rm F}$ &  2.618033988750  &  2.873731249334  &  3.000000000000  &  3.246979603717  &  3.383128531235  &  3.414213562373  &  3.506987965042  \\
\hline
\hline
 Beraha &2.618033988750  &   3 & 3.246979603717  & 3.414213562373  & 3.532088886238  & 3.618033988750  & 3.682507065662  \\
\hline
\end{tabular}
\caption{
   Real zeros of the chromatic polynomials of finite triangular-lattice strips
   with zig-zag boundary conditions, 
   to 12 decimal places. We use the same notation as in
   Table~\protect\ref{table_zeros_cyl}.
}
\protect\label{table_zeros_zig}
\end{table}

\clearpage
%
%
\begin{table}
\small
\hspace*{-1.5cm}
\centering
\begin{tabular}{|r||c|c|c|c|c|l|l|l||c|c|}
\cline{2-11}
\multicolumn{1}{c||}{\mbox{}}&
\multicolumn{8}{|c||}{Eigenvalue-Crossing Curves ${\cal B}$} &
\multicolumn{2}{|c|}{Isolated Points}\\
\hline\hline
Lattice      & \# C & \# E & \# T & \# D & \# ER & 
\multicolumn{1}{|c|}{$\min \real q$} & 
\multicolumn{1}{|c|}{$q_0$}& 
\multicolumn{1}{|c||}{$\max \real q$} & \# RI& \# CI   \\
\hline\hline
$2_{\rm F}$  &      &      &                
                                  &      &       &                & 
                    &                            &  2   &    0    \\
$3_{\rm F}$  &   3  &   6  & \phantom{1}0   
                                  &   0  &   0   & $\phantom{-}1.20474$& 
$2.56984$           & $3.40223$                  &  3   &    0    \\
$4_{\rm F}$  &   2  &   6  & \phantom{1}2 
                                  &   0  &   0   & $\phantom{-}0.81647$& 
$2.75925\pm 0.15444\,i^*$  & $3.63983$           &  4   &    0    \\
$5_{\rm F}$  &   1  &   6  & 12   &   0  &   4   & $\phantom{-}0.55862$&
$3$                        & $3.77830$           &  4   &    0    \\
$6_{\rm F}$  &   1  &   6  & \phantom{1}4     
                                  &   0  &   0   & $\phantom{-}0.37963$&
$3.16093$                  & $3.86641$           &  5   &    0    \\
$7_{\rm F}$  &  $1^\dagger$
                    &  $6^\dagger$  
                           & \phantom{1}$4^\dagger$
                                  &   0  &  $0^\dagger$   
                                                 & $\phantom{-}0.25054$&
$3.27640$                  & $3.92580$           &  6   &    0    \\
$8_{\rm F}$  &  $1^\dagger$  
                    &  $6^\dagger$  
                           & \phantom{1}$4^\dagger$     
                                  &   0  &  $0^\dagger$   
                                                 & $\phantom{-}0.13343$& 
$3.36106$                  & $3.96756$           &  6   &    0    \\
$9_{\rm F}$  &      &      & \phantom{1} 
                                  &      &       & $\phantom{-}       $&
$3.42513$                  & $       $           &  7   &    0    \\
\hline\hline
$2_{\rm P}$  &      &      &                
                                  &      &       &                & 
                    &                            &  2   &    0    \\
$3_{\rm P}$  &      &      &                
                                  &      &       &                & 
                    &             &  3   &    0    \\
$4_{\rm P}$  &   3  &   4  &  \phantom{1}0   
                                  &   0  &   1   & $\phantom{-}1.37053$&
$3.48141$                        & $4$                 &  5   &    0    \\
$5_{\rm P}$  &   3  &   4  &  \phantom{1}0    
                                  &   0  &   1   & $\phantom{-}0.47725$&
$3.20722$                  & $3.87699$           &  5   &    0    \\
$6_{\rm P}$  &   1  &   4  &  \phantom{1}2
                                  &   0  &   0   & $\phantom{-}0.02077$&
$3.25242$                  & $4.28386$           &  6   &    0    \\
$7_{\rm P}$  &   1  &   4  &  \phantom{1}4 
                           &   0  &   1   & $-0.22792$     &
$3.47900$                  & $3.99964$           &  7   &    0    \\
$8_{\rm P}$  &   1  &   4  &  \phantom{1}2
                           &   0  &   0  &  $-0.37137$     & 
$3.51477$                  & $4.04970$           &  7   &    0    \\
$9_{\rm P}$  &   1  &   4  &  \phantom{1}2
                           &   0  &   0  & $-0.45760$     & 
$3.52706$                  & $4.28286$           &  7   &    0    \\
$10_{\rm P}$ &  $1^\dagger$  
                    &  $4^\dagger$
                           & \phantom{1}$2^\dagger$ 
                           &  0   &  $0^\dagger$     
                                         & $-0.51081$     &
$3.63483$                  & $4.12341$           &  9   &    0    \\
$11_{\rm P}$ &  $1^\dagger$
                    &  $4^\dagger$
                           & \phantom{1}$2^\dagger$ 
                           &  0   &  $0^\dagger$
                                         & $-0.54399$&
$3.64414$                  & $4.15609$           &  9   &    0    \\
$12_{\rm P}$ &      &      & \phantom{1}$2^\dagger$
                                  &      &       & $\phantom{-}       $&
$3.64317$                  & $       $           &  9   &    0    \\
\hline\hline
$2_{\rm Z}$  &      &      &
                                  &      &       &                &
                    &                            &  2   &    0    \\
$4_{\rm Z}$  &   2  &   4  & \phantom{1}0
                                  &   0  &   0   & $\phantom{-}2.09914$&
 $2.73717 \pm 0.17233\,i^*$         & $4.00485$           &  5   &    0    \\
$6_{\rm Z}$  &   1  &   4  & \phantom{1}2
                                  &   0  &   0   & $\phantom{-}0.36185$&
$3.17526$                  & $4.25895$           &  6   &    1    \\
$8_{\rm Z}$  &   $1^\dagger$
                    &   $4^\dagger$
                           & \phantom{1}$2^\dagger$
                           & 0    &   $0^\dagger$
                                                 & $-0.21435$&
$3.39410$                  & $4.28991$           &  7   &   $1^\dagger$\\
$10_{\rm Z}$ &      &      & \phantom{1} 
                                  &      &       & $\phantom{-}       $&
$3.52044$                  & $       $           &  9   &         \\
\hline\hline
\end{tabular}

\vspace{1cm}
\caption{
   Summary of qualitative results for the eigenvalue-crossing curves $\scrb$
   and for the isolated limiting points of zeros.
   For each triangular-lattice strip considered in this paper,
   we give the number of connected components of $\scrb$ (\# C),
   the number of endpoints (\# E),
   the number of T points (\# T),
   the number of double points (\# D),
   and the number of enclosed regions (\# ER);
   we also give the minimum value of $\real q$ on $\scrb$,
   the smallest value $q_0$ where $\scrb$ intersects the real axis
   (${}^*$ denotes an almost-crossing),
   and the maximum value of $\real q$ on $\scrb$.
   We also report the number of real isolated limiting points of zeros
   (\# RI) 
   and the number of complex-conjugate pairs of isolated limiting points
   (\# CI).
   The symbol $^\dagger$ indicates uncertain results.
}
\protect\label{table_summary}
\end{table}

\clearpage
%
%
\begin{table}
\centering
\begin{tabular}{|r|r|r|r|r|}
\hline
\multicolumn{1}{|c|}{$m$}         &
\multicolumn{1}{|c|}{TriCyl($m$)} &
\multicolumn{1}{|c|}{\# VA}       & 
\multicolumn{1}{|c|}{TriCyl${}'$($m$)}& 
\multicolumn{1}{|c|}{SqCyl($m$)}  \\
\hline\hline
 1  &    1  &    0 &   1 &    1 \\
 2  &    1  &    0 &   1 &    1 \\
 3  &    1  &    0 &   1 &    1 \\
 4  &    2  &    0 &   2 &    2 \\
 5  &    2  &    0 &   2 &    2 \\
 6  &    5  &    0 &   5 &    5 \\
 7  &    6  &    0 &   6 &    6 \\
 8  &   15  &    2 &  13 &   14 \\
 9  &   28  &   12 &  16 &   22 \\
10  &   67  &   32 &  35 &   51 \\
11  &  145  &  100 &  45 &   95 \\
12  &  368  &  272 &  96 &  232 \\
13  &  870  &  $742^*$ & $126^*$ &  498 \\
14  & 2211  & $1940^*$ & $267^*$ & 1239 \\
\hline\hline
\end{tabular}
\caption{
   Transfer-matrix dimensions for a
   triangular-lattice strip of width $m$ and cylindrical boundary conditions. 
   For each value of the strip 
   width $m$ we give the dimension of the transfer matrix [TriCyl($m$)],  
   the number of vanishing amplitudes (\# VA), and the {\em effective}\/
   dimension of the transfer matrix [TriCyl${}'$($m$) = TriCyl($m$) $-$ \# VA].
   For comparison, we also give the dimensionality of the transfer matrix
   for a square-lattice
   strip of width $m$ and cylindrical boundary conditions [SqCyl($m$)]. 
   The values of TriCyl($m$) and SqCyl($m$) were obtained in 
   \protect\cite{transfer1,Tutte_sq}. 
   An asterisk denotes {\em conjectured}\/ results.
}
\protect\label{table_dimensions}
\end{table}

\clearpage
%
%
\def\kk{\phantom{1}}

\begin{table}
\centering
\begin{tabular}{|c|c|r|r|l|l|r|}
\hline
\multicolumn{1}{|c|}{Curve}            &
\multicolumn{1}{|c|}{Point}            &
\multicolumn{1}{|c|}{$\Arg p_{\rm Fit}/\pi$}&
\multicolumn{1}{|c|}{$\Arg p/\pi$}      &
\multicolumn{1}{|c|}{``discrepancy''}   & 
\multicolumn{1}{|c|}{$\epsilon$}        & 
\multicolumn{1}{|c|}{$\theta/\pi$}     \\
\hline\hline
 $C_1$   &G &$ 1.0000000(3)$ &$1\kk\kk\kk$ 
                                       &$0.00000005$ &$0.0000002$&$ 1/6\kk\kk$\\
         &H &$-0.4999999(3)$ &$-1/2\kk$&$0.0000001$  &$0.000001$ &$ 2/3\kk\kk$\\
\hline
 $C_2$   &I &$ 0.5000001(3)$ &$1/2\kk$ &$0.0000001 $ &$0.000001$ &$ 2/9\kk\kk$\\
         &J &$ 0.4285703(3)$ &$3/7\kk$ &$0.000008  $ &$0.0006$   &$ 7/30\kk$\\
\hline
 $C_3$   &L &$-0.2500003(3)$ &$-1/4\kk$&$0.000001 $  &$0.00004$  &$ 4/9\kk\kk$\\
         &K &$-0.1999993(3)$ &$-1/5\kk$&$0.000004 $  &$0.0001$   &$ 5/12\kk$\\
\hline
 $C_4$   &P &$ 0.2727311(7)$ &$  3/11$ &$0.00004   $ &$0.006$    &$11/42\kk$\\
         &O &$ 0.2499995(3)$ &$1/4\kk$ &$0.000002  $ &$0.00006$  &$ 4/15\kk$\\
\hline
 $C_5$   &M &$-0.1428584(3)$ &$-1/7\kk$&$0.000009 $  &$0.0006$   &$ 7/18\kk$\\
         &N &$-0.1249980(3)$ &$-1/8\kk$&$0.00002  $  &$0.001$    &$ 8/21\kk$\\
\hline
 $C_7$   &T &$-0.3846221(7)$ &$ -5/13$ &$0.00009  $  &$0.02$     &$13/24\kk$\\
         &S &$-0.3749980(3)$ &$-3/8\kk$&$0.00002  $  &$0.001$    &$ 8/15\kk$\\
\hline
 $C_8$   &  &$ 0.1250023(3)$ &$1/8\kk$ &$0.00002   $ &$0.002$    &$ 8/27\kk$\\
         &  &$ 0.1199957(7)$ &$  3/25$ &$0.0001    $ &$0.05$     &$25/84\kk$\\
\hline
 $C_{17}$&  &$-0.4285727(3)$ &$-3/7\kk$&$0.000009  $ &$0.0006$   &$7/12\kk$\\
         &  &$-0.4230929(3)$ &$-11/26$ &$0.0007    $ &$0.04$     &$26/45\kk$\\
\hline
 $C_{18}$&  &$-0.1000025(3)$ &$ -1/10$ &$0.00003   $ &$0.003$    &$10/27\kk$\\
         &  &$-0.0909045(7)$ &$ -1/11$ &$0.00005   $ &$0.007$    &$11/30\kk$\\
\hline
 $C_{19}$&  &$-0.0769285(7)$ &$ -1/13$ &$0.00007   $ &$0.01$     &$13/36\kk$\\
         &  &$-0.0714257(7)$ &$ -1/14$ &$0.00004   $ &$0.008$    &$14/39\kk$\\
\hline\hline
 Others  &  &$-0.5999993(3)$ &$-3/5\kk$&$0.000004  $ &$0.0002$   &$ 5/6\kk\kk$\\
         &  &$-0.7500002(3)$ &$-3/4\kk$&$0.00000009$ &$0.00003$  &$ 4/27\kk$\\
         &  &$ 0.7499995(3)$ &$3/4\kk$ &$0.000002  $ &$0.00006$  &$ 4/21\kk$\\
         &  &$ 0.6000006(3)$ &$3/5\kk$ &$0.000003  $ &$0.0001$   &$ 5/24\kk$\\
         &  &$ 0.2999986(7)$ &$  3/10$ &$0.00001   $ &$0.002$    &$10/39\kk$\\
         &V &$ 0.2000006(3)$ &$1/5\kk$ &$0.000003  $ &$0.0001$   &$ 5/18\kk$\\
         &  &$ 0.1428558(3)$ &$1/7\kk$ &$0.00001   $ &$0.0007$   &$ 7/24\kk$\\
         &  &$ 0.0999991(7)$ &$  1/10$ &$0.000009  $ &$0.001$    &$10/33\kk$\\
\hline\hline
\end{tabular}
\caption{
  Endpoints in the complex $p$-plane with $|p|=1$ of the curves where
  $|g_2/g_1|=1$. For each endpoint we show the ``Curve'' to which it belongs
  (see Figure~\protect\ref{Figure_Baxter_p}), 
  the estimated value of its phase $\Arg p_{\rm Fit}$ (see text), 
  the conjectured exact value $\Arg p$,
  the ``discrepancy'' [cf.\ \protect\reff{def_discrepancy}],
  the ``significance level'' $\epsilon$ [cf.\ \protect\reff{eq.number.3}],
  and the corresponding ``primary'' $\theta$ value. 
  For some selected values we also include a label (``Point'').  
  When we have a curve for which one endpoint is well-determined and the 
  other is not, we include the former point in the category ``Others'' 
  (see text). 
}
\protect\label{table_fits_p}
\end{table}

\clearpage
%
%
\def\kk{\phantom{1}}

\begin{table}
\centering
\begin{tabular}{|c|c|r|r|l|l|r|}
\hline
\multicolumn{1}{|c|}{Curve}            &
\multicolumn{1}{|c|}{Point}            &
\multicolumn{1}{|c|}{$\Arg y_{\rm Fit}/\pi$}&
\multicolumn{1}{|c|}{$\Arg y/\pi$}      &
\multicolumn{1}{|c|}{``discrepancy''}   & 
\multicolumn{1}{|c|}{$\epsilon$}        &
\multicolumn{1}{|c|}{$\theta/\pi$}     \\
\hline\hline
$D_1$   &G&$-0.8000002(3)$&$ -4/5\kk\kk $&$0.000001 $&$0.00004$&$ 1/6\kk\kk$\\
        &H&$ 0.0000000(6)$&$0\kk\kk\kk$ &$0.00000002$&$0.00000006$&$2/3\kk\kk$\\
\hline    
$D_2$   &K&$ 0.8571435(3)$&$  6/7\kk\kk $&$0.000005 $&$0.0003$  &$ 5/12\kk$\\  
        &L&$ 0.7999997(3)$&$  4/5\kk\kk $&$0.000002 $&$0.00008$ &$ 4/9\kk\kk$\\
\hline                                     
$D_4$   & &$-0.7692308(7)$&$-10/13\kk   $&$0.0000001$&$0.00003$ &$ 2/15\kk$\\  
        & &$-0.7272743(3)$&$ -8/11\kk   $&$0.00002  $&$0.002$   &$ 1/12\kk$\\  
\hline     
$D_6$   & &$-0.4615359(7)$&$ -6/13\kk   $&$0.00003  $&$0.006$   &$35/48\kk$\\  
        & &$-0.4000003(3)$&$ -2/5\kk\kk $&$0.000001 $&$0.00006$ &$13/18\kk$\\  
\hline                                     
$D_7$   & &$-0.5714279(3)$&$ -4/7\kk\kk $&$0.000004 $&$0.0003$  &$ 20/27\kk$\\  
        & &$-0.5454561(3)$&$ -6/11\kk   $&$0.00002  $&$0.002$   &$ 31/42\kk$\\  
\hline                                     
$D_9$   &N&$ 0.9230794(3)$&$ 12/13\kk   $&$0.00003  $&$0.006$   &$ 8/21\kk$\\  
        &M&$ 0.9090890(3)$&$ 10/11\kk   $&$0.00002  $&$0.003$   &$ 7/18\kk$\\  
\hline\hline
Others  & &$-0.4705939(7)$&$  -8/17\kk  $&$0.0001   $&$0.03$    &$ 46/63\kk$\\  
        & &$-0.5263122(7)$&$ -10/19\kk  $&$0.00007  $&$0.02$    &$ 53/72\kk$\\  
        & &$-0.6153817(3)$&$ -8/13\kk   $&$0.00004  $&$0.007$   &$38/51\kk$\\  
        & &$-0.7058815(7)$&$ -12/17\kk  $&$0.00001  $&$0.004$   &$  1/18\kk$\\ 
        &I&$-0.8571420(1)$&$ -6/7\kk\kk $&$0.000006 $&$0.0004$  &$ 2/9\kk\kk$\\
        &O&$-0.9090929(3)$&$ -10/11\kk  $&$0.00002  $&$0.003$   &$  4/15\kk$\\ 
        &V&$-0.9230785(7)$&$ -12/13\kk  $&$0.00002  $&$0.004$   &$  5/18\kk$\\ 
        &S&$ 0.5714292(3)$&$   4/7\kk\kk$&$0.000004 $&$0.0003$  &$  8/15\kk$\\  
        &T&$ 0.5454538(7)$&$   6/11\kk  $&$0.000008 $&$0.001$   &$ 13/24\kk$\\  
        & &$ 0.3999997(3)$&$  2/5\kk\kk $&$0.000002 $&$0.00007$ &$ 7/12\kk$\\  
        & &$ 0.2857149(7)$&$   2/7\kk\kk$&$0.000004 $&$0.0003$  &$ 11/18\kk$\\ 
\hline\hline
\end{tabular}
\caption{
  Endpoints in the complex $y$-plane with $|y|=1$ of the curves where
  $|g_3/g_1|=1$. For each endpoint we show the ``Curve'' to which it belongs
  (see Figure~\protect\ref{Figure_Baxter_y}), 
  the estimated value of its phase $\Arg y_{\rm Fit}$ (see text), 
  the conjectured exact value $\Arg p$,
  the ``discrepancy'' [cf.\ \protect\reff{def_discrepancy}],
  the ``significance level'' $\epsilon$ [cf.\ \protect\reff{eq.number.3}],
  and the corresponding ``primary'' $\theta$ value. 
  For some selected values we also include a label (``Point'').  
  When we have a curve for which one endpoint is well-determined and the
  other is not, we include the former point in the category ``Others''
  (see text).
}
\protect\label{table_fits_y}
\end{table}

\clearpage
%
%
\begin{table}
\centering
\begin{tabular}{|c|l|}
\hline
\multicolumn{1}{|c|}{$L$}     &
\multicolumn{1}{|c|}{Beraha numbers}  \\ 
\hline\hline
$2_{\rm F}$ & ${\hbox{\bf\emph{B}}_2}$ ${\hbox{\bf\emph{B}}_3}$ \\ 
$3_{\rm F}$ & ${\hbox{\bf\emph{B}}_2}$ ${\hbox{\bf\emph{B}}_3}$
    ${\hbox{\bf\emph{B}}_4}$ \\ 
$4_{\rm F}$ & ${\hbox{\bf\emph{B}}_2}$ ${\hbox{\bf\emph{B}}_3}$
    ${\hbox{\bf\emph{B}}_4}$ ${\hbox{\bf\emph{B}}_5}$ $B_6$ \\ 
$5_{\rm F}$ & ${\hbox{\bf\emph{B}}_2}$ ${\hbox{\bf\emph{B}}_3}$
    ${\hbox{\bf\emph{B}}_4}$ ${\hbox{\bf\emph{B}}_5}$ $B_6$ \\
$6_{\rm F}$ & ${\hbox{\bf\emph{B}}_2}$ ${\hbox{\bf\emph{B}}_3}$
    ${\hbox{\bf\emph{B}}_4}$ ${\hbox{\bf\emph{B}}_5}$
    ${\hbox{\bf\emph{B}}_6}$ $B_7$ \\
$7_{\rm F}$ & ${\hbox{\bf\emph{B}}_2}$ ${\hbox{\bf\emph{B}}_3}$
    ${\hbox{\bf\emph{B}}_4}$ ${\hbox{\bf\emph{B}}_5}$
    ${\hbox{\bf\emph{B}}_6}$ ${\hbox{\bf\emph{B}}_7}$ $B_8$ \\
$8_{\rm F}$ & ${\hbox{\bf\emph{B}}_2}$ ${\hbox{\bf\emph{B}}_3}$
    ${\hbox{\bf\emph{B}}_4}$ ${\hbox{\bf\emph{B}}_5}$
    ${\hbox{\bf\emph{B}}_6}$ ${\hbox{\bf\emph{B}}_7}$ $B_8$ $B_9$ \\
$9_{\rm F}$ & ${\hbox{\bf\emph{B}}_2}$ ${\hbox{\bf\emph{B}}_3}$
    ${\hbox{\bf\emph{B}}_4}$ ${\hbox{\bf\emph{B}}_5}$
    ${\hbox{\bf\emph{B}}_6}$ ${\hbox{\bf\emph{B}}_7}$ 
    ${\hbox{\bf\emph{B}}_8}$ $B_9$ $B_{10}$ \\
\hline
$2_{\rm P}$ & ${\hbox{\bf\emph{B}}_2}$ ${\hbox{\bf\emph{B}}_3}$ \\                
$3_{\rm P}$ & ${\hbox{\bf\emph{B}}_2}$ ${\hbox{\bf\emph{B}}_3}$
    ${\hbox{\bf\emph{B}}_4}$ \\
$4_{\rm P}$ & ${\hbox{\bf\emph{B}}_2}$ ${\hbox{\bf\emph{B}}_3}$
    ${\hbox{\bf\emph{B}}_4}$ ${\hbox{\bf\emph{B}}_5}$ $B_6$ \\  
$5_{\rm P}$ & ${\hbox{\bf\emph{B}}_2}$ ${\hbox{\bf\emph{B}}_3}$
    ${\hbox{\bf\emph{B}}_4}$ ${\hbox{\bf\emph{B}}_5}$
    ${\hbox{\bf\emph{B}}_6}$ \\
$6_{\rm P}$ & ${\hbox{\bf\emph{B}}_2}$ ${\hbox{\bf\emph{B}}_3}$
    ${\hbox{\bf\emph{B}}_4}$ ${\hbox{\bf\emph{B}}_5}$
    ${\hbox{\bf\emph{B}}_6}$ ${\hbox{\bf\emph{B}}_7}$ $B_{10}$ \\
$7_{\rm P}$ & ${\hbox{\bf\emph{B}}_2}$ ${\hbox{\bf\emph{B}}_3}$
    ${\hbox{\bf\emph{B}}_4}$ ${\hbox{\bf\emph{B}}_5}$
    ${\hbox{\bf\emph{B}}_6}$ ${\hbox{\bf\emph{B}}_7}$
    ${\hbox{\bf\emph{B}}_8}$ $B_{10}$\\
$8_{\rm P}$ & ${\hbox{\bf\emph{B}}_2}$ ${\hbox{\bf\emph{B}}_3}$
    ${\hbox{\bf\emph{B}}_4}$ ${\hbox{\bf\emph{B}}_5}$
    ${\hbox{\bf\emph{B}}_6}$ ${\hbox{\bf\emph{B}}_7}$
    ${\hbox{\bf\emph{B}}_8}$ $B_9$ $B_{10}$ $B_{14}$ \\
$9_{\rm P}$ & ${\hbox{\bf\emph{B}}_2}$ ${\hbox{\bf\emph{B}}_3}$
    ${\hbox{\bf\emph{B}}_4}$ ${\hbox{\bf\emph{B}}_5}$
    ${\hbox{\bf\emph{B}}_6}$ ${\hbox{\bf\emph{B}}_7}$
    ${\hbox{\bf\emph{B}}_8}$ $B_9$ $B_{10}$ $B_{14}$\\
$10_{\rm P}$& ${\hbox{\bf\emph{B}}_2}$ ${\hbox{\bf\emph{B}}_3}$
    ${\hbox{\bf\emph{B}}_4}$ ${\hbox{\bf\emph{B}}_5}$
    ${\hbox{\bf\emph{B}}_6}$ ${\hbox{\bf\emph{B}}_7}$
    ${\hbox{\bf\emph{B}}_8}$ ${\hbox{\bf\emph{B}}_9}$
    ${\hbox{\bf\emph{B}}_{10}}$ $B_{11}$ $B_{14}$ $B_{18}$ \\
$11_{\rm P}$& ${\hbox{\bf\emph{B}}_2}$ ${\hbox{\bf\emph{B}}_3}$
    ${\hbox{\bf\emph{B}}_4}$ ${\hbox{\bf\emph{B}}_5}$
    ${\hbox{\bf\emph{B}}_6}$ ${\hbox{\bf\emph{B}}_7}$
    ${\hbox{\bf\emph{B}}_8}$ ${\hbox{\bf\emph{B}}_9}$
    ${\hbox{\bf\emph{B}}_{10}}$ $B_{11}$ $B_{12}$ $B_{14}$ $B_{18}$ \\
$12_{\rm P}$& ${\hbox{\bf\emph{B}}_2}$ ${\hbox{\bf\emph{B}}_3}$
    ${\hbox{\bf\emph{B}}_4}$ ${\hbox{\bf\emph{B}}_5}$
    ${\hbox{\bf\emph{B}}_6}$ ${\hbox{\bf\emph{B}}_7}$
    ${\hbox{\bf\emph{B}}_8}$ ${\hbox{\bf\emph{B}}_9}$
    ${\hbox{\bf\emph{B}}_{10}}$ 
    $B_{11}$ $B_{12}$ $B_{13}$ $B_{14}$ $B_{18}$ $B_{22}$\\
\hline
$2_{\rm Z}$ & ${\hbox{\bf\emph{B}}_2}$ ${\hbox{\bf\emph{B}}_3}$ \\
$4_{\rm Z}$ & ${\hbox{\bf\emph{B}}_2}$ ${\hbox{\bf\emph{B}}_3}$
    ${\hbox{\bf\emph{B}}_4}$ ${\hbox{\bf\emph{B}}_5}$ $B_6$ \\
$6_{\rm Z}$ & ${\hbox{\bf\emph{B}}_2}$ ${\hbox{\bf\emph{B}}_3}$
    ${\hbox{\bf\emph{B}}_4}$ ${\hbox{\bf\emph{B}}_5}$
    ${\hbox{\bf\emph{B}}_6}$ $B_7$ \\
$8_{\rm Z}$ & ${\hbox{\bf\emph{B}}_2}$ ${\hbox{\bf\emph{B}}_3}$
    ${\hbox{\bf\emph{B}}_4}$ ${\hbox{\bf\emph{B}}_5}$
    ${\hbox{\bf\emph{B}}_6}$ ${\hbox{\bf\emph{B}}_7}$ $B_8$ $B_9$\\
$10_{\rm Z}$& ${\hbox{\bf\emph{B}}_2}$ ${\hbox{\bf\emph{B}}_3}$
    ${\hbox{\bf\emph{B}}_4}$ ${\hbox{\bf\emph{B}}_5}$
    ${\hbox{\bf\emph{B}}_6}$ ${\hbox{\bf\emph{B}}_7}$ 
    ${\hbox{\bf\emph{B}}_8}$ $B_9$ $B_{10}$ $B_{11}$ \\
\hline\hline
\end{tabular}
\caption{
  Beraha numbers $B_n$ that are zeros of $\det D(q)$.
  Those shown in boldface (resp.\ normal face)
  correspond to the vanishing of a dominant (resp.\ subdominant) amplitude. 
}
\protect\label{table_zeros_D}
\end{table}

%
%
\clearpage
%
%
\begin{figure}
  \centering
  \setlength{\unitlength}{1pt}
  \begin{picture}(300,300)(0,0)
  \Thicklines
  \drawline(0,0)(300,0)
  \drawline(0,50)(300,50)
  \drawline(0,0)(0,100)
  \drawline(50,0)(50,100)
  \drawline(100,0)(100,100)
  \drawline(150,0)(150,100)
  \drawline(200,0)(200,100)
  \drawline(250,0)(250,100)
  \drawline(300,0)(300,100)

  \drawline(0,0)(50,50)
  \drawline(0,50)(50,100)
  \drawline(100,0)(150,50)
  \drawline(100,50)(150,100)
  \drawline(200,0)(250,50)
  \drawline(200,50)(250,100)

  \drawline(100,0)(50,50)
  \drawline(100,50)(50,100)
  \drawline(200,0)(150,50)
  \drawline(200,50)(150,100)
  \drawline(300,0)(250,50)
  \drawline(300,50)(250,100)

  \put(0,0){\circle*{10}}
  \put(50,0){\circle*{10}}
  \put(100,0){\circle*{10}}
  \put(150,0){\circle*{10}}
  \put(200,0){\circle*{10}}
  \put(250,0){\circle*{10}}
  \put(300,0){\circle*{10}}
  \put(0,50){\circle*{10}}
  \put(50,50){\circle*{10}}
  \put(100,50){\circle*{10}}
  \put(150,50){\circle*{10}}
  \put(200,50){\circle*{10}}
  \put(250,50){\circle*{10}}
  \put(300,50){\circle*{10}}
  \put(0,100){\circle*{10}}
  \put(50,100){\circle*{10}}
  \put(100,100){\circle*{10}}
  \put(150,100){\circle*{10}}
  \put(200,100){\circle*{10}}
  \put(250,100){\circle*{10}}
  \put(300,100){\circle*{10}}

  \put(143,-50){\makebox(0,0)[lb]{\raisebox{0pt}[0pt][0pt]{\large (b)}}}

  \drawline(0,200)(300,200)
  \drawline(0,250)(300,250)
  \drawline(0,200)(0,300)
  \drawline(50,200)(50,300)
  \drawline(100,200)(100,300)
  \drawline(150,200)(150,300)
  \drawline(200,200)(200,300)
  \drawline(250,200)(250,300)
  \drawline(300,200)(300,300)
  \drawline(0,250)(50,200)
  \drawline(0,300)(100,200) 
  \drawline(50,300)(150,200)
  \drawline(100,300)(200,200)
  \drawline(150,300)(250,200)
  \drawline(200,300)(300,200)
  \drawline(250,300)(300,250)
  \put(0,200){\circle*{10}}
  \put(50,200){\circle*{10}}
  \put(100,200){\circle*{10}}
  \put(150,200){\circle*{10}}
  \put(200,200){\circle*{10}}
  \put(250,200){\circle*{10}}
  \put(300,200){\circle*{10}}
  \put(0,250){\circle*{10}}
  \put(50,250){\circle*{10}}
  \put(100,250){\circle*{10}}
  \put(150,250){\circle*{10}}
  \put(200,250){\circle*{10}}
  \put(250,250){\circle*{10}}
  \put(300,250){\circle*{10}}
  \put(0,300){\circle*{10}}
  \put(50,300){\circle*{10}}
  \put(100,300){\circle*{10}}
  \put(150,300){\circle*{10}}
  \put(200,300){\circle*{10}}
  \put(250,300){\circle*{10}}
  \put(300,300){\circle*{10}}

  \put(143,150){\makebox(0,0)[lb]{\raisebox{0pt}[0pt][0pt]{\large (a)}}}

  \end{picture}
  \vspace*{2cm} 
  \caption{
  Two ways of building a triangular-lattice strip using a transfer-matrix
  approach. (a) Standard method (see e.g. Ref.~\protect\cite{transfer1}).  
  (b) Alternative method (called ``zig-zag'' boundary conditions).
  }
\protect\label{Figure_transfer}
\end{figure}

\clearpage
%
%
\begin{figure}
  \centering
  \epsfxsize=400pt\epsffile{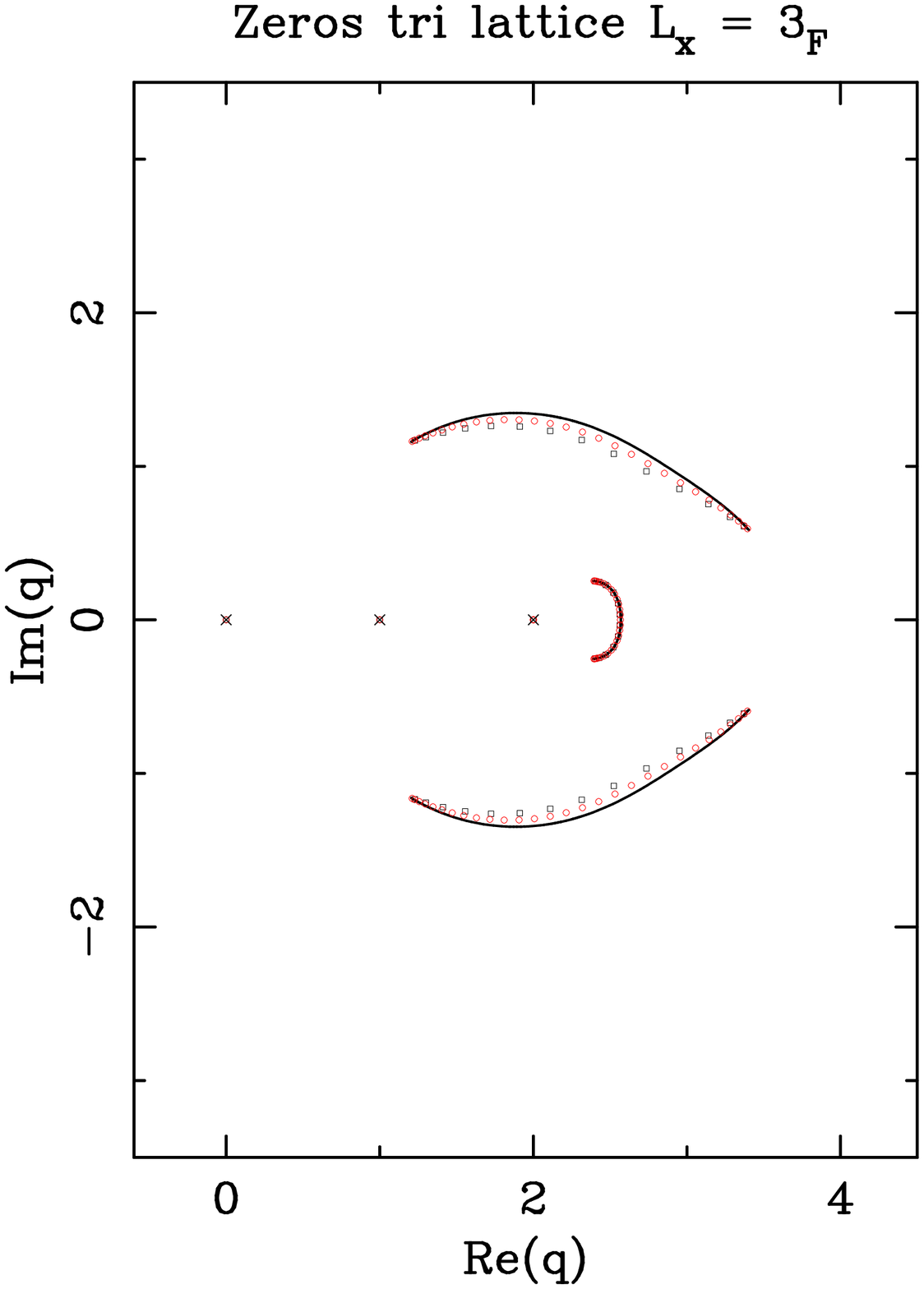}
  \caption{
  Zeros of the partition function of the $q$-state Potts antiferromagnet
  on the triangular lattices $3_F \times 15_F$ (squares), $3_F \times 30_F$ 
  (circles) and $3_F\times\infty_F$ (solid line). The isolated limiting points
  are denoted by a $\times$. 
  }
\protect\label{Figure_tri_3FxInftyF}
\end{figure}

\clearpage
%
%
\begin{figure}
  \centering
  \epsfxsize=400pt\epsffile{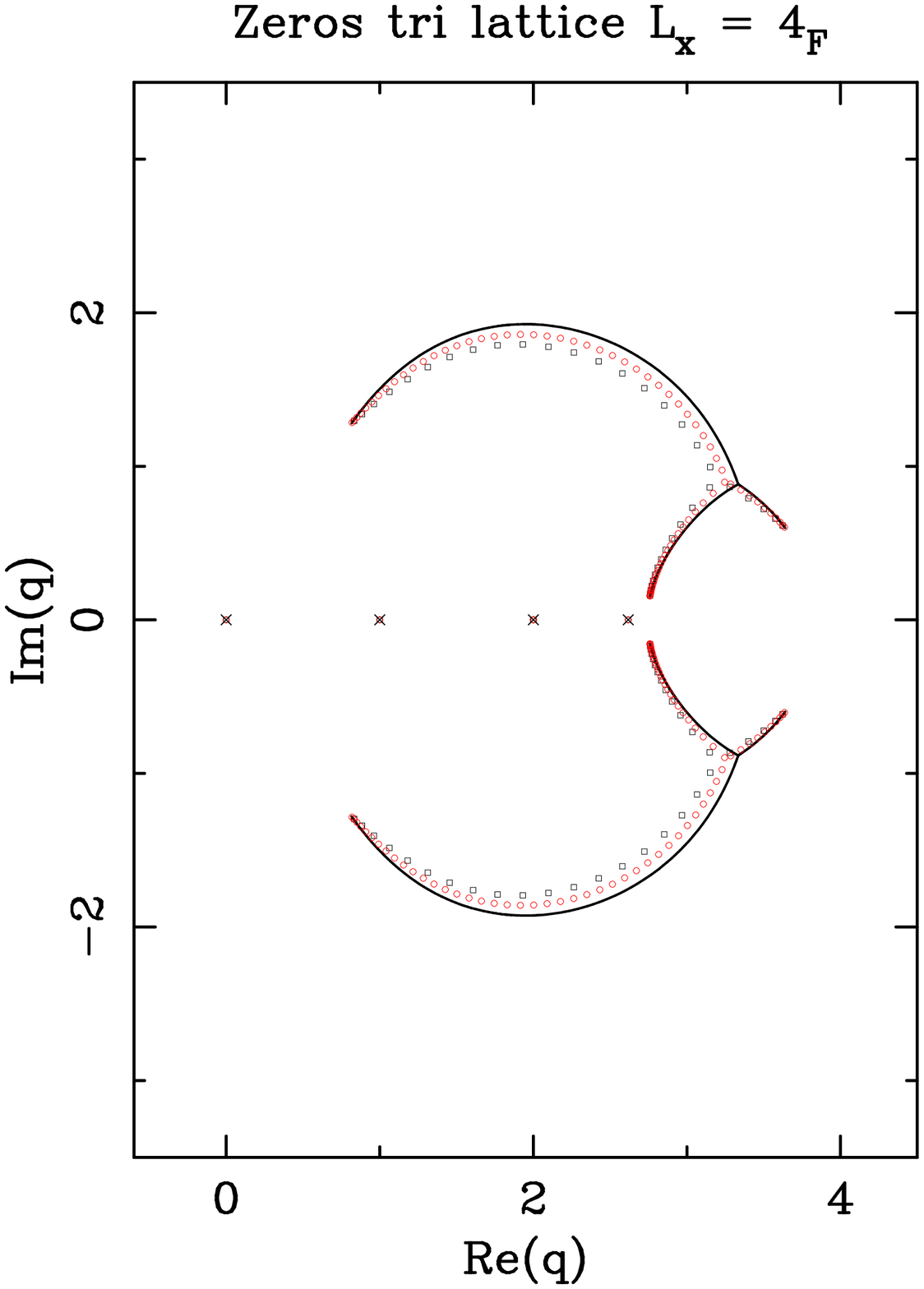}
  \caption{
  Zeros of the partition function of the $q$-state Potts antiferromagnet
  on the triangular lattices $4_F \times 20_F$ (squares), 
  $4_F \times 40_F$ (circles) and $4_F\times\infty_F$ (solid line).
  The isolated limiting points are denoted by a $\times$. 
  }
\protect\label{Figure_tri_4FxInftyF}
\end{figure}

\clearpage
%
%
\begin{figure}
  \centering
  \epsfxsize=400pt\epsffile{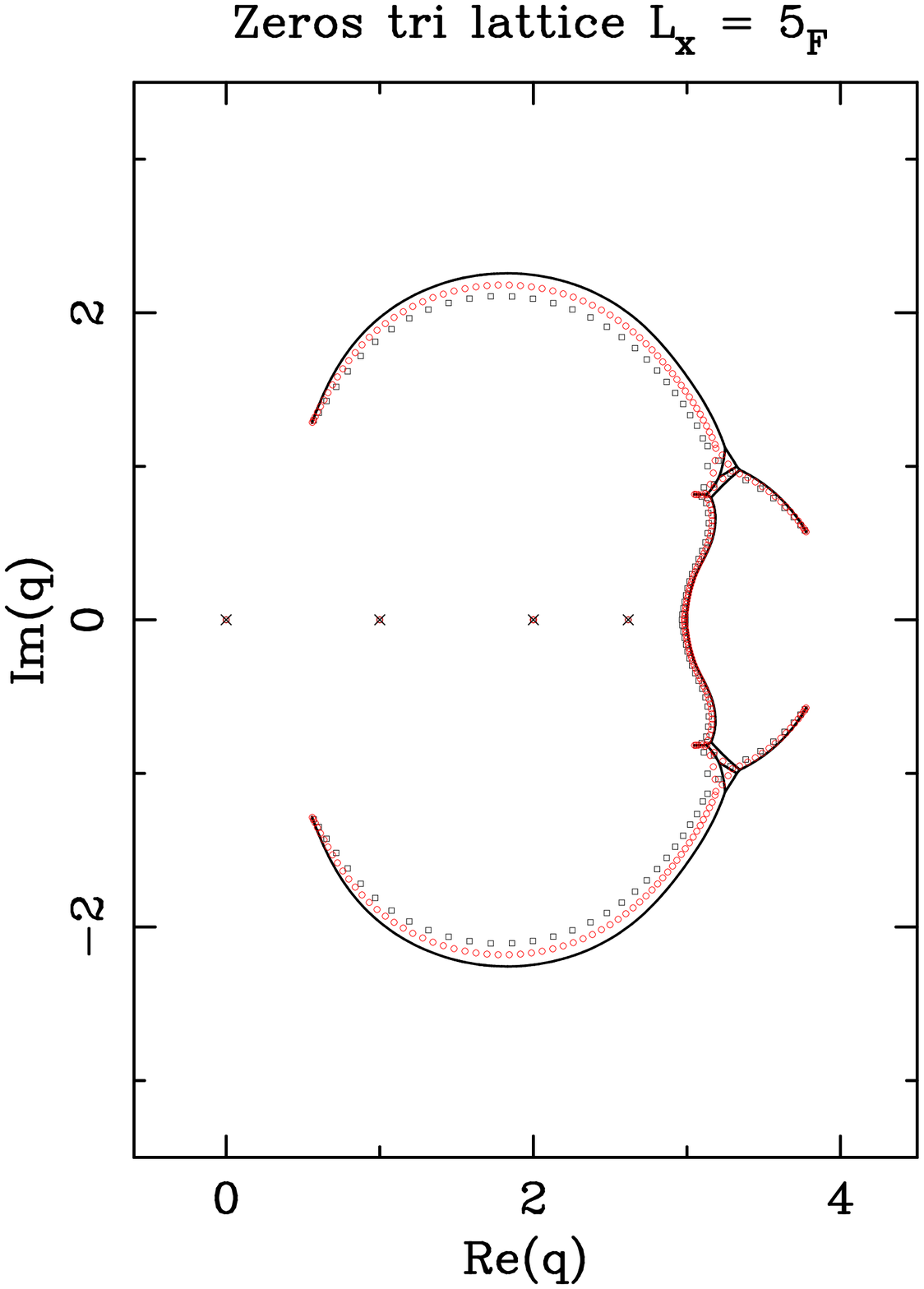}
  \caption{
  Zeros of the partition function of the $q$-state Potts antiferromagnet
  on the triangular lattices $5_F \times 25_F$ (squares), $5_F \times 50_F$ 
  (circles) and $5_F\times\infty_F$ (solid line).
  The isolated limiting points are denoted by a $\times$. 
  }
\protect\label{Figure_tri_5FxInftyF}
\end{figure}

\clearpage
%
%
\begin{figure}
  \centering
  \epsfxsize=400pt\epsffile{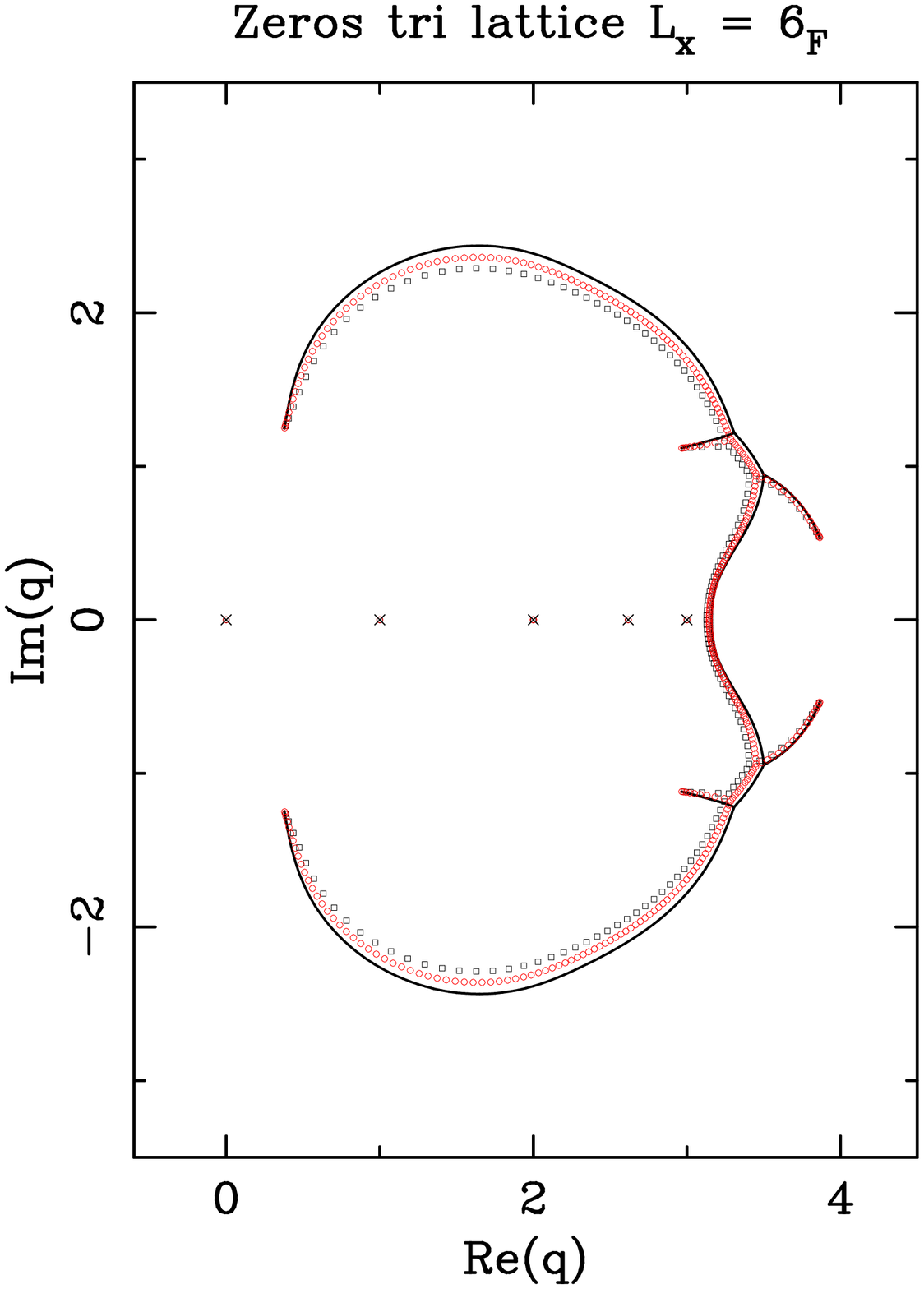}
  \caption{
  Zeros of the partition function of the $q$-state Potts antiferromagnet
  on the triangular lattices $6_F \times 30_F$ (squares), $6_F \times 60_F$ 
  (circles) and $6_F\times\infty_F$ (solid line).
  The isolated limiting points are denoted by a $\times$. 
  }
\protect\label{Figure_tri_6FxInftyF}
\end{figure}

\clearpage
%
%
\begin{figure}
  \centering
  \epsfxsize=400pt\epsffile{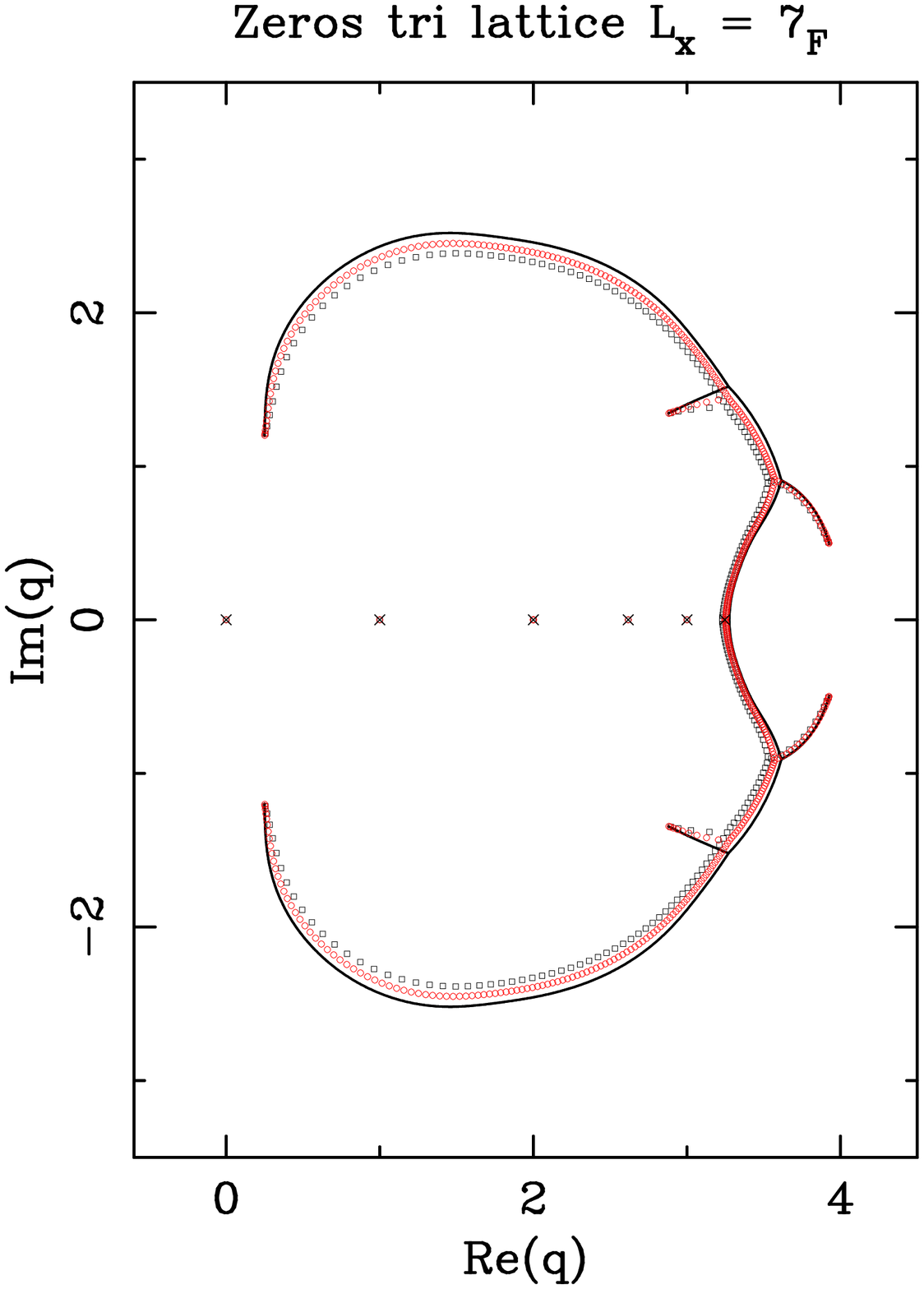}
  \caption{
  Zeros of the partition function of the $q$-state Potts antiferromagnet
  on the triangular lattices $7_F \times 35_F$ (squares), $7_F \times 70_F$ 
  (circles) and $7_F\times\infty_F$ (solid line).
  The isolated limiting points are denoted by a $\times$. 
  }
\protect\label{Figure_tri_7FxInftyF}
\end{figure}

\clearpage
%
%
\begin{figure}
  \centering
  \epsfxsize=400pt\epsffile{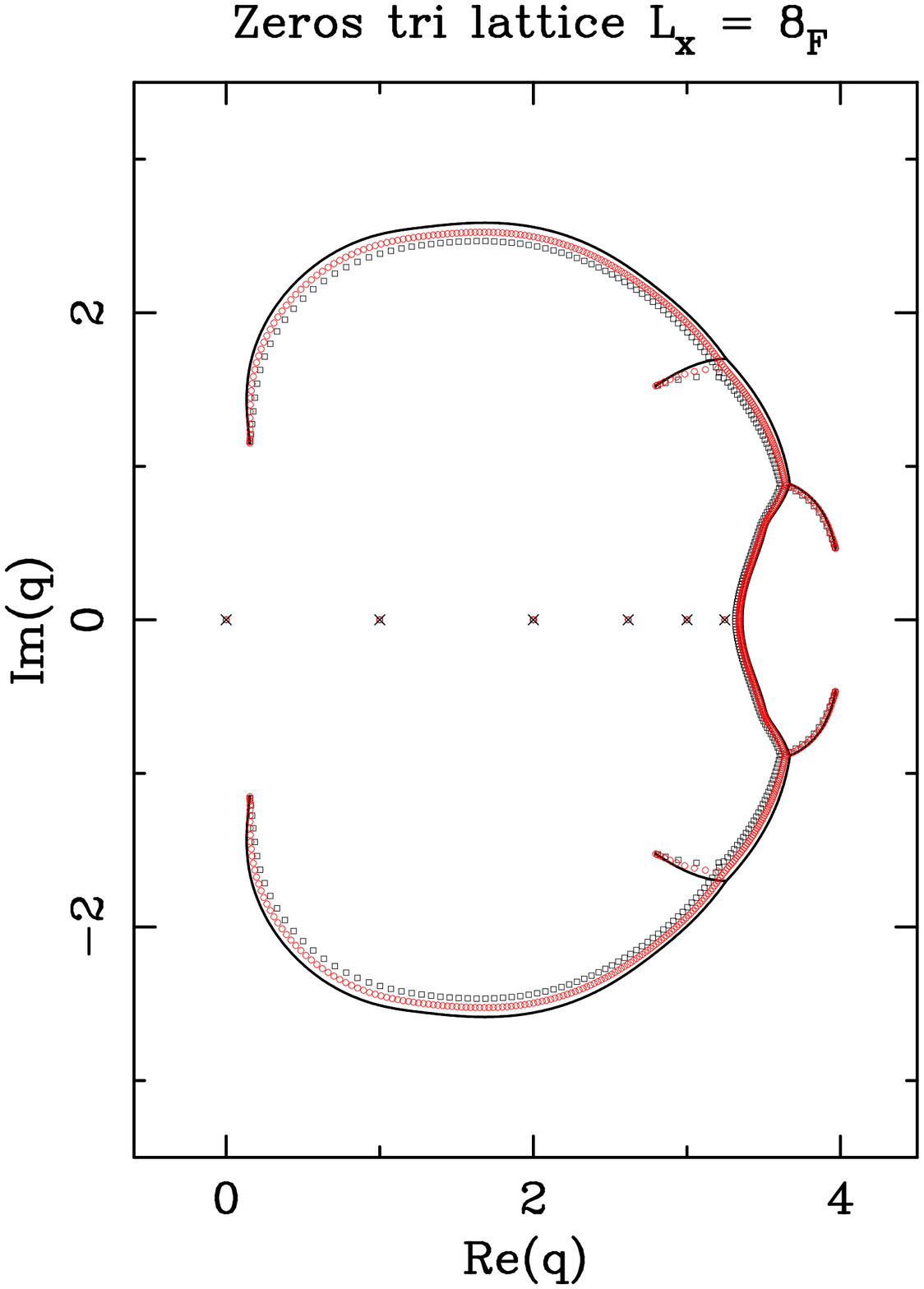}
  \caption{
  Zeros of the partition function of the $q$-state Potts antiferromagnet
  on the triangular lattices $8_F \times 40_F$ (squares), $8_F \times 80_F$ 
  (circles) and $8_F\times\infty_F$ (solid line).
  The isolated limiting points are denoted by a $\times$. 
  }
\protect\label{Figure_tri_8FxInftyF}
\end{figure}

\clearpage
%
%
\begin{figure}
  \centering
  \epsfxsize=400pt\epsffile{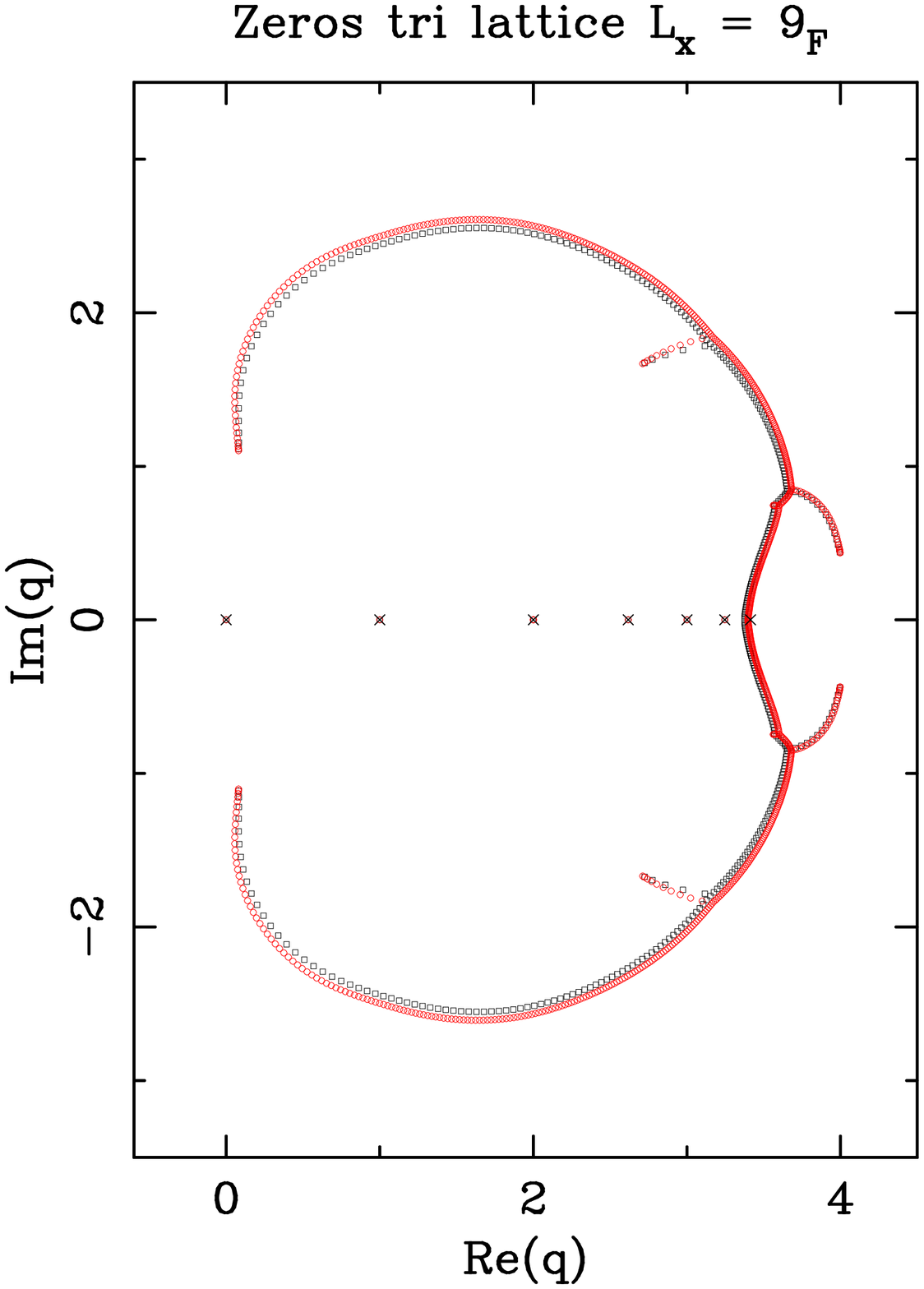}
  \caption{
  Zeros of the partition function of the $q$-state Potts antiferromagnet
  on the triangular lattices $9_F \times 45_F$ (squares) and 
  $9_F \times 90_F$ (circles).  
  The isolated limiting points are denoted by a $\times$.
  }
\protect\label{Figure_tri_9FxInftyF}
\end{figure}

\clearpage
%
%
\begin{figure}
  \centering
  \epsfxsize=400pt\epsffile{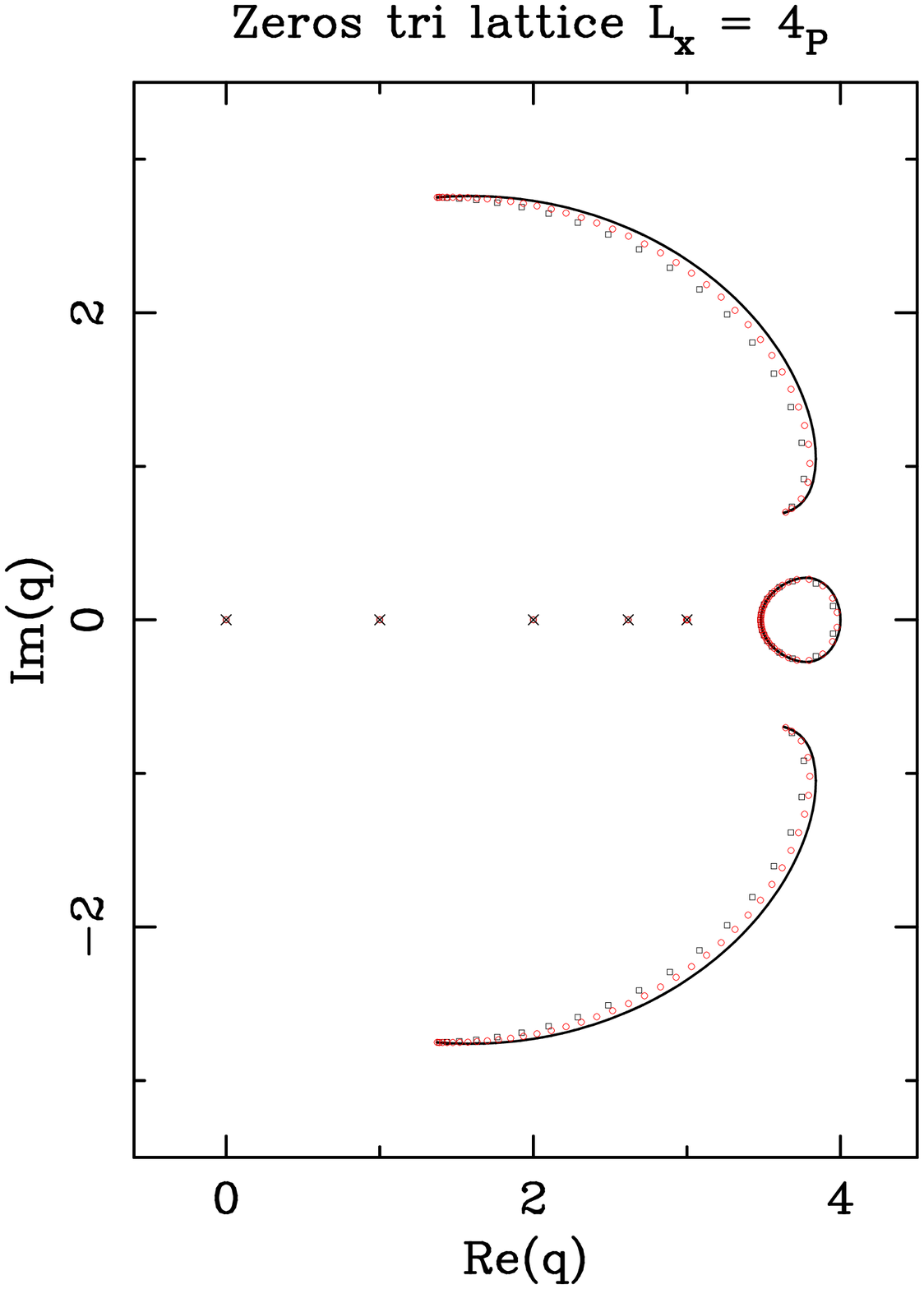}
  \caption{
  Zeros of the partition function of the $q$-state Potts antiferromagnet
  on the triangular lattices $4_P \times 20_F$ (squares), $4_P \times 40_F$  
  (circles) and $4_P\times\infty_F$ (solid line).
  The isolated limiting points are denoted by a $\times$.
  }
\protect\label{Figure_tri_4PxInftyF}
\end{figure}

\clearpage
%
%
\begin{figure}
  \centering
  \epsfxsize=400pt\epsffile{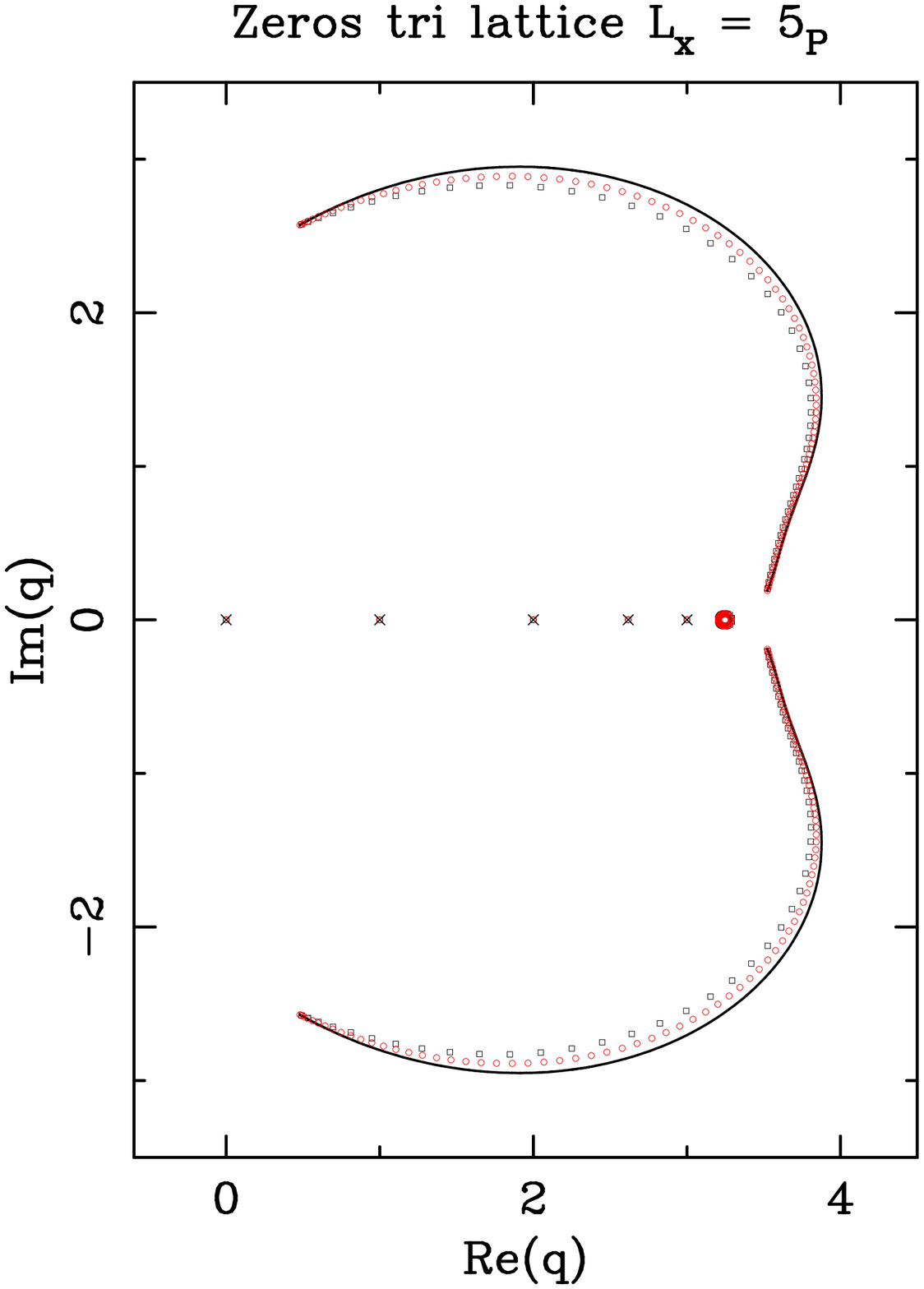}
  \caption{
  Zeros of the partition function of the $q$-state Potts antiferromagnet
  on the triangular lattices $5_P \times 25_F$ (squares), $5_P \times 50_F$ 
  (circles) and $5_P\times\infty_F$ (solid line).
  The isolated limiting points are denoted by a $\times$. 
  }
\protect\label{Figure_tri_5PxInftyF}
\end{figure}

\clearpage
%
%
\begin{figure}
  \centering
  \epsfxsize=400pt\epsffile{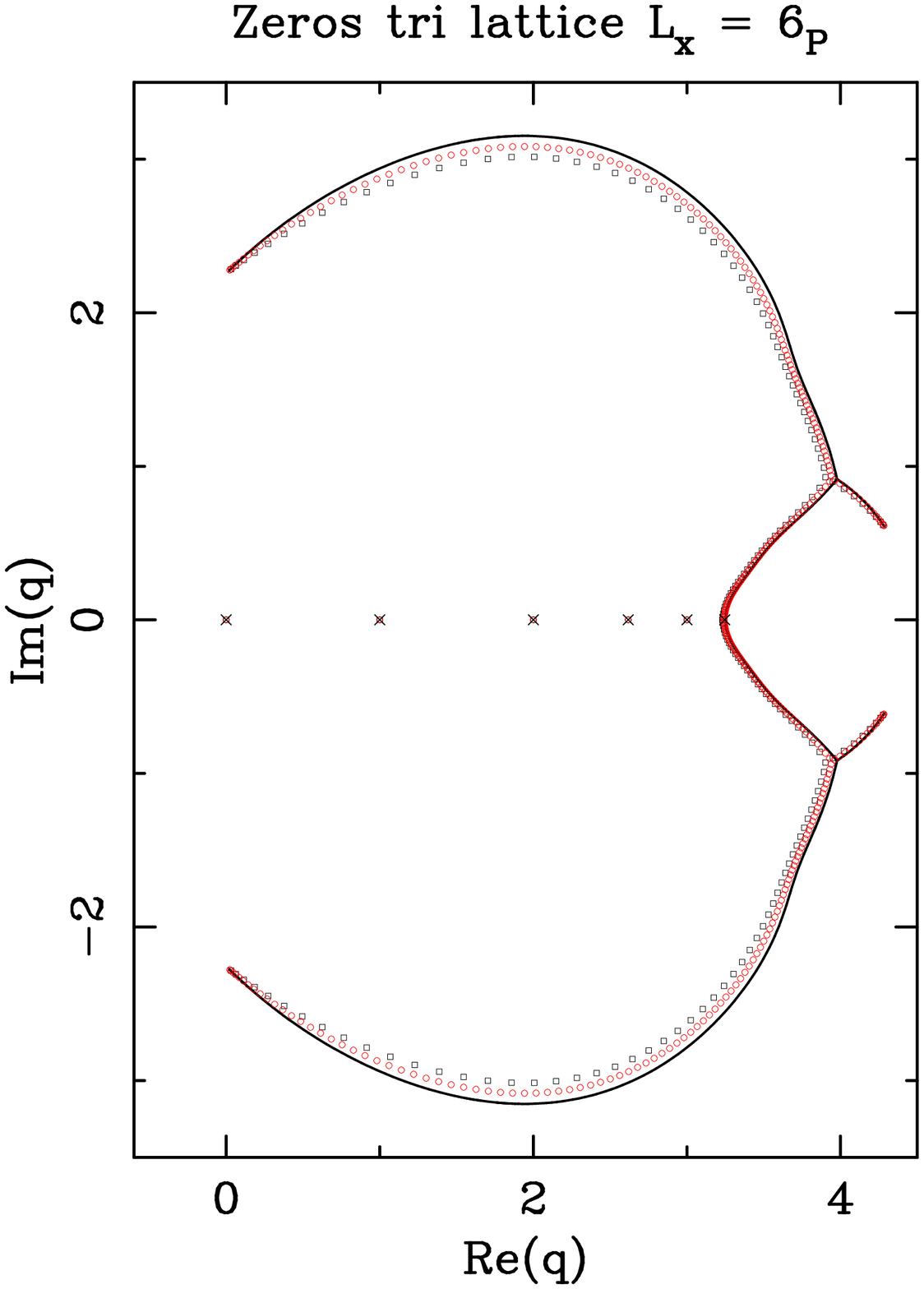}
  \caption{
  Zeros of the partition function of the $q$-state Potts antiferromagnet
  on the triangular lattices $6_P \times 30_F$ (squares), $6_P \times 60_F$  
  (circles) and $6_P\times\infty_F$ (solid line).
  The isolated limiting points are denoted by a $\times$.
  }
\protect\label{Figure_tri_6PxInftyF}
\end{figure}

\clearpage
%
%
\begin{figure}
  \centering
  \epsfxsize=400pt\epsffile{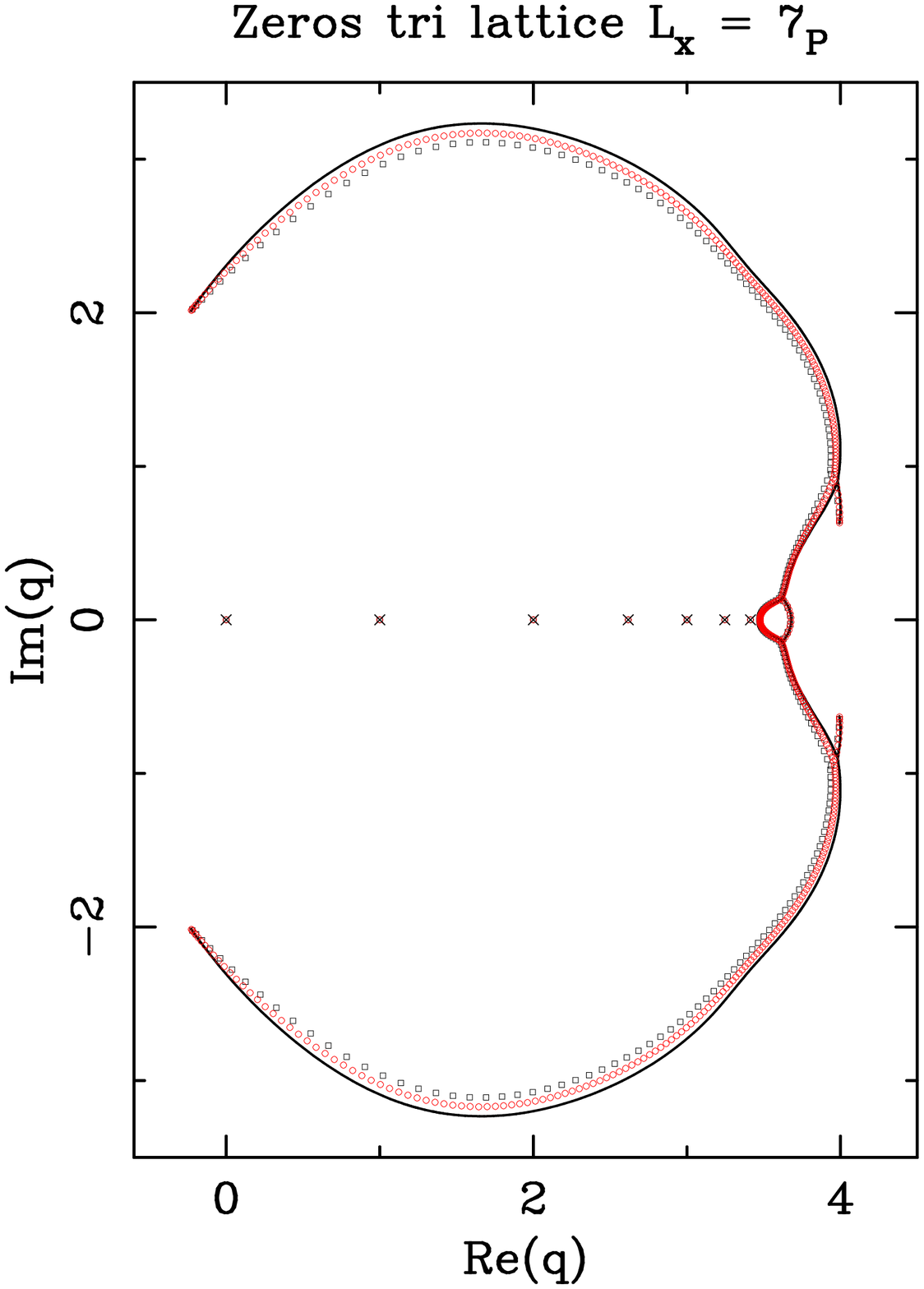}
  \caption{
  Zeros of the partition function of the $q$-state Potts antiferromagnet
  on the triangular lattices $7_P \times 35_F$ (squares), $7_P \times 70_F$ 
  (circles) and $7_P\times\infty_F$ (solid line).
  The isolated limiting points are denoted by a $\times$.
  }
\protect\label{Figure_tri_7PxInftyF}
\end{figure}

\clearpage
%
%
\begin{figure}
  \centering
  \epsfxsize=400pt\epsffile{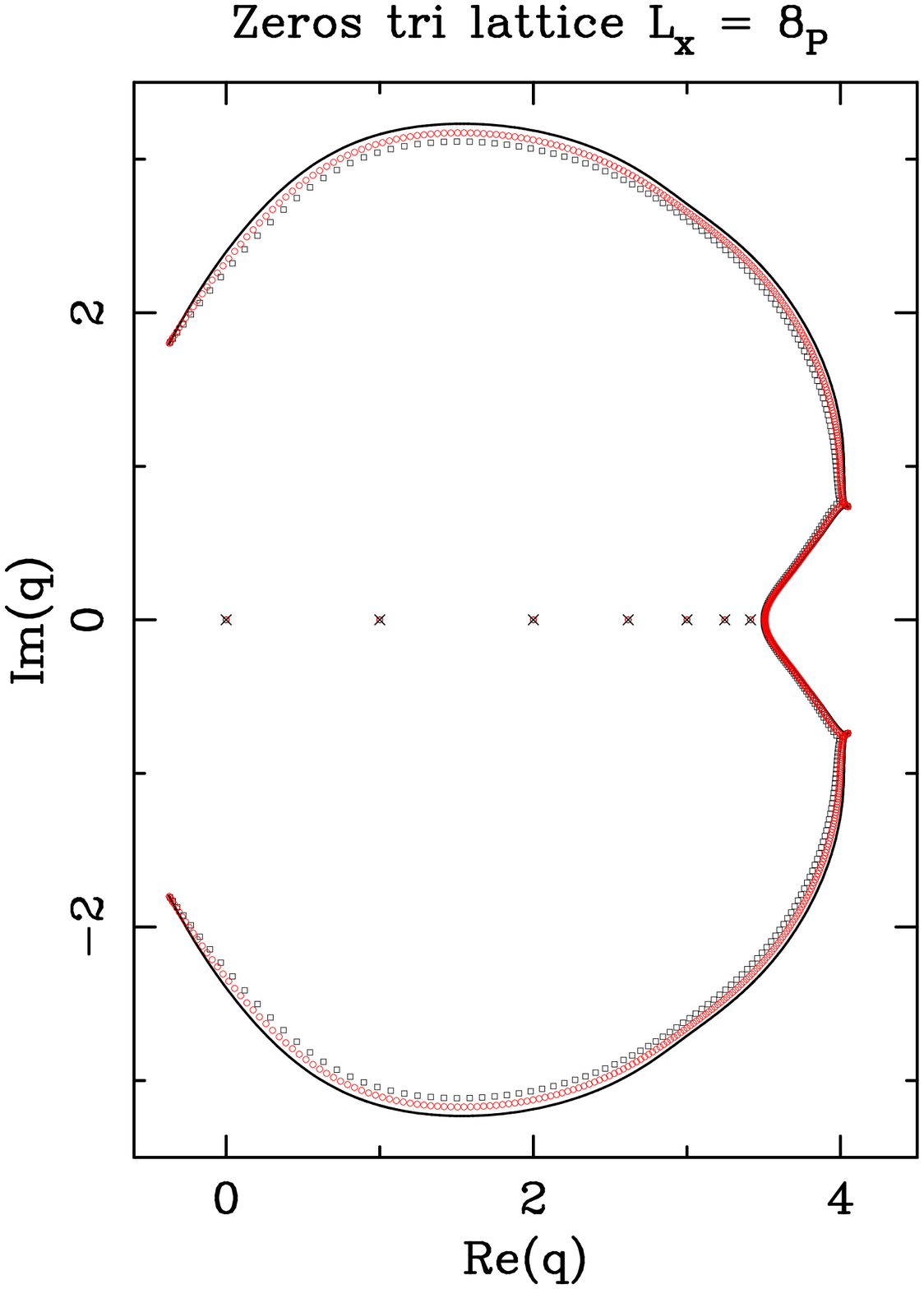}
  \caption{
  Zeros of the partition function of the $q$-state Potts antiferromagnet
  on the triangular lattices $8_P \times 40_F$ (squares), $8_P \times 80_F$  
  (circles) and $8_P\times\infty_F$ (solid line).
  The isolated limiting points are denoted by a $\times$. 
  }
\protect\label{Figure_tri_8PxInftyF}
\end{figure}

\clearpage
%
%
\begin{figure}
  \centering
  \epsfxsize=400pt\epsffile{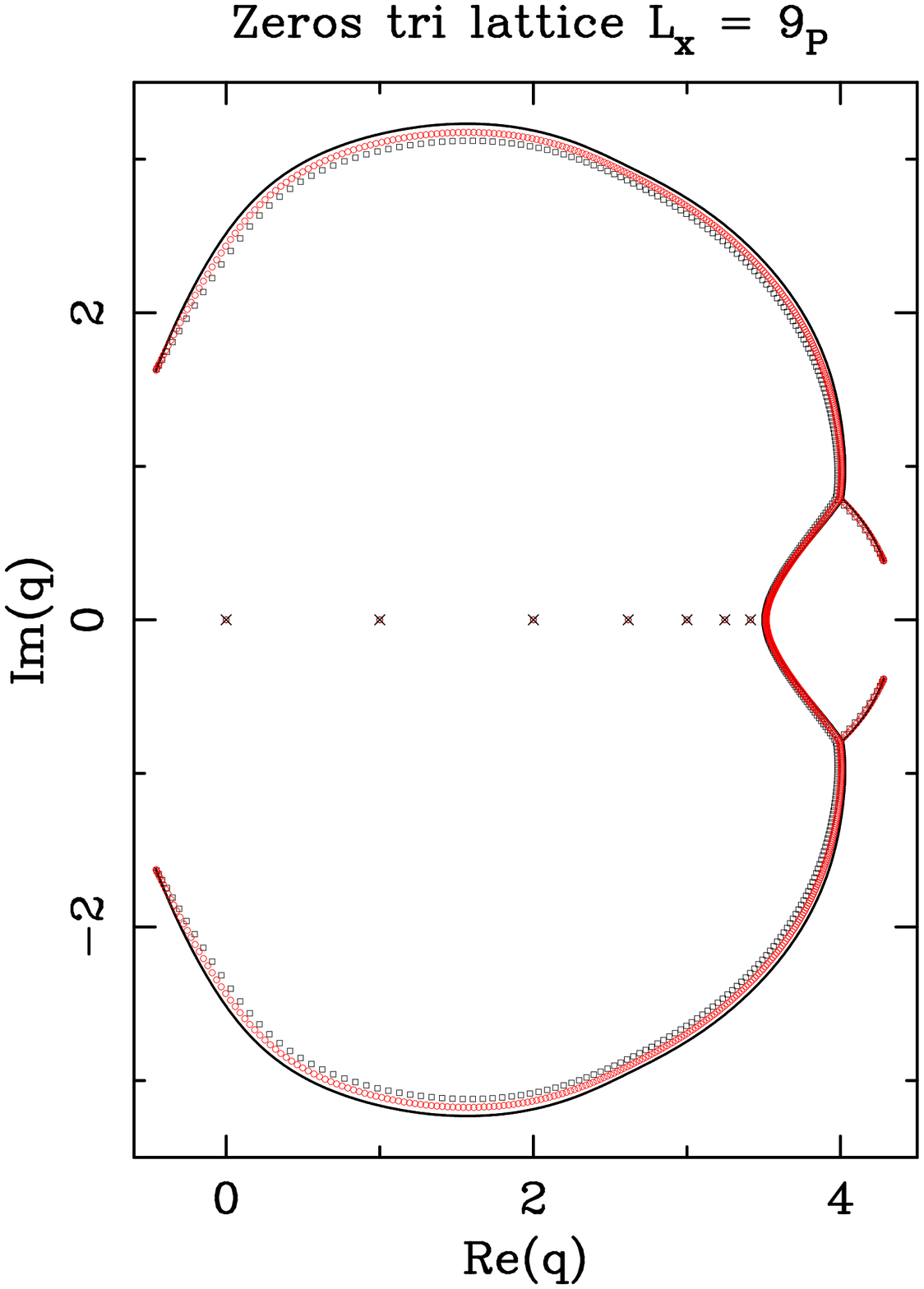}
  \caption{
  Zeros of the partition function of the $q$-state Potts antiferromagnet
  on the triangular lattices $9_P \times 45_F$ (squares), $9_P \times 90_F$
  (circles) and $9_P\times\infty_F$ (solid line).
  The isolated limiting points are denoted by a $\times$.
  }
\protect\label{Figure_tri_9PxInftyF}
\end{figure}

\clearpage
%
%
\begin{figure}
  \centering
  \epsfxsize=400pt\epsffile{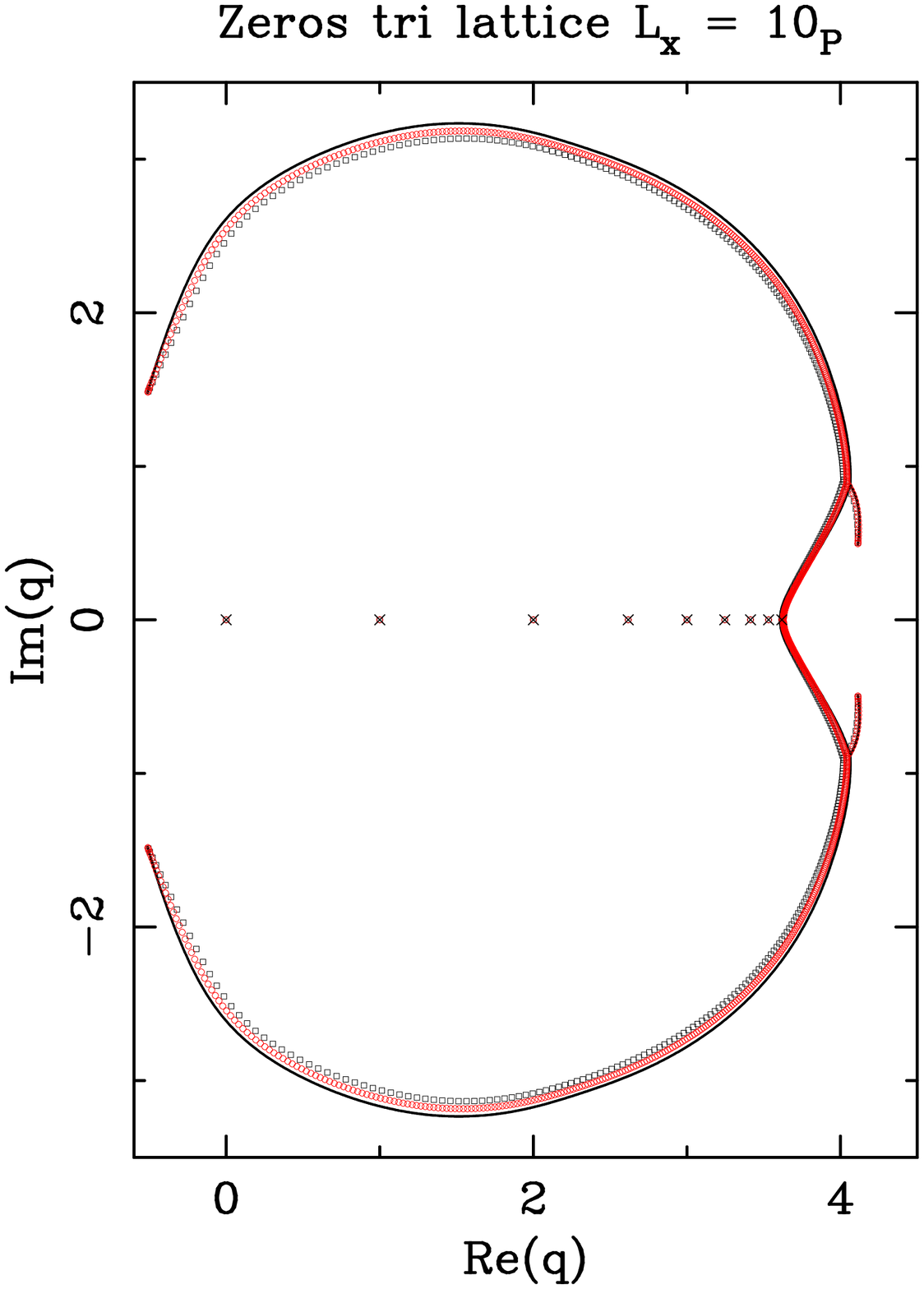}
  \caption{
  Zeros of the partition function of the $q$-state Potts antiferromagnet
  on the triangular lattices $10_P \times 50_F$ (squares), $10_P \times 100_F$
  (circles) and $10_P\times\infty_F$ (solid line).
  The isolated limiting points are denoted by a $\times$.
  }
\protect\label{Figure_tri_10PxInftyF}
\end{figure}

\clearpage
%
%
\begin{figure}
  \centering
  \epsfxsize=400pt\epsffile{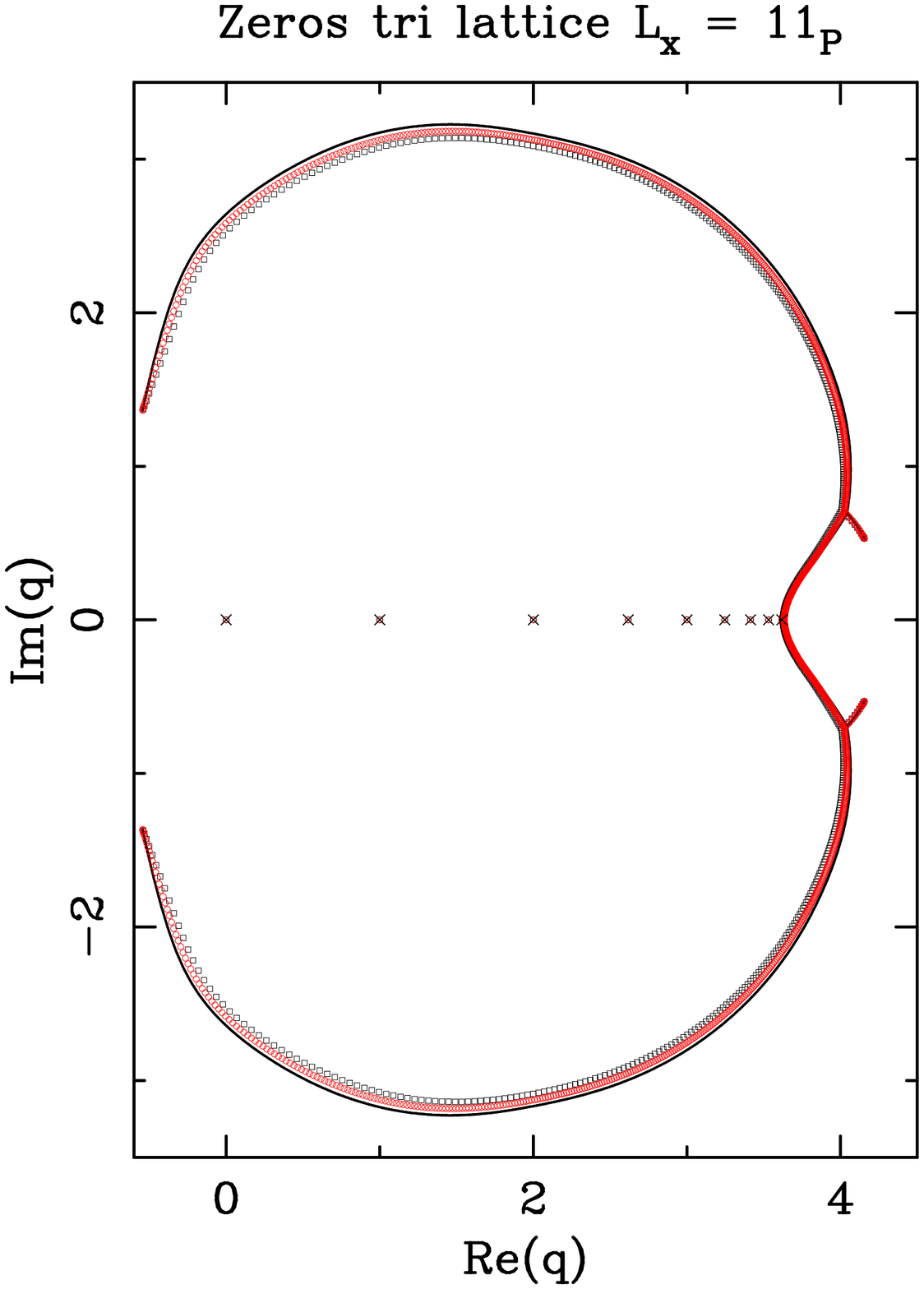}
  \caption{
  Zeros of the partition function of the $q$-state Potts antiferromagnet
  on the triangular lattices $11_P \times 55_F$ (squares), $11_P \times 110_F$
  (circles) and $11_P\times\infty_F$ (solid line).
  The isolated limiting points are denoted by a $\times$.
  }
\protect\label{Figure_tri_11PxInftyF}
\end{figure}

\clearpage
%
%
\begin{figure}
  \centering
  \epsfxsize=400pt\epsffile{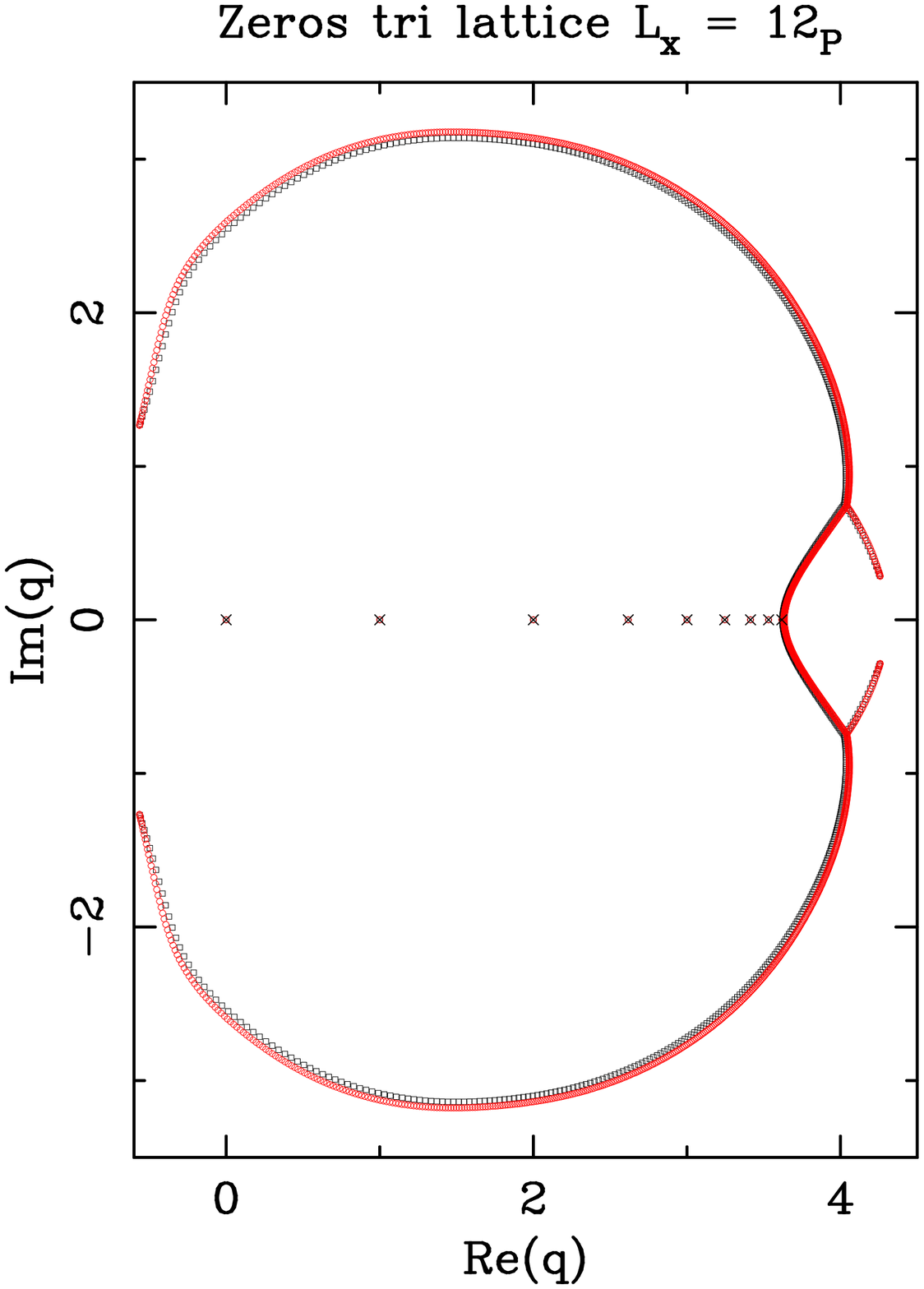}
  \caption{
  Zeros of the partition function of the $q$-state Potts antiferromagnet
  on the triangular lattices $12_P \times 60_F$ (squares) and
   $12_P \times 120_F$ (circles). 
  The isolated limiting points are denoted by a $\times$.
  }
\protect\label{Figure_tri_12PxInftyF}
\end{figure}

\clearpage
%
%
\begin{figure}
  \centering
  \epsfxsize=400pt\epsffile{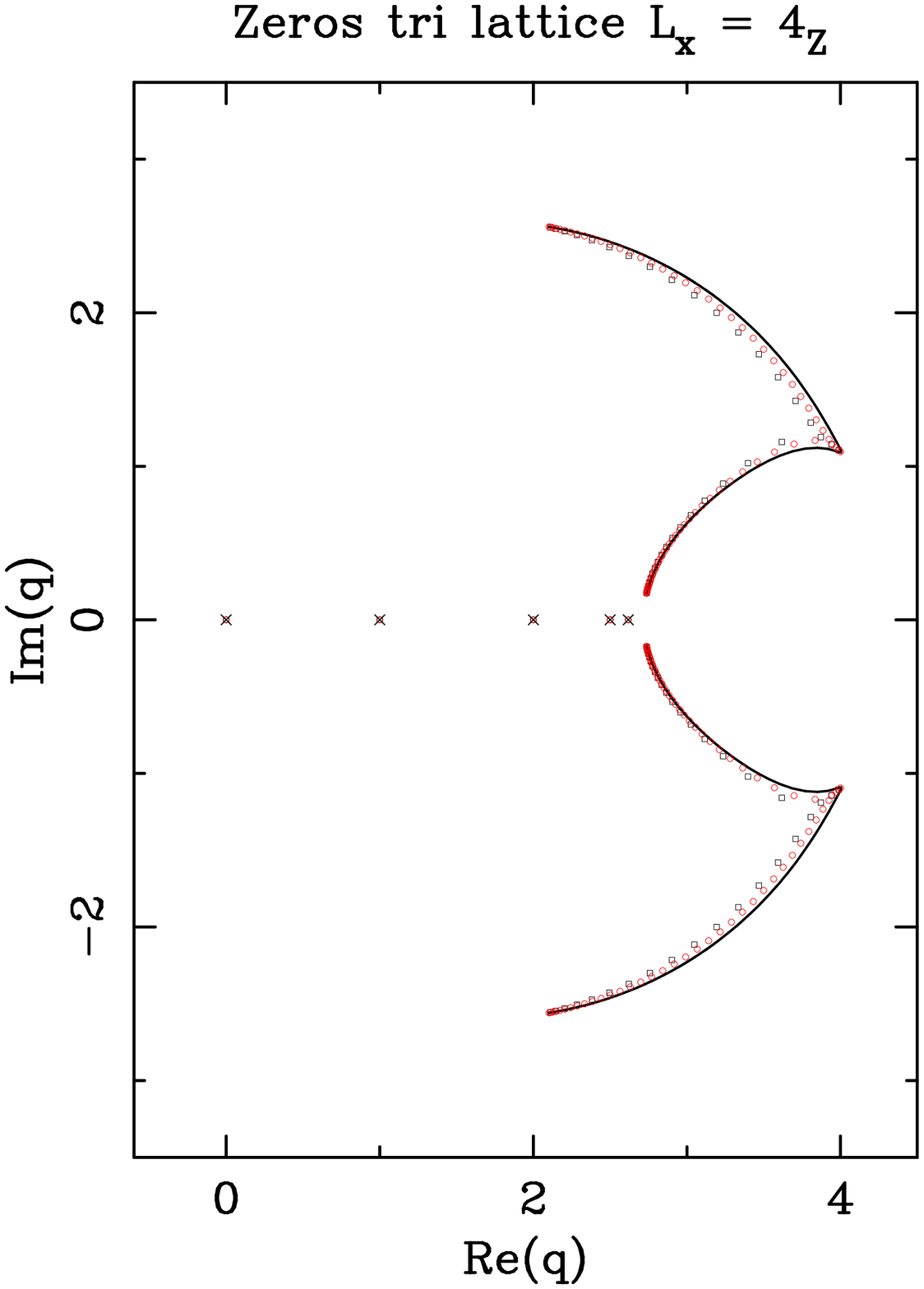}
  \caption{
  Zeros of the partition function of the $q$-state Potts antiferromagnet
  on the triangular lattices $4_P \times 20_Z$ (squares), $4_P \times 40_Z$
  (circles) and $4_P\times\infty_Z$ (solid line).
  The isolated limiting points are denoted by a $\times$.
  }
\protect\label{Figure_tri_4PxInftyZ}
\end{figure}

\clearpage
%
%
\begin{figure}
  \centering
  \epsfxsize=400pt\epsffile{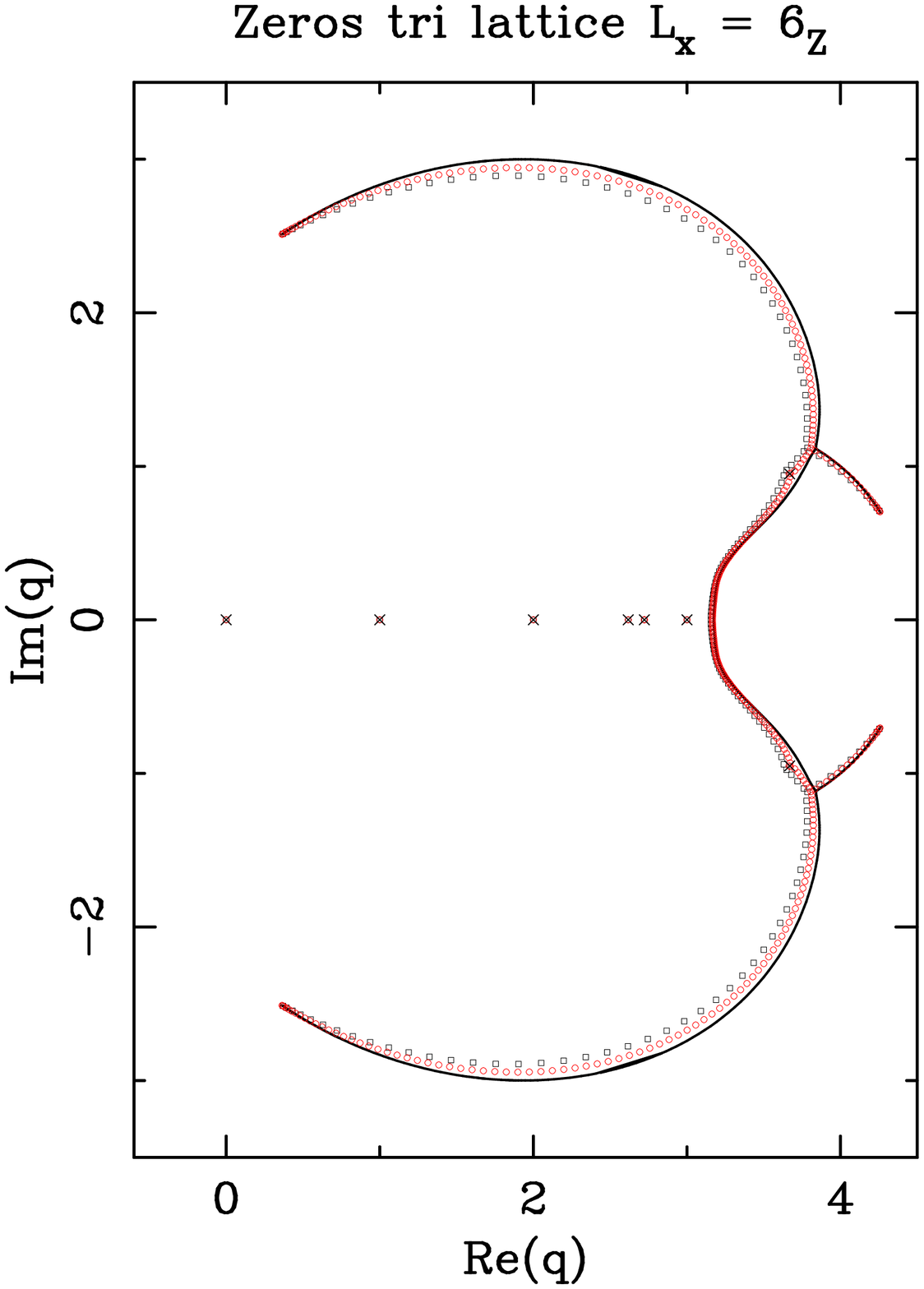}
  \caption{
  Zeros of the partition function of the $q$-state Potts antiferromagnet
  on the triangular lattices $6_P \times 30_Z$ (squares), $6_P \times 60_Z$
  (circles) and $6_P\times\infty_Z$ (solid line).
  The isolated limiting points are denoted by a $\times$.
  }
\protect\label{Figure_tri_6PxInftyZ}
\end{figure}

\clearpage
%
%
\begin{figure}
  \centering
  \epsfxsize=400pt\epsffile{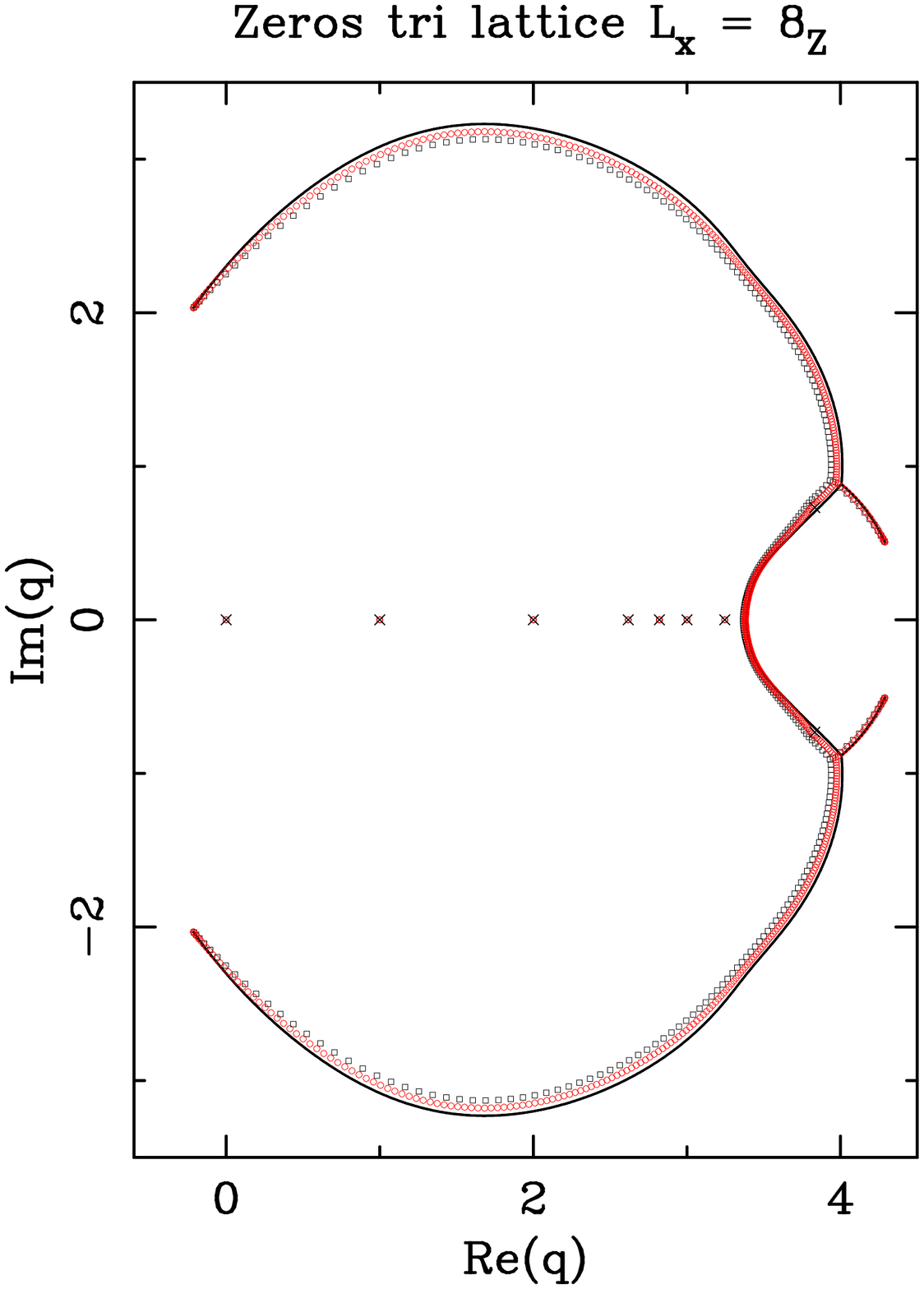}
  \caption{
  Zeros of the partition function of the $q$-state Potts antiferromagnet
  on the triangular lattices $8_P \times 40_Z$ (squares), $8_P \times 80_Z$
  (circles) and $8_P\times\infty_Z$ (solid line).
  The isolated limiting points are denoted by a $\times$.
  }
\protect\label{Figure_tri_8PxInftyZ}
\end{figure}

\clearpage
%
%
\begin{figure}
  \centering
  \epsfxsize=400pt\epsffile{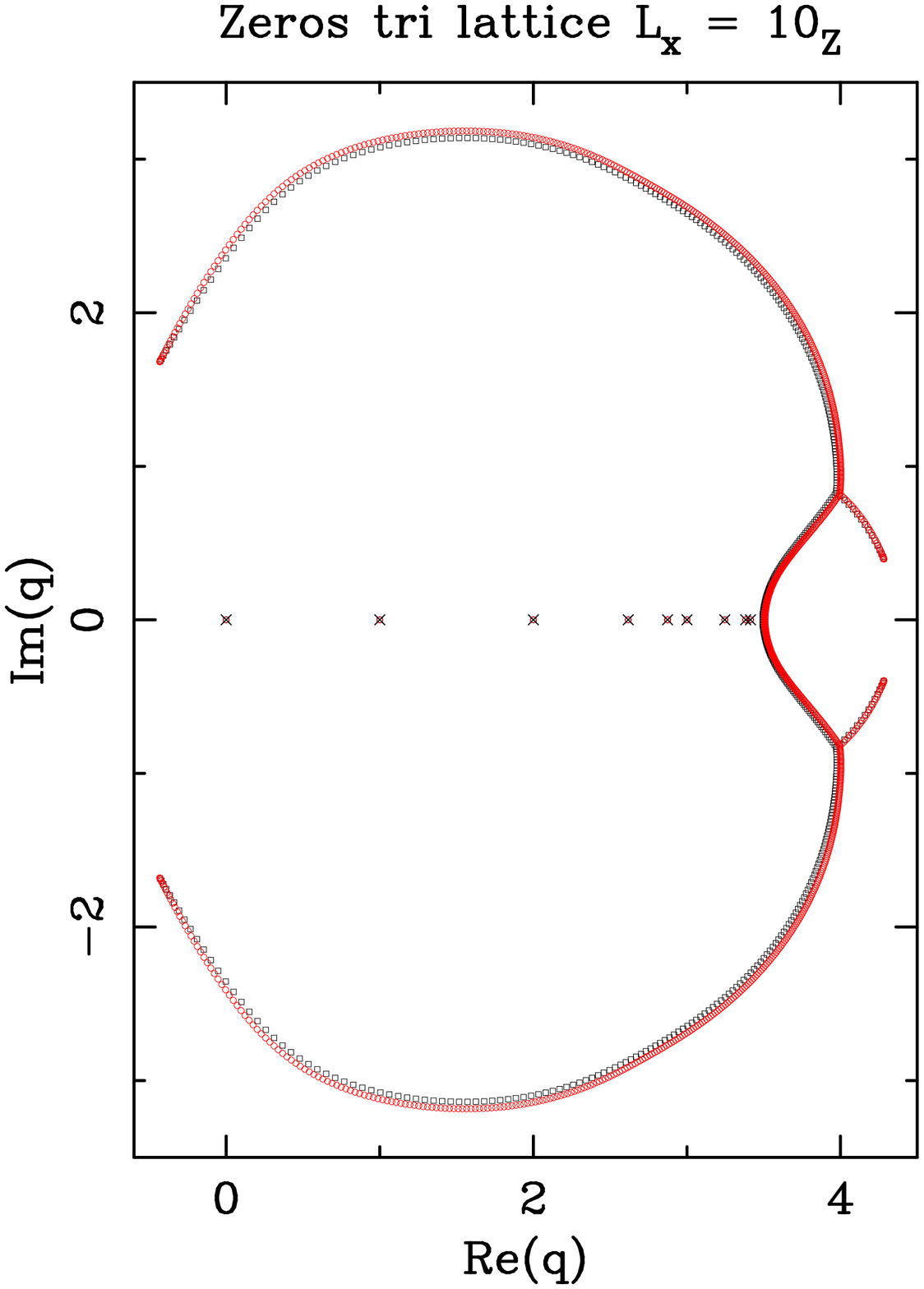}
  \caption{
  Zeros of the partition function of the $q$-state Potts antiferromagnet
  on the triangular lattices $10_P \times 50_Z$ (squares) and 
  $10_P \times 80_Z$ (circles). 
  The isolated limiting points are denoted by a $\times$.
  }
\protect\label{Figure_tri_10PxInftyZ}
\end{figure}

\clearpage

%
%
\begin{figure}
  \centering
  \epsfxsize=400pt\epsffile{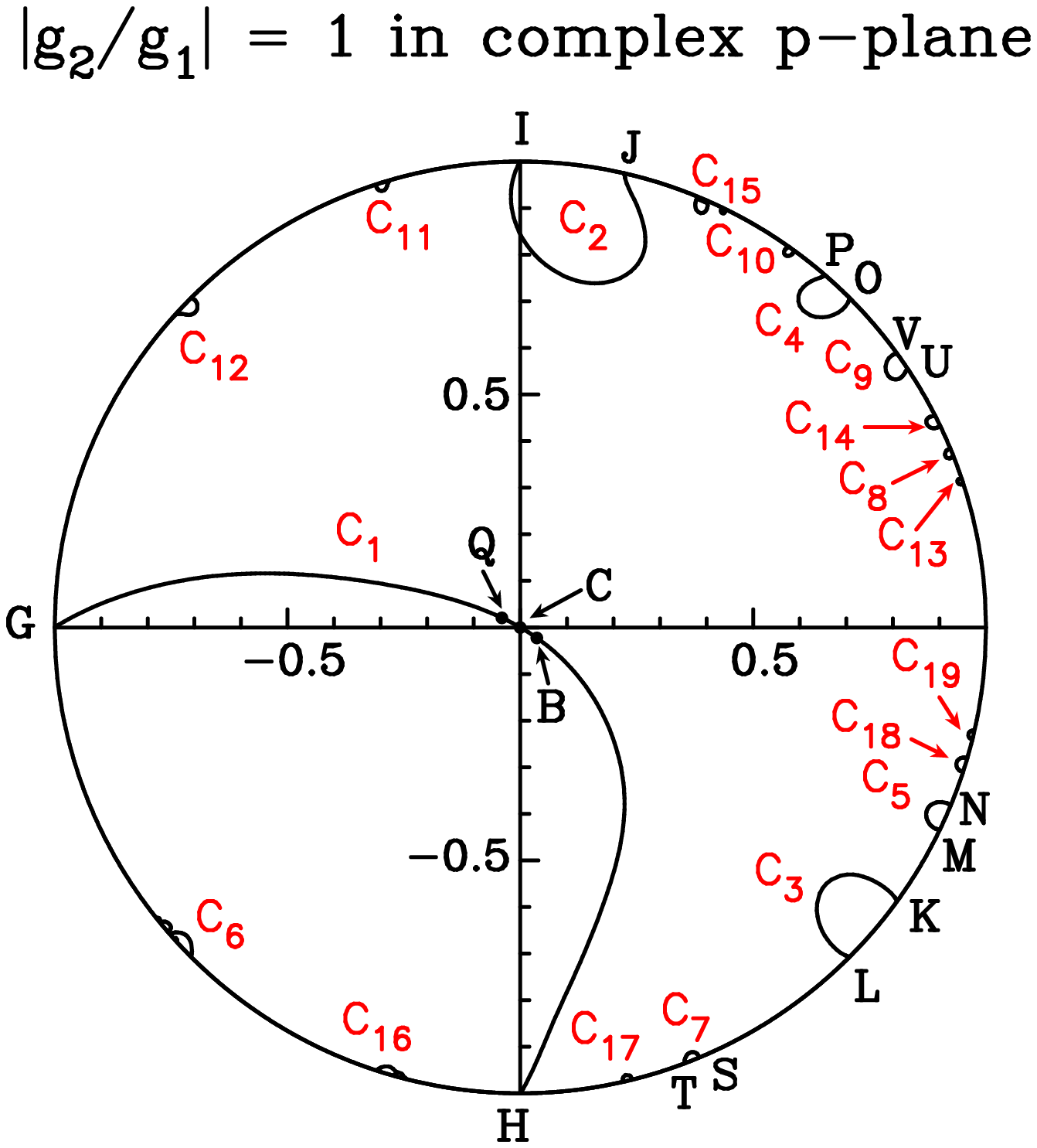}
  \caption{
  Equimodular curves $|g_2/g_1| = 1$ in the complex $p$-plane.
  Several important points are labelled G, H, I, J, \ldots\ (see text).
  }
\protect\label{Figure_Baxter_p}
\end{figure}

\clearpage

%
%
\begin{figure}
  \centering
  \epsfxsize=400pt\epsffile{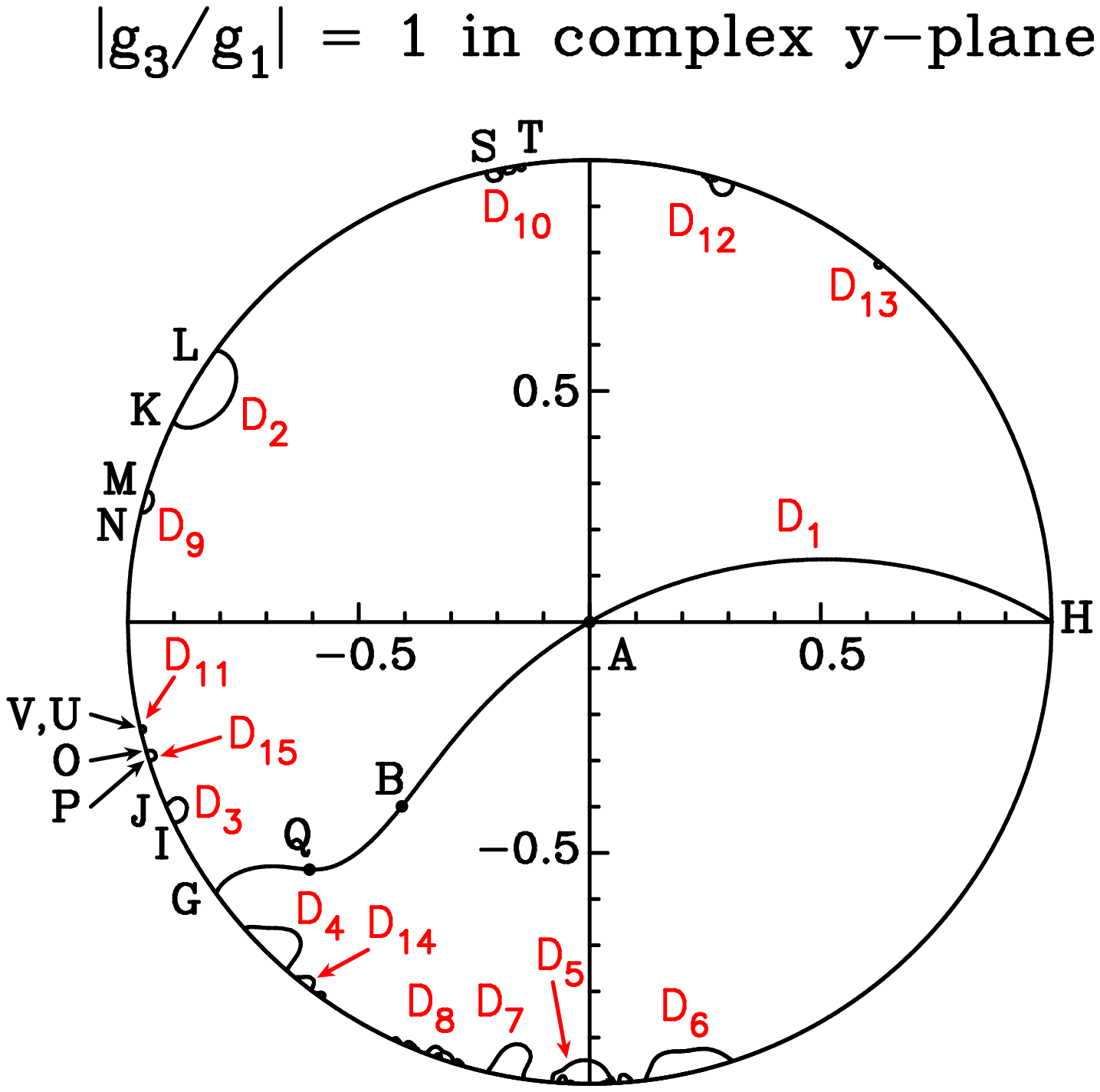}
  \caption{
  Equimodular curves $|g_3/g_1| = 1$ in the complex $y$-plane. 
  Several important points are labelled G, H, I, J, \ldots\ (see text).
  }
\protect\label{Figure_Baxter_y}
\end{figure}

\clearpage

%
%
\begin{figure}
  \centering
  \epsfxsize=400pt\epsffile{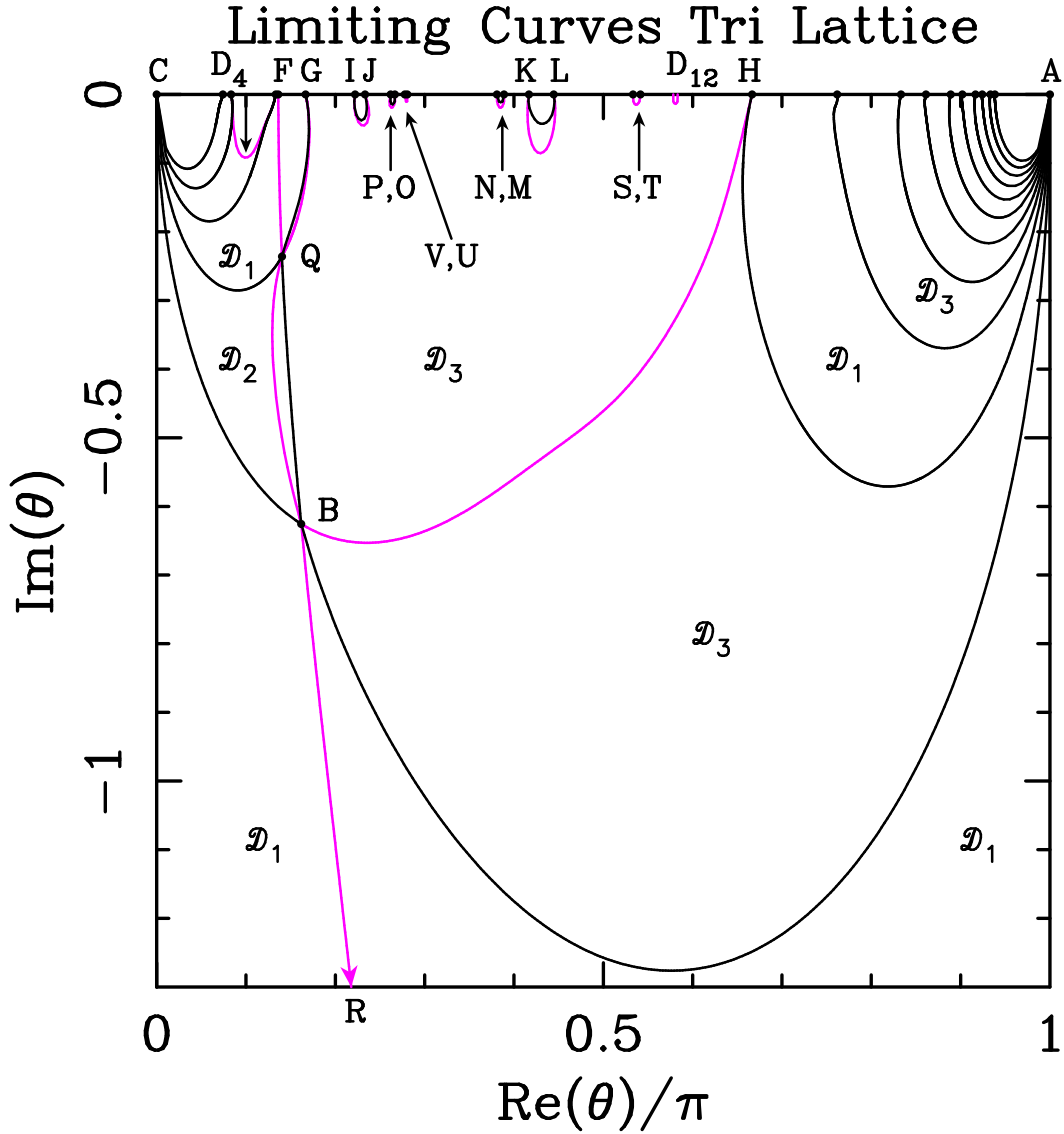}
  \caption{
  Equimodular curves for the eigenvalues $g_i$ in the complex $\theta$-plane. 
  The portions of curves where the equimodular eigenvalues are dominant 
  (resp.\  subdominant) are depicted in black (resp.\  \subdominantcolor). 
  The eigenvalue $g_i$ is dominant in each region labelled ${\cal D}_i$.
  Several important points are labelled A, B, C, \ldots\ (see text).
  To facilitate comparison with Baxter's results
  \protect\cite[Fig.~5]{Baxter_87}, we have used the same labelling of points
  wherever possible.
  }
\protect\label{Figure_Baxter_t}
\end{figure}

\clearpage

%
%
\begin{figure}
  \centering
  \epsfxsize=400pt\epsffile{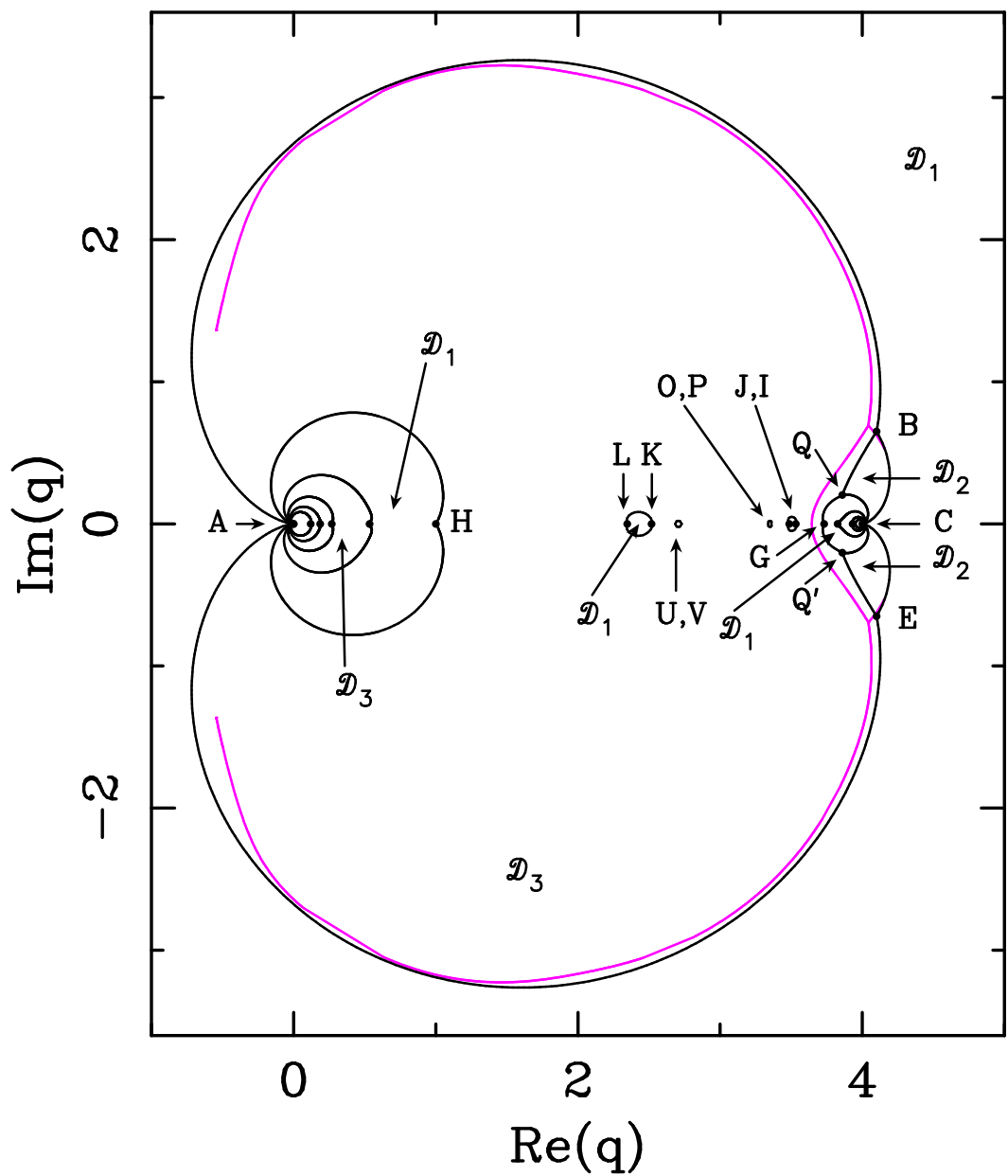}
  \caption{
  Dominant equimodular curves for the eigenvalues $g_i$
  in the complex $q$-plane (in black).
  For comparison, we show (in \subdominantcolor) the limiting 
  curve $\scrb$ for the strip $L_x=11_{\rm P}$. 
  To facilitate comparison with Baxter's results
  \protect\cite[Fig.~5]{Baxter_87}, we have used the same labelling of points
  wherever possible.
  We warn the reader that the presence of the additional curves
  (AH, LK, etc.)\ to the right of $q=0$ has not been definitively established;
  see Sections~\ref{sec6.3} and \ref{sec6.4} for a detailed discussion.
  }
\protect\label{Figure_Baxter_q}
\end{figure}

\clearpage
%
%
\begin{figure}
  \centering
  \epsfxsize=300pt\epsffile{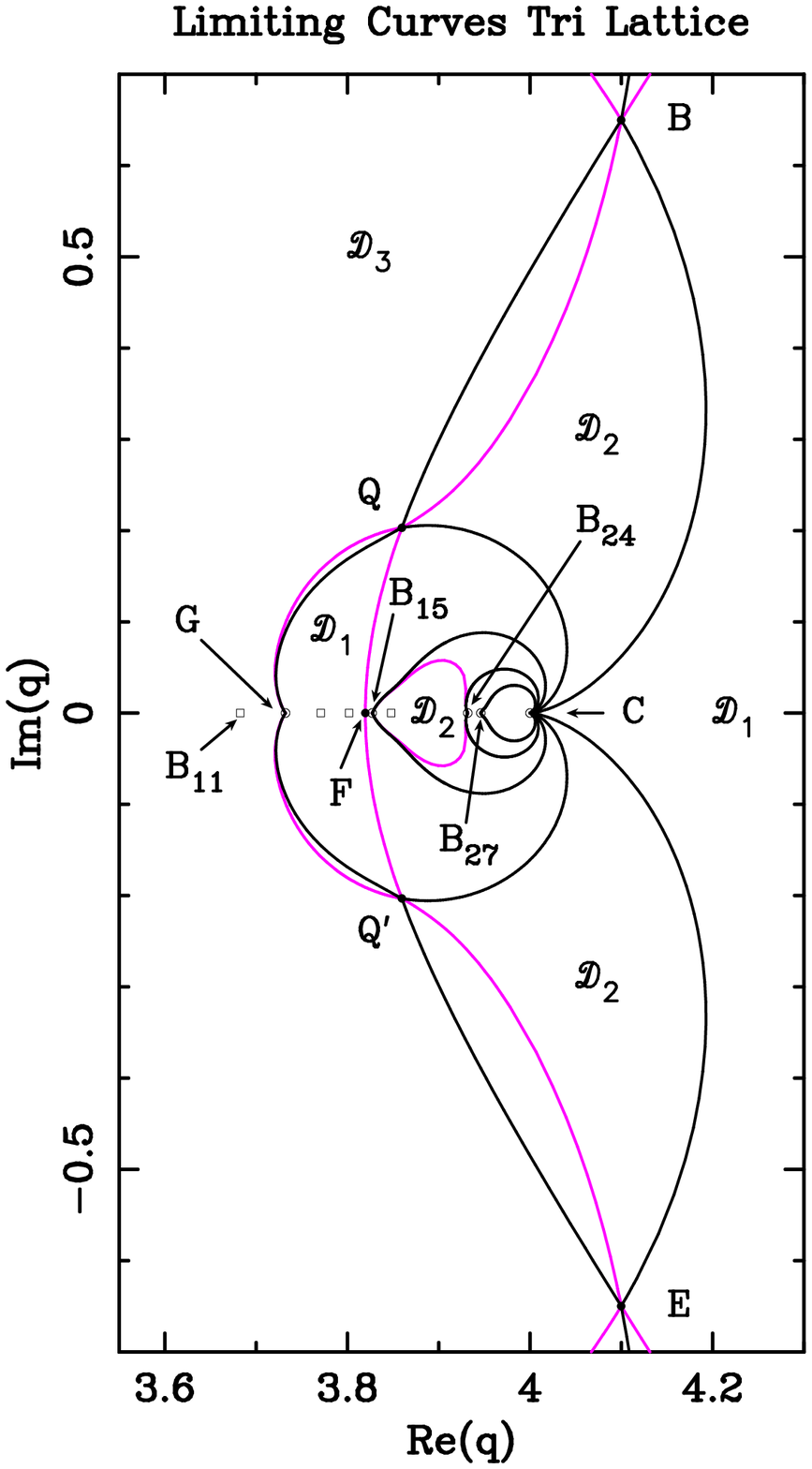}
  \caption{
  Detail of Figure~\protect\ref{Figure_Baxter_q} around the point $q=4$. 
  We depict the dominant (resp.\  subdominant) equimodular curves in black 
  (resp.\  \subdominantcolor). The solid circles ($\bullet$) denote 
  special points discussed in the text, while the squares ($\square$) and empty
  circles ($\circ$) denote the Beraha numbers 
  $q=B_{11},\ldots,B_{16}$, $B_{24}$, $B_{27}$ and 
  $B_\infty = 4$ (point C). The empty circles denote those Beraha numbers 
  which belong to any equimodular curve. 
  We denote by ${\cal D}_i$ the regions where
  the eigenvalue $g_i$ is dominant.  
  }
\protect\label{Figure_Baxter_q_zoom}
\end{figure}

\clearpage
%
%
\begin{figure}
  \centering
  \epsfxsize=400pt\epsffile{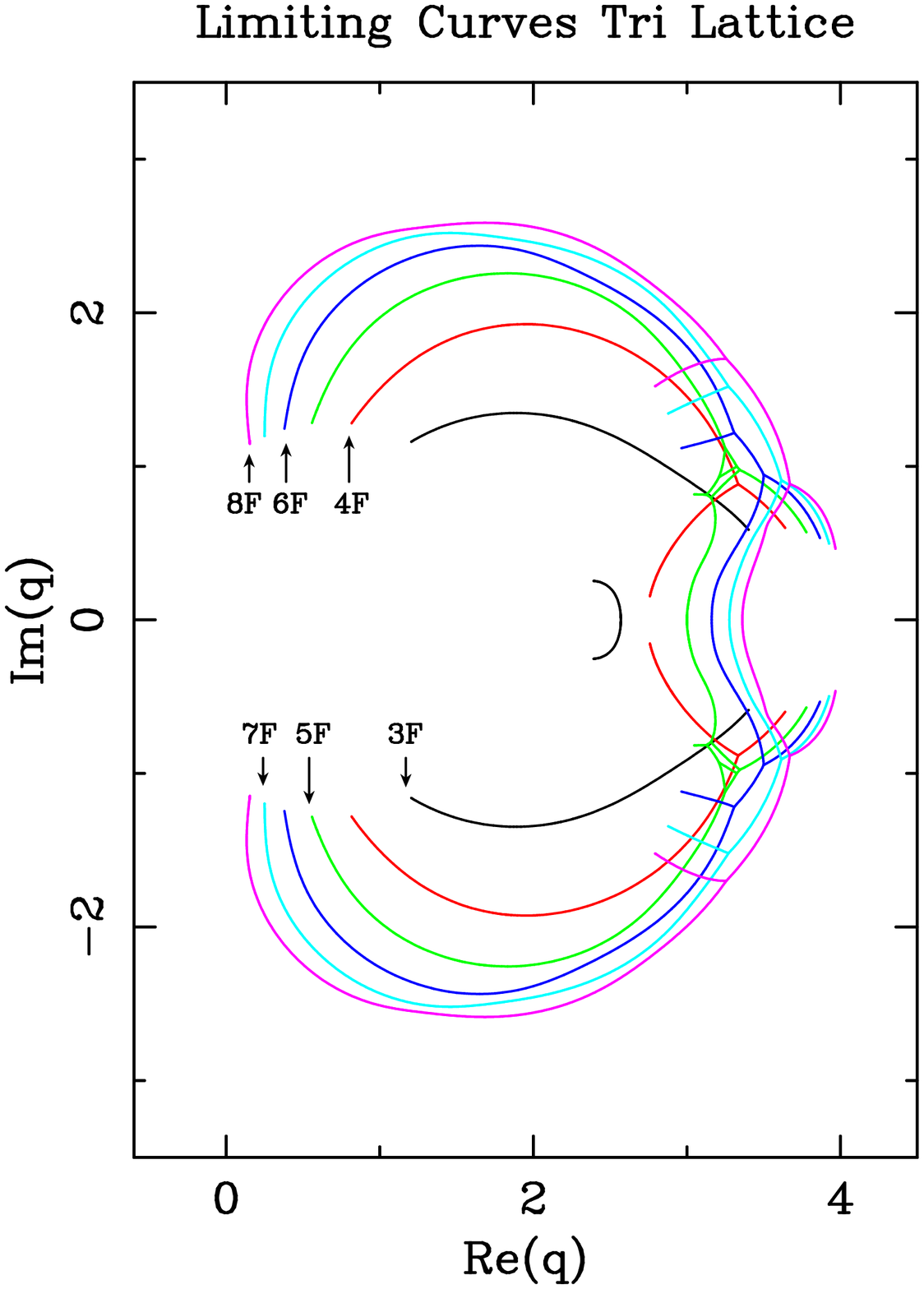}
  \caption{
  Limiting curves for the triangular-lattice strips 
  $L_{\rm F}\times\infty_{\rm F}$ with $3 \leq L \leq 8$.  
  }
\protect\label{Figure_tri_allF}
\end{figure}

\clearpage
%
%
\begin{figure}
  \centering
  \epsfxsize=400pt\epsffile{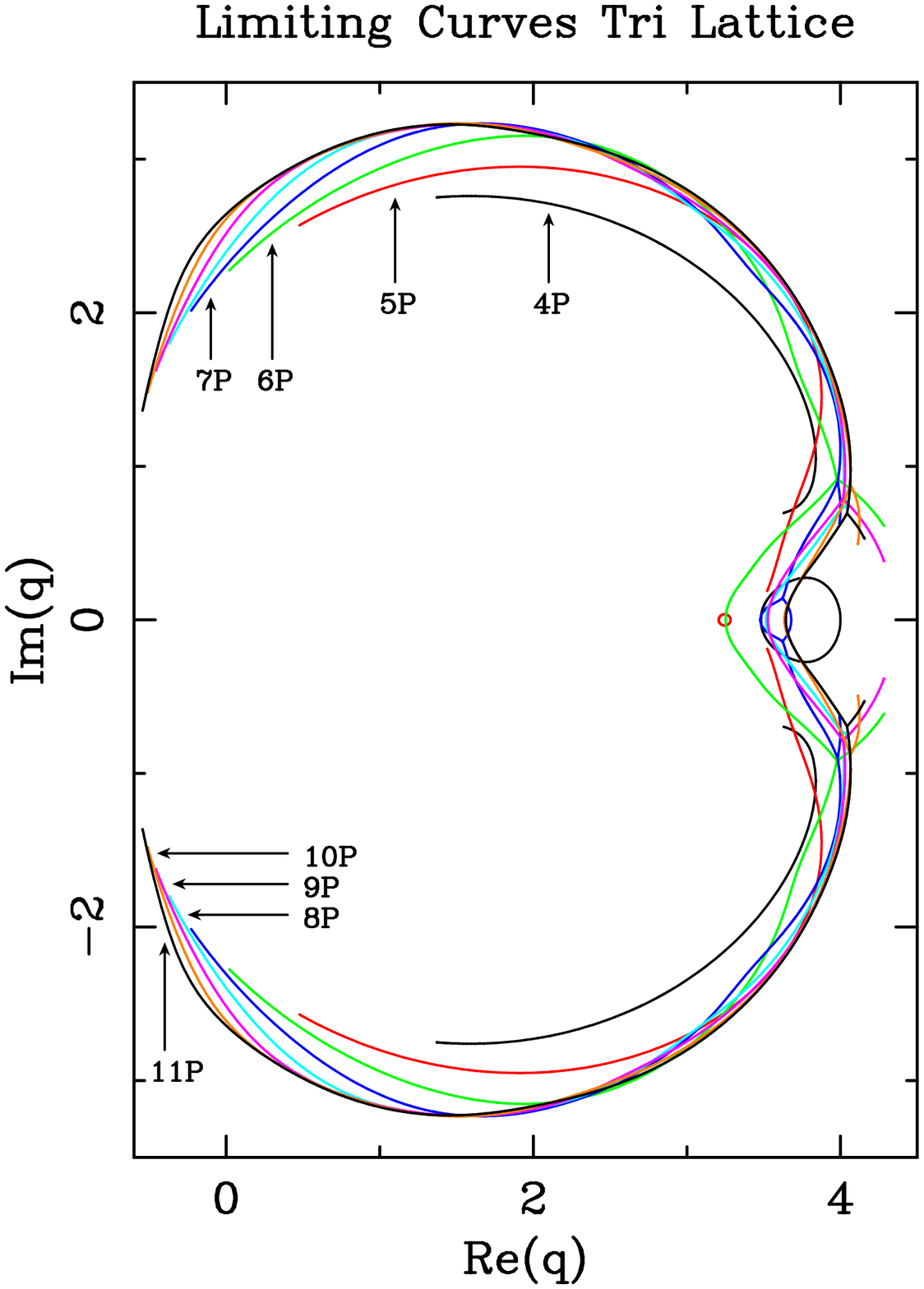}
  \caption{
  Limiting curves for the triangular-lattice strips 
  $L_{\rm P}\times\infty_{\rm F}$ with $4 \leq L \leq 11$. 
  }
\protect\label{Figure_tri_allP}
\end{figure}

\clearpage
%
%
\begin{figure}
  \centering
  \epsfxsize=400pt\epsffile{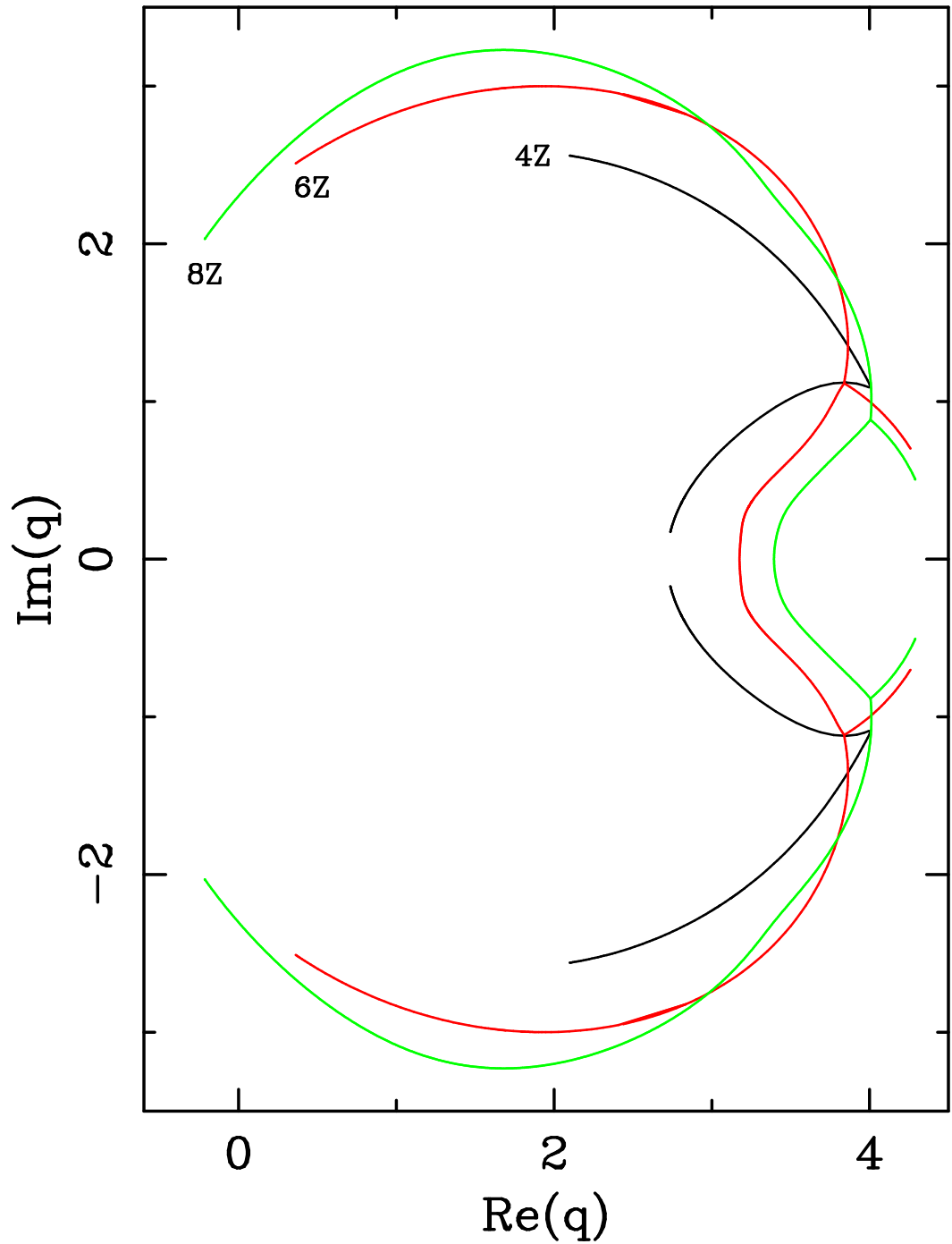}
  \caption{
  Limiting curves for the triangular-lattice strips
  $L_{\rm Z}\times\infty_{\rm F}$ with $L=4,6,8$.
  }
\protect\label{Figure_tri_allZ}
\end{figure}

\clearpage
%
%
\begin{figure}
  \centering
  \epsfxsize=400pt\epsffile{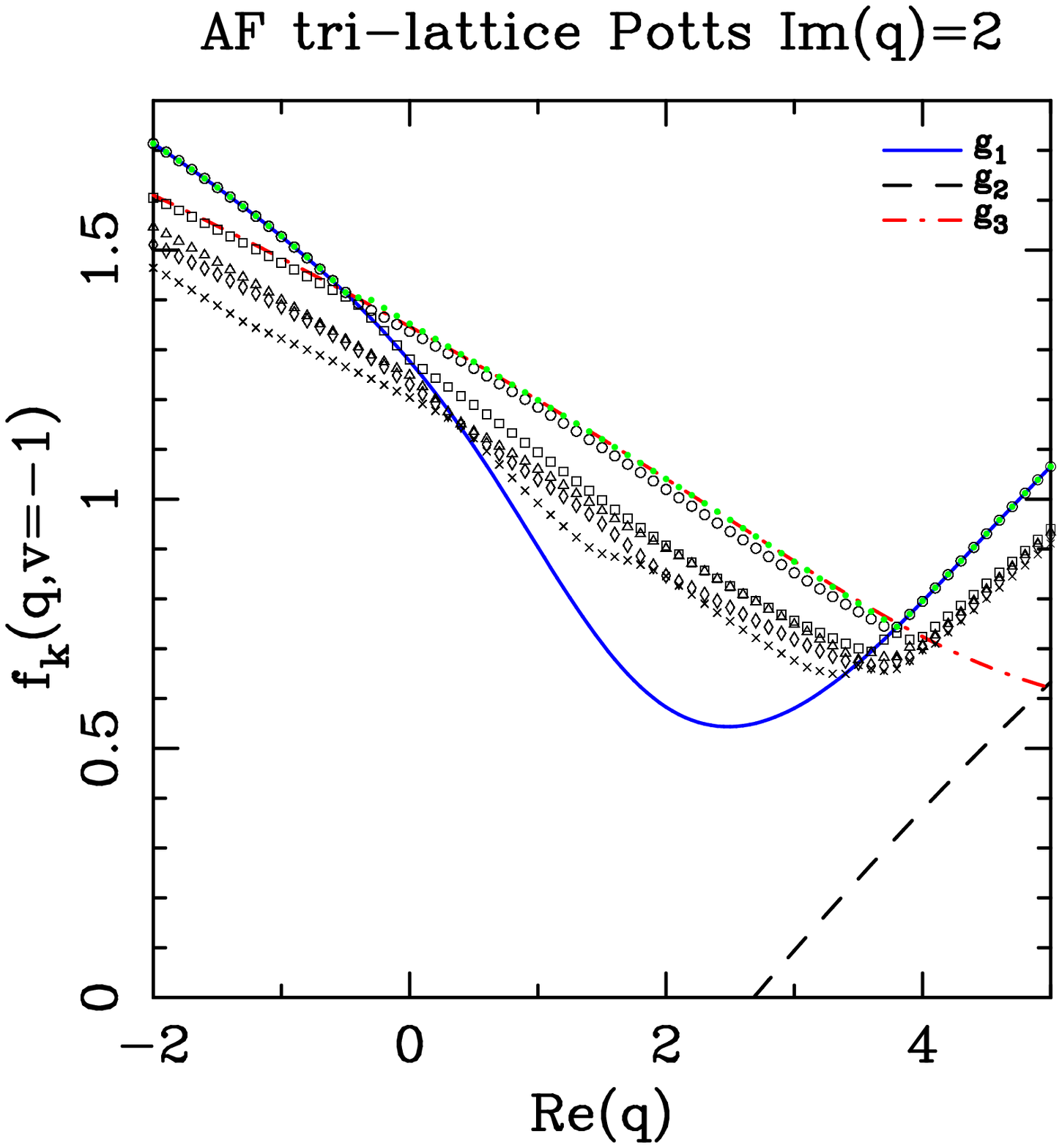}
  \caption{
  Comparison of Baxter's eigenvalues \protect\reff{def_gi} with  
  the most dominant eigenvalues of the transfer matrix for $\imag q = 2$. 
  Curves show the logarithm of Baxter's eigenvalues $\log |g_i|$:
  the solid curve corresponds to $g_1$, the dashed curve to $g_2$,
  and the dot-dashed curve to $g_3$.
  Points show the free energy $f_k = (1/L) \log |\lambda_k|$ associated to 
  the five largest eigenvalues (in modulus) of the transfer matrix for 
  a triangular-lattice strip of width $L=12_{\rm P}$. The index $k$ is coded
  as follows: $k=1$ (the dominant eigenvalue, $\circ$), 2 ($\square$), 
  3 ($\triangle$), 4 ($\diamond$), and 5 ($\times$).
  The solid green dots represent the extrapolation of the dominant
  eigenvalue to $L=\infty$.
  }
\protect\label{fig_Imq=2}
\end{figure}

\clearpage
%
%
\begin{figure}
  \centering
  \epsfxsize=400pt\epsffile{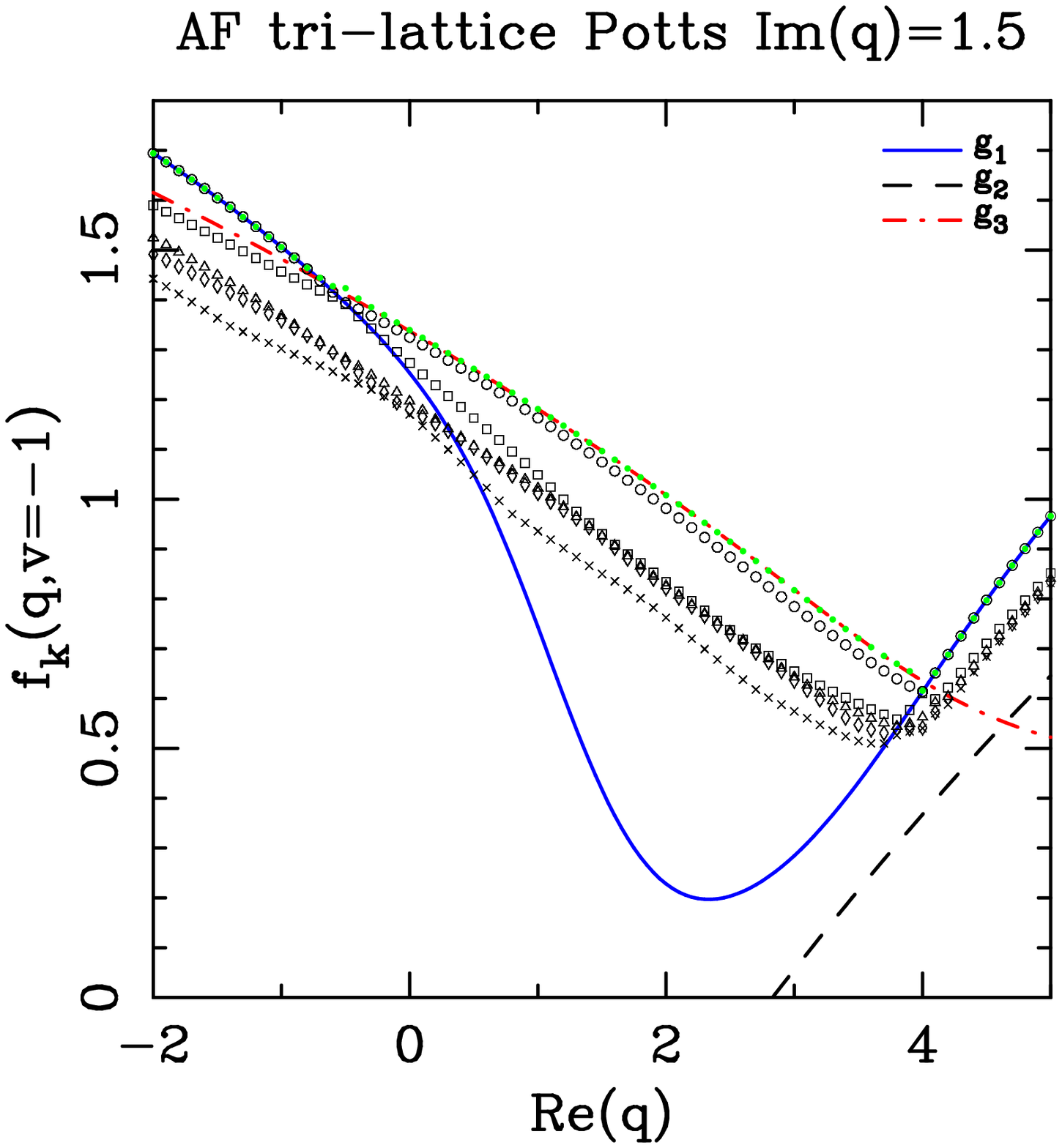}
  \caption{
  Comparison of Baxter's eigenvalues \protect\reff{def_gi} with
  the most dominant eigenvalues of the transfer matrix for $\imag q = 1.5$.
  The notation is as in Figure~\ref{fig_Imq=2}.
  }
\protect\label{fig_Imq=1.5}
\end{figure}

\clearpage
%
%
\begin{figure}
  \centering
  \epsfxsize=400pt\epsffile{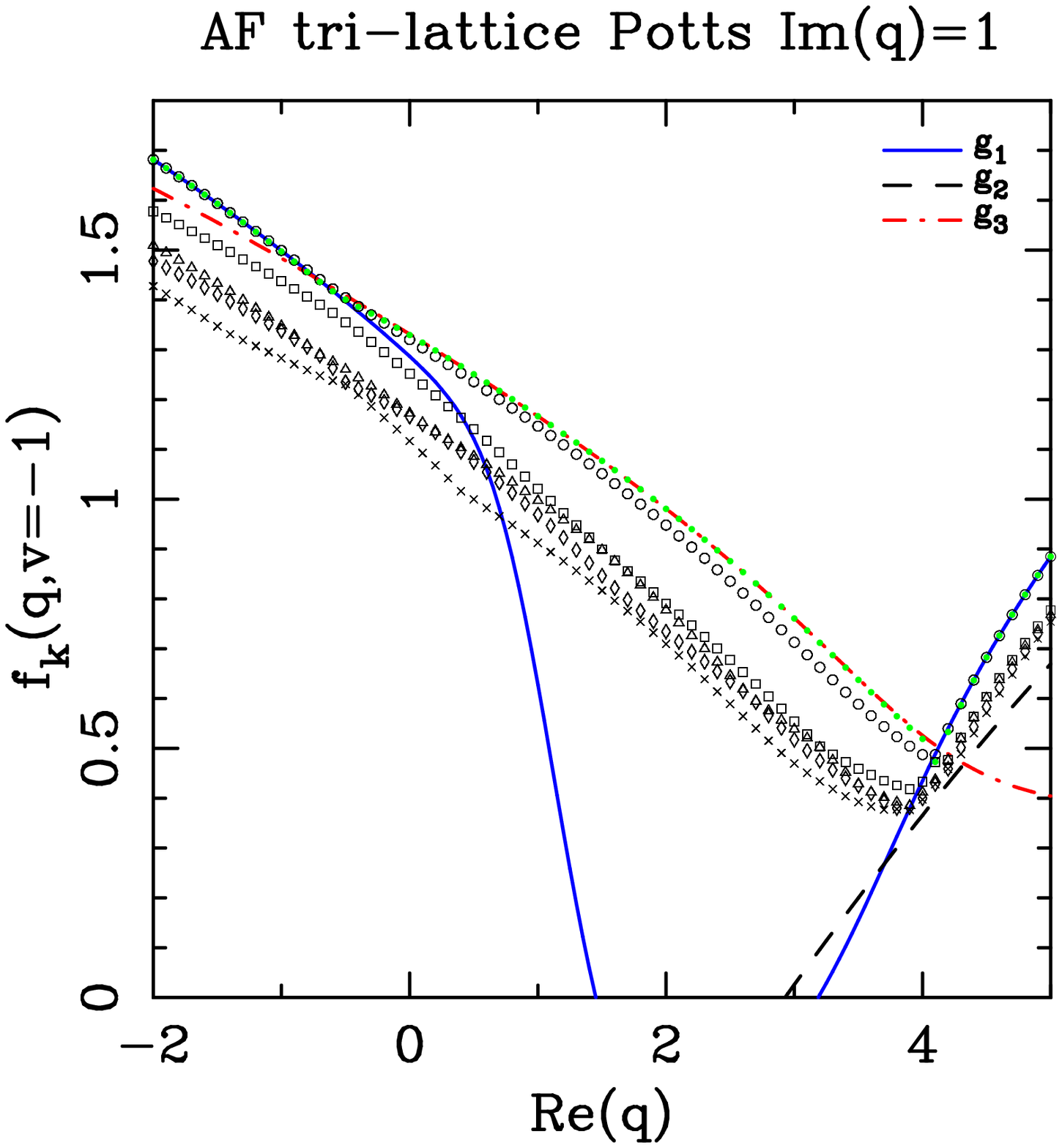}
  \caption{
  Comparison of Baxter's eigenvalues \protect\reff{def_gi} with
  the most dominant eigenvalues of the transfer matrix for $\imag q = 1$.
  The notation is as in Figure~\ref{fig_Imq=2}.
  }
\protect\label{fig_Imq=1}
\end{figure}

\clearpage
%
%
\begin{figure}
  \centering
  \epsfxsize=400pt\epsffile{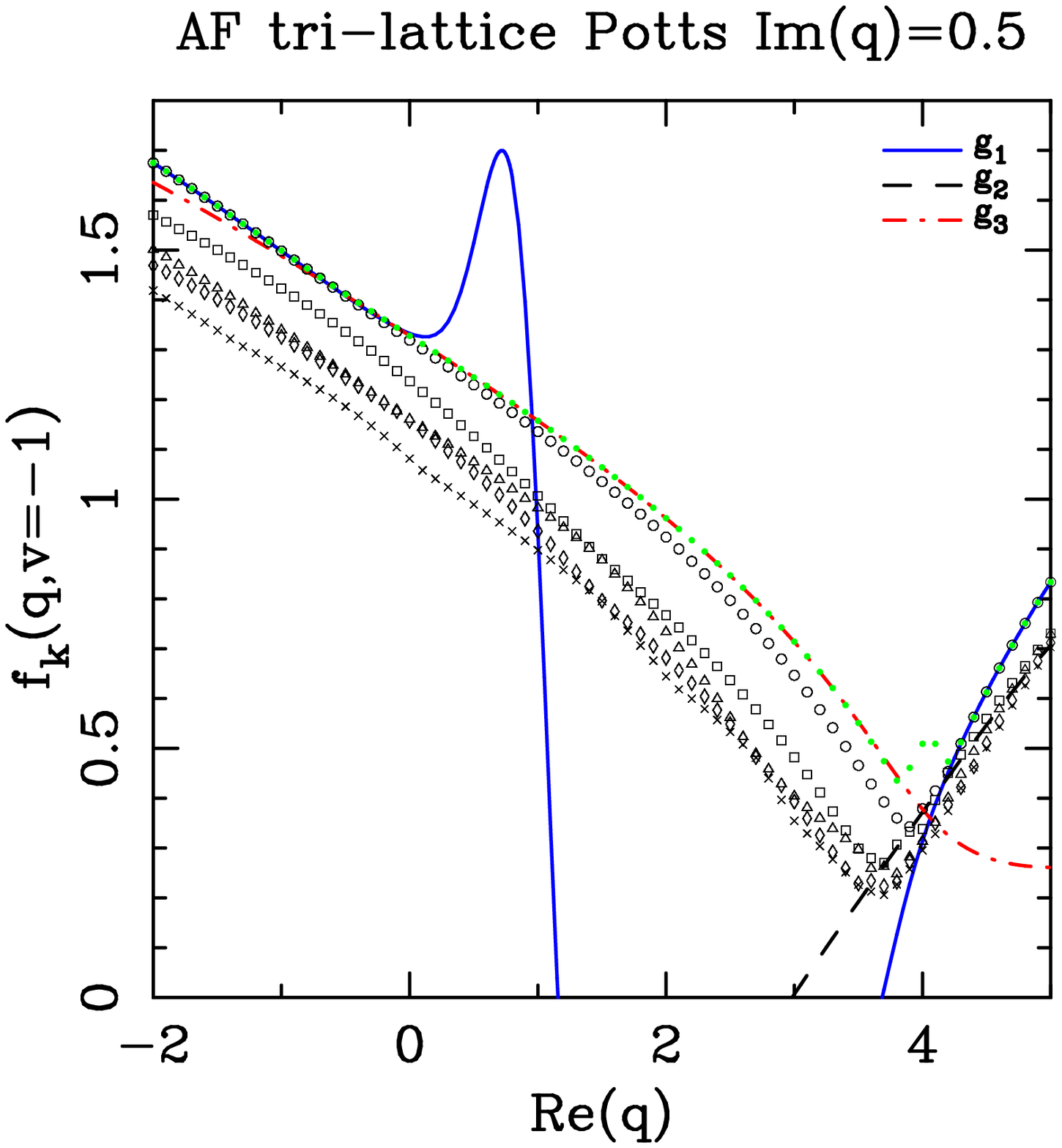}
  \caption{
  Comparison of Baxter's eigenvalues \protect\reff{def_gi} with
  the most dominant eigenvalues of the transfer matrix for $\imag q = 0.5$.
  The notation is as in Figure~\ref{fig_Imq=2}.
  }
\protect\label{fig_Imq=0.5}
\end{figure}

\clearpage
%
%
\begin{figure}
  \centering
  \epsfxsize=400pt\epsffile{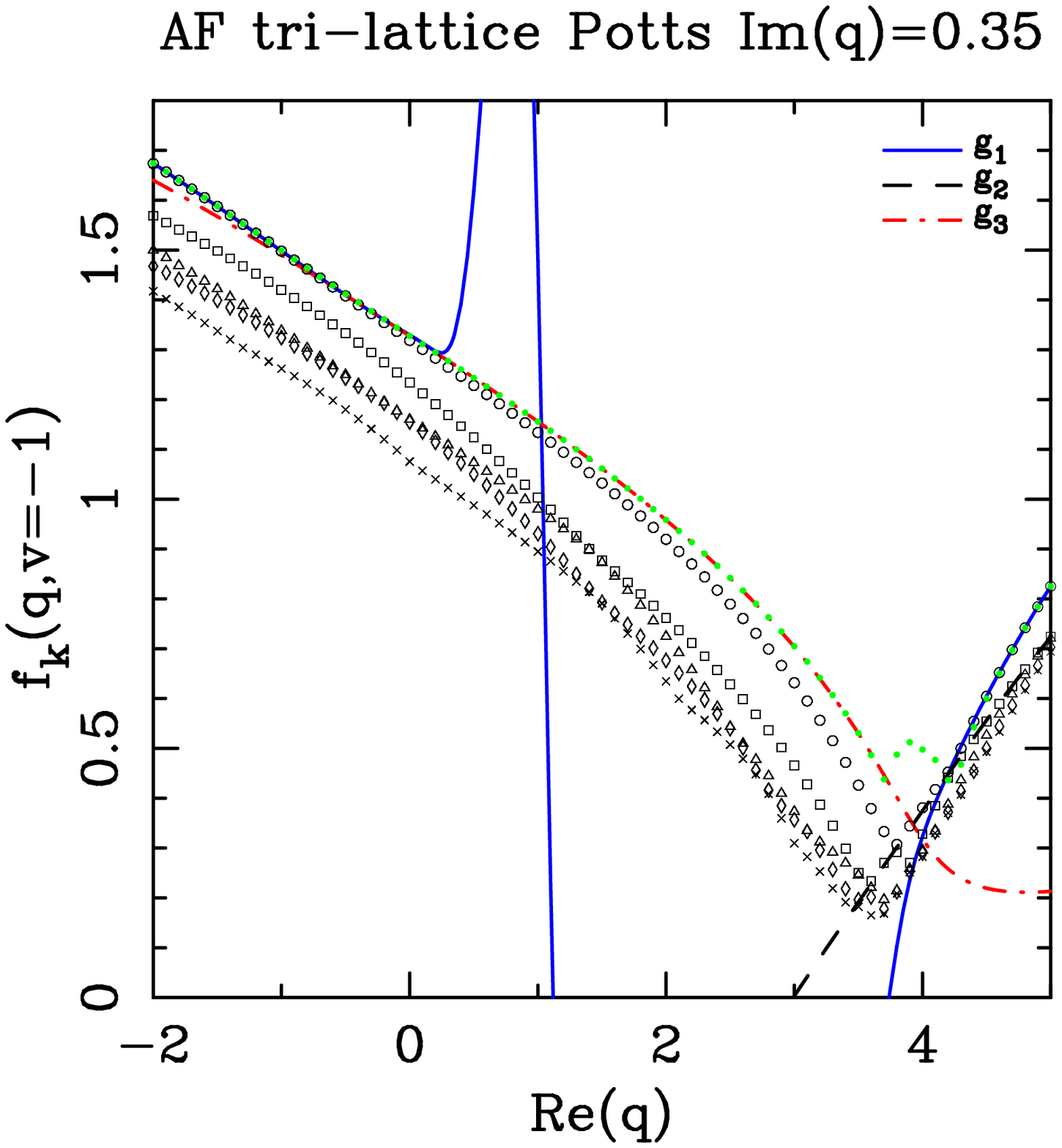}
  \caption{
  Comparison of Baxter's eigenvalues \protect\reff{def_gi} with
  the most dominant eigenvalues of the transfer matrix for $\imag q = 0.35$.
  The notation is as in Figure~\ref{fig_Imq=2}.
  }
\protect\label{fig_Imq=0.35}
\end{figure}

\clearpage
%
%
\begin{figure}
  \centering
  \epsfxsize=400pt\epsffile{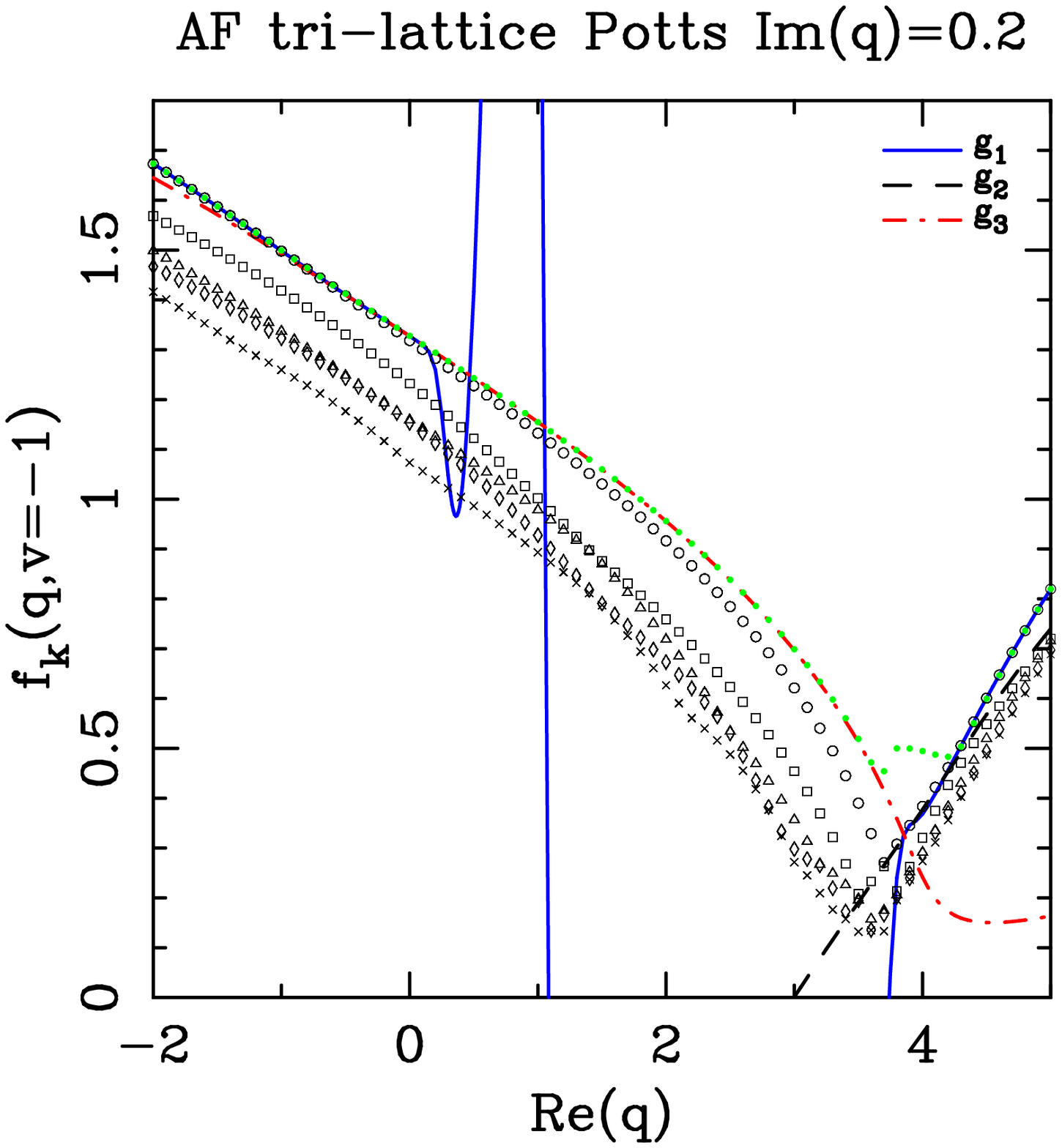}
  \caption{
  Comparison of Baxter's eigenvalues \protect\reff{def_gi} with
  the most dominant eigenvalues of the transfer matrix for $\imag q = 0.2$.
  The notation is as in Figure~\ref{fig_Imq=2}.
  }
\protect\label{fig_Imq=0.2}
\end{figure}

\clearpage
%
%
\begin{figure}
  \centering
  \epsfxsize=400pt\epsffile{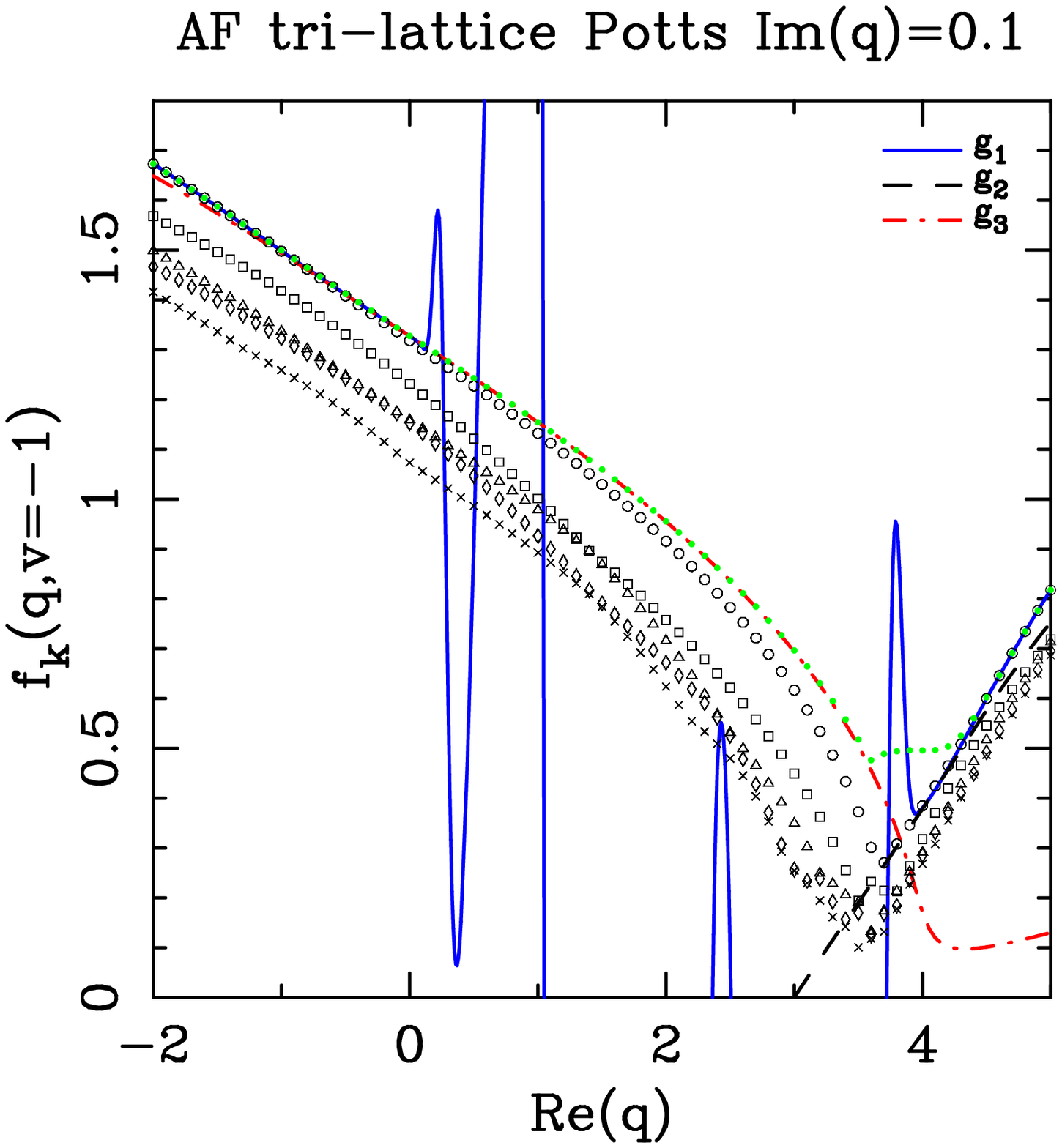}
  \caption{
  Comparison of Baxter's eigenvalues \protect\reff{def_gi} with
  the most dominant eigenvalues of the transfer matrix for $\imag q = 0.1$.
  The notation is as in Figure~\ref{fig_Imq=2}.
  }
\protect\label{fig_Imq=0.1}
\end{figure}

\clearpage
%
%
\begin{figure}
  \centering
  \epsfxsize=400pt\epsffile{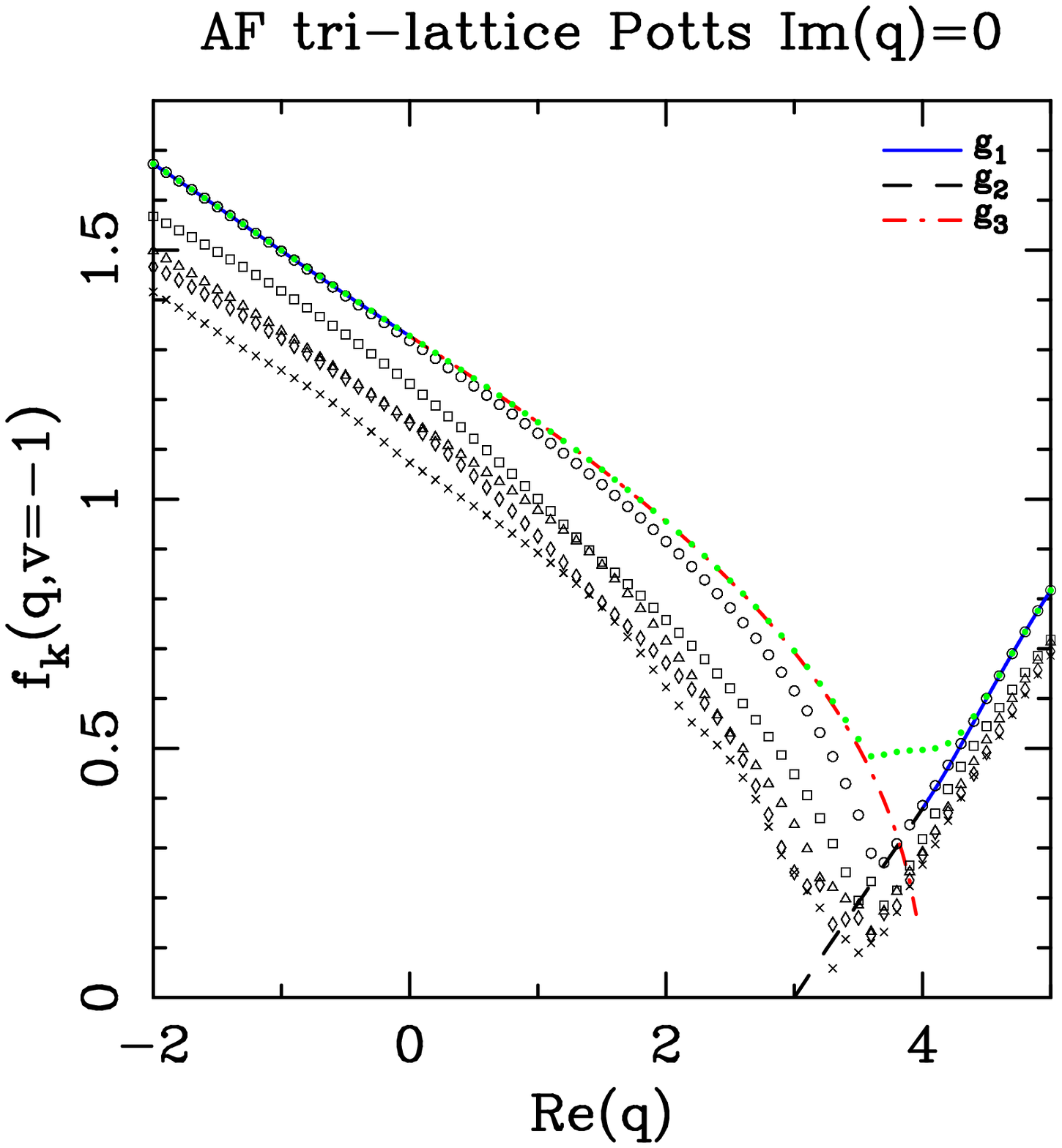}
  \caption{
  Comparison of Baxter's eigenvalues \protect\reff{def_gi} with
  the most dominant eigenvalues of the transfer matrix for $\imag q = 0$.
  The notation is as in Figure~\ref{fig_Imq=2}.
  Baxter's eigenvalues $g_2$ [cf.\ \protect\reff{def_g2}]
  and $g_3$ [cf.\ \protect\reff{def_g3}]
  are depicted only on the interval $0 \le q \le 4$,
  while $g_1$ [cf.\ \protect\reff{def_g1}] is shown only outside that interval
  (see text for details).
  }
\protect\label{fig_Imq=0}
\end{figure}


\begin{thebibliography}{199}

\bibitem{Wu_82}  F.Y. Wu, Rev. Mod. Phys. {\bf 54}, 235 (1982);
   {\bf 55}, 315 (E) (1983).

\bibitem{Wu_84}  F.Y. Wu, J. Appl. Phys. {\bf 55}, 2421 (1984).

\bibitem{Baxter_82}  R.J. Baxter, {\em Exactly Solved Models in Statistical
   Mechanics}\/ (Academic Press, London--New York, 1982).

\bibitem{Martin_91}  P.P. Martin,
   {\em Potts Models and Related Problems in Statistical Mechanics}\/.
   (World Scientific, Singapore, 1991).

\bibitem{Baxter_70_TRI}  R.J. Baxter, J. Math. Phys. {\bf 11}, 784 (1970).

\bibitem{Baxter_82b} R.J. Baxter, Proc. Roy. Soc. London A {\bf 383}, 
  43 (1982).

\bibitem{WSK_90}  J.-S. Wang, R.H. Swendsen and R. Koteck\'y,
  Phys. Rev. B {\bf 42}, 2465 (1990).  

\bibitem{Saleur_90}  H. Saleur,
   Commun. Math. Phys. {\bf 132}, 657 (1990).

\bibitem{Saleur_91}  H. Saleur,
   Nucl. Phys. B {\bf 360}, 219 (1991).

\bibitem{Adler_95}  J. Adler, A. Brandt, W. Janke and S. Shmulyan,
   J. Phys. A {\bf 28}, 5117 (1995).

\bibitem{Salas_97}  J. Salas and A.D. Sokal,
    J. Stat. Phys. {\bf 86}, 551 (1997), cond-mat/9603068.

\bibitem{Salas_98}   J. Salas and A.D. Sokal,
   J. Stat. Phys. {\bf 92}, 729 (1998), cond-mat/9801079.

\bibitem{Ferreira_99}  S.J. Ferreira and A.D. Sokal,
   J. Stat. Phys. {\bf 96}, 461 (1999), cond-mat/9811345.

\bibitem{Cardy_01}  J. Cardy, J.L. Jacobsen and A.D. Sokal,
   J. Stat. Phys. {\bf 105}, 25 (2001), cond-mat/0101197.

\bibitem{Read_88}  R.C. Read and W.T. Tutte,
   in {\em Selected Topics in Graph Theory 3}\/,
   ed.\ L.W. Beineke and R.J. Wilson
   (Academic Press, London, 1988).

\bibitem{transfer1}  J. Salas and A.D. Sokal, J. Stat. Phys. {\bf 104}, 609
   (2001), cond-mat/0004330. 

\bibitem{transfer2}  J.L. Jacobsen and J. Salas, J. Stat. Phys. {\bf 104}, 
   701 (2001), cond-mat/0011456.  

\bibitem{Baxter_86}  R.J. Baxter, J. Phys. A {\bf 19}, 2821 (1986).

\bibitem{Baxter_87}  R.J. Baxter, J. Phys. A {\bf 20}, 5241 (1987).

\bibitem{Yang-Lee_52}  C.N. Yang and T.D. Lee, Phys. Rev. {\bf 87}, 404 (1952).

\bibitem{Beraha_79}  S. Beraha and J. Kahane,
   J. Combin. Theory B {\bf 27}, 1 (1979).

\bibitem{Beraha_80}  S. Beraha, J. Kahane and N.J. Weiss,
   J. Combin. Theory B {\bf 28}, 52 (1980).

\bibitem{Shrock_97a}  R. Shrock and S.-H. Tsai,
   Phys. Rev. E {\bf 55}, 5165 (1997), cond-mat/9612249.

\bibitem{Shrock_98a}  M. Ro\v{c}ek, R. Shrock and S.-H. Tsai,
   Physica A {\bf 252}, 505 (1998), cond-mat/9712148.

\bibitem{BKW_75}  S. Beraha, J. Kahane and N.J. Weiss,
   Proc. Nat. Acad. Sci. USA {\bf 72}, 4209 (1975).

\bibitem{BKW_78}  S. Beraha, J. Kahane and N.J. Weiss,
   in {\em Studies in Foundations and Combinatorics}\/
   (Advances in Mathematics Supplementary Studies, Vol.~1),
   ed.~G.-C. Rota (Academic Press, New York, 1978).

\bibitem{Sokal_chromatic_roots}   A.D. Sokal, Chromatic roots are dense
   in the whole complex plane, Combin. Probab. Comput. (to appear),
   cond-mat/0101197.

\bibitem{Shrock_97b}  R. Shrock and S.-H. Tsai,
   Phys. Rev. E {\bf 56}, 1342 (1997), cond-mat/9703249.

\bibitem{Beraha_unpub}  S. Beraha, unpublished, circa 1974.

\bibitem{Baxter_78}  R.J. Baxter, H.N.V. Temperley and S.E. Ashley,
   Proc. Roy. Soc. London A {\bf 358}, 535 (1978).

\bibitem{Nienhuis_82}  B. Nienhuis, Phys. Rev. Lett. {\bf 49}, 1062 (1982).

\bibitem{Stephenson_64}  J. Stephenson, J. Math. Phys. {\bf 5}, 1009 (1964).

\bibitem{Blote_82b} H.W.J. Bl\"ote and H.J. Hilhorst,
   J. Phys. A {\bf 15}, L631 (1982).

\bibitem{Nienhuis_84b}  B. Nienhuis, H.J. Hilhorst and H.W.J. Bl\"ote,
   J. Phys. A {\bf 17}, 3559 (1984).

\bibitem{Henley_unpublished} C.L. Henley, private communications.

\bibitem{Salas_TRI4state}  J. Salas and A.D. Sokal, unpublished.

\bibitem{vEFS_unpub} A.C.D. van Enter, R. Fern\'andez and A.D. Sokal,
   unpublished (1996).

\bibitem{Salas-Sokal_TRI56}  J. Salas and A.D. Sokal, in preparation.

\bibitem{Kasteleyn_69}  P.W. Kasteleyn and C.M. Fortuin,
   J. Phys. Soc. Japan {\bf 26} (Suppl.), 11 (1969).

\bibitem{Fortuin_72}  C.M. Fortuin and P.W. Kasteleyn,
   Physica {\bf 57}, 536 (1972).

\bibitem{Shrock_98c}  M. Ro\v{c}ek, R. Shrock and S.-H. Tsai,
   Physica A {\bf 259}, 367 (1998), cond-mat/9807106.

\bibitem{Shrock_98f}  R. Shrock and S.-H. Tsai,
   Phys. Rev. E {\bf 58}, 4332 (1998), cond-mat/9808057.

\bibitem{Shrock_99g} R. Shrock, Discrete Math. {\bf 231}, 421 (2001),  
   cond-mat/9908307.

\bibitem{Shrock_00c} S.-C. Chang and R. Shrock, 
  Annals Phys. {\bf 290}, 124 (2001), cond-mat/0004129.

\bibitem{Shrock_00a} R. Shrock and S.-H. Tsai,
  Physica A {\bf 275}, 429 (2000), cond-mat/9907403.

\bibitem{Shrock_01a} S.-C. Chang and R. Shrock, 
 Physica A {\bf 292}, 307 (2001), cond-mat/0007491.   

\bibitem{Shrock_01b} S.-C. Chang and R. Shrock,
 Physica A {\bf 296}, 131 (2001), cond-mat/0005232.

%
%
\bibitem{Shrock_00e} S.-C. Chang and R. Shrock, 
     Physica A {\bf 286}, 189 (2000), cond-mat/0004181. 

\bibitem{Shrock_inprep} S.-C. Chang, J.L. Jacobsen, J. Salas and R. Shrock, 
 Exact Potts model partition functions for strips of the triangular lattice,
 cond-mat/0211623. 

%
%
\bibitem{Shrock_02a} S.-C. Chang and R. Shrock, 
   Physica A {\bf 316}, 335 (2002), cond-mat/0201223.

\bibitem{Bini_package}  D.A. Bini and G. Fiorentino,
   Numerical computation of polynomial roots using MPSolve version 2.2
   (January 2000).
   Software package and documentation available for download at
   {\tt ftp://ftp.dm.unipi.it/pub/mpsolve/}.

\bibitem{Bini-Fiorentino}  D.A. Bini and G. Fiorentino,
   Numer. Algorithms {\bf 23}, 127 (2000).

\bibitem{Tutte_sq} S.-C. Chang, J. Salas and R. Shrock,
 J. Stat. Phys. {\bf 107}, 1207 (2002), cond-mat/0108144.

\bibitem{Niven_63}  I. Niven, {\em Diophantine Approximation}\/
   (Wiley--Interscience, New York, 1963).

\bibitem{Rademacher_77}  H. Rademacher, {\em Lectures on Elementary
   Number Theory}\/ (Robert E. Krieger, Huntington NY, 1977).

\bibitem{Cassels_57}  J.W.S. Cassels, {\em An Introduction to
   Diophantine Approximation}\/ (Cambridge University Press, Cambridge, 1957).

\bibitem{Graham_94}  R.L. Graham, D.E. Knuth and O. Patashnik,
  {\em Concrete Mathematics: A Foundation for Computer Science}\/,
  2nd ed.~(Addison-Wesley, Reading, Mass., 1994).

\bibitem{Baxter-Kelland-Wu}  R.J. Baxter, S.B. Kelland and F.Y. Wu,
   J. Phys. A {\bf 9}, 397 (1976).

\bibitem{Baxter_JSP82}  R.J. Baxter, J. Stat. Phys. {\bf 28}, 1 (1982).

\bibitem{Euler}  L. Euler, {\em Introduction to Analysis of the Infinite}\/
  [{\em Introductio in Analysin Infinitorum}\/, 1748], 2 vols.,
   translated by John D. Blanton (Springer-Verlag, New York, 1988/1990).

\bibitem{Sokal_infprod}  A.D. Sokal, Numerical computation of
       $\prod\limits_{n=1}^\infty (1 - tx^n)$,
       preprint (December 2002), math.NA/0212035.

\bibitem{Andrews_98}  G.E. Andrews, {\em The Theory of Partitions}\/
   (Cambridge University Press, Cambridge, 1998).

\bibitem{Knopp_70}  M.I. Knopp,
   {\em Modular Functions in Analytic Number Theory}\/
   (Markham, Chicago, 1970).  

\bibitem{Remmert_98}  R. Remmert,
   {\em Classical Topics in Complex Function Theory}\/
   (Springer-Verlag, New York--Berlin--Heidelberg, 1998).
                                                            
\end{thebibliography}
\end{document}